# Interdiffusion study in group IVB, VB and VIB refractory metal-silicon systems

A thesis
Submitted For the degree of
Doctor of Philosophy
2014

Soumitra Roy

Department of Materials Engineering
Indian Institute of Science
Bangalore – 560 012, INDIA

***To***
***My Parents***

# Synopsis

The knowledge of diffusion parameters provides important understanding of many physical and mechanical properties of materials. In most of the applications silicides are grown by a diffusion controlled process mainly in thin film condition. Because of this reason, most of the studies till date are available in thin film condition. Although more than one phase is present in all these systems, mainly disilicides were found at the interface. In this thesis bulk interdiffusion studies are conducted by coupling pure refractory metals (group IVB, VB and VIB elements) with single crystal Si.

Several phase layers grow between binary refractory metal and Si systems. The layer thicknesses of the phases are measured from the microstructures. Composition profiles were measured in electron probe microanalyzer. Different diffusion parameters are estimated such as parabolic growth constants, integrated diffusion coefficients, activation energy for diffusion and ratio of tracer diffusivities of the components are estimated. Growth mechanisms of the phases are discussed with the help of diffusion parameters. The atomic mechanism of the diffusion is discussed considering crystal structure of the phases along with possible defects present.

Solid diffusion couple experiments are conducted to analyse the growth mechanism of the phases and the diffusion mechanism of the components in the Ti-Si system. The calculation of the parabolic growth constant and of the integrated diffusion coefficients substantiates that the analysis is intrinsically prone to erroneous conclusions if it is based just on the parabolic growth constants determined for a multiphase interdiffusion zone. The location of the marker plane is detected based on the uniform grain morphology in the $TiSi_2$ phase, which indicates that this phase grows mainly because of Si diffusion. The growth mechanism of the phases and morphological evolution in the interdiffusion zone are explained with the help of imaginary diffusion couples. The activation enthalpies for the integrated diffusion coefficient of $TiSi_2$ and the Si tracer diffusion are calculated as 190±9 and 170±12 kJ/mol, respectively. The crystal structure, details on the nearest neighbours of the elements and the relative mobilities of the components indicate that the vacancies are mainly present on the Si sublattice.

Diffusion controlled growth of the phases in the Hf-Si and Zr-Si are studied by bulk diffusion couple technique. Only two phases grow in the interdiffusion zone, although several phases are present in both the systems. The location of the Kirkendall marker plane detected based on the grain morphology indicates that the disilicides grow by the diffusion of Si. Diffusion of the metal species in these phases is negligible. This indicates that vacancies are present mainly on the Si sublattice. The activation energies for integrated diffusion coefficients in the $HfSi_2$ and $ZrSi_2$ are estimated as 394 ± 37 and 346 ± 34 kJ/mol, respectively. The same is calculated for the HfSi phase as 485±42 kJ/mol. The activation energies for Si tracer diffusion in the $HfSi_2$ and $ZrSi_2$ phases are estimated as 430 ± 36 and 348 ± 34 kJ/mol, respectively.

We conducted interdiffusion studies to understand the atomic mechanism of the diffusing species and the growth mechanism of the phases. Integrated diffusion coefficients and the ratio of tracer diffusion coefficients were estimated for these analyses. The activation energies for the integrated diffusion coefficients were calculated as 550 ± 70 and 410 ± 39 kJ/mol in the $TaSi_2$ and the $Ta_5Si_3$ phases, respectively. In the $TaSi_2$ phase, Ta has a slightly lower but comparable diffusion rate with respect to Si, although no Ta–Ta bonds are present in the crystal. In the $Ta_5Si_3$ phase, Si has higher diffusion rate, which is rather unusual, if we consider the atoms in the nearest-neighbor positions for both the elements. The ratio of Si to Ta tracer diffusion coefficients is found to be lower in the Si-rich phase, $TaSi_2$, compared to the Si-lean phase, $Ta_5Si_3$, which is also unusual. This indicates the type of structural defects present. An analysis on the growth mechanism of the phases indicates that duplex morphology and the Kirkendall marker plane should only be present in the $TaSi_2$ phase. This is not present in the $Ta_5Si_3$ phase because of the very high growth rate of the $TaSi_2$ phase, which consumes most of the $Ta_5Si_3$ phase layer. The problems in the calculation method used previously by others in this system are also explained.

Experiments are conducted in the W-Si system to understand the diffusion mechanism of the species. The activation energies for integrated diffusion are found to be 152±7 and 301±40 kJ/mol in the $WSi_2$ and $W_5Si_3$ phases, respectively. In both the phases, Si has a much higher diffusion rate compared to W. The result found in the $WSi_2$ phase is not surprising, if we consider the nearest neighbors in the crystal. However, it is rather unusual to find that Si has higher diffusion rate in the $W_5Si_3$ phase, indicating the presence of high concentration of Si antisites in this phase.

In the group IVB, VB and VIB M-Si systems are considered to show an interesting pattern in diffusion of components with the change in atomic number in a particular group. $MSi_2$ and $M_5Si_3$ are considered for this discussion. Except in the Ta-Si system, activation energy for integrated diffusion of $MSi_2$ is always lower than $M_5Si_3$. Interestingly, in both the phases, the relative mobilities measured by the ratio of tracer diffusion coefficients, $\frac{D_{Si}^*}{D_M^*}$ decreases with the increase in atomic number in both the groups. Both the phases have similar crystal structures in a particular group in which these parameters are calculated. In both the phases Si has higher diffusion rate compared to M. Absence of any M-M bonds in $MSi_2$ and increase in the diffusivities of M with the increase in atomic number substantiates the increasing concentration of M anti-sites and higher interactions of M with vacancies. Only one or two Si-Si bonds are present in $M_5Si_3$, however, the higher diffusion rate of Si indicates the presence of vacancies mainly on its sublattice. On the other hand, increase in $\frac{D_{Si}^*}{D_M^*}$ with increasing atomic number in both the groups substantiates increasing interactions of M and vacancies.

## Acknowledgement

I am thankful to all those who helped and supported me directly or indirectly from the beginning to end of my research work and to the shaping of my thesis.

First, I am grateful to the Indian Institute of Science, the premier institute of the country, (the Department of Materials Engineering) for giving me the opportunity and confidence.

I would like to show my greatest appreciation to Prof. Aloke Paul. I can't thank him enough for his tireless support and help. I feel motivated and encouraged every time I attend his meeting. Without his encouragement and guidance this project would not have materialized.

I thank Prof. V. Jayaram and Prof. K. Chattopadhyay, present and former Chairman, Department of Materials Engineering, Indian Institute of Science, for extending the facilities of the Department. Specially I would like to thank Prof. S. Suwas and Dr. R. Ravi for their valuable suggestions and discussions during my research work. I would like to thank all the faculty and the staff of the department for their helping hand. Special thank to Krishnamurty sir (XRD) and Vandana for (EPMA).

I thank Dr. S .V. Divinski (Institute of Materials Physics, University of Münster, Germany) and Dr. T. Laurila (Department of Electronics, School of Electrical Engineering, Aalto University, Espoo, Finland ) for their cooperation and collaborative research projects. I thank Prof. A. Garg, Department of Materials Science and Engineering, for allowing me to work in his laboratory.

I am extremely thankful to my labmates Soma, Divya, Sangeeta, Kiruthika, Varun, Bivu, Avik, Sarfaraj and Tanaji for their technical help, support and valuable friendly assistance in the lab.

I express my deepest regards and reverence to my parents and my sister Satabdi, They always extended their support and encouraged me with their best wishes. I would not have reached this stage without their love, care, and the faith they showed in me.

I thank all my friends for their love and support. The immense trust they showed in me has helped me to build up my confidence.

Finally I would like to thank my wife, Kanika. She was always there cheering me up and stood by me through the good times and bad, whose love and encouragement allowed me to finish this journey.

# Declaration

I hereby declare that the thesis entitled **"Interdiffusion studies in group IVB, VB and VIB refractory metal- silicon systems"** is the result of my own research work and effort carried out under Prof. Aloke Paul supervision in the Department of Materials Engineering, Indian Institute of Science, Bangalore, India.

Any contributions where others are involved, mentioned with due reference to the literature, and clear acknowledgement of collaborative research and discussions are given. I declare that this thesis has not been submitted anywhere for the award for any University of Institution of higher learning.

Date Soumitra Roy

# Contents

# Chapter 1

# Introduction

## 1.1 Motivation

First known application of metal silicides, which is $MoSi_2$, was as heating element in electrical furnaces by Moissan (1904) [1]. In 1950`s, silicides were used to protect exposed parts of the rocket and jet engines in oxidizing atmosphere [2]. Later on silicides were considered for various innovative applications because of their high temperature stability, excellent oxidation resistance, and low temperature coefficient of electrical resistivity. Combination of these properties and metal like behavior, attracted them in wide range of applications such as (a) high temperature structural applications (b) Schottky barrier and ohomic contacts in integrated circuit technology (c) Low resistivity metallization for gates and interconnects (d) diffusion barrier in Cu based interconnects. For example, vanadium disilicide is used as a protective coating in combination with $MoSi_2$ [3] and Ohmic contact for Schottky diodes [4]. Mo- and Nb-based silicides are being developed to replace Ni-based superalloys in jet engine applications [5-9]. Extensive research is in progress to improve the fracture toughness and the oxidation resistance. Ta and W are proposed as the diffusion barrier layer in Cu interconnects [10-13], since these are immiscibile with Cu even at elevated temperature [14]. Growth of silicides at the interface of the barrier metal and the substrate influence the performance of the product. Molybdenum-silicon system is important for many applications, such as Schottky contacts [15, 16], VLSI (very large scale integration) interconnects and photomask for VLSI fabrication [17, 18.], soft X-ray mirror [19, 20] etc. Heating elements for furnaces and high temperature coatings are produced from $MoSi_2$ because of its excellent oxidation resistance [21]. Growth of silicides controls the overall performance of the structure when Mo is used as interlayer to diffusion bond $Si_3N_4$ [21]. Tungsten disilicides, which is grown by reactive diffusion, could be important for making integrated circuits because of its low electrical resistivity and good thermal stability [13, 22, 23]. $TiSi_2$ is used as contacts and interconnects in metal oxide semiconductor (MOS) [24].Semiconducting rhenium silicide $ReSi_{1.8}$ is promising for thermo electric application [13, 25-30] because of their high career mobilities (370 $cm^2/V$) at room temperature [31].

Wide range of applications motivated to study diffusion and growth behavior of various refractory metal silicides. In solid state diffusion, the compound phase or phases grow in between two end members which can be predicted by equilibrium phase diagrams. Moreover, in many applications, silicides are grown by a diffusion controlled process mainly in thin film condition [32-43]. Because of this reason, most of the studies till date are available in this condition [10, 11, 22, 44-54]. Although more than one phase is present in all the systems, mainly disilicides phase were found at the interface. As speculated, the growth mechanism of the phases in thin film could be very complicated [55]. Because of stress and nucleation problems, a few equilibrium phases might not grow. In some cases, a metastable phase could be found. Many a times, sequential instead of simultaneous growth of the phases are reported [55]. Kirkendall marker experiments were conducted in some systems entrapping gas bubble at the interface as inert Kirkendall marker. It should be noted here that such small bubbles could be dragged by grain boundary and might not act as suitable markers. Overall, higher diffusion rate of Si compared to the refractory metal species in the disilicides phase in all the systems were reported [22, 23, 49, 56-58]. However, no quantitative analysis is available for estimating the ratio of diffusivities, which could shed light on the atomic mechanism of diffusion.

To understand the growth mechanism of the phases and diffusion mechanism of the species, experiments in bulk conditions are required. It is much easier to estimate the relative mobilities of the species using the inert particles as the kirkendall markers in bulk diffusion couple. Although there are reasonable numbers of publications are related to study of diffusion in bulk condition, mainly the growth kinetics of the phases were calculated, which is not a material constant. Parabolic growth constant depends on the end member compositions of the diffusion couples. There are very few studies are related to estimation of the interdiffusion/integrated diffusion coefficients [47, 49, 54]. Few studies are also available using inter markers to detect the Kirkendall plane, however, the mobilities of the species were not calculated. It should be noted here that the diffusion couple technique can be followed to estimate the mobilities of the species indirectly.

It is already known from the experimental analysis and theoretical calculations that diffusion in ordered phases is assisted mainly by two types of defects; vacancies and antisites. Moreover, different concentration of defects might be present on different sublattices. Many a times, this is the reason to find very different diffusion rates of the species of different elements in a phase.

The aim of this study is to conduct interdiffusion studies following the solid state diffusion couple technique. The parabolic growth of the phases was examined for the present conditions. Time dependent experiments have been conducted to examine the diffusion controlled growth of the phases, and temperature dependent experiments to calculate the activation energies. Integrated diffusion coefficients and the activation energies have been calculated from the composition profiles. Grain morphology in the interdiffusion zone indicates relative mobilities of the species, which help to predict defects and diffusion of elements.

## 1.2 Outline of the thesis

There could be many silicides present in the phase diagram, however, all the phases do not grow with reasonable thickness for quantitative analysis. We have listed below the phases we considered for extensive analysis.

| Group | Systems | Silicides discussed |
|---|---|---|
| IVB | Ti/Si | $TiSi_2$, TiSi, $Ti_5Si_4$ |
| | Zr/Si | $ZrSi_2$ |
| | Hf/Si | $HfSi_2$, HfSi |
| VB | Ta/Si | $TaSi_2$, $Ta_5Si_3$ |
| VIB | W/Si | $WSi_2$, $W_5Si_3$ |

Table 1.1: The systems and the phases considered for present study.

The intermetalic phases of these silicides are purely line compounds. The calculation methodology of different diffusion parameters are discussed in chapter 3.

Bulk diffusion couple experiments are conducted and the diffusion parameters are determined for all the phases mentioned in the Table 1.1. The integrated diffusion coefficient, $\tilde{D}_{int}$ of the phases and the tracer diffusion coefficient, $D^*$ of the components are determined. Further the growth mechanism of the phases is discussed based on the data calculated in this study. Based on the tracer diffusion coefficient of the species in each phase, the atomic mechanism of diffusion is discussed for all the systems.

In Chapter 4.1, bulk diffusion couple experiments are studied in the Ti-Si system. The Significance of the parabolic growth constant in multiphase interdiffusion is discussed. The location of the marker plane is detected by etching the sample with acid solution. Activation energy and integrated diffusion coefficients of the phases are calculated. The Si tracer diffusivity in the $TiSi_2$ phase is calculated with the help of the available thermodynamic data in the literature. Presence of defects in the crystal structure is discussed with help of relative mobilities of the diffusing components.

Similar studies have been conducted with Zr/Si and Hf/Si systems. In Chapter 4.2 these two systems are discussed together because of their similar behavior in growth of the phases. Diffusion controlled growth of the phases in the Hf-Si and Zr-Si are studied by bulk diffusion couple technique. The activation energies for integrated diffusion coefficients in the $HfSi_2$, HfSi and $ZrSi_2$ are estimated. The activation energies for Si tracer diffusion in the $HfSi_2$ and $ZrSi_2$ phases are estimated.

Interdiffusion studies conducted to understand the atomic mechanism of the diffusing species and the growth mechanism of the phases in Ta/Si system are explained in Chapter 5. Incremental diffusion couple studies are conducted in $TaSi_2$/Ta system for the growth of $Ta_5Si_3$ phase. Integrated diffusion coefficients and the ratio of tracer diffusion coefficients were estimated for these analyses. Structural defects in $TaSi_2$ and $Ta_5Si_3$ phases are discussed with the ratio of Si to Ta tracer diffusion coefficients. Growth mechanism of the phases and the presence of the Kirkendall marker plane in $TaSi_2$ phase discussed. The problems in the calculation method used previously by others in this system are also explained. Similar studies on W/Si system is discussed in Chapter 6.

Group IVB, VB and VIB M-Si systems show an interesting pattern in diffusion of components with the change in atomic number in a particular group, which are explained in Chapter 7. $MSi_2$ in group IVB, $MSi_2$ and $M_5Si_3$ in group VB and VIB are considered for this discussion. The position of the Kirkendall marker plane in the disilicides shows a common trend in their respective groups. The type of defects present in the crystal is discussed with nearest neighbor of the species. The change of defect concentration also shows a systematic trend with increasing atomic number for a particular phase in the periodic table, which is discussed with the help of homologous temperature plot.

# Chapter 2

# Diffusion Concepts

## 2.1 Introduction

Diffusion has an important role on microstructural evolution, processing and heat treatment of metals and alloys, phase transformation, precipitation, homogenisation, recrystallization, high temperature creep and oxidation. Many device performance such as turbine blades, flip chip bonding and intermetallic superconductors depends on diffusion.

The Scotish chemist, Thomas Graham conducted the first systematic study on diffusion [1,2]. Most of his works are on gases and liquids. His first published work is found in 1833. In 1855 Adolf Fick [3] proposed the quantitative laws of diffusion, which was inspired from the work of Graham. Fick's perception was diffusion could be described with the same mathematical formalism as Fourier's law of heat conduction or Ohm's law of electricity. Following the same analogy, he proposed that the flux of matter and gradient of its concentration are directly proportional. Latter on the study of diffusion in solids, was found by W.C. Roberts-Austen [4] in the Au/Pb system in 1896.

There are essentially two different ways to understand the basic concept of the diffusion theory, a phenomenological approach and an atomistic approach. Phenomenological approach based on the basic diffusion law and its mathematical solution which proposed by Adolf Fick. Atomistic approach is based on the 'random walk' of particles which developed by Robert Brown and mathematically formulated by Albert Einstein.

For our explanation, we shall mostly deal with the phenomenological approach. The equations and derivations used this chapter are taken from the Refs. [5-7].

## 2.2 Fick's Law

The phenomenological diffusion theory is proposed by Adolf Fick [3]. The flux of particles (atoms, molecules, ions etc) is directly proportional to its concentration gradient in one dimensional system. The relation can be expressed as,

$$J = -D\left(\frac{\partial C}{\partial x}\right) \tag{2.1}$$

where $J$ ($mole/m^2/s$) is the flux, $D$ ($m^2/s$) is the diffusion coefficient, $C$ ($mole/m^3$) is the concentration and $x$ ($m$) is the position parameter. The negative sign in this relation shows that diffusion occurs in the direction opposite to the increasing concentration gradient. In the steady state condition, the flux and the diffusion coefficients are directly related by Eqn 2.1. But, in most of the practical applications, concentration always changes with annealing time. To explain that Fick's second law is derived, as below.

Usually, in a simplified diffusion process, number of diffusing particles is conserved. This means that the diffusing components are neither created nor destroyed. Hence, in such a case, if we consider two planes, 1 and 2 separated by distance $\Delta x$ (as shown in Fig 2.1 and the corresponding flux through each plane as $J(x)$ and $J(x+\Delta x)$, from the conservation of mass, the change in concentration in time $\partial t$ can be written as,

$$J(x+\Delta x) - J(x) = -\frac{\partial C}{\partial t}\Delta x \tag{2.2}$$

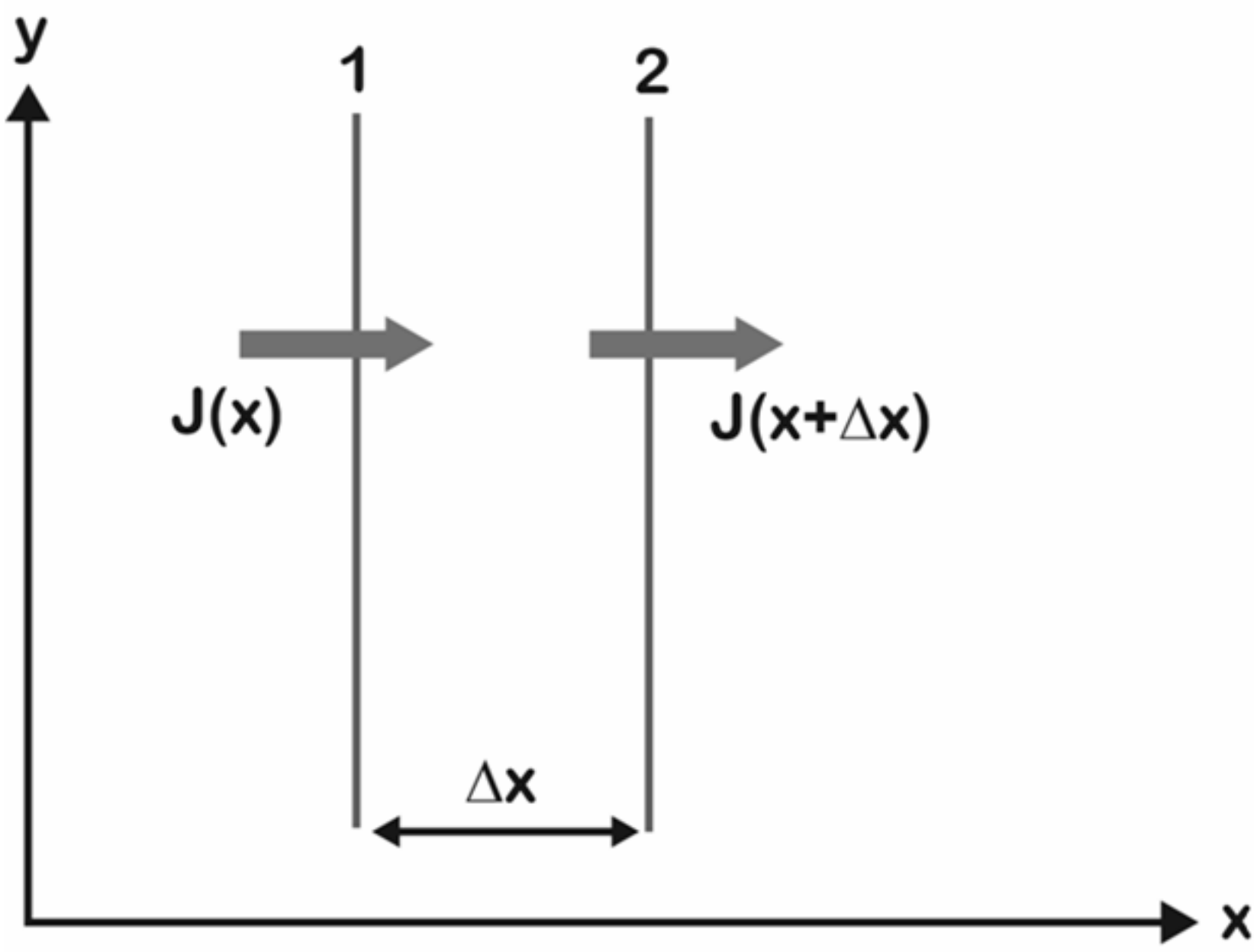

Fig 2.1: Two planes, 1 and 2 separated by distance $\Delta x$. Corresponding flux through each plane is *J(x)* and *J(x+Δx)*.

Combining the Fick's first law (Eqn 2.1) and law of conservation of mass (Eqn 2.2) we can write an expression which is called the "Diffusion equation" or the Fick's second law,

$$\frac{\partial C}{\partial t} = -\frac{\partial J}{\partial x} = \frac{\partial}{\partial x}\left( D\frac{\partial C}{\partial x} \right) \tag{2.3}$$

where *t* is time(s).

If we consider diffusion coefficient, *D*, is independent of composition, Fick's second law can be simplified and rewritten as

$$\frac{\partial C}{\partial t} = D\frac{\partial^2 C}{\partial x^2} \tag{2.4}$$

Using error function analysis the above expression can be solved. In most of the practical cases, where concentration changes with position, the diffusing atoms in a system shows different diffusion coefficient, which is called the interdiffusion coefficient, $\tilde{D}$. The interdiffusion coefficient, which is a function of composition, in most of the cases is not constant. Therefore Fick's second law can be expressed as

$$\frac{\partial C}{\partial t} = \frac{\partial}{\partial x}\left(\tilde{D}\frac{\partial C}{\partial x}\right) = \frac{\partial \tilde{D}}{\partial x}\frac{\partial C}{\partial x} + \tilde{D}\frac{\partial^2 C}{\partial x^2} \qquad (2.5)$$

The term $\partial \tilde{D}/\partial x$ makes the equation inhomogeneous, and it is not possible to solve this non-linear partial differential equation. However, it is very much possible to calculate the composition dependent $\tilde{D}$ using the classical Matano-Boltzmann analysis and Sauer-Freise which also follows a similar approach.

## 2.3 Determination of the Interdiffusion Coefficient, $\tilde{D}$

From early 20th century, it was evident that the diffusion coefficient in solid-state is not constant, but is a function of composition and temperature [3, 8]. In structural applications very often two different materials are in contact with each other at a temperature when interdiffusion takes place. The composition changes with time in a particular position because of the concentration gradient. Therefore, the interdiffusion coefficient, $\tilde{D}$, also changes at a particular position. The interdiffusion coefficient relates the interdiffusion flux $\tilde{J}_i$ with the concentration gradient by $\tilde{J}_i = -\tilde{D}\frac{\partial C_i}{\partial x}$; $i$ = components A or B. The interdiffusion fluxes are defined in the laboratory-fixed frame of reference [5]. The interdiffusion coefficient $\tilde{D}$ is, in general, a function of composition and, therefore, of $x$ $\left(\tilde{D} = \tilde{D}(x)\right)$ and Fick's second law can be written as

$$\frac{\partial C}{\partial t} = \frac{\partial}{\partial x}\left(\tilde{D}\frac{\partial C}{\partial x}\right) = \frac{\partial \tilde{D}}{\partial x}\frac{\partial C}{\partial x} + \tilde{D}\frac{\partial^2 C}{\partial x^2} \tag{2.6}$$

The term $\partial\tilde{D}/\partial x$ makes the equation in homogeneous, and the solution in closed form is not possible.

We can estimate $\tilde{D} = \tilde{D}(C)$ and this treatment is known as Matano-Boltzmann analysis. By this method interdiffusion coefficients, $\tilde{D}$ can be measured, at different compositions from the concentration profile, which is measured by microprobe analysis. Considering a diffusion couple of $C_B^-$ and $C_B^+$ and annealed for reasonably short time, *t*, such that still end of these materials are not affected by diffusion Fig. (2.2). Boundary conditions can be written as bellow

$$\begin{array}{llll} C_B = C_B^- & for & x\langle 0 & at \quad t = 0 \\ C_B = C_B^+ & for & x\rangle 0 & at \quad t = 0 \end{array} \tag{2.7}$$

where "-" and "+" represents the left- and right-hand end of the reaction couple.

In 1894, Boltzmann [8] introduced a variable λ, with the help of which Matano [9] was able to transform the complicated non-linear partial differential equation, Eqn 2.5 to a simpler form of differential equation with an assumption that $\tilde{D}$ is only a function of concentration, *C*. The Boltzmann parameter, λ is defined by the relation,

$$\lambda = \lambda(C_B) = \frac{x - x_o}{t^{1/2}} \tag{2.8}$$

This above relation states that every composition will have one fixed value of λ irrespective of the position $x$ with respect to initial contact plane, $x_o$ in a diffusion couple. It further states that one particular composition will move with respect to the $x_o$ such that it will have a fixed value after any annealing time, $t$.

From Eqn. 2.6, we can write,

$$\frac{\partial \lambda}{\partial t} = -\frac{1}{2}\frac{x}{t^{3/2}} \quad \text{and} \quad \frac{\partial \lambda}{\partial x} = \frac{1}{t^{1/2}} \tag{2.9}$$

Now $$\frac{\partial C}{\partial t} = \frac{\partial C}{\partial \lambda}\frac{\partial \lambda}{\partial t} = -\frac{1}{2}\frac{x}{t^{3/2}}\frac{\partial C}{\partial \lambda} \quad \text{and} \quad \frac{\partial C}{\partial x} = \frac{\partial C}{\partial \lambda}\frac{\partial \lambda}{\partial x} = \frac{1}{t^{1/2}}\frac{\partial C}{\partial \lambda} \tag{2.10}$$

Fick's second law, as given by Eqn 2.3 can be rewritten as

$$-\frac{1}{2}\frac{x}{t^{3/2}}\frac{\partial C}{\partial \lambda} = \frac{\partial}{\partial x}\left(\frac{\tilde{D}}{t^{1/2}}\frac{\partial C}{\partial \lambda}\right) \tag{2.11}$$

From Eqn 2.6 and 2.7, we can rewrite Eqn 2.9 as,

$$-\frac{1}{2}\frac{\lambda}{t}\frac{\partial C}{\partial \lambda} = \frac{\partial}{\partial \lambda}\left(\frac{\tilde{D}}{t}\frac{\partial C}{\partial \lambda}\right) \quad \text{or} \quad -\frac{1}{2}\lambda\frac{\partial C}{\partial \lambda} = \frac{\partial}{\partial \lambda}\left(\tilde{D}\frac{\partial C}{\partial \lambda}\right) \tag{2.12}$$

Eqn 2.10 is also known as Boltzmann transformed version of the Fick's second law. Matano in 1933 [9] used this transformation to derive the interdiffusion coefficient, $\tilde{D}$, based on the concentration profile by specifying appropriate boundary conditions which are

$C = C_B^-$ at $(x < 0, t = 0)$ that is, $\lambda = -\infty$

$C = C_B^+$ at $(x > 0, t = 0)$ that is, $\lambda = +\infty$

Typical concentration profile has been shown in Fig 2.3. Now, Eqn 2.10 contains only total differentials and $\partial\lambda$ can be cancelled from both sides. So we integrate both sides over the range varying between the initial concentration, $C_B^-$ and concentration under consideration, $C_B^*$, at which the interdiffusion coefficient is to be measured.

$$-\frac{1}{2}\int_{C_B^-}^{C_B^*}\lambda dC_B = \tilde{D}\frac{dC_B}{d\lambda}\bigg|_{C_B^-}^{C_B^*} \tag{2.13}$$

In a semi infinite diffusion couple, it is assumed that the ends of the couple are unaffected by the diffusion process. Hence, from the geometry of the composition profile, we can say that $dC_B/dx = 0$, at $C = C_B^-$ and $C = C_B^+$. Also, the interdiffusion coefficient is always measured at some particular annealing time and hence we can assume $t$ to be constant. Hence, Eqn 2.13 can be rewritten as,

$$-\frac{1}{2}\int_{C_B^-}^{C_B^*} x dC_B = \tilde{D}t\frac{dC_B}{dx}\bigg|_{C_B^-}^{C_B^*} = \tilde{D}t\left(\frac{dC_B}{dx}\right)_{C_B^*}$$

or $$\tilde{D}(C_B^*) = -\frac{1}{2t}\left(\frac{dx}{dC_B}\right)_{C_B^*}\int_{C_B^-}^{C_B^*} x dC_B \tag{2.14}$$

This relation is more popularly known as Matano-Boltzmann relation. Integrating the above equation over the entire range of composition, that is, $C_B^-$ and $C_B^+$, we get

$$\int_{C_B^-}^{C_B^+} x dC_B = 0 \tag{2.15}$$

This equation defines the Matano plane or the initial contact plane between the end members. This plane can be experimentally determined by equalising the areas P and Q in Fig 2.2 which is based on the conservation principle.

Applying integration by parts to Eqn 2.14, we get,

$$\tilde{D}(C_B^*) = -\frac{1}{2t}\left(\frac{\partial x}{\partial C_B}\right)_{C_B^*}\left[x^*(C_B^* - C_B^-) - \int_{-\infty}^{x^*}(C_B - C_B^-)dx\right]$$

$$= \frac{1}{2t}\left(\frac{\partial x}{\partial C_B}\right)_{C_B^*}[M + N] \qquad (2.16)$$

Areas M and N is shown in Fig 2.2 below.

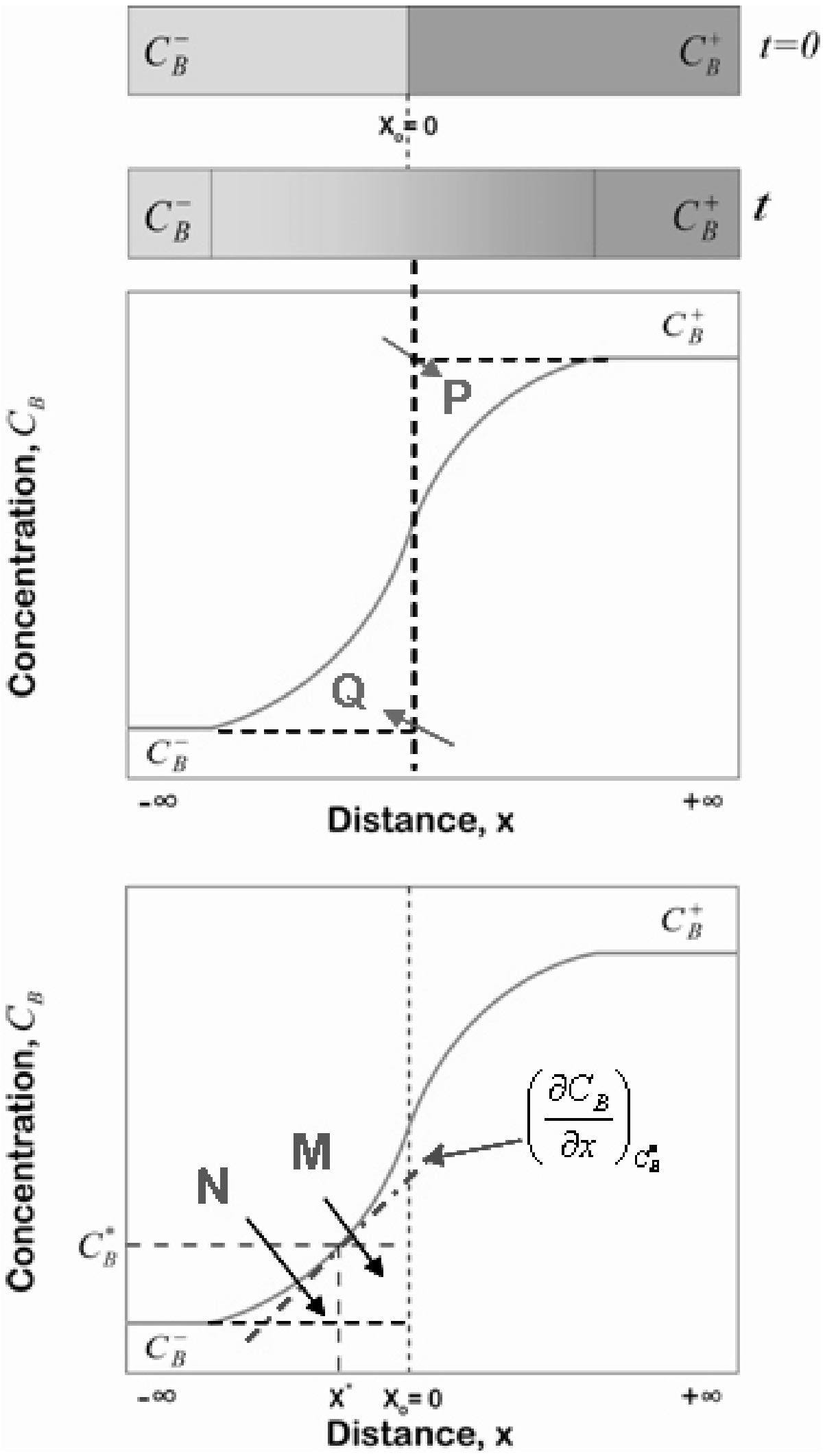

Fig 2.2: Schematic composition profile explains the calculation methodology of the interdiffusion coefficient following the Matano-Boltzmann relation.

There are few points that should be kept in mind when using the Matano-Boltzmann analysis,

- The analysis refers to an infinite system which means that during the annealing, the concentration change should not reach the end of the system
- The analysis shows huge error near the end member composition. This is due to the fact that as we move farther away from the Matano plane, the integral part of the Eqn 2.14 as well as the slope term, $\partial C_B / \partial x$ term approaches zero.
- The analysis can only be applied to the systems where there is no change in the volume after reaction/mixing during the interdiffusion mixing. This is so because there is a mismatch in the position of the Matano plane when determined from the left hand side or the right hand side of the diffusion couple.

To overcome the problem of molar volume change during interdiffusion, Ballufi [10] for the first derived the solution for $\tilde{D}$ for the case of molar volume deviation from ideality. Later on Sauer-Freise [11] generalised the Matano-Boltzmann relation while Den Boerder [12] developed the same theory based on graphical representation. Wagner [13] also derived a solution for the determination of interdiffusion coefficient utilising the Sauer-Freise variable so that there is no need for determining the Matano plane. The Wagner's theory is discussed below,

Before moving ahead, first we shall consider few simple thermodynamical relation for a binary system of species A and B,

$$N_A V_A + N_B V_B = V_m$$

$$C_A = N_A / V_m \quad \text{and} \quad C_B = N_B / V_m$$

(2.17)

Sauer-Freise normalisation variable *Y*, is defined as,

$$Y = \frac{N_B - N_B^-}{N_B^+ - N_B^-} \tag{2.18}$$

Hence, we can write,

$$N_A = \left(1 - N_B^+\right) Y + \left(1 - N_B^-\right)\left(1 - Y\right) \tag{2.19a}$$

$$N_B = N_B^+ \, Y + N_B^- \, \left(1 - Y\right) \tag{2.19b}$$

From Fick's second law as expressed by Eqn 2.3, we know that

$$\frac{\partial C_A}{\partial t} = -\frac{\partial J_A}{\partial x} \text{ and } \frac{\partial C_B}{\partial t} = -\frac{\partial J_B}{\partial x} \tag{2.20}$$

Also, from Eqn 2.6 and 2.7, we know that

$$\frac{\partial \lambda}{\partial t} = -\frac{1}{2}\frac{x}{t^{3/2}} = -\frac{\lambda}{2t} \tag{2.21}$$

Hence by substituting for C from Eqn 2.15, N from Eqn 2.17 and δt from Eqn 2.19, we can rewrite Eqn 2.18 as,

$$\frac{\lambda}{2t}\left[\left(1 - N_B^+\right)\frac{\partial}{\partial \lambda}\left(\frac{Y}{V_m}\right) + \left(1 - N_B^-\right)\frac{\partial}{\partial \lambda}\left(\frac{1-Y}{V_m}\right)\right] = \frac{\partial \tilde{J}_A}{\partial x} \tag{2.22a}$$

$$\frac{\lambda}{2t}\left[N_B^+ \frac{\partial}{\partial \lambda}\left(\frac{Y}{V_m}\right) + N_B^- \frac{\partial}{\partial \lambda}\left(\frac{1-Y}{V_m}\right)\right] = \frac{\partial \tilde{J}_B}{\partial x} \tag{2.22b}$$

Now, by solving for $N_B^- \times Eqn\ 3.20a - \left(1 - N_B^-\right) \times Eqn\ 2.20b$, we get,

$$-\frac{\lambda}{2t}\left(N_B^+ - N_B^-\right)\frac{d}{d\lambda}\left(\frac{Y}{V_m}\right) = N_B^- \frac{\partial \tilde{J}_A}{\partial x} - \left(1 - N_B^-\right)\frac{\partial \tilde{J}_B}{\partial x} \tag{2.23a}$$

Similarly by solving for $N_B^+ \times Eqn\ 3.20a - \left(1 - N_B^+\right) \times Eqn\ 2.20b$, we get,

$$\frac{\lambda}{2t}\left(N_B^+ - N_B^-\right)\frac{d}{d\lambda}\left(\frac{1-Y}{V_m}\right) = N_B^+ \frac{\partial \tilde{J}_A}{\partial x} - \left(1 - N_B^+\right)\frac{\partial \tilde{J}_B}{\partial x} \tag{2.23b}$$

Also, as already shown in Eqn 2.7,

$$\frac{\partial \lambda}{\partial x} = \frac{1}{t^{1/2}} \text{ or } \partial \lambda = \frac{\partial x}{t^{1/2}}$$

Hence, multiplying $\partial \lambda$ to the left hand side and $\frac{\partial x}{t^{1/2}}$ to the right side of Eqn 2.21a, followed by integrating the expression by parts within the limits $\lambda = -\infty$ to $\lambda = \lambda^*$, ($\lambda^*$ corresponds to the position at which the interdiffusion coefficient is to be calculated), we get,

$$\frac{1}{2t}\left(N_B^+ - N_B^-\right)\left[-\frac{\lambda^* Y^*}{V_m^*} + \int_{-\infty}^{\lambda^*} \frac{Y}{V_m} d\lambda\right] = \frac{1}{t^{1/2}}\left[N_B^- \tilde{J}_A^* - \left(1 - N_B^-\right)\tilde{J}_B^*\right] \tag{2.24a}$$

Similarly, applying same treatment to Eqn 2.21b and integrating within the limits $\lambda = \lambda^*$ to $\lambda = +\infty$, we get,

$$\frac{1}{2t}\left(N_B^+ - N_B^-\right)\left[-\frac{\lambda^*\left(1 - Y^*\right)}{V_m^*} - \int_{\lambda^*}^{+\infty} \frac{\left(1 - Y\right)}{V_m} d\lambda\right] = \frac{1}{t^{1/2}}\left[-N_B^+ \tilde{J}_A^* - \left(1 - N_B^+\right)\tilde{J}_B^*\right] \tag{2.24b}$$

In Eqns 2.22, $\tilde{J}_A^*$ and $\tilde{J}_B^*$ are the fluxes of the diffusion species A and B at position $\lambda = \lambda^*$.

Now, by solving for $\left(1 - Y^*\right) \times Eqn\ 2.22a - Y^* \times Eqn\ 2.22b$, we get,

$$\frac{1}{2t}\left(N_B^+ - N_B^-\right)\left[\left(1 - Y^*\right)\int_{-\infty}^{\lambda^*} \frac{Y}{V_m} d\lambda + Y^* \int_{\lambda^*}^{+\infty} \frac{1 - Y}{V_m} d\lambda\right] = \frac{1}{t^{1/2}}\left[N_B^* \tilde{J}_A^* - \left(1 - N_B^*\right)\tilde{J}_B^*\right] \tag{2.25}$$

Now, writing Fick's first law in terms of intrinsic flux,

$$\tilde{J}_A = -\tilde{D}\left(\frac{\partial C_A}{\partial x}\right) = \frac{V_B}{V_A}\left(\frac{\partial C_B}{\partial x}\right) = -\tilde{J}_B \frac{V_B}{V_A} \tag{2.26}$$

since, $V_A dC_A + V_B dC_B = 0$

Also, $\tilde{D}\frac{dC_B}{dx} = \tilde{D}\frac{V_A}{V_m^2}\frac{dN_B}{dx} = -\tilde{J}_B = -\frac{(N_B V_B + N_A V_A)}{V_m}\tilde{J}_B$ (2.27)

since, $V_A N_A + V_B N_B = V_m$

Hence, from Eqn 2.24 and 2.25, we can write,

$$\tilde{D} = \frac{V_m\left(N_B\tilde{J}_A - N_A\tilde{J}_B\right)}{\partial N_B/\partial x} \tag{2.28}$$

Now, substituting Eqn 2.26 in Eqn 2.23 for $N_B = N_B^*$ and also for $d\lambda = \frac{dx}{t^{1/2}}$, Eqn 2.23

can be expressed in terms of the interdiffusion coefficient $\tilde{D}$ as,

$$\tilde{D}(N_B^*) = \frac{(N_B^+ - N_B^-)V_m^*}{2t\,(\partial N_B/\partial x)_{x=x^*}*}\left[(1-Y^*)\int_{-\infty}^{x^*}\frac{Y}{V_m}dx + Y^*\int_{x^*}^{\infty}\frac{(1-Y)}{V_m}dx\right] \tag{2.29}$$

Now, by differentiating Eqn 2.16 and substituting it in Eqn 2.27, we get,

$$\tilde{D}(Y^*) = \frac{V_m^*}{2t\left(\frac{dY^*}{dx}\right)}\left[(1-Y^*)\int_{-\infty}^{x^*}\frac{Y}{V_m}dx + Y^*\int_{x^*}^{+\infty}\frac{(1-Y)}{V_m}dx\right] \tag{2.30}$$

We can also show schematically, the above expression for the calculation of the interdiffusion coefficient (Fig 2.3)

From Fig. 2.3, we can rewrite Eqn 2.28 as,

$$\tilde{D}(Y^*) = \frac{V_m^*}{2t\left(\frac{dY}{dx}\right)_*}\left[(1-Y^*)P + Y^*Q\right] \tag{2.31}$$

Den Broeder [11] also solved for the interdiffusion coefficient by modifying the Matano-Boltzmann equation by graphically treating it and gave the expression,

$$\tilde{D}(C_B^*) = \left(\frac{\partial x}{\partial C_B}\right)_{C_B^*}\left[(1-Y^*)\int_{-\infty}^{x^*}(C_B - C_B^-)\,dx + Y^*\int_{x^*}^{+\infty}(C_B^+ - C_B)dx\right] \tag{2.32}$$

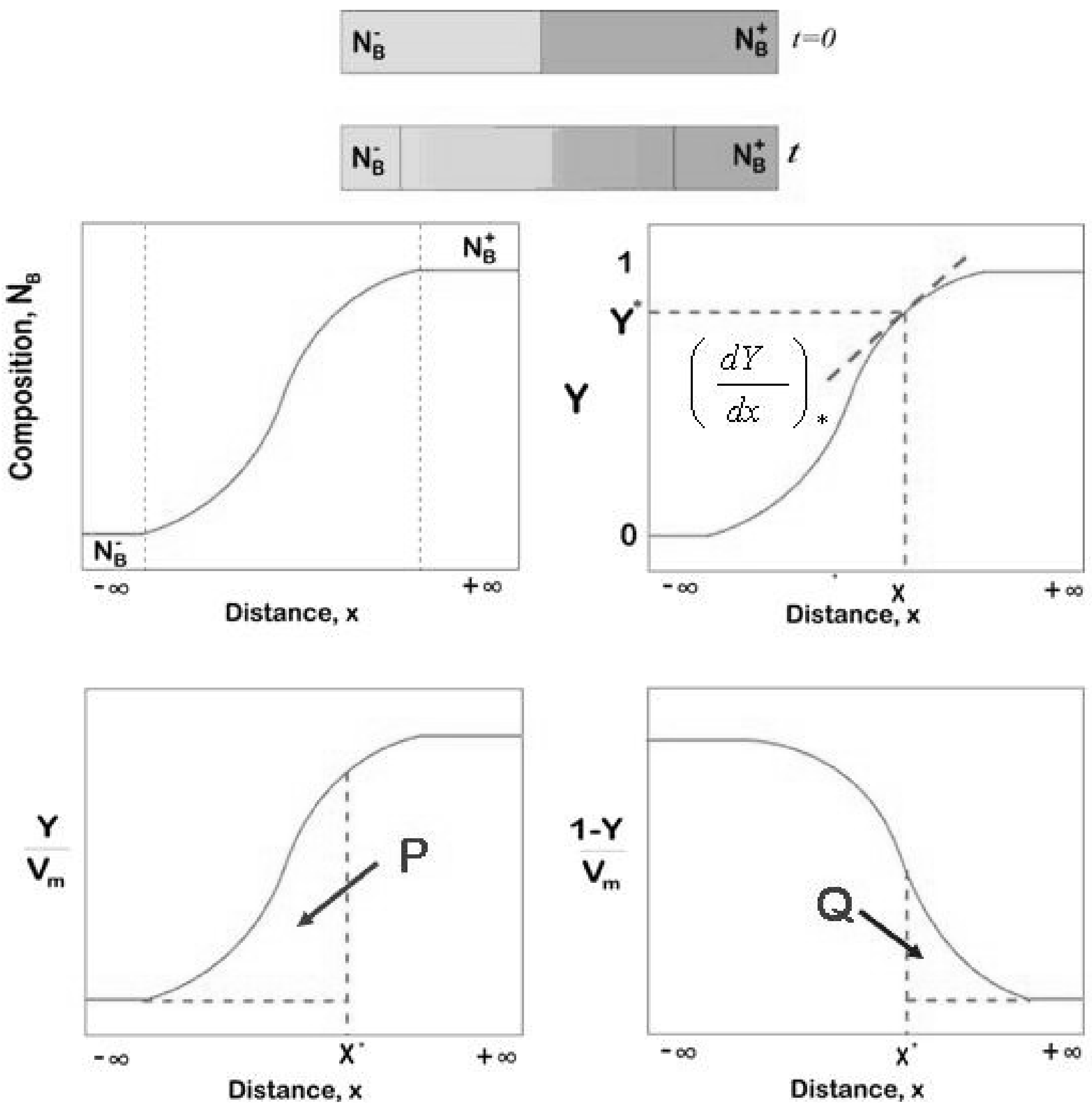

Fig 2.3: Schematic representation explains the calculation methodology of the interdiffusion coefficient following the Wagner's method.

## 2.4 Determination of the integrated diffusion coefficient, $\tilde{D}_{int}$

In the case of line compounds or compounds with a narrow homogeneity range, it is not possible to calculate the vanishingly small concentration gradient. Hence, the expression derived for the interdiffusion coefficient in the previous section, Eqn 2.28, can not be applied in such cases. Hence, Wagner introduced a new variable, the integrated diffusion coefficient, $\tilde{D}_{int}$ to show the diffusion in line compounds. If we consider β phase, as a line compound with a very narrow homogeneity range of $N_B^{\beta 1}$ and $N_B^{\beta 2}$, such that $N_B^{\beta 1} \le N_B^{\beta} \le N_B^{\beta 2}$, in a particular system as shown in the schematic phase diagram Fig. 2.4a. Two end members $N_B^-$ and $N_B^+$ are coupled and annealed at elevated temperature for time *t*. The resulting concentration profile is shown in Fig. 2.4b. The total thickness of β is $\Delta x_\beta \left(= x_{\beta 1} - x_{\beta 1}\right)$. The integrated diffusion coefficient can be expressed for β as,

$$\tilde{D}_{\text{int}}^{\beta} = \int_{N_B^{\beta 1}}^{N_B^{\beta 2}} \tilde{D} dN_B = \frac{\left(N_B^{\beta} - N_B^-\right)\left(N_B^+ - N_B^{\beta}\right)}{N_B^+ - N_B^-}\frac{\Delta x^2}{2t}$$
$$+ \frac{\Delta x}{2t}\left[\frac{N_B^+ - N_B^{\beta}}{N_B^+ - N_B^-} \times \int_{-\infty}^{x^{\beta 1}} \frac{V_m^{\beta}}{V_m}\left(N_B - N_B^-\right)dx + \frac{N_B^{\beta} - N_B^-}{N_B^+ - N_B^-} \times \int_{x^{\beta 2}}^{\infty} \frac{V_m^{\beta}}{V_m}\left(N_B^+ - N_B\right)dx\right]$$

(2.33)

$\tilde{D}_{\text{int}}^{\beta}$ can be related to the average interdiffusion coefficient as,

$$\tilde{D}_{av}^{\beta} = \frac{\tilde{D}_{\text{int}}^{\beta}}{N_B^{\beta 2} - N_B^{\beta 1}} = \frac{1}{N_B^{\beta 2} - N_B^{\beta 1}} \int_{N_B^{\beta 1}}^{N_B^{\beta 2}} \tilde{D} dN_B$$

which cannot be used because of the unknown value of $\left(N_B^{\beta 2} - N_B^{\beta 1}\right)$.

In the case of a diffusion system where several line compounds exist and where no solubility in the end members of the components occur as shown in Fig. 2.5a and 2.5b, Eqn 2.31 can be written as,

$$\tilde{D}_{\text{int}}^{\beta} = \frac{\left(N_B^{\beta} - N_B^-\right)\left(N_B^+ - N_B^{\beta}\right)}{N_B^+ - N_B^-}\frac{\Delta x_\beta^{\ 2}}{2t} + \frac{\Delta x_\beta}{2t} \text{x}$$
$$\left[\frac{\left(N_B^+ - N_B^{\beta}\right)\sum_{\nu=2}^{\nu=\beta-1} \frac{V_m^{\beta}}{V_m^{\nu}}\left(N_B^{\nu} - N_B^-\right)\Delta x_\nu + \left(N_B^{\beta} - N_B^-\right)\sum_{\nu=\beta+1}^{\nu=n-1} \frac{V_m^{\beta}}{V_m^{\nu}}\left(N^+ - N^{\nu}\right)\Delta x_\nu}{N_B^+ - N_B^-}\right]$$

(2.34)

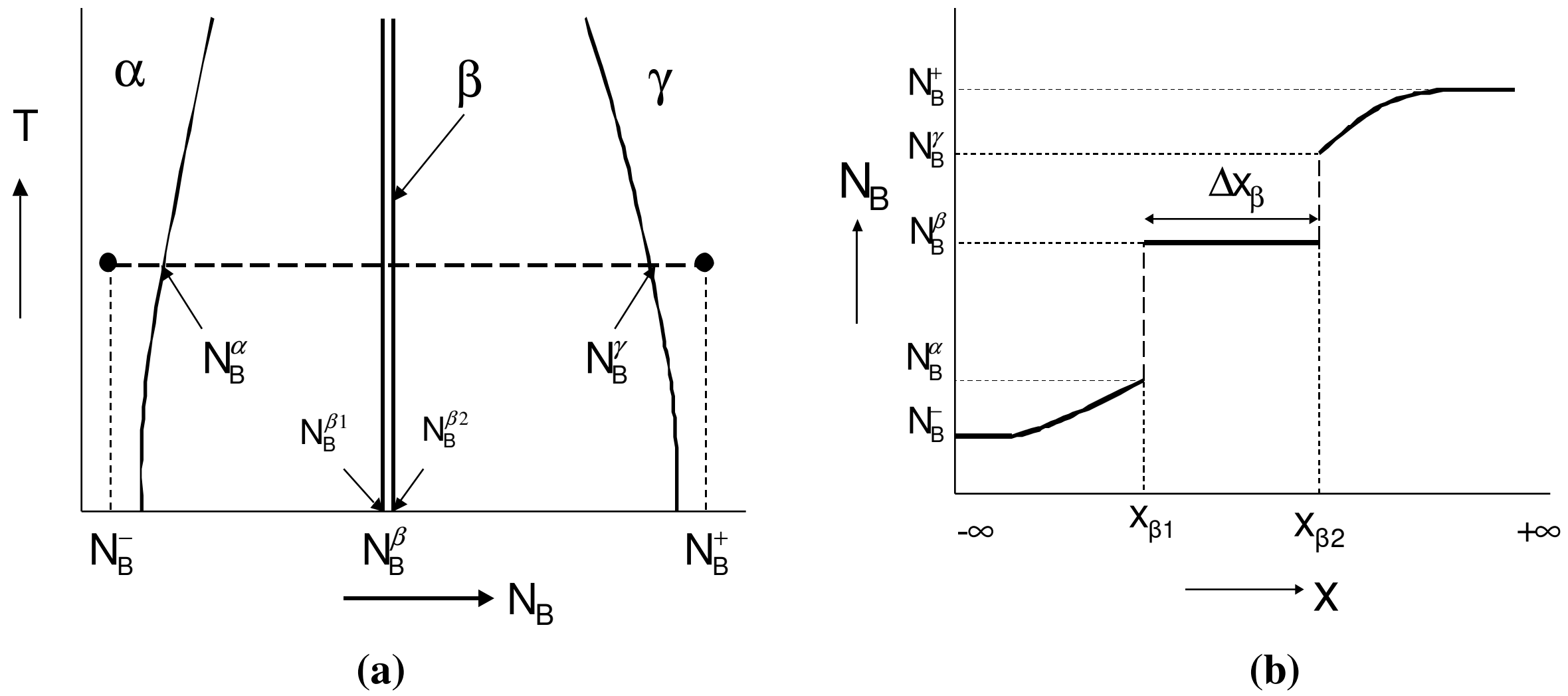

Fig 2.4: (a) Presence of only one line compound β phase between two end members $N_B^-$ and $N_B^+$ shown schematically. (b) Schematically shown a composition profile of a diffusion couple $N_B^-$ and $N_B^+$ after annealing time *t*.

Following Fig. 2.5b, Eqn. 2.32 can be expressed as,

$$\tilde{D}_{\text{int}}^{\beta} = \frac{ab}{a+b}\frac{\Delta x_\beta^2}{2t} + \frac{\Delta x_\beta}{2t}\left[\frac{b\left(\dfrac{V_m^\beta}{V_m^\varepsilon}P + \dfrac{V_m^\beta}{V_m^\alpha}Q\right) + a\left(\dfrac{V_m^\beta}{V_m^\gamma}R + \dfrac{V_m^\beta}{V_m^\delta}S\right)}{a+b}\right] \tag{2.35}$$

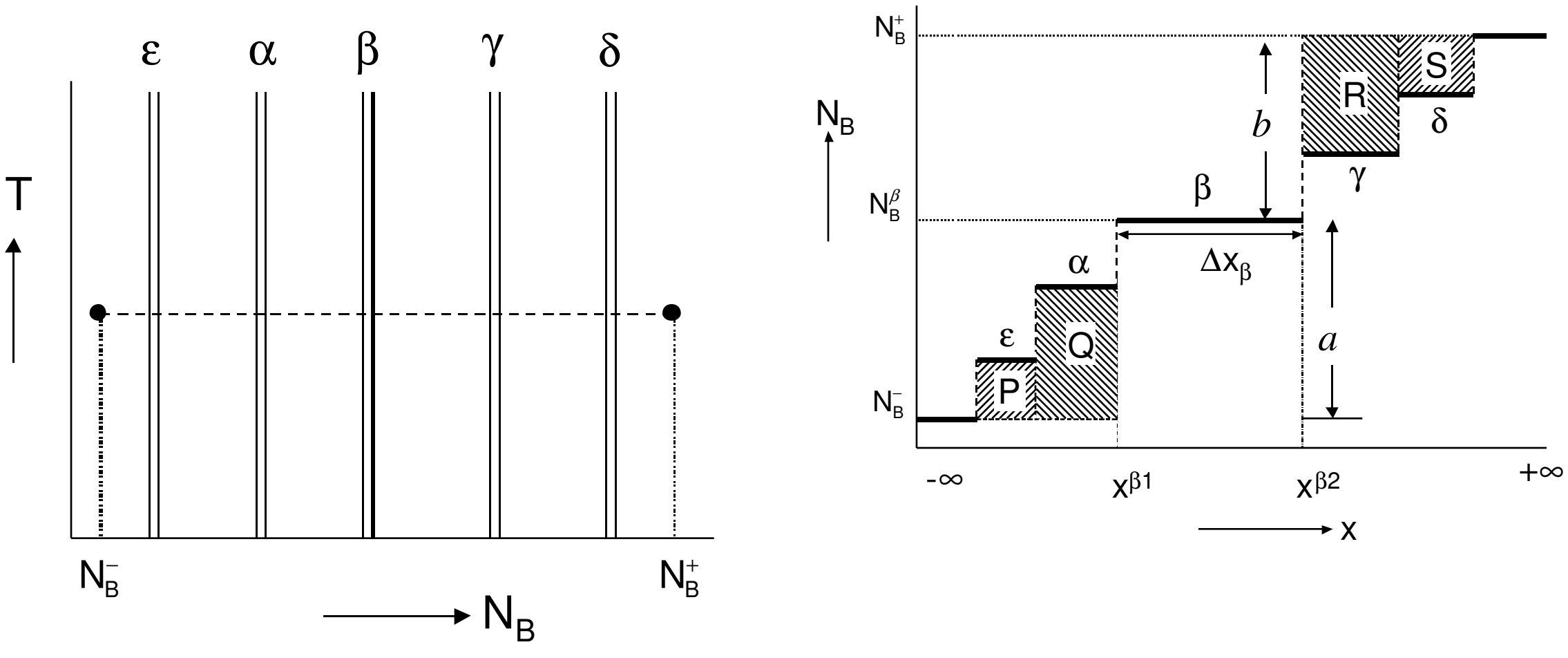

(a) (b)

Fig 2.5: (a) Schematic representation of a phase diagram having several line compounds. (b) Schematic composition profile of a diffusion couple, after annealing time *t*. Compositions correspond to the phase diagram shown in 2.5a.

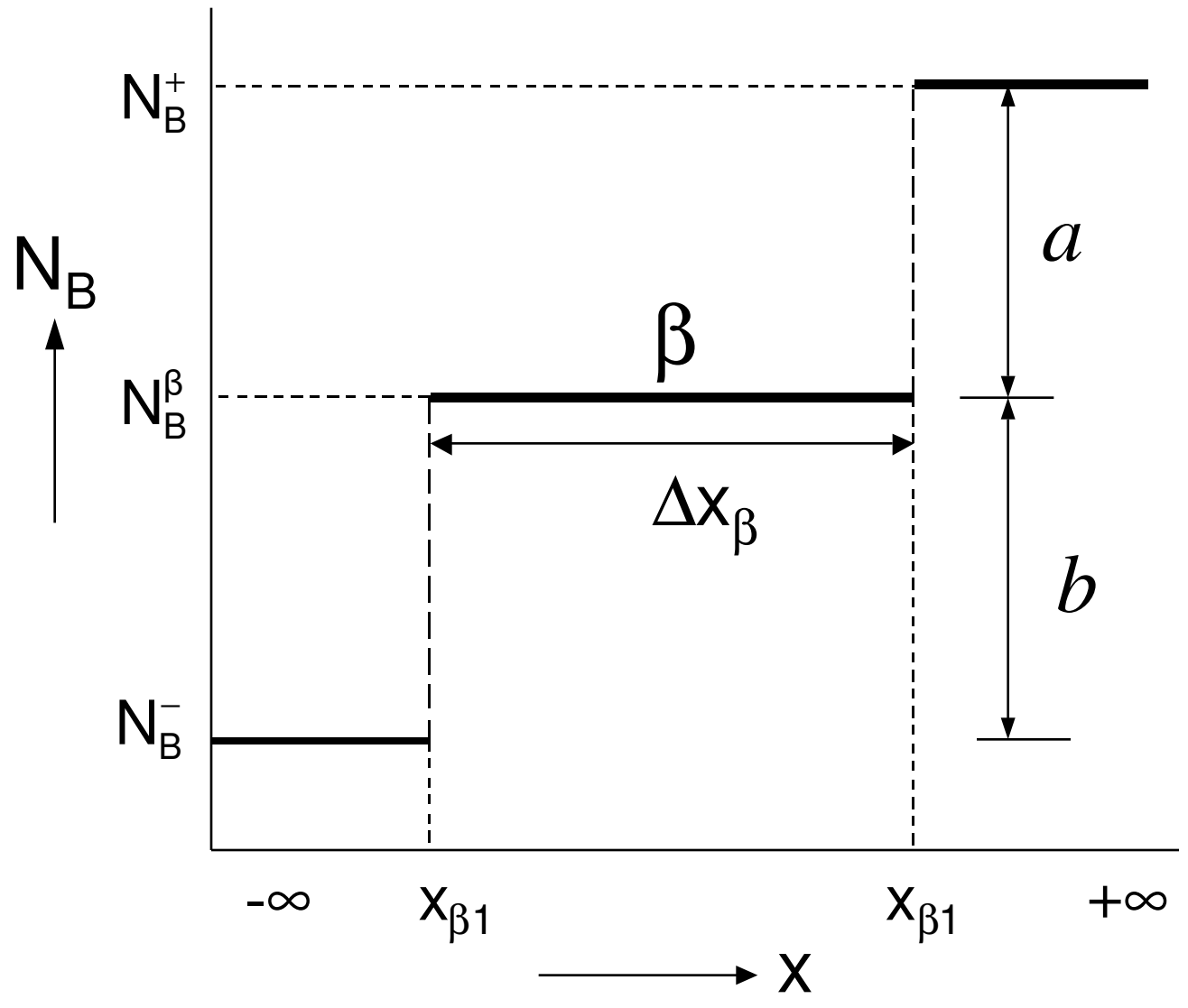

Fig 2.6: Schematic representation of a composition profile of a diffusion couple

In the special case that only one intermetallic compound β, exists between two end members, $N_B^-$ and $N_B^+$, as shown in Fig. 2.6, then Eqn. 2.32 reduces to,

$$\tilde{D}_{\text{int}}^{\beta} = \frac{a.b}{a+b}\frac{\Delta x_{\beta}^{2}}{2t} = \frac{ab}{a+b}k_{p} \tag{2.36}$$

where $k_p$ is the parabolic growth constant.

**2.5 Intrinsic Diffusion Coefficient and the Kirkendall Effect**

Till now we have discussed only about the composition dependent interdiffusion coefficient to describe the diffusion process, which represents average diffusivity. However, in a real system, the diffusing species have unequal transfer rate that can be expressed in terms of individual *intrinsic diffusion coefficient* of the species, which is again composition dependent. Due to this inequality of the flux of each species, there is net mass flow in the interdiffusion process. Because of this reason the diffusion couple swells on one side and shrinks on the other side. This effect is popularly called *'Kirkendall effect'* and was discovered by Smigelkas and Kirkendall [14] in 1947. However, such effect was earlier reported by Pfiel [15] in 1929 in the oxidation study of iron and steel.

Smigelkas and Kirkendall made use of inert marker, Molybdenum wire, to show the inequality of the diffusing species in Cu-Zn system. The experimental setup is shown below, Fig 2.7. In the experiment, Mo wire was placed on both sides of brass bar (Cu-30 wt % Zn) and Cu was coated over it. This couple was then subjected to annealing at 785 $^{o}$C. After annealing for certain duration the couple was section and examined to get the distance between the Mo wires. The process was repeated to get the data at different annealing times. The results showed that the Zn atoms moved faster outwards than Cu inwards, that is, $D_{Zn} > D_{Cu}$. This resulted in shrinking of the core brass which in turn caused the Mo wires to move inwards and the distance between the wire decreases parabolically with annealing time.

Hence, $x_K = k\sqrt{t}$, where $x_K$ is the position of the Kirkendall plane which moves parabolically with the annealing time, *t*. This confirms the process to be diffusion controlled and the Kirkendall plane is the only plane which starts moving from the beginning of the diffusion process with a velocity given as $v_K = x_K/2t$.

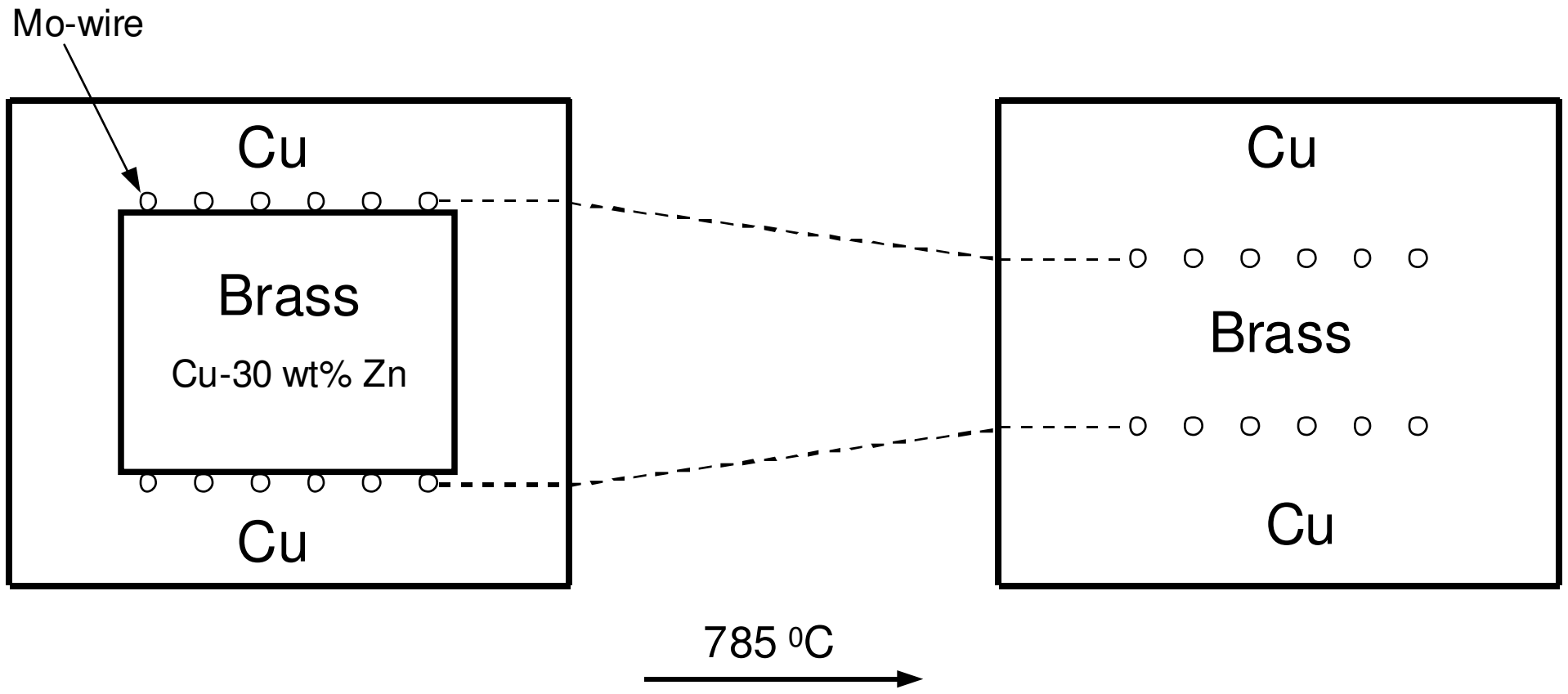

Fig 2.7: Schematically shown the cross-section of the diffusion couples prepared by Smigelkas and Kirkendall [14] The diffusion couple was annealed at 785 °C. Schematically shown that the molybdenum wires moved closer to each other after annealing.

Prior to the establishment of the Kirkendall effect, it was a common belief that diffusion in solids occur by direct exchange or the ring mechanism which implies that diffusivity of both the species are equal. Subsequently, the vacancy mechanism of substitutional diffusion got an acceptance.

After the establishment of the Kirkendall effect, it was quite apparent that the diffusion process cannot be described completely with the help of only single diffusion coefficient and that diffusivity of both the species is required in a binary system. In 1948,

Darken [16] for the first time attempted to mathematically relate the interdiffusion coefficient with intrinsic diffusion coefficient.

Suppose the intrinsic diffusivity of B is higher than that of A, that is, $D_B > D_A$ at the Kirkendall plane. If we consider that the Kirkendall plane, $x_K$ moves from the initial contact plane $x_{M/o}$ with a velocity $v_K$, then the relation between the interdiffusion fluxes, $\tilde{J}_A$ and $\tilde{J}_B$ (measured with respect to the initial contact plane) and intrinsic diffusion fluxes, $J_A$ and $J_B$ (measured with respect to the Kirkendall frame of reference measured at $x_K$) can be written as

$$\tilde{J}_i = J_i + v_K C_i \quad \text{where } i=A, B \tag{2.37}$$

Now, in terms of volume flux, we can write,

$$J_i^{vol} = V_i \tilde{J}_i = V_i J_i + v_K V_i C_i \tag{2.38}$$

Now, for an infinite binary diffusion couple, $J_A^{vol} + J_B^{vol} = 0$ (2.39)

From the Fick's first law, intrinsic flux is related to the intrinsic diffusivity, that is,

$J_i = -D_i \left( \frac{\partial C_i}{\partial x} \right)_K$, hence from Eqn 2.35 and 2.36, we can write,

$$v_K = V_B (D_B - D_A) \left( \frac{\partial C_B}{\partial x} \right)_K = \frac{V_A V_B}{V_m^2} (D_B - D_A) \left( \frac{\partial N_B}{\partial x} \right)_K = -(V_B J_B + V_A J_A) \tag{2.40}$$

Now, substituting Eqn 2.37 in Eqn 2.34 written for species B, we get

$$-\tilde{D} \frac{\partial C_B}{\partial x} = -D_B \frac{\partial C_B}{\partial x} - C_B \left( -V_B D_B \frac{\partial C_B}{\partial x} - V_A D_A \frac{\partial C_A}{\partial x} \right)$$

But, we know that $V_A C_A + V_B C_B = 1$ and hence, $V_A \partial C_A = -V_B \partial C_B$

Hence, $\tilde{D}\frac{\partial C_B}{\partial x} = (1 - V_B C_B) D_B \frac{\partial C_B}{\partial x} + V_B C_B D_A \frac{\partial C_B}{\partial x}$

or, $\tilde{D} = V_A C_A D_B + V_B C_B D_A$ (2.41)

In the special case when partial molar volumes of the components are equal and do not change with the composition, $V_m = V_A = V_B$, we can rewrite Eqn 2.38 as,

$$\tilde{D} = N_A D_B + N_B D_A \tag{2.42}$$

This relation of interdiffusion coefficient with the intrinsic diffusion coefficient is known as Darken's relation.

Intrinsic diffusion coefficient of the species in an interdiffusion zone can be calculated only at the location of the Kirkendall marker plane, since only this plane moves parabolically with time starting from *t=0* (contrary to markers introduced at other positions in the couple at *t=0*). This is the the only plane, which is immediately affected by the diffusion process at the start of the annealing and once the markers are trapped in a certain composition, they stay at that same composition for the entire annealing time and this plane acts as a reference to determine the intrinsic diffusivities. The derivation given below was done by Paul [5].

Solving for $N_B^+ \times Eqn\ 2.22a + N_B^- \times Eqn\ 2.22b$, we get,

$$\frac{1}{2t}\left(N_B^+ - N_B^-\right)\left[-\lambda_K\left\{\frac{Y_K}{V_m^K}N_B^+ + \frac{1-Y_K}{V_m^K}N_B^-\right\} + \left\{N_B^+\int_{-\infty}^{\lambda_K}\frac{Y}{V_m}d\lambda - N_B^-\int_{\lambda_K}^{+\infty}\frac{1-Y}{V_m}d\lambda\right\}\right]$$
$$= \frac{1}{t^{1/2}}\left[N_B^-\left(1-N_B^+\right)\tilde{J}_A^K - N_B^+\left(1-N_B^-\right)\tilde{J}_B^K\right]$$

(2.43)

After rearranging, we get,

$$\tilde{J}_B^K = \frac{t^{1/2}}{2t}\left[\lambda_K \frac{N_B^K}{V_m^K} - N_B^+ \int_{-\infty}^{\lambda_K} \frac{Y}{V_m} d\lambda + N_B^- \int_{\lambda_K}^{+\infty} \frac{1-Y}{V_m} d\lambda\right] \quad (2.44)$$

Now, substituting for $t^{1/2} d\lambda = dx$ and then rearranging, we get,

$$\tilde{J}_B^K = v_K C_B^K - \frac{1}{2t}\left[N_B^+ \int_{-\infty}^{\lambda_K} \frac{Y}{V_m} dx - N_B^- \int_{\lambda_K}^{+\infty} \frac{1-Y}{V_m} dx\right] \quad (2.45)$$

Since, $\tilde{J}_B = J_B + v_K C_B = -D_B \dfrac{\partial C_B}{\partial x} + v_K C_B$, we can rewrite Eqn 2.42 as,

$$D_B = \frac{1}{2t}\left(\frac{\partial x}{\partial C_B}\right)_K \left[N_B^+ \int_{-\infty}^{x_K} \frac{Y}{V_m} dx - N_B^- \int_{x_K}^{+\infty} \frac{(1-Y)}{V_m} dx\right] \quad (2.46a)$$

Similarly, for the diffusing species A, we can write,

$$D_A = \frac{1}{2t}\left(\frac{\partial x}{\partial C_A}\right)_K \left[N_A^+ \int_{-\infty}^{x_K} \frac{Y}{V_m} dx - N_A^- \int_{x_K}^{+\infty} \frac{(1-Y)}{V_m} dx\right] \quad (2.46b)$$

and the ratio of intrinsic diffusivity can be written as,

$$\frac{D_B}{D_A} = \frac{V_B}{V_A}\left[\frac{N_B^+ \int_{-\infty}^{x_K} \frac{Y}{V_m} dx - N_B^- \int_{x_K}^{\infty} \frac{(1-Y)}{V_m} dx}{-N_A^+ \int_{-\infty}^{x_K} \frac{Y}{V_m} dx + N_A^- \int_{x_K}^{\infty} \frac{(1-Y)}{V_m} dx}\right] \quad (2.47)$$

This was also derived by van Loo [17] differently. However, in the case of line compound, it is not possible to calculate the intrinsic diffusion coefficients of elements, because of the vanishingly small concentration gradient at the position of the Kirkendall plane for a compound with a very narrow homogeneity range. Nevertheless, one always calculate the ratio of the intrinsic diffusivities following Eqn 2.44, since it does not require the concentration gradient term. In the case of a diffusion couple as shown in Fig.

2.8a, if we consider growth of β phase and the Kirkendall marker plane situated at $x_K$, the ratio of intrinsic diffusivities can be written as

$$\frac{D_B}{D_A}=\frac{V_B}{V_A}\left[\frac{N_B^{+}\Phi-N_B^{-}\Psi}{-N_A^{+}\Phi+N_A^{-}\Psi}\right] \tag{2.48}$$

where,

$$\Phi=\left(\int_{-\infty}^{x_{\beta 1}}\frac{\left(N_B-N_B^{-}\right)}{V_m}dx+\frac{\left(N_B^{\beta}-N_B^{-}\right)}{V_m^{\beta}}\Delta x_1\right) \text{ and } \Psi=\left(\int_{x_{\beta 2}}^{+\infty}\frac{\left(N_B^{+}-N_B\right)}{V_m}dx+\frac{\left(N_B^{+}-N_B^{\beta}\right)}{V_m^{\beta}}\Delta x_2\right)$$

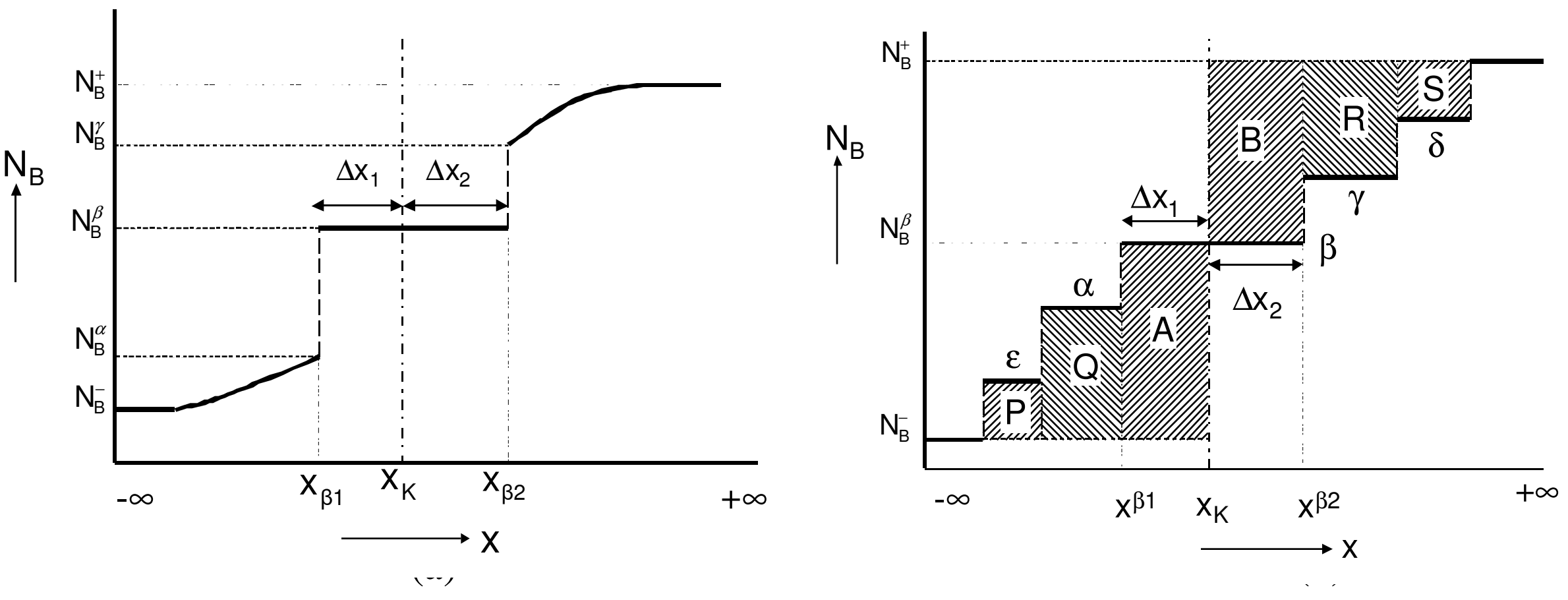

Fig 2.8: (a) Schematic composition profile for the case when only one line compound β grows between the end members α and γ. The Kirkendall marker plane is in the $\beta$ phase at position $x_K$. (b) Schematic composition profile for the case when several line compounds exist. The Kirkendall marker plane is in the β phase at position $x_K$.

If we consider the diffusion couple in Fig. 2.8b, where no solubility exists in the end members, all the compounds have a narrow homogeneity range and the Kirkendall plane is located in β-phase at $x_K$, then Φ and Ψ in Eqn. 2.18 can be written as,

$$\Phi = \left( \sum_{\nu=2}^{\beta-1} \frac{\left(N_B^{\nu} - N_B^{-}\right)}{V_m^{\nu}} \Delta x^{\nu} + \frac{\left(N_B^{\beta} - N_B^{-}\right)}{V_m^{\beta}} \Delta x_1 \right) \text{and}$$

$$\Psi = \left( \sum_{\nu=\beta+1}^{n-1} \frac{\left(N_B^{+} - N_B^{\nu}\right)}{V_m^{\nu}} \Delta x^{\nu} + \frac{\left(N_B^{+} - N_B^{\beta}\right)}{V_m^{\beta}} \Delta x_2 \right) \tag{2.49}$$

From Fig 2.8b, we can rewrite Eqn 2.46 as

$$\Phi = \left( \frac{P}{V_m^{\varepsilon}} + \frac{Q}{V_m^{\alpha}} + \frac{A}{V_m^{\beta}} \right) \text{ and } \Psi = \left( \frac{S}{V_m^{\delta}} + \frac{R}{V_m^{\gamma}} + \frac{B}{V_m^{\beta}} \right)$$

In the special case when there is only one layer, β, growing between two end members, and the Kirkendall plane position $x_K$ is present in that phase, as shown in Fig. 2.9, the equations reduce to,

$$\Phi = \frac{\left(N_B^{\beta} - N_B^{-}\right)}{V_m^{\beta}} \Delta x_1 = \frac{A}{V_m^{\beta}} \text{ and } \Psi = \frac{\left(N_B^{+} - N_B^{\beta}\right)}{V_m^{\beta}} \Delta x_2 = \frac{B}{V_m^{\beta}}$$

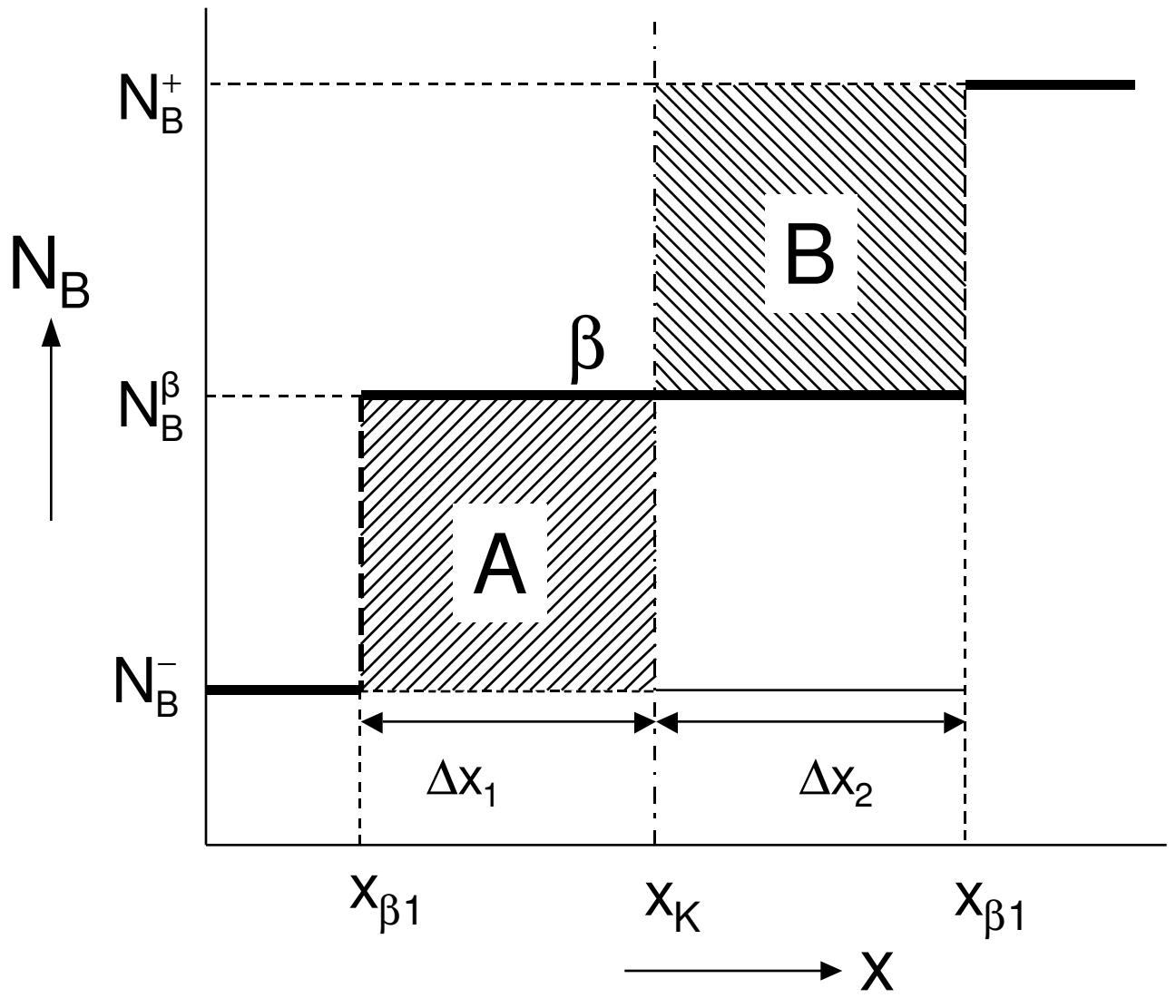

Fig 2.9: Schematic composition profile of a diffusion couple when only one line compound exists. The Kirkendall marker plane is inside the β phase at position $x_K$.

In the case of pure end members, $N_B^- = 0$ and $N_B^+ = 1$, Eqn. 2.48 reduces to

$$\frac{D_B}{D_A} = \frac{V_B}{V_A}\left[\frac{N_B^+ \Phi}{N_A^- \Psi}\right] = \frac{V_B}{V_A}\left[\frac{\Phi}{\Psi}\right] \tag{2.50}$$

and if there is only one compound between two pure end member then,

$\Phi = \frac{N_B^\beta}{V_m^\beta}\Delta x_1$ ; $\Psi = \frac{1 - N_B^\beta}{V_m^\beta}\Delta x_2$ and Eqn. 2.45 further reduces to

$$\frac{D_B}{D_A} = \frac{V_B}{V_A}\left[\frac{N_B^\beta}{\left(1 - N_B^\beta\right)}\frac{\Delta x_1}{\Delta x_2}\right] \tag{2.51}$$

## 2.6 Tracer Diffusion Coefficient

Till now we have discussed about the intermixing of the atoms taking place under concentration gradient, that is, in presence of a driving force. However, even in the absence of any driving force, intermixing of atoms takes place and this phenomenon of atom transport is referred to as *self diffusion*. It is possible to monitor the mobility of atoms by use of radioactive isotopes or *tracer* of the element under consideration. That is, they are chemically same with only one or two neutron mass difference. Another requirement of the tracer material is that the half life should be neither to short nor too long.

Experimentally, the tracer diffusion coefficient is measured by depositing a thin layer of the tracer element over the element/alloy under consideration followed by annealing at temperature of interest, *T*, and annealing time, *t*. Subsequently, thin sections are made at different values of distance, *x* from the surface on which the tracer material was deposited. The concentration can then be evaluated as function of distance from the intensity of the radioactive radiation from each of the section. Hence, the tracer diffusion coefficient, $D^*$ can be calculated following the relation developed as a solution of Fick's second law for thin films where the diffusion distance is very small,

$$C_B^*(x,t) = \frac{N_B^*}{\sqrt{\pi D_B^* t}} \exp\left(-\frac{x^2}{4D_B^* t}\right) \tag{2.52}$$

The experimental procedure for determining the tracer diffusion coefficient is quite cumbersome. However, it is also possible to indirectly calculate $D^*$ from the results of Kirkendall marker experiment. Here we shall discuss indirect methodology for determining the tracer diffusion coefficient.

It is fundamentally incorrect to assume that concentration gradient as the driving force in Fick's laws and more realistically, actual driving force is thechemical potential gradient. Hence Fick's law can be written as

$$J_B = -M_B C_B \frac{d\mu_B}{dx} = -D_B \frac{dC_B}{dx} \tag{2.53}$$

where, mobility $M_B = v_B / F$, $v_B$ is the velocity of the species B and $F$ is the force on the species B due to chemical potential, $\mu_B$. Hence we can write,

$$D_B = M_B C_B \frac{d\mu_B}{dC_B} \tag{2.54}$$

Using standard thermodynamic relation, we can rewrite Eqn 2.51 as,

$$D_B = \frac{V_m}{V_A} M_B N_B \frac{d\mu_B}{dN_B} = \frac{V_m}{V_A} M_B \frac{d\mu_B}{d\ln N_B} \tag{2.55}$$

We also know that,

$$\mu_B = \mu_B^o(T,P) + RT \ln a_B = \mu_B^o(T,P) + RT(\ln N_B + \ln \gamma_B) \tag{2.56}$$

Where, the activity of the species B is $a_B = \gamma_B N_B$, $\gamma_B$ is the activity coefficient of B and R is the gas constant. Hence, from Eqn 2.53, Eqn 2.52 can be rewritten as

$$D_B = \frac{V_m}{V_A} M_B RT \frac{d\ln a_B}{d\ln N_B} = \frac{V_m}{V_A} M_B RT \left(1 + \frac{d\ln \gamma_B}{d\ln N_B}\right) \tag{2.57}$$

In case of diffusion of an infinitely thin layer of radioactive tracer of B, the volume terms and the non ideality term can be ignored and thus Eqn 2.54 can be written as,

$$D_B^* = M_B^* RT \tag{2.58}$$

And this relation is called the Nernst-Einstein Equation.

Thus form Eqn 2.55, Eqn 2.54 can be rewritten as,

$$D_B = D_B^* \frac{V_m}{V_A} \frac{d\ln a_B}{d\ln N_B} = D_B^* \frac{V_m}{V_A} \left(1 + \frac{d\ln \gamma_B}{d\ln N_B}\right) \tag{2.59a}$$

Similarly, we can write Equation for species A as,

$$D_A = D_A^* \frac{V_m}{V_B} \frac{d\ln a_A}{d\ln N_A} = D_A^* \frac{V_m}{V_B} \left(1 + \frac{d\ln \gamma_A}{d\ln N_A}\right) \tag{2.59b}$$

Now, as per the Gibbs-Duhem Relation,

$$\frac{d\ln a_A}{d\ln N_A} = \frac{d\ln a_B}{d\ln N_B} \tag{2.60}$$

Now, from Eqns 2.56 and 2.57, the Darken's relation for interdiffusion coefficient (Eqn 2.38), can be rewritten as,

$$\tilde{D} = \left(N_A D_B^* + N_B D_A^*\right)\left(\frac{d\ln a_B}{d\ln N_B}\right) \tag{2.61}$$

This relation was first time proposed by Darken [16]. The ratio of diffusivity can be expressed by,

$$\frac{D_A}{D_B} = \frac{V_A}{V_B} \frac{D_A^*}{D_B^*} \tag{2.62}$$

Eqn 2.58 changes a little when we consider growth of a line compound or a compound with a narrow homogeneity range. Now if we consider formation of β phase as shown in Fig 2.10.

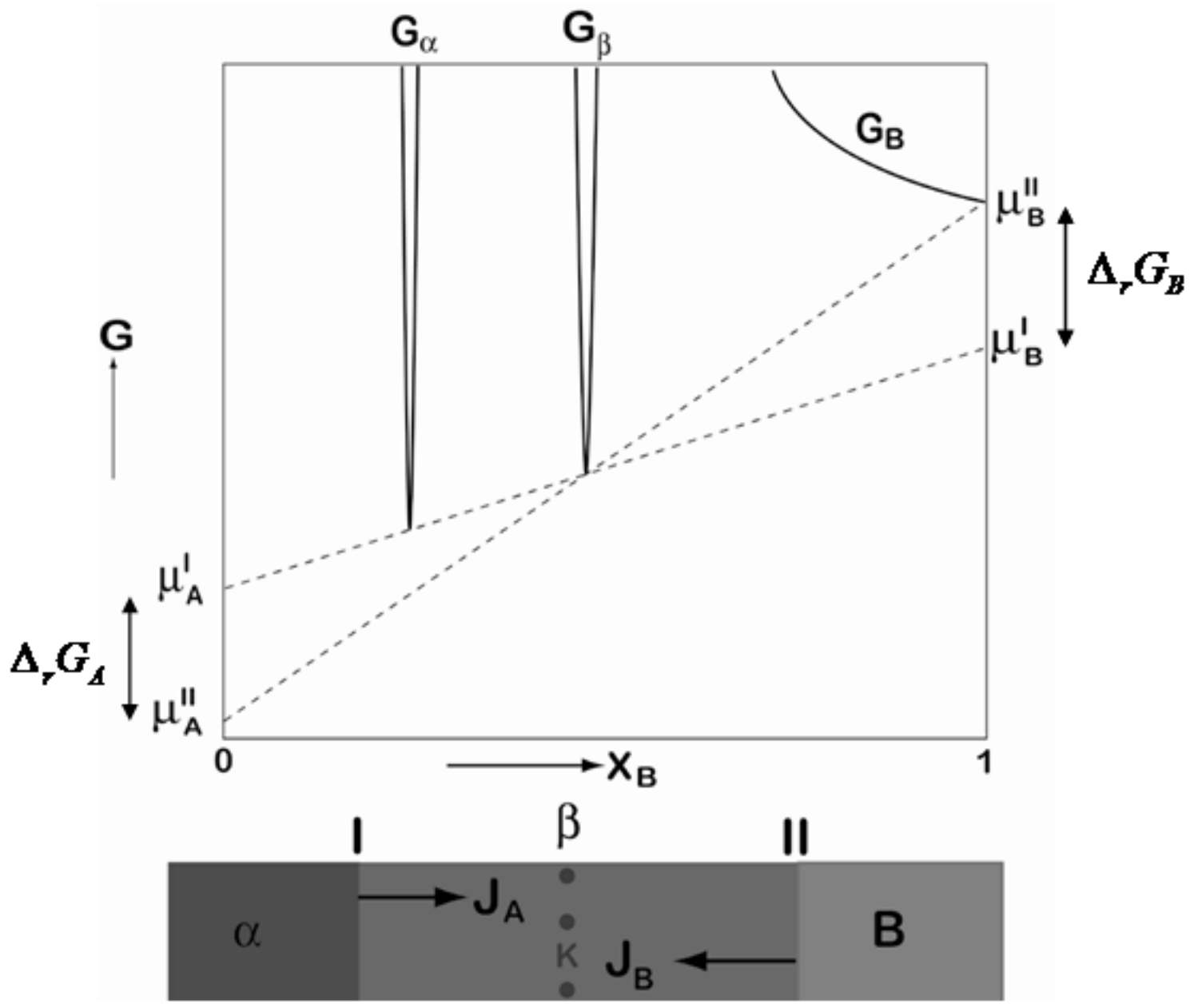

Fig 2.10: The driving force calculation for the growth of the phase β phase formed in the α/B diffusion couple is shown schematically

By definition of the integrated diffusion coefficient of phase β as shown in previous section, we also can write,

$$\tilde{D}_{\text{int}}^{\beta} = \int_{N_{\beta I}}^{N_{\beta II}} \tilde{D} dN_B \tag{2.63}$$

where, $N_{\beta I}$ and $N_{\beta II}$ are the compositions of β phase at interface I and II such that $N_{\beta I} \le N_{\beta} \le N_{\beta II}$.

Now, from Eqn 2.58, Eqn 2.60 can be rewritten as,

$$\tilde{D}^{\beta}_{\text{int}} = \int_{I}^{II} \left(N_A D_B^* + N_B D_A^*\right) N_B \partial \ln a_B = \left(N_A D_B^* + N_B D_A^*\right) N_B \left[\ln a_B^{II} - \ln a_B^{I}\right]$$

(2.64)

Here, the energy per mole of B for dissociation at interface II can be represented by the Gibbs Energy, $\Delta G^o = -RT \ln a_B^{II}$ where, $\mu_B = \mu_B^o + RT \ln a_B$. Similarly, energy per mole of B for dissociation at interface I to form β can be represented by $\Delta G^o = -RT \ln a_B^{I}$. So the total energy to form β by diffusion of one mole of B (by dissociation or reaction) can be written as $\Delta_r G_B^o = -RT\left(\ln a_B^{II} - \ln a_B^{I}\right)$. So Eqn 2.61 can be written as,

$$\tilde{D}^{\beta}_{\text{int}} = -\left(N_A D_B^* + N_B D_A^*\right) \frac{N_B \Delta_r G_B^o}{RT} \tag{2.65}$$

It should be noted here that, $N_B \Delta_r G_B^o = N_A \Delta_r G_A^o$. Also, if β phase forms between pure end members, that is, ( $N_B^- = 0$ and $N_B^+ = 1$ ), Eqn 2.62 can also be written as,

$$\tilde{D}^{\beta}_{\text{int}} = -\left(N_A D_B^* + N_B D_A^*\right) \frac{\Delta_f G_\beta^o}{RT} \tag{2.66}$$

where, $\Delta_f G_\beta^o$ is the Gibbs free energy for formation per mole of particles for the β phase.

## 2.7 Vacancy Wind Effect and Manning's Correction

As the Kirkendall effect was established, Seitz [18] and Bardeen [19] identified that the original Darken equations are only approximation. If we look at the diffusion process at atomic level, substitutional interdiffusion occurs by vacancy exchange mechanism. Darken's equation assumes that vacancy concentration is in thermal equilibrium. From the Kirkendall effect we know that vacancies are created on one side and annihilated on the other side, resulting in a net flux of vacancies to maintain the local equilibrium. This also means there are sufficient sources and sinks of vacancy in the diffusion couple. Manning [20, 21] proposed that the net flux of vacancy causes an effect which he called *vacancy wind effect* which enhances the intrinsic diffusivity of faster diffusing species and lowers that of slower species. He proposed a *correction factor*, $W_i$, to take care of the vacancy wind effect.

The intrinsic diffusion coefficient can then be related to the tracer diffusion coefficient as

$$D_A = \frac{V_m}{V_B} D_A^* \Theta (1 + W_A) \tag{2.67a}$$

$$D_B = \frac{V_m}{V_A} D_B^* \Theta (1 - W_B) \tag{2.67b}$$

where the relation of vacancy wind factor $W_i = \frac{2N_i (D_A^* - D_B^*)}{M_o (N_A D_A^* + N_B D_B^*)}$, $M_o$ is a constant, dependent on the crystal structure of the system, $N_i$ is the mole fraction of species *i* and $\Theta$ is the thermodynamic factor $\partial \ln a_A / \partial \ln N_A = \partial \ln a_B / \partial \ln N_B$. Further the relation for interdiffusion coefficient in terms of tracer diffusion coefficient of elements can be written as

$$\tilde{D} = (N_A D_B^* + N_B D_A^*) \; \Theta \; W_{AB} \tag{2.68}$$

where, $$W_{AB} = 1 + \frac{2N_A N_B (D_A^* - D_B^*)^2}{M_o (N_A D_B^* + N_B D_A^*)(N_A D_A^* + N_B D_B^*)} \tag{2.69}$$

However, in general in most of the systems the contribution of vacancy wind effect is found to be negligible [22].

**Further read**

# Chapter 3

# Experimental procedure

## 3.1 Experimental Technique

The most commonly used procedure to study solid state diffusion is diffusion couple technique. In this study the diffusion couple technique was followed as the whole study is focused on solid state diffusion in bulk materials. In this method two different materials are ground and polished flat, coupled together, clamped and annealed at the temperature of interest in the vacuum furnace. The cross sections of the diffusion bonded samples were analyzed in microscope to identify and measure the phases grown in the interdiffusion zone.

## 3.2 Starting materials

Pure materials as well as alloys are used in this study. The specifications of pure materials used are shown in Table 3.1. Pure Zr ingot and other alloys are prepared using the vacuum arc melting furnace. The alloys required for the experiments are prepared using the vacuum arc melting furnace. First, a desired amount of material is cleaned with acetone, weighed and placed on the copper base of the arc melting furnace. The furnace is evacuated to a pressure of ~$10^{-7}$ kPa. Then, the furnace is purged with Ar gas to eliminate the presence of oxygen. Again, the furnace is evacuated and filled with Ar gas. Finally, the arc is produced by striking the tungsten electrode on the copper base and placing it appropriately on the samples. Current is adjusted so that all the metals in the alloy melt properly. In order to ensure better mixing, the samples are melted for four to five times turning them upside down. Further, the ingots are homogenized just below the melting point of the alloys in a vacuum tube furnace for 2-4 days. The ingots are checked for a composition using the energy dispersive X-ray spectroscopy (EDS) technique attached to the electron microscope and/or EPMA. Ingots prepared are cut into 1-2 mm thick slices using the electro-discharge machine (EDM) cutting.

| **Material** | **Purity (%)** | **Supplier** |
|---|---|---|
| Ti | 99.7 | Alfa Aesar |
| Zr | 99.8 | Alfa Aesar |
| Hf | 99.95 | Alfa Aesar |
| Ta | 99.95 | Sigma Aldrich |
| W | 99.95 | Alfa Aesar |
| Re | 99.98 | Sigma Aldrich |
| TaN | 99.5 | Cathay Advanced Materials |
| $SiO_2$ | 99.995 | Angstrom Sciences Inc. |
| Ni | 99.98 | Alfa Aesar |
| Pt | 99.99 | Arora Matthey |
| Si (100) single crystal wafer | 99.999 | Semiconductor wafer Inc. |
| Mo | 99.95 | Alfa Aesar |
| Cr | 99.997 | Alfa Aesar |

Table 3.1: Specifications of the pure elements used for the present study.

## 3.3 Analysis of the diffusion couples

1 mm thick foils of refractory metal element and 0.7 mm thick (1 0 0) oriented one side polished silicon wafers were used to produce the diffusion couples. Samples were cut into pieces with the dimension of approximately 5 mm × 5 mm and the standard grinding and polishing method was followed. Immediately after that, the couples' halves were bonded in a special fixture and annealed in the temperature range of our interest in vacuum (~$10^{-4}$ Pa) for a fixed annealing time. The annealing temperature was controlled within ±5 °C. In most of the M/Si diffusion couples, the thickness of $M_5Si_3$ was much smaller compared to $MSi_2$. Even the Kirkendall marker plane was also present in the $MSi_2$ phase and the relative mobilities of the species are possible to calculate by the relation developed by F.J.J. van Loo [1] in this phase only. Therefore, to grow the $M_5Si_3$ phase with reasonable thickness and for the determination of the relative mobilities of the species in this phase, incremental diffusion couple experiments were conducted [2]. First M/Si couples were annealed at desired temperature, in which mainly the $MSi_2$ phase grew with very small thickness of the $M_5Si_3$ phase. Following, Si was separated from the couple. It was done by hitting lightly since the Si/$MSi_2$ interface was very weak. Many a times, it was separated during cooling because of the difference in the thermal coefficient of expansion with $MSi_2$. This does not damage the relatively stronger M/$M_5Si_3$ interface. After removal of Si, couples were annealed in the temperature range of interest such that mainly $M_5Si_3$ could grow in the interdiffusion zone.

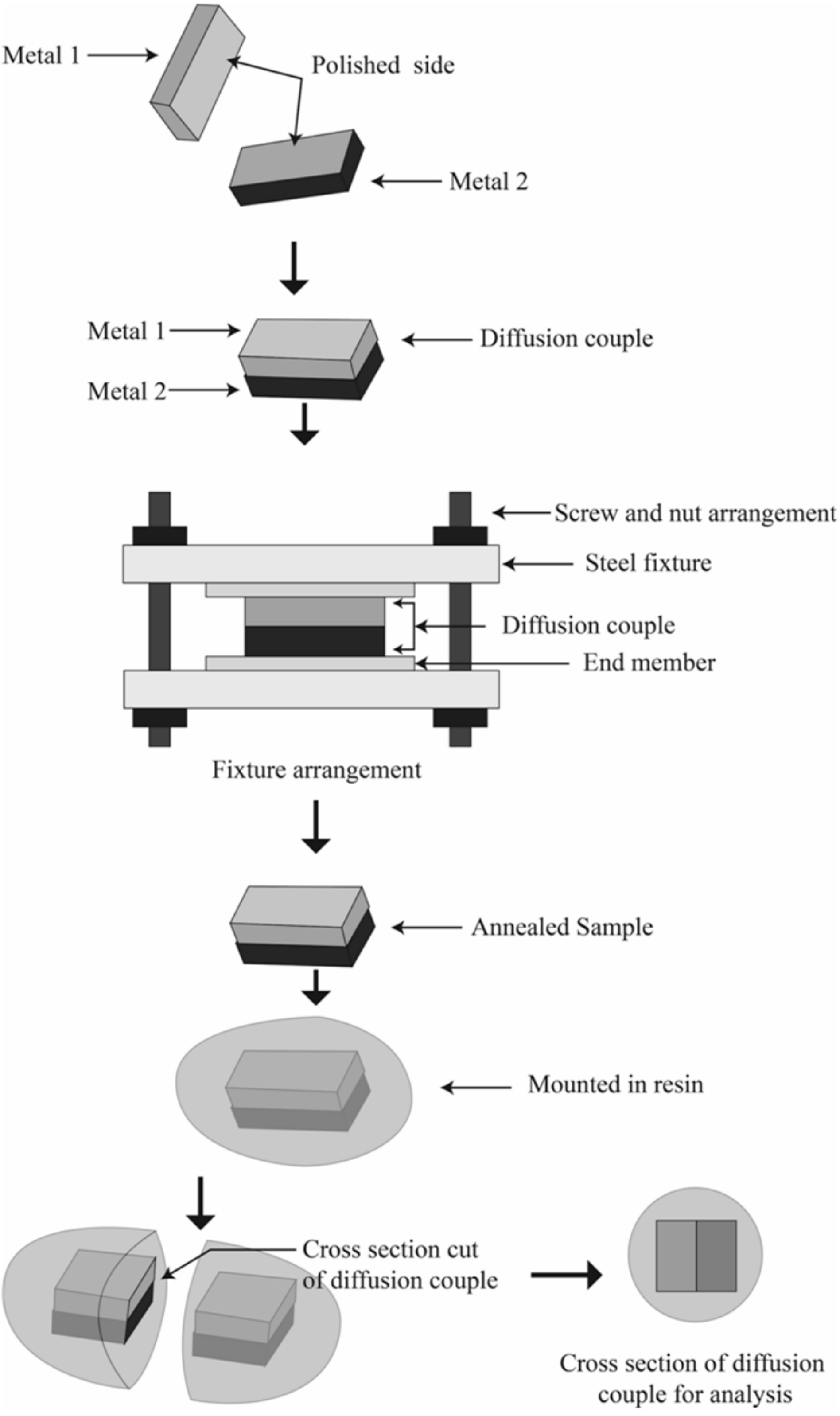

Fig. 3.1: Schematic diagram of the sample preparation method

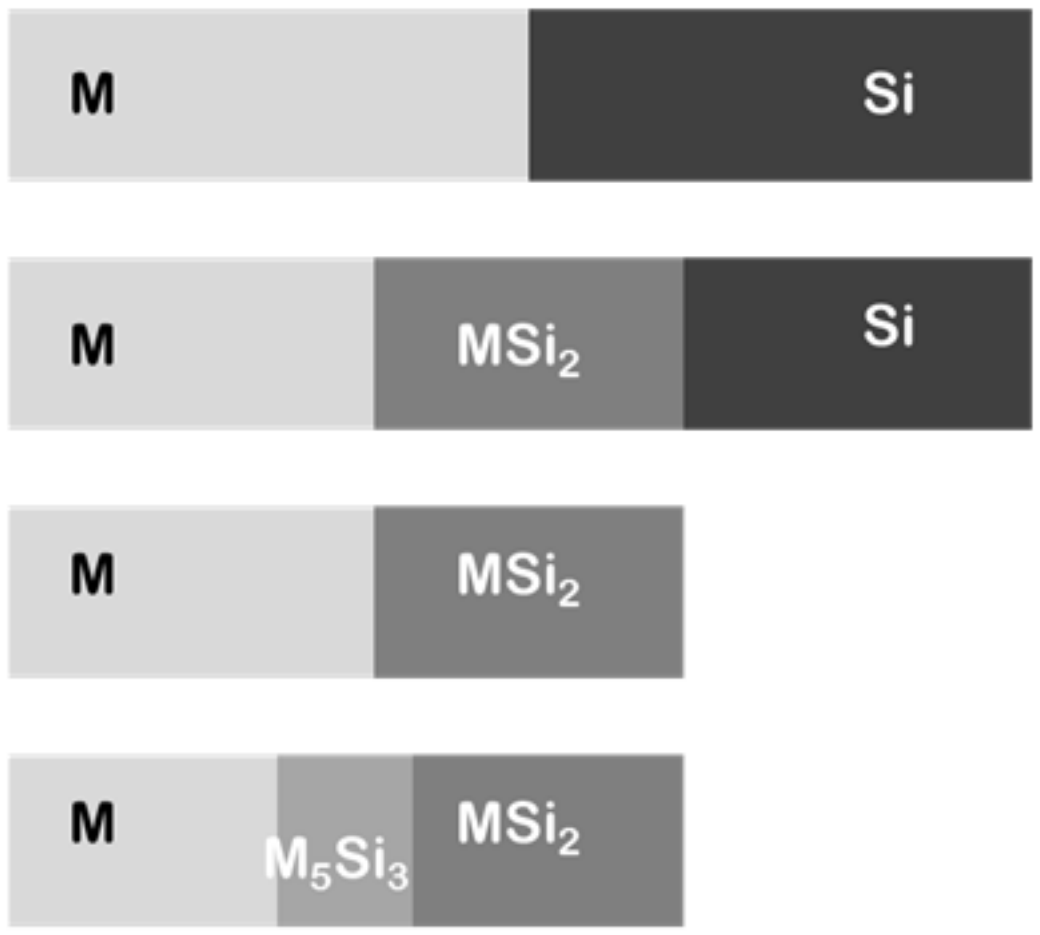

Fig 3.2: Schematic diagram of incremental diffusion couple used in the Ta/Si and W/Si system.

The experimental temperature range for the different systems is listed in the table 3.2.

| Group | Systems | Annealing Temperatures | Reasonable phase formation |
|---|---|---|---|
| | Ti/Si | 1150 – 1250 °C | $TiSi_2$, $TiSi$, $Ti_5Si_4$ |
| IVB | Zr/Si | 1150 – 1250 °C | $ZrSi_2$ |
| | Hf/Si | 1175 – 1275 °C | $HfSi_2$ |
| VB | Ta/Si | 1200 – 1275 °C | $TaSi_2$ |
| | Ta/$TaSi_2$ (incremental) | 1200 – 1350 °C | $Ta_5Si_3$ |
| VIB | W/Si | 1150 – 1275 °C | $WSi_2$ |
| | W/$WSi_2$ (incremental) | 1150 – 1350 °C | $W_5Si_3$ |

Table 3.2: The range of annealing temperature for the systems in this study.

This incremental diffusion couple technique is shown in the Fig. 3.2. After the experiments, samples were mounted in resin and cross-sectioned by a slow-speed diamond saw. Following, the final polishing was done with 0.04 μm colloidal silica.

Images were captured in a scanning electron microscope (SEM) and the compositions of the phases were measured by an electron probe microanalyzer (EPMA).

### 3.4 The Kirkendall marker position

The interdiffusion coefficient which can be calculated from the concentration profile is only the average diffusivity of the species or the measure of the intermixing of the species involved. It does not give any information about the mobilities of the individual species involved in the diffusion process. Kirkendall [3] first experimentally showed this behavior and latter Darken [4] described that in a solid-solid interdiffusion process the two independent component fluxes in opposite direction need not necessarily be equal. Because of this inequality, net mass flow in the interdiffusion process causes the diffusion couple to shrink on one side and swell on the other side. This process can be made to visible in the microstructure by the presence of the inert marker in the interdiffusion zone. These markers move towards the faster diffusing species.

Sometime scratches, pores or voids present in the interdiffusion zone act as a marker, which always not possible to detect or sometimes can draw the wrong conclusion. Thefore inclusion of the inert marker in the in the diffusion coupe during the experiment is a common practice. These markers should not react with neither of the diffusion couple nor the phases they form. Generally stable oxides $ThO_2$, $TiO_2$, $Y_2O_3$ are used as inert markersOne should be careful about the particle size of the inert markers. Too large particle can hinder the diffusion process on other hand too small particles can be dragged by the moving grain boundary [4, 5]. Further the Kirkendall marker plane in the microstructure can be located by revealing the grain morphology. Interestingly, sometimes duplex grain morphology develops inside the diffusion grown compound layers. The product layer which has the same chemical compound looks like it consists of two different sublayers marked by distinct boundary. These two sublayers differ by size, shape or orientation of the grains and apparently this boundary is called the Kirkendall marker plane [7]. Further the mobilities of the species and growth of the phases is studied by A. Paul [8, 9].

To reveal grain morphology, polished samples were etched by an acid mixture of $HNO_3$ and HF. Sometimes, electron back-scattered diffraction (EBSD) images were taken to reveal the grain morphology. In the Fig 3.3 schematically is shown the duplex grain morphology and the Kkirkendal marker position.

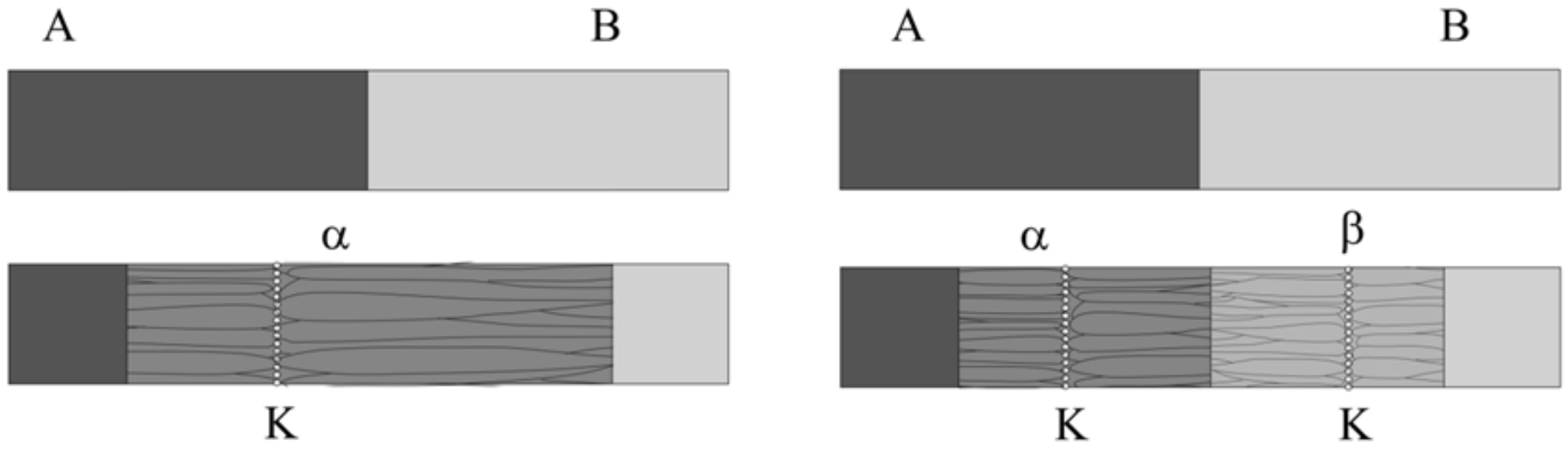

Fig. 3.3: Schematically shown how the Kirkendall marker plane is detected in the single phase and double phase layer.

# Chapter 4.1

# Reactive diffusion in the Ti-Si system and the significance of the parabolic growth constant

## 4.1.1 Introduction

Ti-Si is one of the most important refractory metal-silicon systems used extensively in very large scale integrated (VLSI) industry. The C54 phase $TiSi_2$ is used as a contact because of its low resistivity and excellent thermal stability. It is grown by depositing a thin film of Ti on Si. Latter, self-aligned silicide (salicide) is grown at the interface by a diffusion controlled process. It is reported that the metastable orthorhombic base centered C49 phase grows before transformation to the desirable orthorhombic face centered C54 phase with much lower resistivity [1-3]. To date, many diffusion studies have been conducted in thin film conditions. Hung et al. [4] studied the growth kinetics of $TiSi_2$ in the temperature range of 475-550 °C after depositing Ti on amorphous and single crystal silicon. They found a parabolic growth of the phase with the activation energy of 174±10 kJ/mol. Bentini et al. [5] reported that the growth of the phase(s) depends on the annealing temperature and time. They found that only the $TiSi_2$ phase grows at 550 °C, whereas above 600 °C a TiSi phase was also observed. With increasing annealing time, only the $TiSi_2$ phase was found. They also reported the parabolic growth of $TiSi_2$ and similar activation energy to that found by Hung et al. [4] based on the experiments in the

---

This chapter is written based on

S. Roy, S. Divinski and A. Paul, Reactive diffusion in the Ti –Si system and the significance of the parabolic growth constant, Philosophical Magazine, 94 92014) 683 – 699.

range of 550-650 °C. Additionally, they predicted that the layer grows mainly by the diffusion of Si. Murarka and Fraser [6] did experiments in the temperature range of 400–1100 °C and reported the growth of two phases, TiSi and $TiSi_2$ when the annealing temperature was less than 700 °C. Above this, only the $TiSi_2$ phase grew. Révész et al. [7] found that this phase grows parabolically with time even on $SiO_2$.

A tracer diffusion study of different elements was also conducted in the $TiSi_2$ phase. Importantly, the activation energy of the tracer diffusion of Ge, which is supposed to mimic the diffusion behaviour of Si, was reported to be 290 kJ/mol [8]. The similarity of the tracer diffusion properties of Si and Ge in silicides has been experimentally confirmed for the $MoSi_2$ phase [9, 10]. The significant difference of the activation energies of Ge tracer diffusion and of the growth rate of the $TiSi_2$ phase indicates an existing inconsistency of the diffusion database. As it will be shown in this manuscript for the case of a single phase growth in a couple, the activation energy of the phase growth has to correspond to that of the interdiffusion coefficients, which in the case of the $TiSi_2$ phase is determined by diffusion of Si. Therefore, further study is required to investigate the growth mechanism of the phases and the diffusion mechanism of the components in this system.

Because of their relevance, most recent studies have been conducted in thin film conditions. However, this might not be suitable to analyze the growth mechanism of the phases. It is known that the stress generated during the deposition of the thin film affects the growth of the phase. A metastable phase might grow in the beginning before the growth of the equilibrium phases. Sequential growth of the phases is also found very frequently [11]. On the other hand, the simultaneous growth of the phases is found in a solid diffusion couple although one or more phases might grow with very small thickness because of a much slower growth rate. A study of solid diffusion couples will highlight the diffusion mechanism of the components based on the determination of the activation energy and the relative mobilities of the components. This is an indirect but reliable technique to study the relative mobilities of the components using inert particles as the Kirkendall markers [12-14]. It should be noted here, that, frequently, growth and diffusion mechanisms are discussed based on the activation energies calculated from the parabolic growth constants. We shall show that this can be done only when a single phase layer grows (within a narrow homogeneity range) in the interdiffusion zone. However, in multiphase growth, this will lead to a wrong calculation, since the parabolic growth constant is not a material constant but depends on end member compositions. One should calculate activation energies based on the calculation of diffusion parameters, which are material constants. To the best of our knowledge, to date, only one diffusion study in this system is available in the bulk conditions. Cockeram and Wang [15] studied multiphase growth in the temperature range of 700-1150 °C. They calculated the parabolic growth constant and then the average interdiffusion coefficients with the help of the activity data available in the literature. It should be noted here that the diffusion coefficients can be calculated directly from the composition profiles without knowledge of the thermodynamic data [16, 17].

The aim of this study is to conduct interdiffusion studies following the solid diffusion couple technique. The parabolic growth of the phases will be examined for the present conditions. Integrated diffusion coefficients and the activation energies are calculated from the composition profiles. The tracer diffusion coefficient of fast-diffusion component is calculated in the $TiSi_2$ phase and the diffusion mechanism is discussed in view of atomic arrangements in the crystal.

### 4.1.2. Results and discussion

### 4.1.3 Growth of the phases

In the interdiffusion zone of a bulk diffusion couple, all the equilibrium phases are expected to grow with different thicknesses depending on their growth rates. According to the phase diagram [18], as shown in Fig. 4.1.1, four phases $TiSi_2$, TiSi, $Ti_5Si_4$ and $Ti_5Si_3$ with a narrow homogeneity range should grow in the temperature range of our interest (1150–1250 °C). Additionally, at temperatures below 1170 °C, $Ti_3Si$ also must be present.

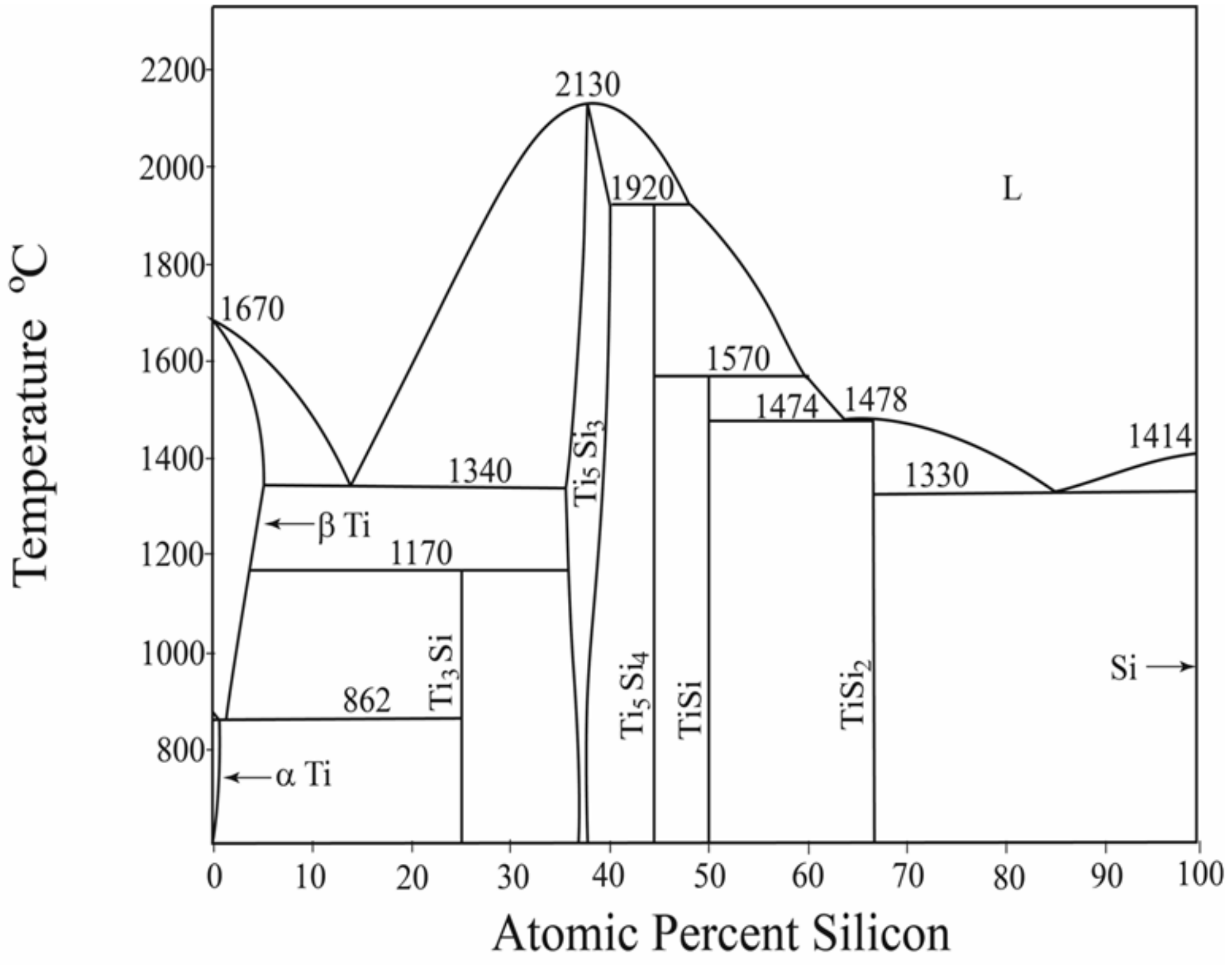

Fig. 4.1.1: Ti - Si phase diagram [18].

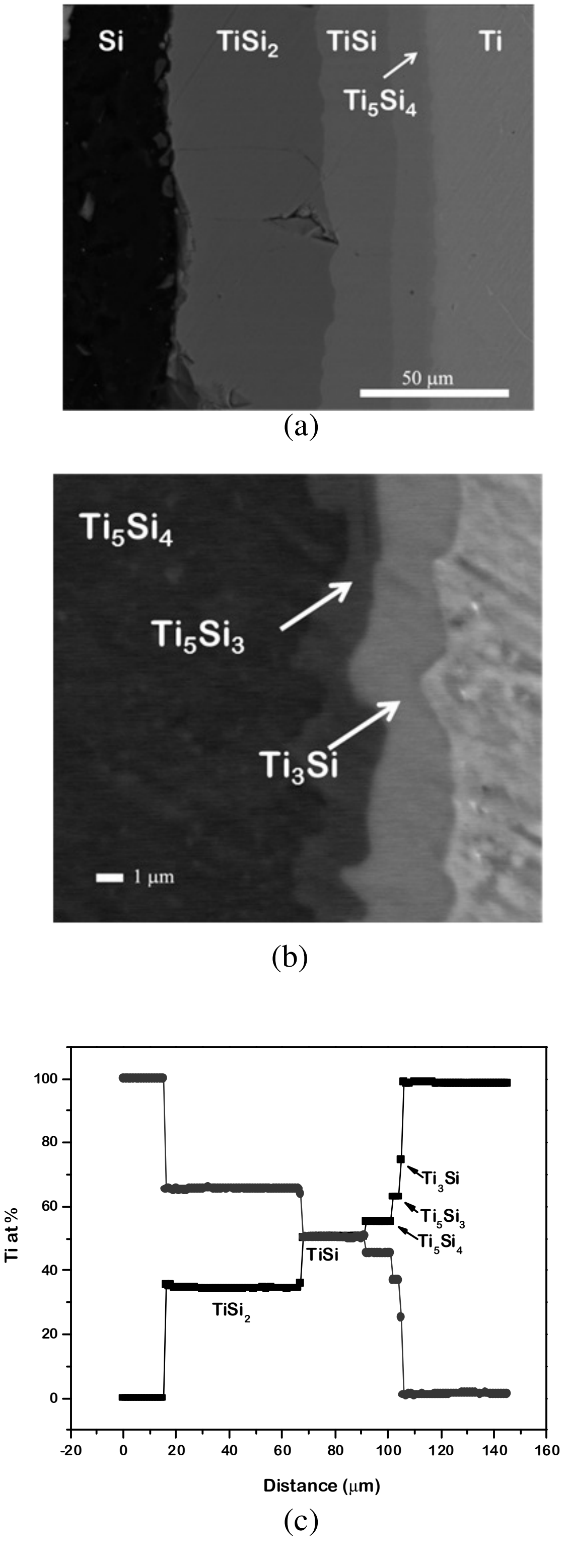

Fig. 4.1.2: (a) Scanning electron micrograph of Ti/Si diffusion couple annealed at 1200 °C for 16 hrs, (b) magnified image of the Ti-rich part showing the presence of $Ti_5Si_3$ and $Ti_3Si$ phases and (c) the composition profile of the interdiffusion zone.

As shown in Fig. 4.1.2a, in the interdiffusion zone grown at 1200 °C, three phases $TiSi_2$, TiSi, $Ti_5Si_4$ are clearly found. A close examination reveals the presence of two other phases, $Ti_5Si_3$ and $Ti_3Si$.

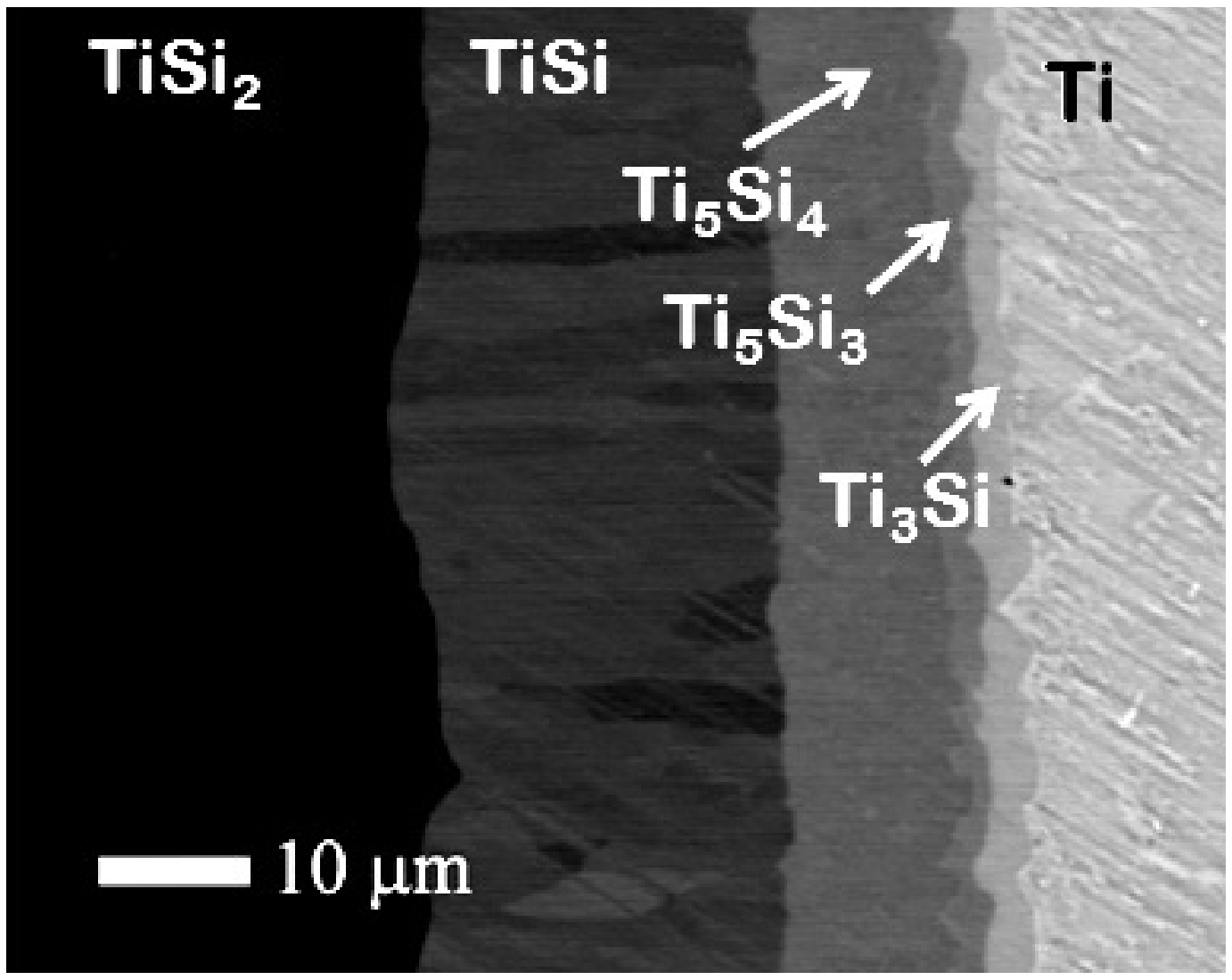

(a)

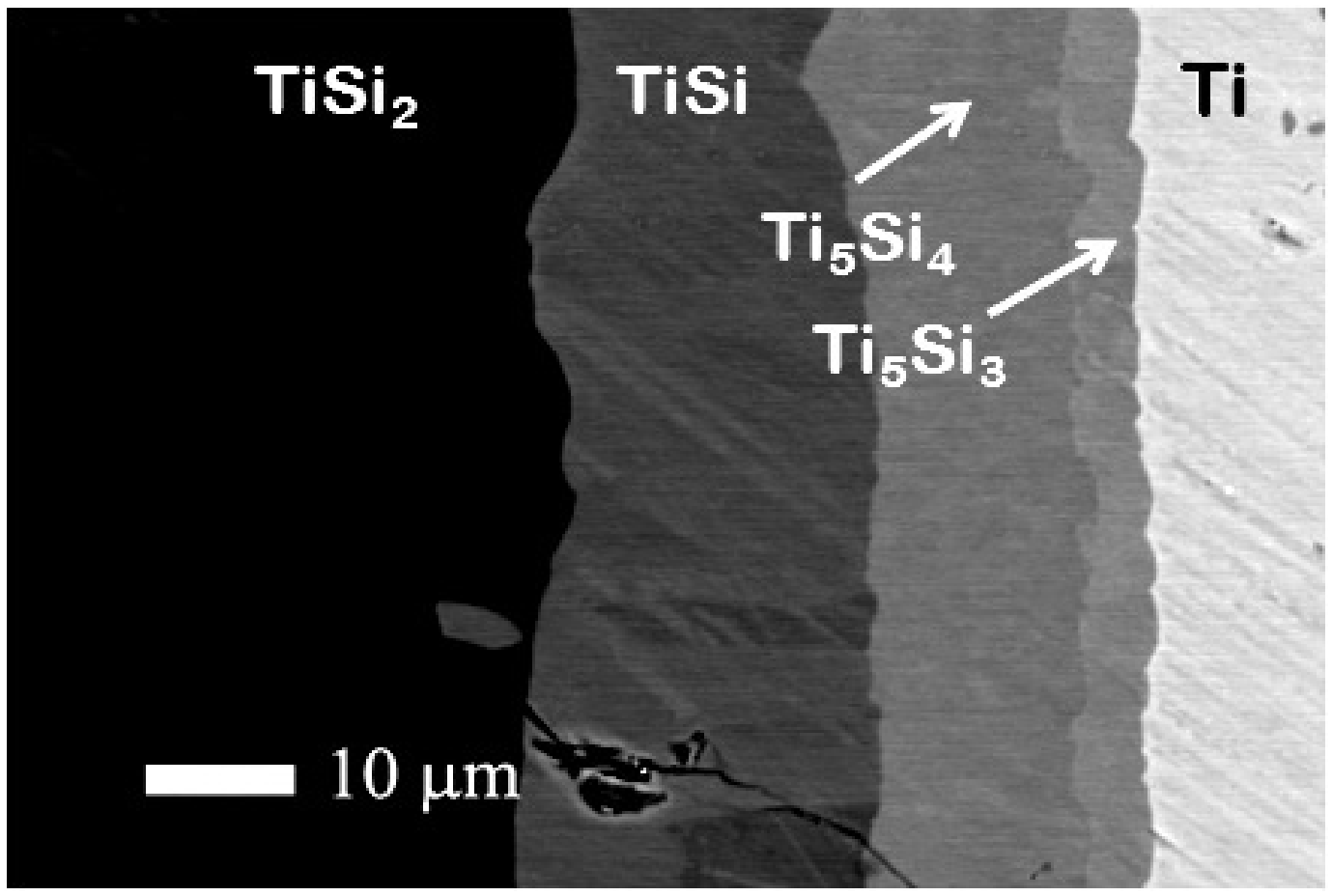

(b)

Fig. 4.1.3: (a) Magnified image of the Ti/Si diffusion couples annealed at (a) 1225 °C for 16 hrs and (b) 1250 °C for 16 hrs showing the presence of Ti-rich phases.

A measured composition profile of the interdiffusion zone, as shown in Fig. 4.1.2c, indicates no dissolution of the species in the end members, although a β Ti(Si) solid solution phase is present. It should be noted here that according to the phase diagram, $Ti_3Si$ should not be present in the interdiffusion zone above 1170 °C. To investigate further, a close examination of the interdiffusion zone near the Ti end member of the couples annealed at 1225 and 1250 °C is done, as shown in Fig. 4.1.3. We found the presence of this phase in the diffusion couple annealed at 1225 °C. However, it is not present at 1250 °C. Therefore, the maximum temperature up to which this phase is stable lies between 1225–1250 °C instead of 1170 °C as reported in the phase diagram. This indicates the need to revisit the phase diagram.

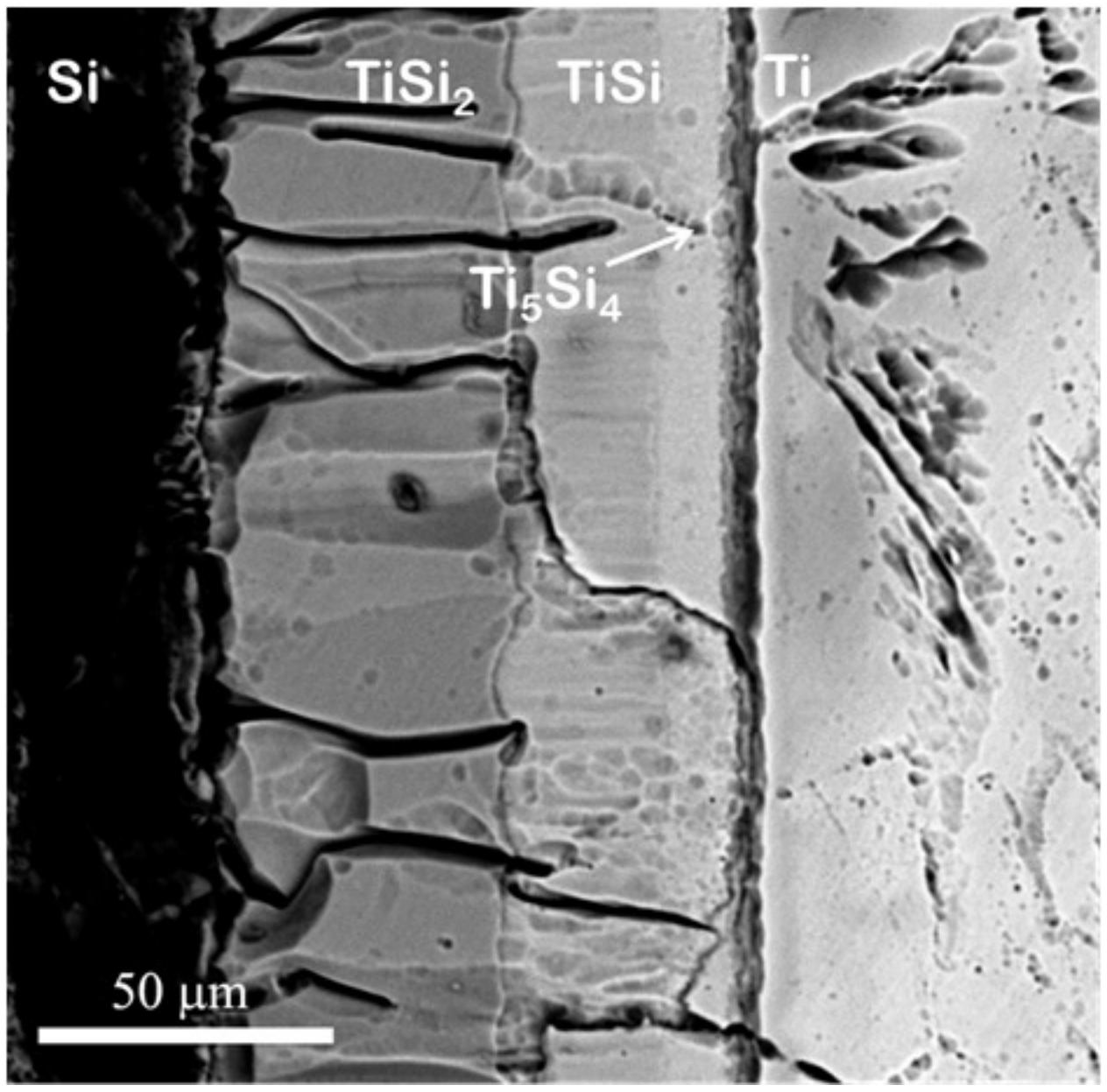

Fig. 4.1.4: Etched micrograph annealed at 1200 °C for 16 hrs showing grains covering the whole phase layers in both the $TiSi_2$ and TiSi phases.

The grain morphology, as revealed by etching, indicates the relative mobilities of the species [19-22]. In general, the Kirkendall marker planes are present at the Si/$MSi_2$ interface or inside the $MSi_2$ phase in refractory metal (M)−silicon systems [23, 24]. Since Si is the fast diffusing component in most disilicides, uniform grain morphology in the $TiSi_2$ phase, as shown in Fig. 4.1.4 substantiates that the Kirkendall marker plane must be present at the Si/$TiSi_2$ interphase interface. Similar behaviour was found in the V-Si and Mo-Si systems. Further, TiSi also has uniform grain morphology. Other phases might also have uniform morphology, but this could not be detected.

The resulting grain morphology is illustrated in Fig. 4.1.5 by a schematic diagram of an imaginary diffusion couple. Let us consider two phases, $\alpha$ and $\beta$, in the interdiffusion zone. The discussion is though applicable to any number of phases. The detailed phenomenological treatment can be found elsewhere [19, 25]. When inert particles are used as the Kirkendall markers, depending on the relative components' mobilities in different phases, three situations might arise. As explained in the diagram, the phases grow via several processes. The element A diffuses through $\alpha$ to interface II and reacts with $\beta$ to produce $\alpha$. B dissociates from $\beta$ to produce $\alpha$ at interface II. The same B diffuses to interface I and reacts with A to produce $\alpha$. A similar dissociation reaction process occurs at interfaces II and III. At interface II, A dissociates from $\alpha$ to produce $\beta$. The same A diffuses to interface III to react with B and produce $\beta$. B, from interface III, diffuses through $\beta$ to reach interface II and reacts with $\alpha$ to produce $\beta$.

(a)

(b)

(c)

(d)

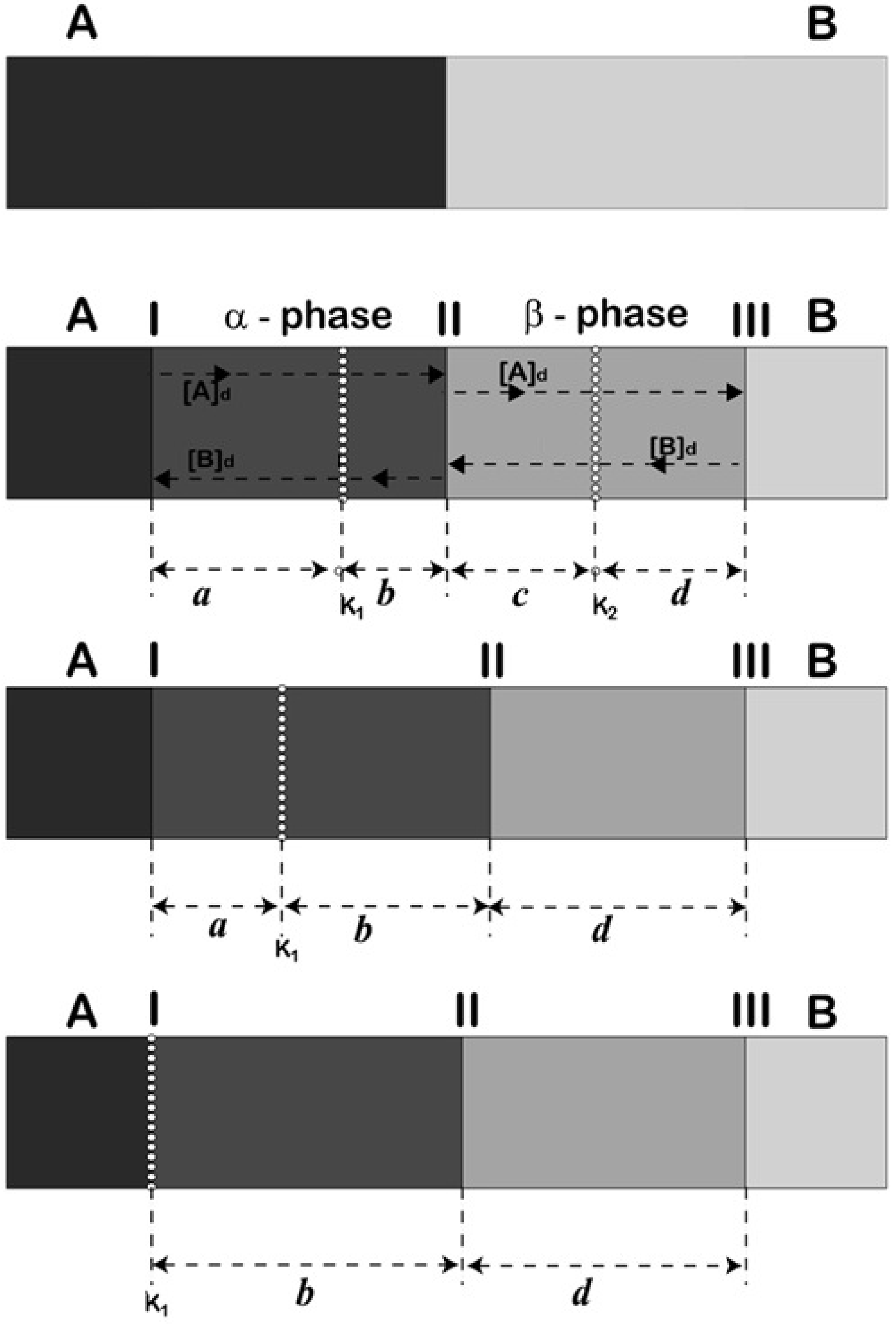

Fig. 4.1.5: (a) Schematic diagram of imaginary diffusion couple of A and B (b) two phases α and β are grown and the Kirkendall marker planes are present in both the phases (c) The Kirkendall marker plane is present only in the α phase and (d) ) the Kirkendall marker plane is present at the A/α interface

Therefore, at interface II, α consumes β for its own growth. At the same time, it is consumed because of the growth of β. The thicknesses of the phase layers depend on the integrated diffusion coefficients of the phases. On the other hand, the thicknesses of the sublayers on either side of the marker plane in a given phase depend on the relative mobilities of the components. For example, in certain conditions, both sublayers *b* and *c* could be positive such that both phases will have a marker plane, as shown in Fig. 4.1.5b. This was found in the Co-Si system [19]. There could be a situation in a particular system such that *b* is positive; however, *c* is negative. This means that even after the consumption of α by β at interface II, sublayer *b* can grow. On the other hand, from the same interface, β cannot grow because of the high rate of consumption by α. Therefore, the Kirkendall marker plane will be present inside the α phase and no marker plane will be found in the β phase, as shown in Fig. 4.1.5c. This behaviour was found in the Nb-Si [16], Ta-Si [26] and W-Si [27] systems. The marker plane was found in the $MSi_2$ (M = Nb, Ta, W) phase and no marker plane was found in the $M_5Si_3$ phase. Another situation is sketched in Fig. 4.1.5d. The marker plane is found at one end of the interdiffusion zone. In this particular case, it indicates that element A has a diffusion rate many orders of magnitude higher than B in α phase. Since the diffusion of B is negligible, sublayer *a* cannot grow. That means that the whole phase is build by sublayer *b*. This also indicates that there is a very high consumption of β at interface II because of the growth of α. Therefore, not only *c* but a part of β growing from interface III also gets consumed. That means that the β phase will have only the sublayer *d*. As a result, the marker plane will be missing in the β phase. It should be noted here that this does not indicate a much higher diffusion rate of A compared to B in the β phase also [25]. Both the elements might have comparable diffusion rates in this phase; however, both the sublayers could not grow because of the very high growth rate of $\alpha$ at interface II. This situation was found in the V-Si [24], and Mo-Si [23] systems previously, where the marker plane was located at the Si/$MSi_2$ (M = V, Mo) interface. This is also found in the $TiSi_2$ system. This must be the reason for the uniform grain morphology (continuous columnar grain morphology throughout the phase) found also in the TiSi phase along with the $TiSi_2$ phase, as shown in Fig. 4.1.4.

### 4.1.4 The parabolic growth constant and the diffusion parameters

Before any analysis on the diffusion process, the diffusion controlled growth of the phases should be checked by time dependent experiments. Since many articles have already shown the parabolic growth of the $TiSi_2$ phase (although in thin film conditions), it was not necessary to repeat it. Therefore, the temperature dependent experiments were conducted in the beginning for the calculation of the activation energy for diffusion. The thickness of the phase layers are tabulated in Table 4.1.1a. We have considered only three phases, $TiSi_2$, TiSi and $Ti_5Si_4$, since these grow with reasonable thicknesses. As expected, $TiSi_2$ and $Ti_5Si_4$ show the change in layer thickness with a change in temperature. However, interestingly, the variation in the layer thickness of the TiSi phase is insignificant.

There could be two reasons for this behaviour. In an incremental couple, when only one phase layer grows in the interdiffusion zone, it is not affected by any other phases. On the other hand, the growth kinetics of a phase in a multiphase interdiffusion depends on the growth of other neighbouring phases, as already explained in the phenomenological process in Fig. 4.1.5. Ideally, all the phases should grow according to their own diffusion parameters since these are the material constants. It means that the thicknesses of the phase layers are adjusted depending on end member compositions such that the diffusion parameters remain the same. Therefore, there is a possibility that the growth rate of the other phase change with temperature in such a way that the growth rate of TiSi did not vary significantly with temperature.

| **Temperature ($^oC$)** | **Δx $TiSi_2$ (μm)** | **Δx TiSi (μm)** | **Δx $Ti_5Si_4$ (μm)** |
|---|---|---|---|
| 1150 | 39 ± 1.1 | 23 ± 0.5 | 9 ± 0.3 |
| 1175 | 43 ± 0.6 | 24 ± 0.4 | 12 ± 0.3 |
| 1200 | 52 ± 0.8 | 24 ± 0.6 | 14 ± 0.4 |
| 1225 | 60 ± 0.5 | 25 ± 0.6 | 15 ± 0.7 |
| 1250 | 67 ±0.9 | 25 ± 0.7 | 18 ± 0.4 |

(a)

| **Temperature ($^oC$)** | **Time (hrs)** | **Δx $TiSi_2$ (μm)** | **Δx TiSi (μm)** | **Δx $Ti_5Si_4$ (μm)** |
|---|---|---|---|---|
| 1200 | 4 | 34 ± 1.1 | 18 ± 0.7 | 7 ± 0.4 |
| 1200 | 9 | 40 ± 0.6 | 20 ± 0.4 | 9 ± 0.3 |
| 1200 | 16 | 50 ± 0.8 | 24 ± 0.6 | 12 ± 0.4 |
| 1200 | 25 | 56 ± 1 | 28 ± 0.8 | 14 ± 0.3 |
| 1200 | 36 | 65 ± 0.5 | 33 ± 0.9 | 19 ± 0.3 |

(b)

Table 4.1.1 (a) Thicknesses of phases grown at different temperatures after 16 hrs of annealing and (b) Thickness of phases annealed for different times at 1200 $^oC$.

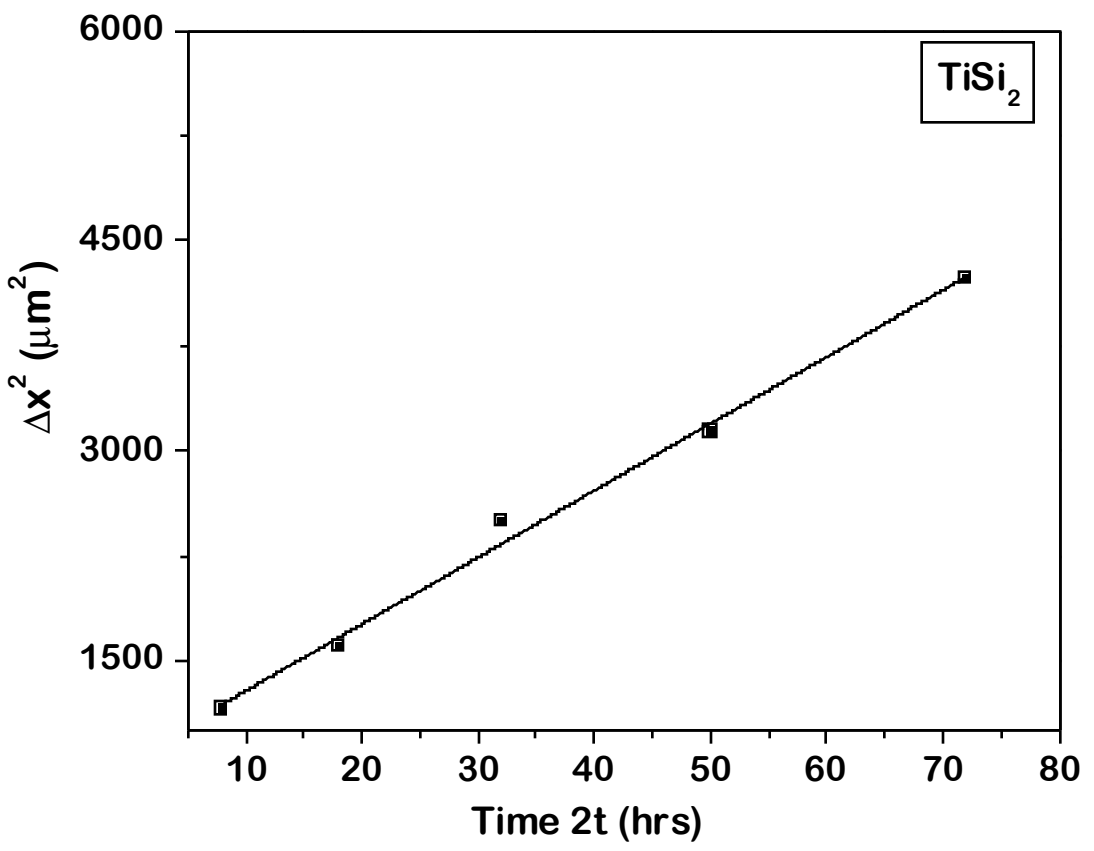

(a)

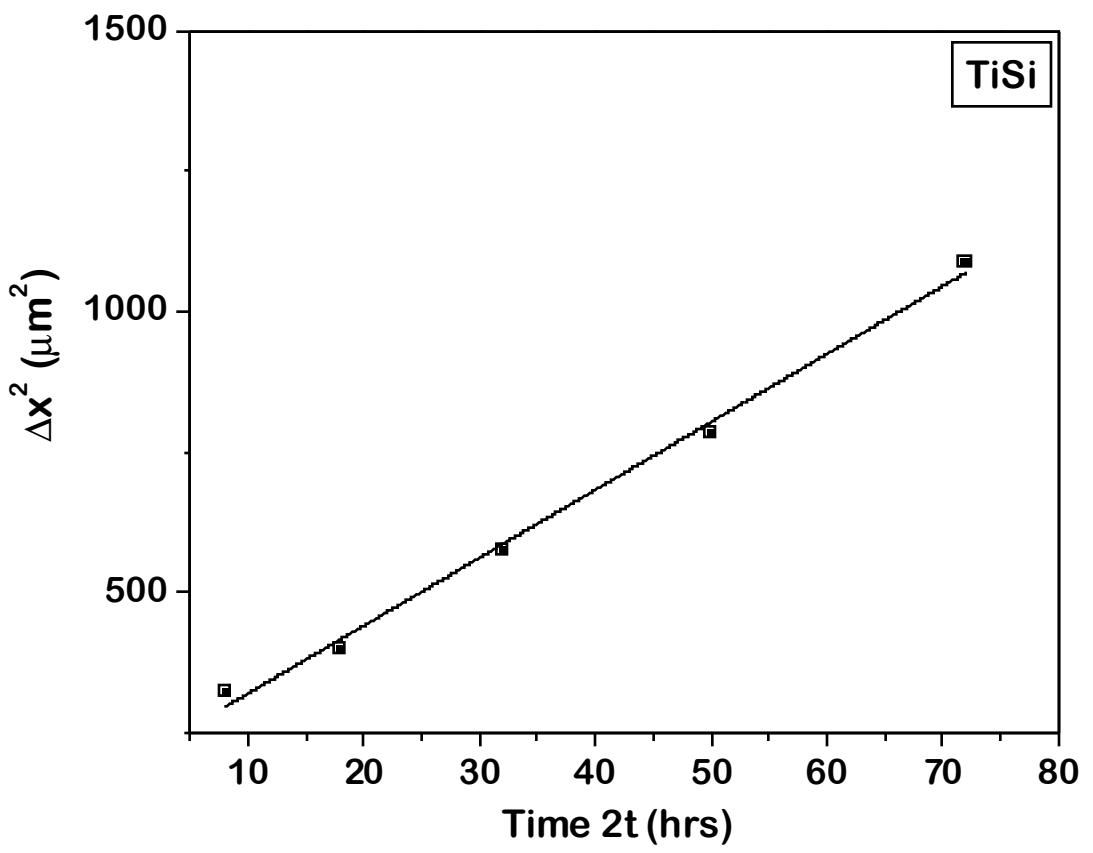

(b)

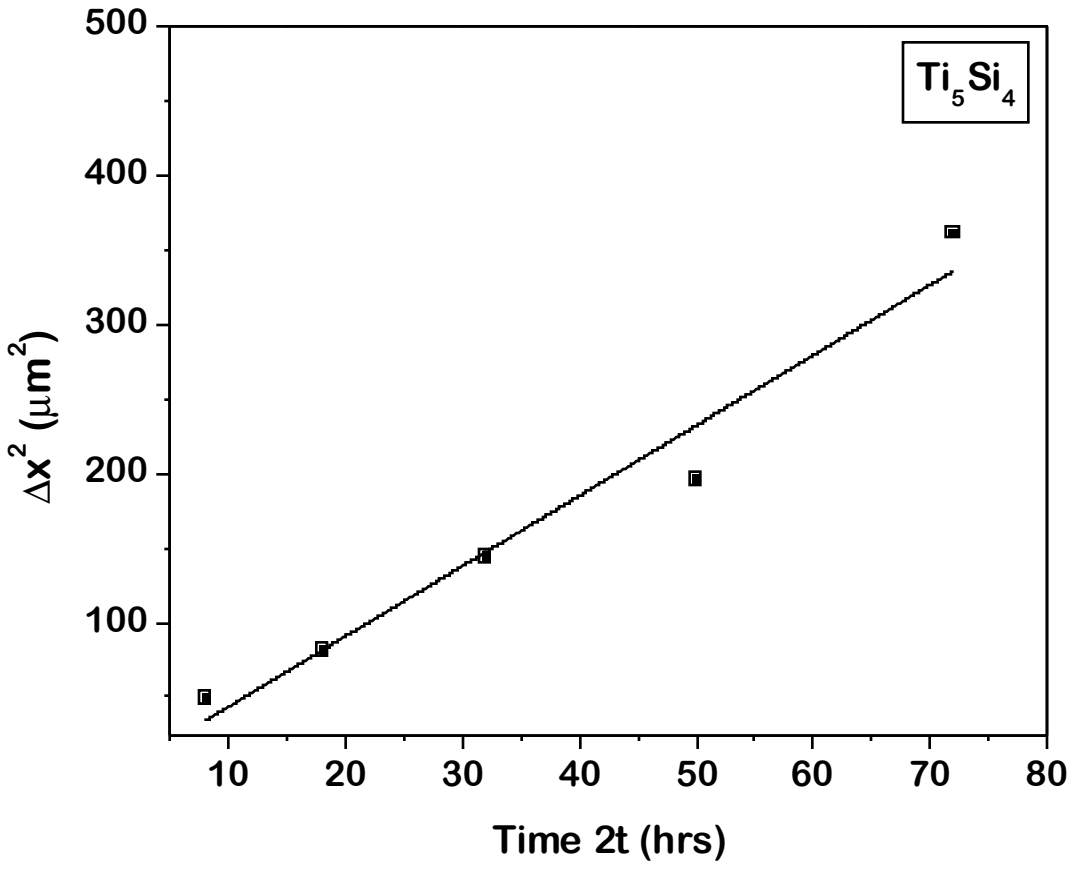

(c)

Fig. 4.1.6: Time dependant experiments at 1200 °C for 4, 9, 16, 25 and 36 hrs are shown by $\Delta x^2$ *vs.* 2t plots for (a) $TiSi_2$ (b) TiSi and (c) $Ti_5Si_4$.

This means also that the activation energy for the growth of this phase is relatively low compared to the other phases. Secondly, it is possible that this phase could not grow ideally, that is, parabolically with time, in a multiphase interdiffusion zone, as was found, e.g., in the Ti-Al system [28]. To investigate this, the time dependent experiments were conducted at 1200 °C. It can be seen clearly in Fig. 4.1.6 that all the three phases grow by a diffusion controlled process since a linear dependence of the phase thickness squares, $\Delta x^2$ vs. time $t$ is found. Therefore it is necessary to estimate the activation energy for diffusion to understand the growth mechanism of the phases .

The parabolic growth constant, for example, for the phase $\beta$ , $k_p^\beta$ is estimated using the relation

$$k_p^\beta = \frac{(\Delta x_\beta)^2}{2t} \tag{4.1.1}$$

where $\Delta x_\beta$ is the phase layer thickness of the phase of interest. $\beta$ , and $t$ is the annealing time. Further, the temperature dependence and the activation energy for the growth can be calculated from the Arrhenius equation expressed as

$$k_p^\beta = k_p^o \exp\left(-\frac{Q}{RT}\right) \tag{4.1.2}$$

where $k_p^o$ is the pre-exponential factor, $Q$ is the activation energy for growth and $T$ is the temperature in K.

The calculated values are plotted in Fig. 4.1.7a for the $TiSi_2$ and $Ti_5Si_4$ phases. Since the thickness variation of TiSi with temperature is very small, the slope is negligible and it makes no sense to estimate the activation energy in this phase from the estimation of the parabolic growth constants.

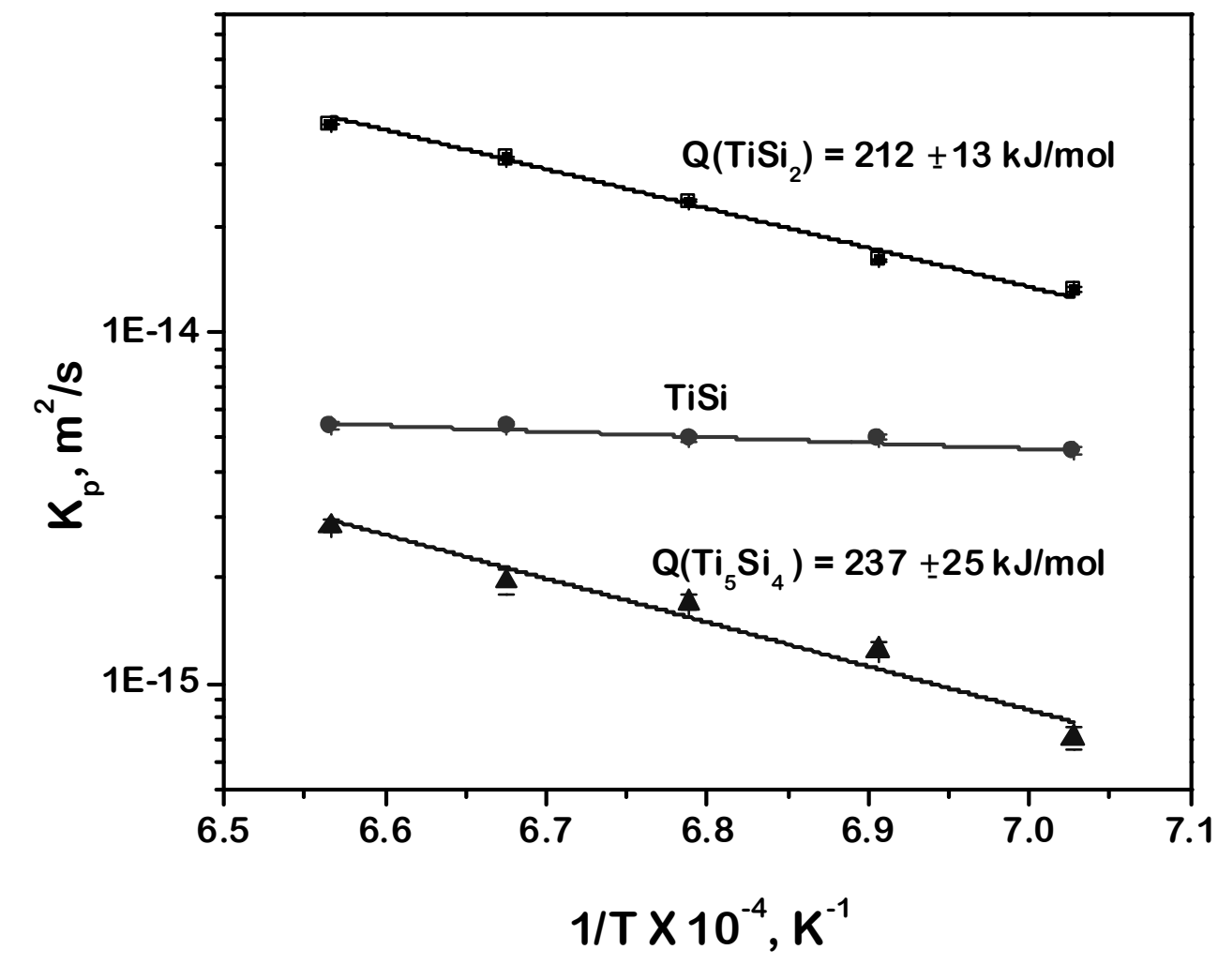

(a)

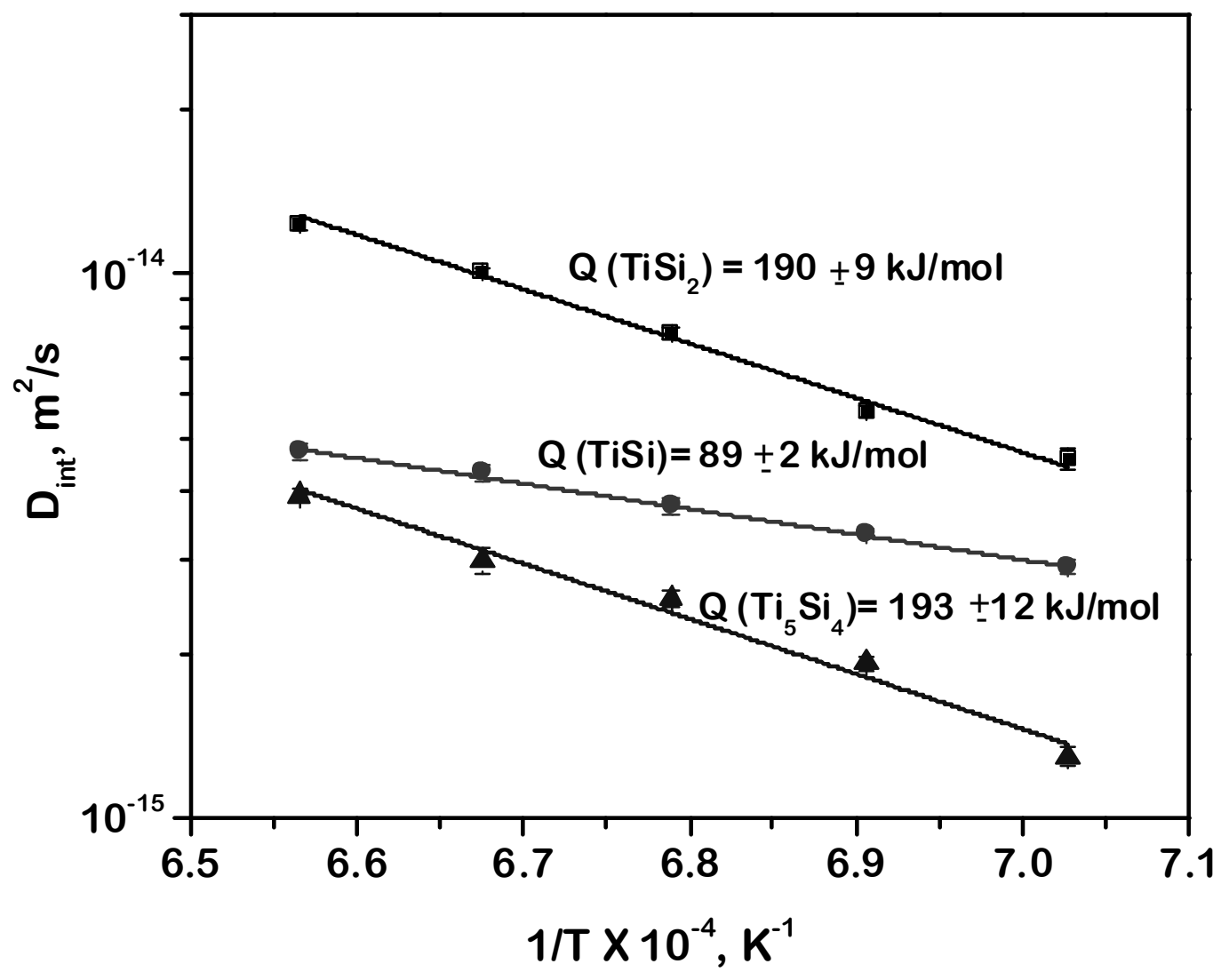

(b)

Fig. 4.1.7: (a) The parabolic growth constants for $TiSi_2$ and $Ti_5Si_4$ and (b) the integrated diffusion coefficients for $TiSi_2$ , TiSi and $Ti_5Si_4$ plotted with respect to Arrhenius equation.

For further clarifications, it is necessary to calculate the diffusion parameters. Since the phases grow within a narrow homogeneity range, we can calculate the integrated diffusion coefficient, $\tilde{D}_{\mathrm{int}}$, let us say for the $\beta$ phase, as [29]

$$\tilde{D}_{\mathrm{int}}^{\beta} = \int_{N_i'}^{N_i''} \tilde{D} dN_i \,, \qquad (4.1.3)$$

where $\tilde{D}^{\beta}$ ($m^2/s$) is the interdiffusion coefficient of the phase β, $N_i$ is the mol fraction of component $i$ and $N_i'$ and $N_i''$ are the mol fractions of the phase boundaries. Since the composition profiles developed in the solid solution phases are negligible, $\tilde{D}_{\mathrm{int}}$ can be calculated directly from the expression

$$\tilde{D}_{\mathrm{int}}^{\beta} = \frac{\left(N_i^{\beta} - N_i^{-}\right)\left(N_i^{+} - N_i^{\beta}\right)}{N_i^{+} - N_i^{-}} \frac{\Delta x_{\beta}^{2}}{2t} + \frac{\Delta x_{\beta}}{2t}\left[\frac{\left(N_i^{+} - N_i^{\beta}\right)\sum\limits_{\nu=2}^{\nu=\beta-1} \frac{V_m^{\beta}}{V_m^{\nu}}\left(N_i^{\nu} - N_i^{-}\right)\Delta x_{\nu} + \left(N_i^{\beta} - N_i^{-}\right)\sum\limits_{\nu=\beta+1}^{\nu=n-1} \frac{V_m^{\beta}}{V_m^{\nu}}\left(N_i^{+} - N_i^{\nu}\right)\Delta x_{\nu}}{N_i^{+} - N_i^{-}}\right] \qquad (4.1.4)$$

where $N_i^{-}$ and $N_i^{+}$ are the mole fractions of the unreacted left and right hand side of the end member, respectively, with respect to element $i$, $V_m^{\nu}$ and $\Delta x_{\nu}$ are the molar volume and the layer thickness of the $\nu$th phase and $t$ is the annealing time. It should be noted here that there is only one integrated diffusion coefficient in a binary system, and it is the same when estimated using the composition profile of any one component. Moreover just by definition, the integrated diffusion coefficient is measured in mol fractions time meter squared per second. It can be seen in the above equation that the first part is directly related to the parabolic growth constant of the phase of interest. The second part is related to the thickness of the other phase layers. Therefore, although there is a very small change in the layer thickness of the TiSi phase, there could be a reasonable change in the integrated diffusion coefficients because of the variation in the layer thickness of the other phases. Molar volumes of the phases are listed in Table 4.1.2. The calculated data are plotted in Fig. 4.1.7b. Since the thicknesses of $Ta_5Si_3$ and $Ti_3Si$ are very small, we have not calculated the diffusion parameters because of the chance of a high error in calculation [24]. It can be seen that, indeed, there is a variation in $\tilde{D}_{int}$ for TiSi with the annealing temperature. The small change in the layer thickness of this phase with the annealing temperature is found because of low activation energy of this phase. Further, the neighbouring phases influence the growth of this phase.

| **Phase** | **Prototype** | **Pearson Symbol** | **Space group** | **Lattice parameters** | | | **Molar Volume ($\times 10^{-6}$ m$^3$/mol)** |
|---|---|---|---|---|---|---|---|
| | | | | **a (Å)** | **b (Å)** | **c (Å)** | |
| $TiSi_2$ | $TiSi_2$ | oF24 | Fddd | 8.236 | 4.773 | 8.523 | 8.41 |
| TiSi | FeB | oP8 | Pnma | 6.54 | 3.63 | 4.99 | 8.92 |
| $Ti_5Si_4$ | $Zr_5Si_4$ | tP36 | $P4_12_12$ | 6.713 | - | 12.171 | 9.18 |

Table 4.1.2: Crystal Structures, lattice parameters and molar volumes of the phases

The importance of the calculation of the diffusion parameters can be understood from the calculations above. It is a common practice to analyze data on the calculation of the parabolic growth constants [30]; however, it might lead to a wrong conclusion if the diffusion parameters are not analyzed in a multiphase diffusion couple. This can be clarified with the help of Eqs. 4.1.1 and 4.1.4. When only a single phase grows in the interdiffusion zone, the second term in Eq. 4.1.4 is zero. Then $\tilde{D}_{\text{int}}$ is proportional to $k_p$. As defined by Wagner [29], the parabolic growth constant in this condition is defined as the parabolic growth constant of the second type, $k_P^{II}$ . Therefore, the activation energy for both these parameters are the same and it is safe to discuss the diffusion mechanism based on the calculation of $k_p$. However, in a multiphase diffusion, the calculation of the parabolic growth constant does not consider the growth of other phases although the diffusion of elements is affected by the number and growth rate of these phases. The parabolic growth constant in this condition is defined as the parabolic growth constant of the first type, $k_P^I$ [30]. Therefore, the activation energies for $\tilde{D}_{\text{int}}$ and $k_p$ are different. This is clear from the values mentioned in Fig. 4.1.7a and 4.1.7b. According to our experience, the activation energies for relatively thicker phases are less affected than those for the thinner phases [31]. If the discussion on diffusion mechanisms is based on the calculation of the parabolic growth constants, one should calculate first $k_P^{II}$ from $k_P^I$. On the other hand, it is always safe to have a discussion based on the calculation of the diffusion parameters. Further, in many systems, phases might have a wide homogeneity range. In this range, the diffusion coefficients and the activation energy might vary significantly. For example, in β-NiAl, the interdiffusion coefficients vary by three orders of magnitude with the change in composition [32]. Parabolic growth constants calculated based on the total layer thickness cannot bring out all these details. The main importance of the calculation of the parabolic growth constant is to examine whether the phases grow by a diffusion controlled process, as has been done in this study. In some cases, it is difficult to calculate the diffusion parameters and authors are forced to discuss the diffusion process on the calculation of the parabolic growth constants [33]. In these cases, one should be aware of the disadvantages and limitations before making any statement.

The activation energy for the integrated diffusion coefficient in the $TiSi_2$ phase is calculated as 190±9 kJ/mol. This is similar to the values found in a few other disilicides [16, 23, 24, 27]. Moreover, it is close to the activation energy for parabolic growth calculated by Hung et al. [4] where a single phase layer was grown in the thin film diffusion couple. As has been already mentioned, the Kirkendall marker plane is located at the Si/ $TiSi_2$ interface. This indicates that $TiSi_2$ grows mainly by the diffusion of Si. In a line compound, we cannot calculate the intrinsic diffusion coefficients since we are unable to estimate the concentration gradient in a phase within a narrow homogeneity range. We can instead calculate the ratio of components' diffusivities using the relation [11, 34]

$$\frac{V_{Ti} D_{Si}}{V_{Si} D_{Ti}} = \frac{D_{Si}^*}{D_{Ti}^*} = \frac{\left[ N_{Si}^+ \int_{x^{-\infty}}^{x_K} \left( N_{Si} - N_{Si}^- \right) dx - N_{Si}^- \int_{x_K}^{x^{+\infty}} \left( N_{Si}^+ - N_{Si} \right) dx \right]}{\left[ -N_{Ti}^+ \int_{x^{-\infty}}^{x_K} \left( N_{Si} - N_{Si}^- \right) dx + N_{Ti}^- \int_{x_K}^{x^{+\infty}} \left( N_{Si}^+ - N_{Si} \right) dx \right]} \qquad (4.1.5)$$

where $D_i$ and $D_i^*$ are the intrinsic and the tracer diffusion coefficients of element *i*. $x_K$ is the Kirkendall marker plane location. $x^{-\infty}$ and $x^{+\infty}$ correspond to the unaffected ends of the diffusion couple. This relation does not consider the vacancy wind effect [35]. As typically done, we have assumed that it is about unity in the present case. Moreover, the partial molar volumes of the components are also not known in a phase within a narrow homogeneity range. Therefore, we actually calculate the tracer diffusion coefficients indirectly following the diffusion couple technique. Since the diffusion rate of Ti is very small compared to Si, the above relation gives the value of $\frac{D_{Si}^*}{D_{Ti}^*} = \infty$. Although it gives the value of infinity, in an actual case, the diffusion rates of the elements could differ by few orders of magnitude. This has already been confirmed by the calculations based on diffusion couple and radio tracer experiments [9, 10, 23] in the Mo-Si system. The integrated and tracer diffusion coefficients are related by [36]

$$\begin{aligned}
\tilde{D}_{\text{int}}^{TiSi_2} &= -\left(N_{Ti} D_{Si}^* + N_{Si} D_{Ti}^*\right) \frac{N_{Ti}\left(\mu_{Ti}^{I} - \mu_{Ti}^{II}\right)}{RT} \\
&= -\left(N_{Ti} D_{Si}^* + N_{Si} D_{Ti}^*\right) \frac{N_{Si}\left(\mu_{Si}^{II} - \mu_{Si}^{I}\right)}{RT} \\
&= -\left(N_{Ti} D_{Si}^* + N_{Si} D_{Ti}^*\right) \frac{N_{Ti} \Delta_d G_{Ti}}{RT} \\
&= -\left(N_{Ti} D_{Si}^* + N_{Si} D_{Ti}^*\right) \frac{N_{Si} \Delta_d G_{Si}}{RT}
\end{aligned} \qquad (4.1.7a)$$

$\mu_i$ is the chemical potential of element *i*, and $\Delta_d G_{Si}$ is the driving force for the diffusion of Si in the $TiSi_2$ phase. It should be noted that $N_{Ti} \Delta_d G_{Ti} = N_{Si} \Delta_d G_{Si}$ (the Gibbs-Duhem relation). Equation (4.1.7a) resembles the standard Darken-Manning relation [DM]. Since the diffusion rate of Ti is negligible, the tracer diffusion coefficient of Si can be estimated by

$$\tilde{D}_{\text{int}}^{TiSi_2} = -N_{Ti} D_{Si}^* \frac{N_{Si} \Delta_d G_{Si}}{RT} \qquad (4.1.7b)$$

Again the vacancy-wind effect is neglected for simplicity. Note here that this relation is applicable when a single phase is grown between two neighbouring phases in the phase diagram. The diffusion parameters are material constants and one can use these parameters calculated from a multiphase interdiffusion zone. Therefore, for the calculation of the driving force, we need to consider as if the $TiSi_2$ phase is grown between Si and TiSi, as shown in Fig. 4.1.8. The free energies of these phases are calculated from the details available in Refs. [37, 38] and listed in Table 4.1.3.

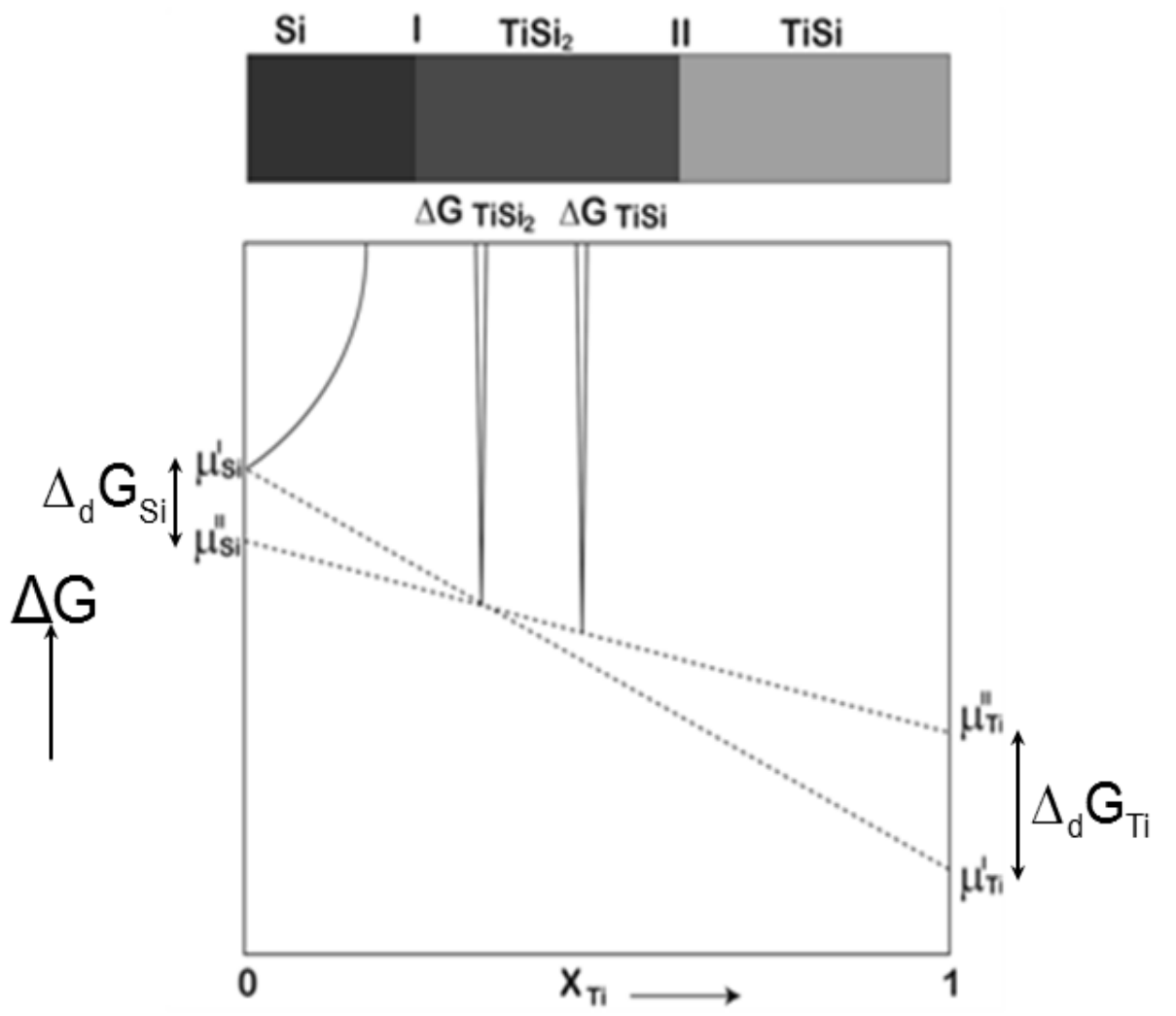

Fig. 4.1.8: Schematic diagram of the driving force calculation for the growth of $TiSi_2$ phase

| T °C | $\Delta G_{si}$–diamond J/mol | $\Delta G$ ($TiSi_2$) J/mol | $\Delta G$ (TiSi) J/mol | $\Delta_d G_{Si}$ J/mol |
|---|---|---|---|---|
| 1250 | -58329.1 | -119422.6 | -137100.6 | -25900 |
| 1225 | -56863.6 | -117889.8 | -135536.6 | -25900 |
| 1200 | -55410.1 | -116370.2 | -133986.4 | -25800 |
| 1175 | -53968.5 | -114863.9 | -132450.2 | -25800 |
| 1150 | -52539.2 | -113371.2 | -130928.1 | -25800 |

Table 4.1.3 : Free energy of the $\Delta G_{si}$ (diamond), $\Delta G$ ($TiSi_2$), $\Delta G$ (TiSi) and $\Delta_d G_{Si}$ at the temperature of our interest.

The Arrhenius plot of $D_{Si}^{*}$ is shown in Fig. 4.1.9 and the activation is calculated as 197±8 kJ/mol. Note that this value is much lower than the value reported by Gas et al. [8] based on the measurements of Ge tracer diffusion. However, it is close to the value determined in the $MoSi_2$ phase, and the activation energy for growth established by Hung et al. [4] in this system. We have explained already that the activation energy for growth found by Hung et al. should be close to the value of the activation energy tracer diffusion coefficient since a single $TiSi_2$ phase layer grows in the interdiffusion zone in a thin film couple. This phase grows mainly by the diffusion of Si. Even the driving force for the diffusion of this element with temperature does not change drastically as can be seen in Table 4.1.3.

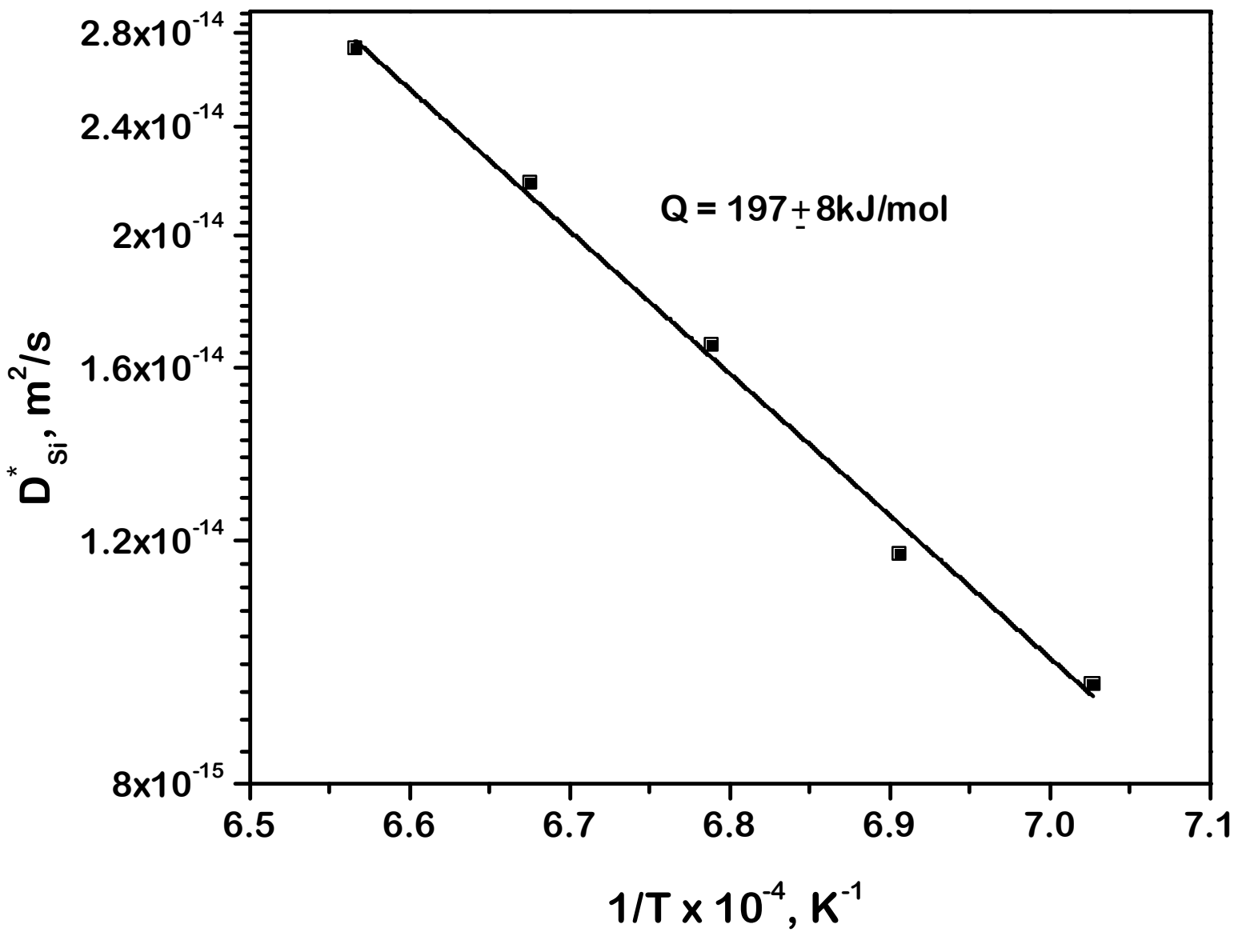

Fig. 4.1.9: Arrhenius plot of the Si tracer diffusion coefficients in the $TiSi_2$ phase.

### 4.1.5 Atomic mechanism of diffusion

Atomic mechanisms of diffusion in intermetallic compounds require a special consideration. In an ordered compound, the elements occupy their own sublattices. Therefore, the migration of substitutional atoms is restricted by a general demand to maintain the given level of the atomic order [32, 39, 40]. Typically, diffusion in a substitutional alloy proceeds by a vacancy-mediated mechanism [40, 41]. Two fundamental types of vacancy-mediated mechanisms can be distinguished with respect to jumps to nearest or next nearest neighbours [40]. In view of diffusion data for $MoSi_2$ [9, 10], the nearest neighbour jumps of vacancies are most probable for the $TiSi_2$ phase, too. Generally, four basic defects can be distinguished in an ordered binary compound, namely the vacancies on the two sublattices and the corresponding anti-sites – i.e. the atoms on a "wrong" sublattice. In fact, the diffusion mechanism of the given component depends on many factors – connectivity of the own sublattice, availability of vacancies on the different sublattices, "energy costs" for the anti-site formation, migration barriers, and the corresponding correlation effects which may even penalize specific types of atom jumps [40]. In view of these factors, a large asymmetry in the different rates of components in a binary alloy could be found, as it was the case of the $MoSi_2$ phase [9, 10, 23]. This diffusion asymmetry correlates with the vacancy concentrations on different sublattices. As it was determined in Ref. [42], mainly Si vacancies are formed in the tI6 structure of $MoSi_2$. A detailed inspection of lattice structure reveal that a Mo atom is surrounded by 10 Si atoms, whereas, a Si atom neighbours to 5 Mo and 5 Si. Therefore, Si can diffuse via its own sublattice and directly profit from the vacancy availability on the Si sublattice. On the other hand, diffusion of Mo is hindered both by absence of own vacancies and high energy costs to create a Mo antisites. This is the reason that Si diffuses by many orders of magnitude faster as Mo [9, 11]. On the other hand, in the $WSi_2$ phase with a similar crystal structure, the significant diffusion of W indicates the presence of a high concentration of W antisites [27].

Based on this understanding, we can analyse the defects present in the $TiSi_2$ phase. It has a face centered orthorhombic structure (oF24) with 24 atoms in the unit cell, as shown in Fig. 4.1.10.

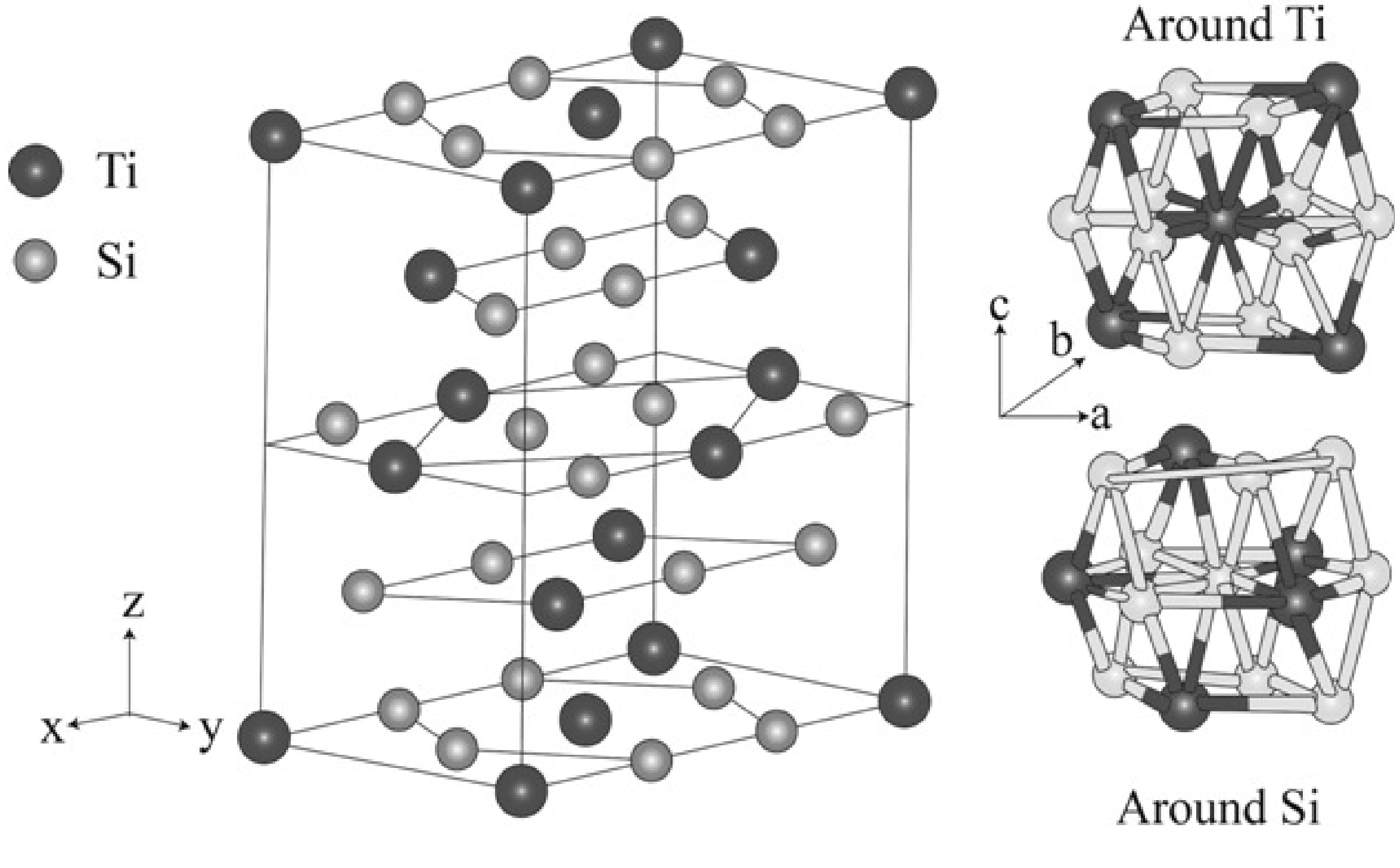

Fig. 4.1.10: Crystal structure of the C54 $TiSi_2$ Phase (oF24) showing nearest neighbours of Ti and Si.

The nearest neighbours for both the components are also shown. Each Ti atom is surrounded by 4 Ti and 10 Si, whereas Si atoms are surrounded by 5 Ti and 9 Si. Therefore, both the elements could potentially diffuse over their own sublattice provided that the corresponding vacancies are present. The fact that the growth of the $TiSi_2$ phase layer occurs mainly because of Si diffusion substantiates that vacancies are present mainly on the Si sublattice. This conclusion agrees with recent computation of the defect formation energies in $TiSi_2$ by tight binding molecular dynamics [43].

### 4.1.6. Conclusion

We have studied the Ti-Si system to understand the growth mechanism of the phases and the diffusion mechanism of the species. The following conclusions are drawn from this study:

- The diffusion couple experiments indicate that the maximum stability temperature for the $Ti_3Si$ phase is higher than the temperature reported in the Ti-Si phase diagram.
- The variation in layer thickness with the change in the annealing temperature of the TiSi phase is very small. On the other hand, time dependent experiments indicate the parabolic diffusion controlled growth of the phase. This indicates that the study on the parabolic growth constant might draw a wrong conclusion on the growth mechanism of the phases. The calculation of the diffusion parameters is always reliable in discussing the diffusion mechanism.
- Uniform grain morphology in the $TiSi_2$ phase indicates the location of the Kirkendall marker plane at the Si/$TiSi_2$ interface. Therefore, $TiSi_2$ grows mainly because of the diffusion of Si. The grain morphology developed in the interdiffusion zone is explained with the help of imaginary diffusion couples.
- In the $TiSi_2$ phase, Ti atoms are surrounded by 4 Ti and 10 Si. Similarly, Si atoms are surrounded by 5 Ti and 9 Si. Therefore both the elements can diffuse via their own sublattice on the condition that vacancies are present. Since the phase grows because of Si diffusion, vacancies must be present mainly on the Si sublattice. Negligible diffusion rate of Ti indicates the low concentration of Ti antistes and vacancies on the Ti sublattice.

# Chapter 4.2

# Growth of hafnium and zirconium silicides by reactive diffusion

## 4.2.1. Introduction

Transition metal silicides draw special attention because of their use in microelectronic devices as Ohmic contacts and Schottky barriers [1]. Low resistivity, metal like behavior and high temperature stability make them attractive from application point of view. [2-5]. Hf and Zr silicates are being considered for replacing silica in sub 100 nm complementary metal oxide semiconductor (CMOS) technology [6]. Advantages of introduction of Hf and Zr films at the Si/$SiO_2$ interface are being examined [5, 7].

To date, several studies have been conducted to examine the growth of the phases in Hf/Si in thin film couples [8-11]. There are five phases in this system [12], however, only two phases, $HfSi_2$ and HfSi are found to grow in the interdiffusion zone. Unlike bulk diffusion couple, sequential growth of phases is very common in thin film couples [13, 14]. In this system, HfSi is found to grow first, followed by $HfSi_2$ which grows very rapidly at the Si/$HfSi_2$ interface by nucleating on HfSi [11, 15]. Parabolic growth has been found, which indicates diffusion controlled growth of the phases [15]. Ziegler et al. [11] studied the growth of the HfSi and $HfSi_2$ phases in the temperature range of 550–750 °C and 750–900 °C respectively, in the He atmosphere, and estimated the activation energy for growth of HfSi to be 2.5 eV (241 kJ/mol).

---

This chapter is written based on
S. Roy and A. Paul, Growth of hafnium and zirconium silicides by reactive diffusion, Materials Chemistry and Physics, 143 (2014) 1309-1314

Movement of the impurities during the growth of this phase indicated that Si is the faster diffusing species. So et al. [15] studied the growth of the $HfSi_2$ phase in the temperature range of 600–650 °C in vacuum and calculated the activation energy as 3.5±0.3 eV. (337±29 kJ/mol). No diffusion studies have been conducted in bulk condition to date, hence no reliable information is available understanding the growth mechanism of phases compared to the study in thin film condition.

Unlike the Hf-Si system, studies on diffusion controlled growth of the phases in the Zr/Si system are very limited [16, 17]. In one of the studies [16], growth of the phases in bulk diffusion couples was reported; however, analysis on diffusion process was not done. Although there are six phases present in the system [18], only two phases, $ZrSi_2$ and ZrSi are found to grow. After the consumption of Si, other phases were found between $ZrSi_2$ and Zr.

Therefore, the aim of this present study is to conduct bulk diffusion couple experiments in these two systems to analyze the growth mechanism of the phases and the diffusion mechanism of the species. Time dependent experiments are conducted to examine the diffusion controlled growth of the phases, and temperature dependent experiments to calculate the activation energies. Grain morphology in the interdiffusion zone indicates relative mobilities of the species, which help to predict defects and diffusion of elements.

### 4.2.2 Results and discussion

Backscattered electron images (BSE) of interdiffusion zones of Hf/Si and Zr/Si diffusion couples annealed at 1200 and 1250 °C for 16 hrs, respectively, are shown in Fig. 4.2.1a and b. X-ray diffractions taken in these diffusion couples are shown in Fig. 4.2.1c and d. Only disilicide and monosilicide phases grew in the interdiffusion zone, although there are many phases present in both the systems. Moreover, the thickness of the disilicides is much higher than the monosilicide, which indicates difference in the growth rates.

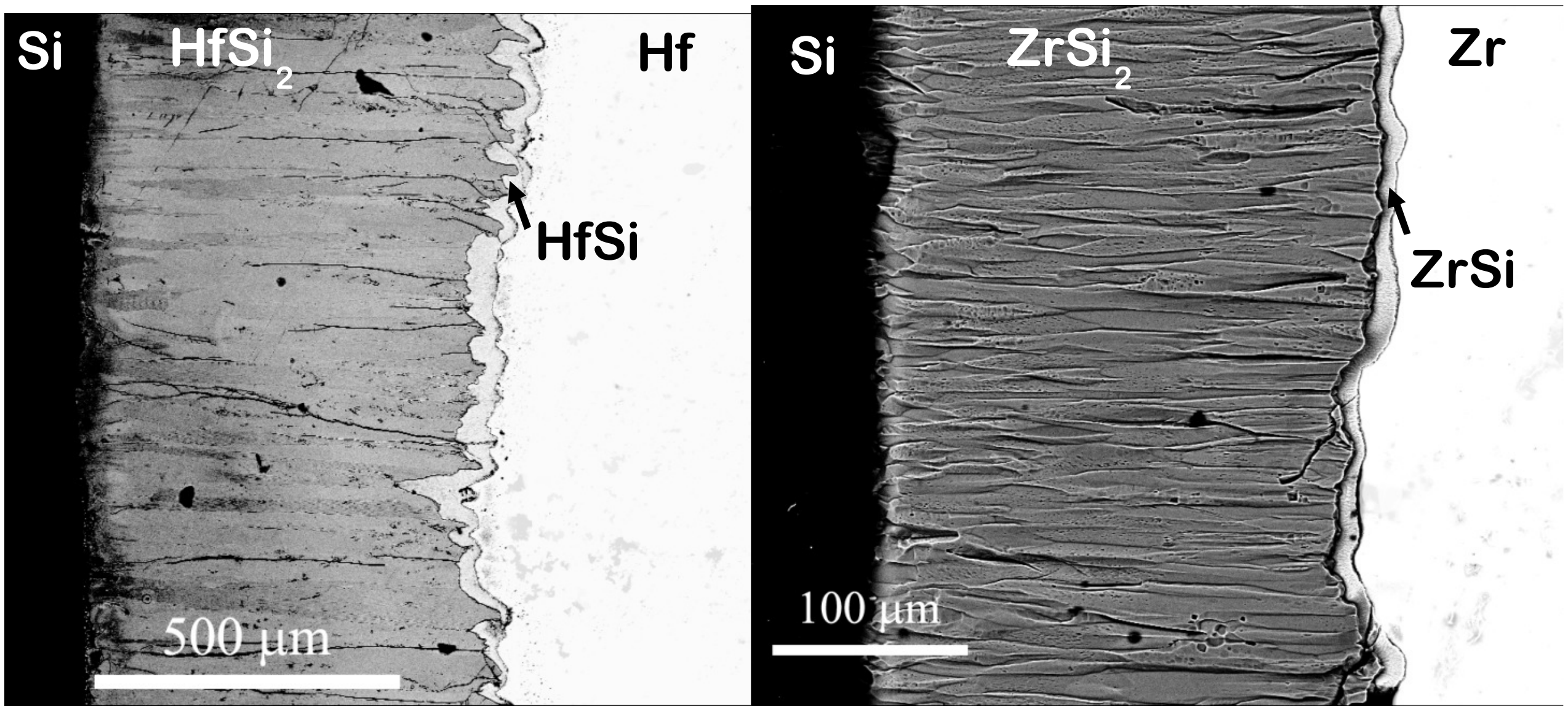

(a) (b)

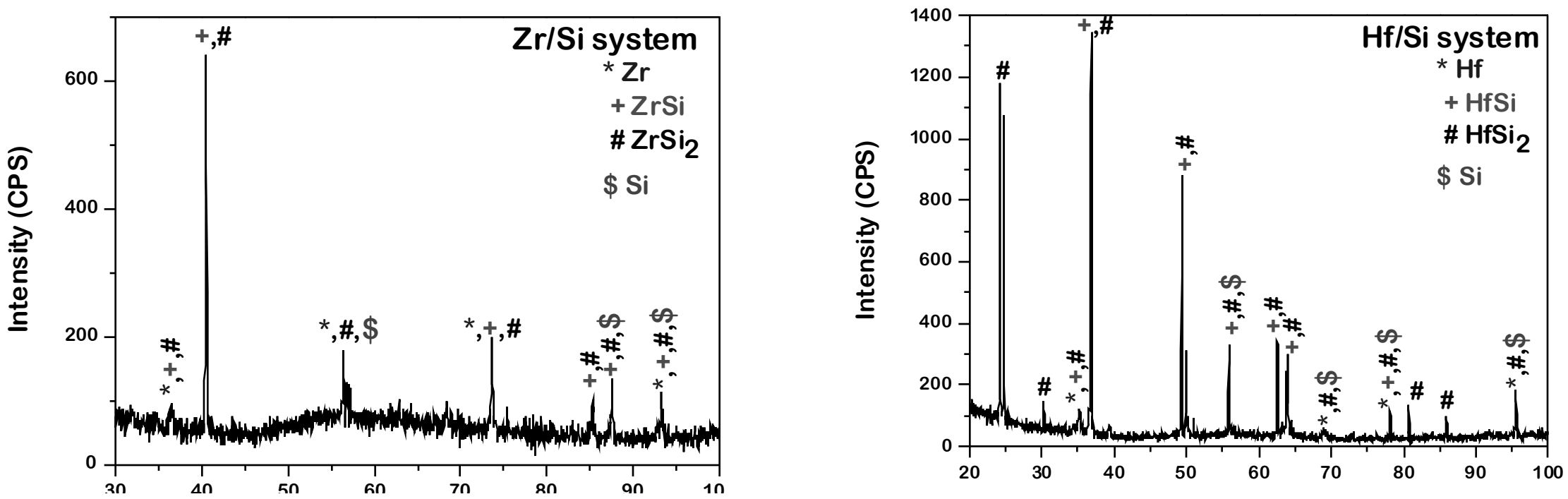

(c) (d)

Fig. 4.2.1: Scanning electron micrograph of etched (a) Hf/Si diffusion couple annealed at 1250 °C for 16 hrs (b) Zr/Si diffusion couple annealed at 1200 °C for 16 hrs. (c) and (d) are the X-ray analysis confirming the phases detected in the interdiffusion zone.

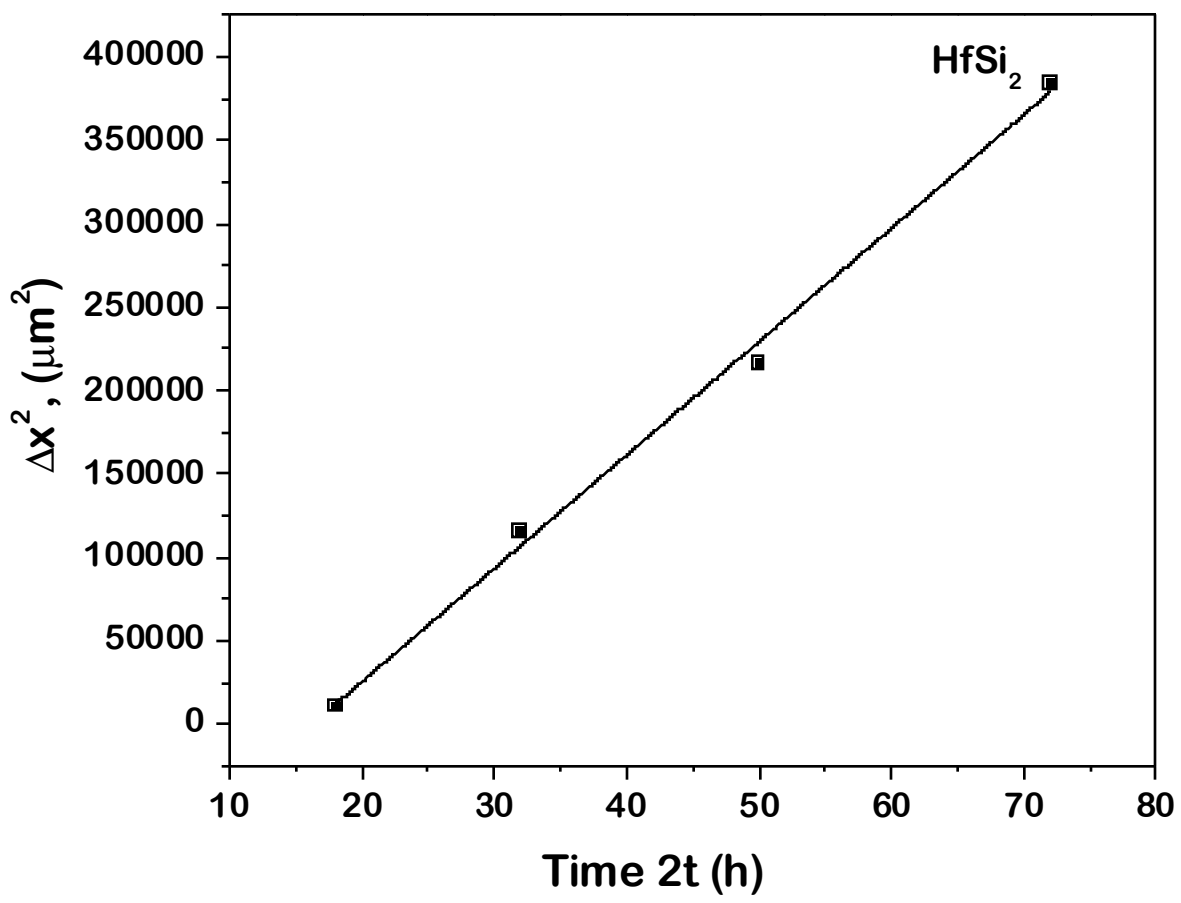

(a)

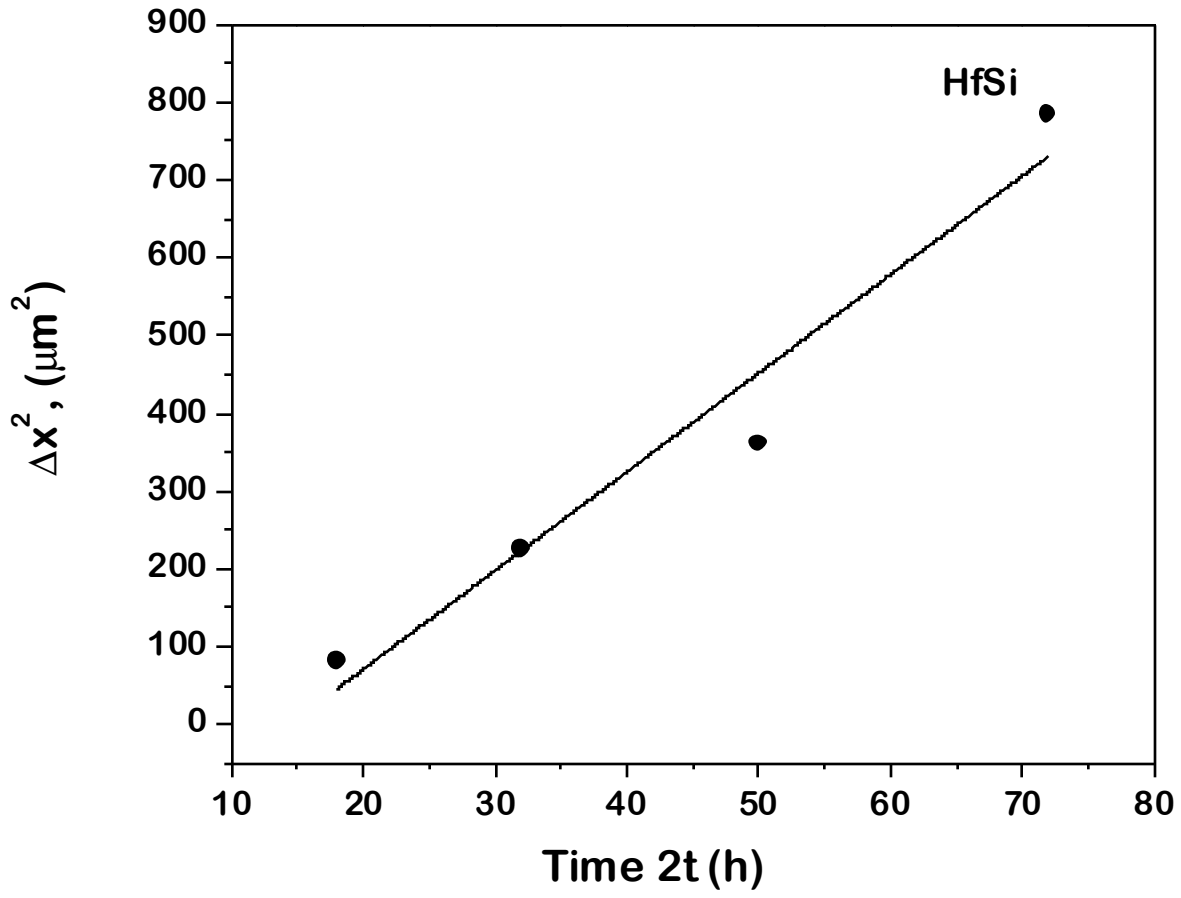

(b)

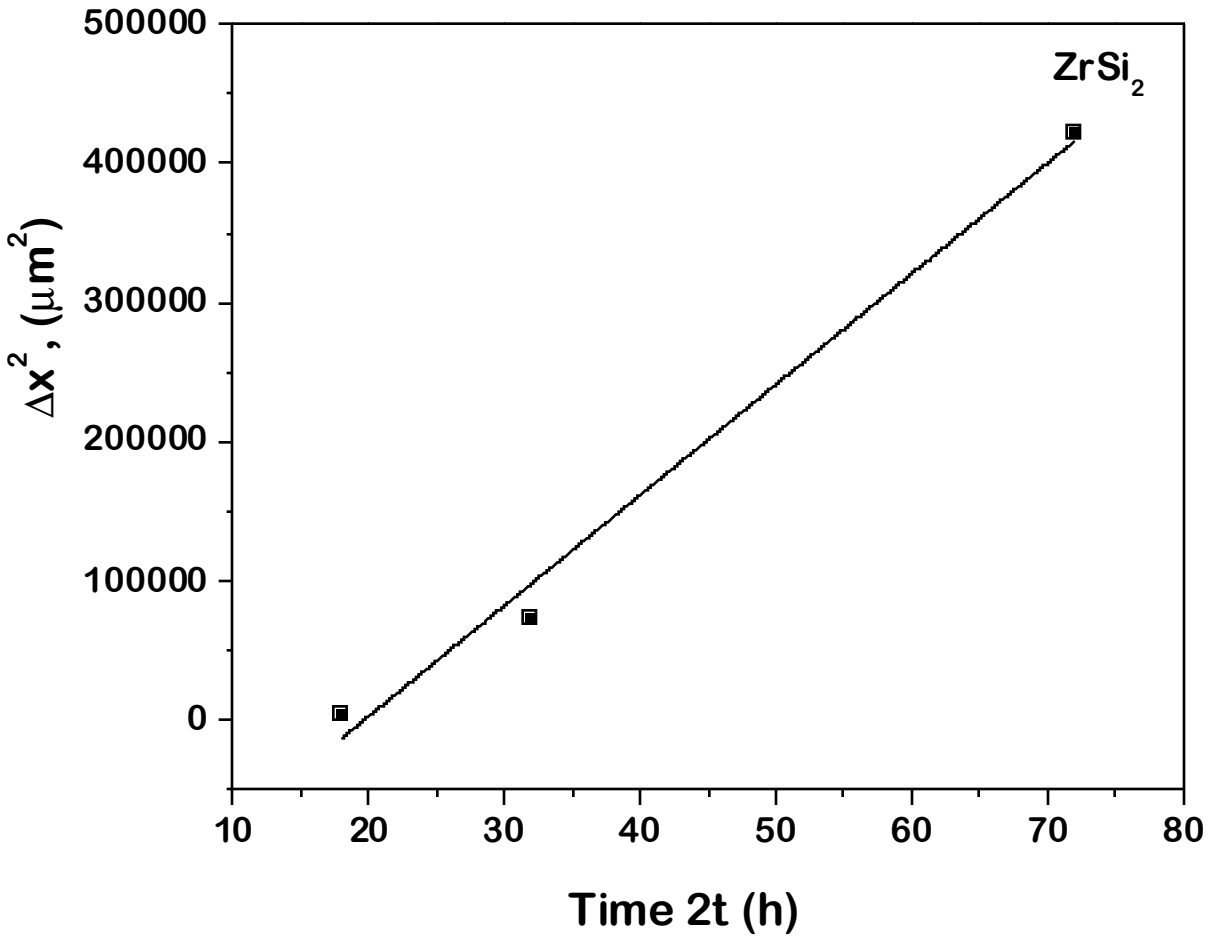

(c)

Fig. 4.2.2: Time dependant experiments done at 1200 °C for 9, 16, 25 and 36 hours. Δx2 vs. 2t plots of (a) $HfSi_2$, (b) HfSi and (c) $ZrSi_2$ phases.

Bertolino et al., [16] reported the growth of other phase after consumption Si between 1698 to 1869 °C (1971 to 2142 K). This indicates that the growth rates of other phases are much smaller compared to the phases found in the interdiffusion zone. Although time dependent experiments at different temperatures conducted in thin film condition indicated the diffusion controlled growth of the phases, we conducted these experiments at one particular temperature (1200 °C) in both the systems, as shown in Fig. 4.2.2a, 2b and 2c. Linear relationships of the data of $\Delta x^2$ vs. 2t plots indicate the diffusion controlled growth. Thicknesses of the phases at different temperatures and annealing times are listed in Table 4.21.

| Temperature (°C) | Δx $HfSi_2$ (μm) | Δx HfSi (μm) | Δx $ZrSi_2$ (μm) |
|---|---|---|---|
| **1150** | - | - | 155 |
| **1175** | 268 | 9 | - |
| **1200** | 340 | 15 | 275 |
| **1225** | - | - | 349 |
| **1250** | 650 | 34 | 390 |
| **1275** | 720 | 40 | - |

(a)

| Temperature (°C) | Time (hrs) | Δx $HfSi_2$ (μm) | Δx HfSi (μm) | Δx $ZrSi_2$ (μm) |
|---|---|---|---|---|
| **1200** | 9 | 103 | 9 | 73 |
| **1200** | 16 | 340 | 15 | 275 |
| **1200** | 25 | 465 | 19 | - |
| **1200** | 36 | 620 | 28 | 650 |

(b)

Table 4.2.1: (a) Layer thicknesses of the phases grown after annealing for 16 hours at different temperatures (b) Layer thicknesses of the phases at 1200 °C annealed for different times.

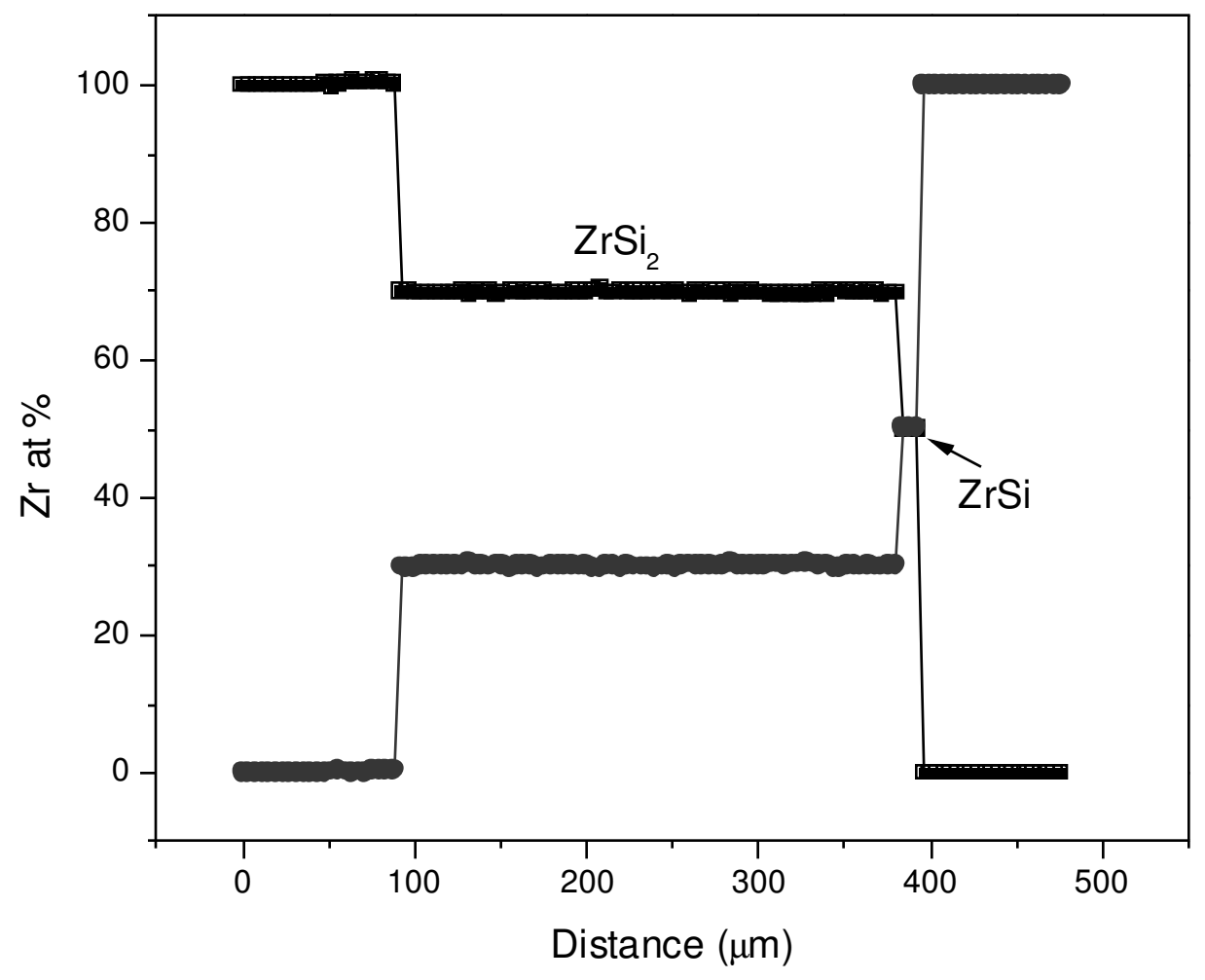

(a)

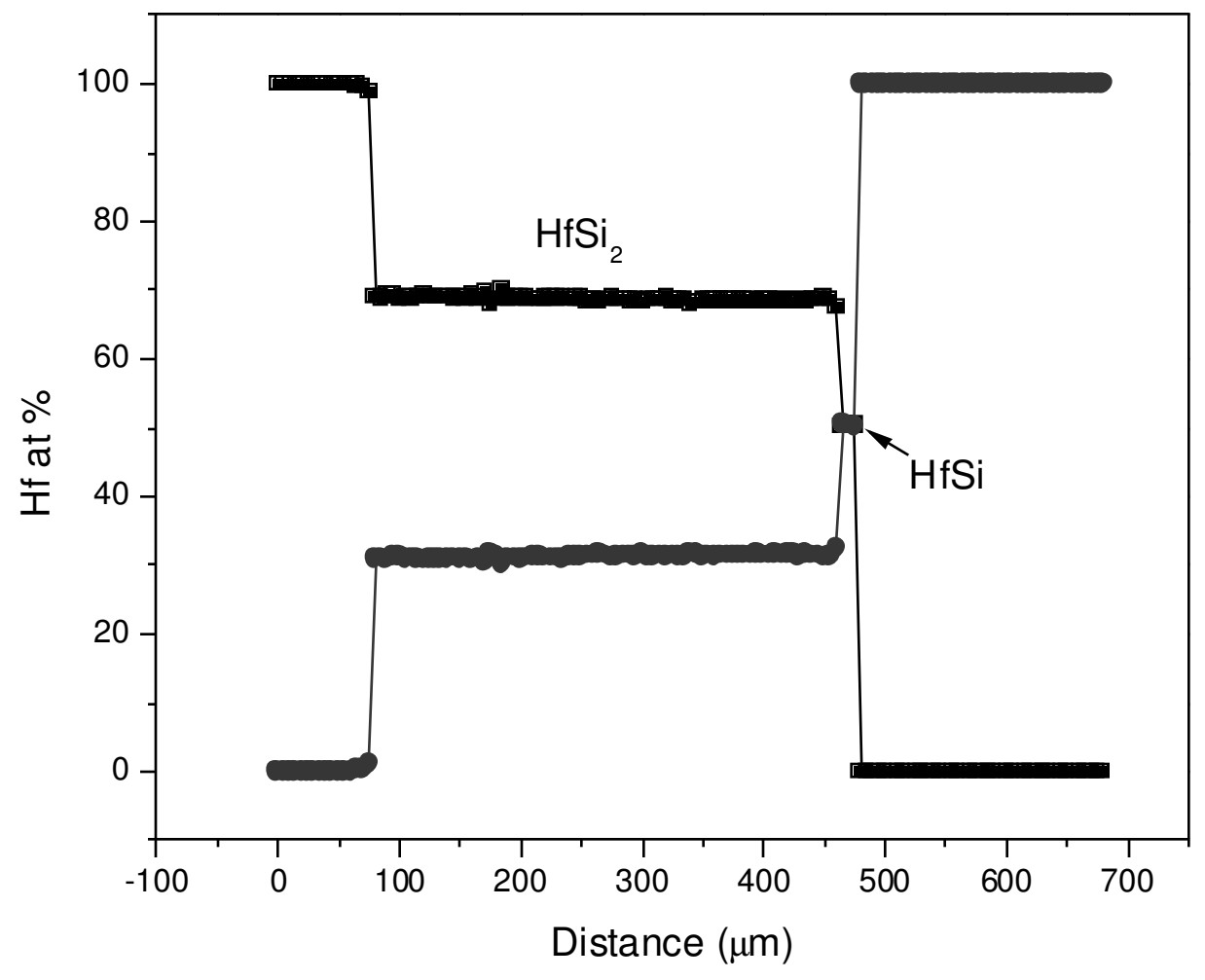

(b)

Fig 4.2.3: The composition profile of the interdiffusion zone (a) Zr/Si diffusion couple annealed at 1200 °C for 16 hrs and (b) Hf/Si diffusion couple annealed at 1250 °C for 16 hrs .

Composition profiles in Fig. 4.2.3 indicate negligible diffusion of elements in the Hf or Zr and Si end members. Since all the phases in the interdiffusion zone are line compounds, we can calculate the integrated diffusion coefficient $\tilde{D}_{int}$ using the Wagner's relation [19]. This relation for the β phase can be written as Eqn. 2.34[20] Although the thickness of HfSi is much smaller compared to $HfSi_2$, which vary in the range of 9 – 40 µm in the temperature range of our interest, we can calculate the integrated diffusion coefficients for both the phases without introducing much error. However, the maximum thickness of the ZrSi phase is less than 10 µm at the highest temperature and decreases rapidly with the decrease in annealing temperature. Therefore, the diffusion parameters are not calculated for this phase in order to avoid error in the calculations. This is discussed in detail based on the results in the V-Si system [21]. Molar volumes of the phases, as listed in Table 4.2.2, are calculated from the lattice parameter data available in the literature [22, 23].

| **Phase** | **Prototype** | **Pearson Symbol** | **Space group** | **Strukturbericht Designation** | **a (Å)** | **b (Å)** | **c (Å)** | **Molar volume ($\times 10^{-6}$ $m^3$/mol)** |
|---|---|---|---|---|---|---|---|---|
| **$HfSi_2$** | $ZrSi_2$ | oC12 | Cmcm | C49 | 3.7 | 14.6 | 3.6 | 9.8 |
| **HfSi** | FeB | oP8 | Pnma | B27 | 6.8 | 3.7 | 5.2 | 9.9 |
| **$ZrSi_2$** | $ZrSi_2$ | oC12 | Cmcm | C49 | 3.7 | 14.8 | 3.7 | 10.0 |

**Table 4.2.2: Details on crystal structure, lattice parameters and molar volume of the phases.**

The calculated $\tilde{D}_{int}$ are shown in Fig. 4.2.4a and b with respect to the Arrhenius equation

$$\tilde{D}_{int}^{\beta} = \tilde{D}_{int}^{o} \exp\left(-\frac{Q}{RT}\right) \tag{4.2.1}$$

where $\tilde{D}^{o}_{\text{int}}$ ($m^2/s$) is the pre-exponential factor, $Q$ (J/mol) is the activation energy and $R$ the universal gas constant (8.314 J/mol.K). The activation energies for $HfSi_2$ and $ZrSi_2$ are estimated as 394±37 and 346±34 kJ/mol, respectively. The activation energy for HfSi is 485±42 kJ/mol.

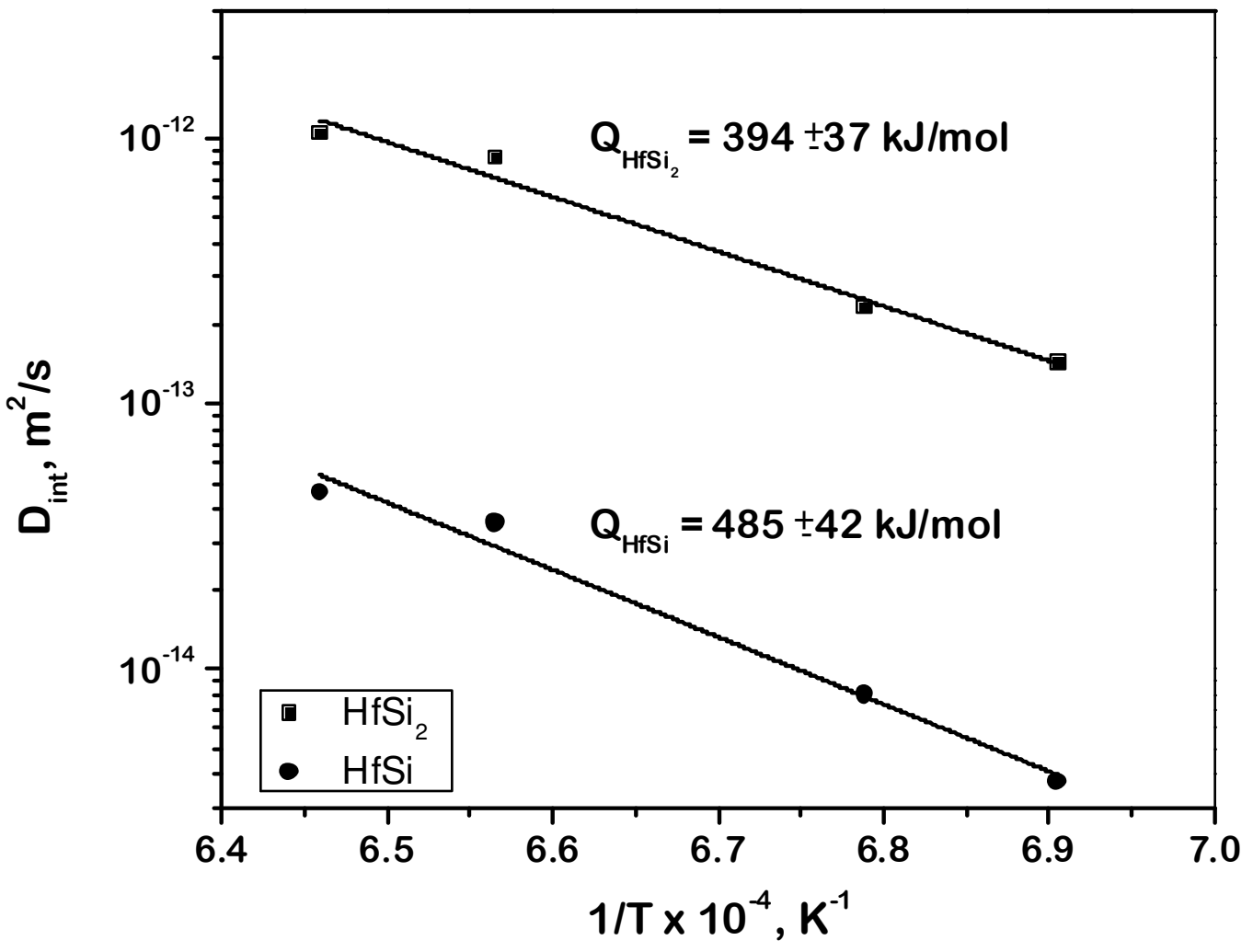

(a)

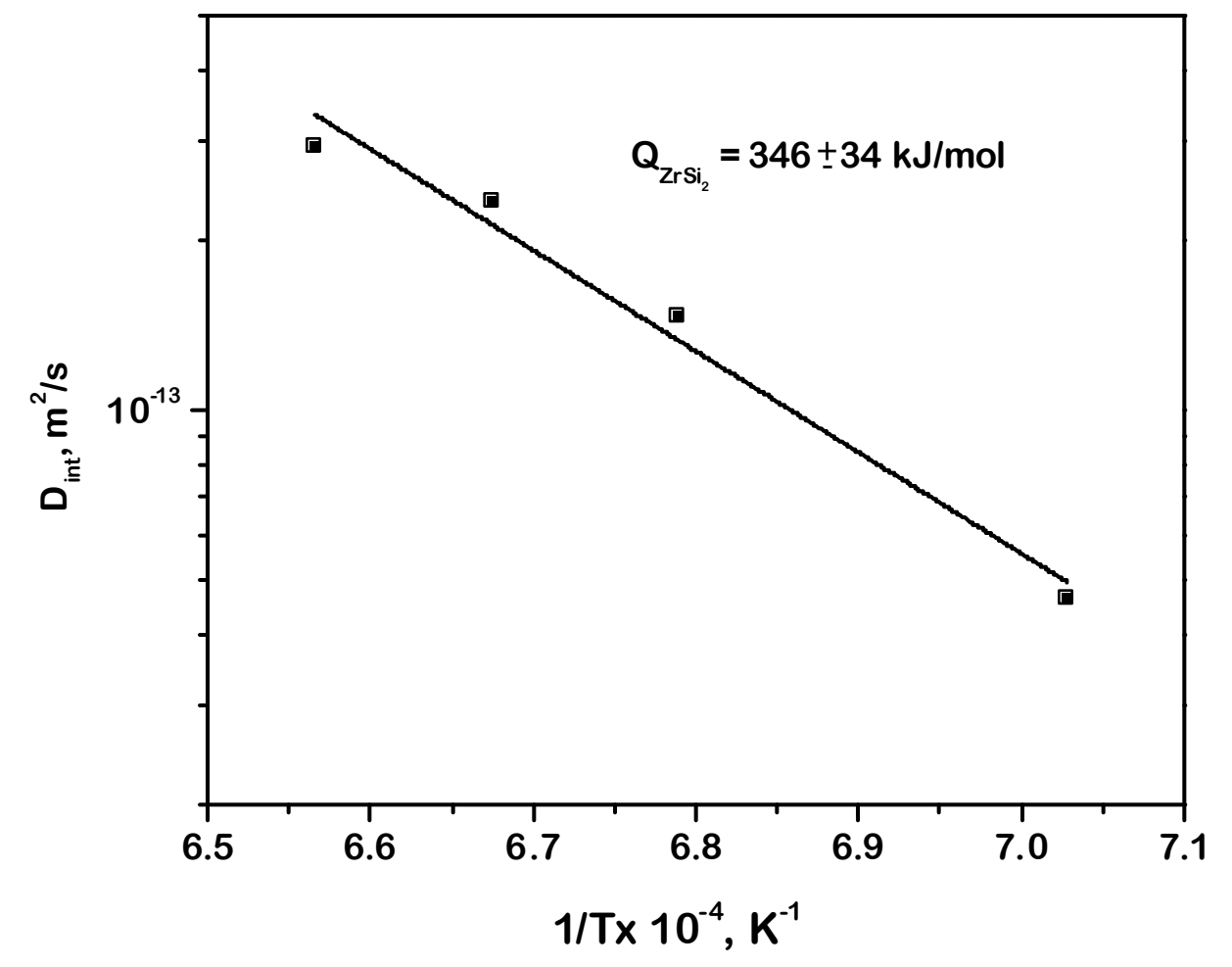

(b)

Fig 4.2.4: Arrhenius plot of the integrated diffusion coefficients calculated for (a) $HfSi_2$ and HfSi, and (b) $ZrSi_2$.

The Kirkendall marker experiments could not be conducted, because, the material halves did not join when inert particles were used. However, location of the marker plane could be detected by revealing the grain morphology, based on which physico-chemical approach is developed [24, 25]. Uniform grain morphologies in the disilicide phases, as shown in Fig. 4.2.1, indicate that one of the species must have much higher diffusion rate compared to the other. Similar behavior was found in the V-Si [26], Mo-Si [27] and Ti-Si [28] systems. According to extensive studies on tracer and interdiffusion in the Mo-Si system, Si has few orders of magnitude higher diffusion rate compared to Mo. In fact, Si has higher diffusion rate compared to the metal species in the disilicides [21, 26-31]. This indicates that the marker plane must be present at the Si/$MSi_2$ interface (M = Hf, Zr). In line compounds, we cannot determine the ratio of intrinsic diffusion coefficients because of unknown partial molar volumes, $V_i$; however, we can determine the ratio of the tracer diffusion coefficients using the relation [20]

$$\frac{V_M D_{Si}}{V_{Si} D_M} = \frac{D^*_{Si}}{D^*_M} = \frac{\left[ N^+_{Si} \int_{x^{-\infty}}^{x_K} \left(N_{Si} - N^-_{Si}\right) dx - N^-_{Si} \int_{x_K}^{x^{+\infty}} \left(N^+_{Si} - N_{Si}\right) dx \right]}{\left[ -N^+_M \int_{x^{-\infty}}^{x_K} \left(N_{Si} - N^-_{Si}\right) dx + N^-_M \int_{x_K}^{x^{+\infty}} \left(N^+_{Si} - N_{Si}\right) dx \right]} \tag{4.2.2}$$

where $D_i$ and $D^*_i$ are the intrinsic and the tracer diffusion coefficient respectively, of element *i*. This relation ignores the vacancy wind effect proposed by Manning [32, 33], since the structure factors required are not known. The marker plane is present at the Si/$MSi_2$ interface and the ratio of the tracer diffusivities estimated by the above relation is $\frac{D^*_{Si}}{D^*_M} = \infty$, because of negligible diffusion rate of M. Further, we can determine the tracer diffusion coefficient by using the relation [34]

$$\begin{aligned}\tilde{D}_{\text{int}}^{MSi_2} &= -\left(N_M D_{Si}^* + N_{Si} D_M^*\right)\frac{N_M\left(\mu_M^I - \mu_M^{II}\right)}{RT}\\ &= -\left(N_M D_{Si}^* + N_{Si} D_M^*\right)\frac{N_{Si}\left(\mu_{Si}^{II} - \mu_{Si}^{I}\right)}{RT}\\ &= -\left(N_M D_{Si}^* + N_{Si} D_M^*\right)\frac{N_M \Delta_d G_M}{RT}\\ &= -\left(N_M D_{Si}^* + N_{Si} D_M^*\right)\frac{N_{Si} \Delta_d G_{Si}}{RT}\end{aligned} \tag{4.2.3a}$$

where $\mu_i$ the chemical potential of element *i*, and $\Delta_d G_{Si}$ the driving force for the diffusion of Si in the $MSi_2$ phase. It should be noted that $N_M \Delta_d G_M^o = N_{Si} \Delta_d G_{Si}^o$. Considering $D_M^*$ as negligible, we get

$$\tilde{D}_{\text{int}}^{MSi_2} = -N_M D_{Si}^* \frac{N_{Si} \Delta_d G_{Si}^o}{RT} \tag{4.2.3b}$$

This relation is applicable only when a phase layer is developed between two neighboring phases in the phase diagram. On the other hand, diffusion parameters are materials constant and one can use these parameters calculated from a multiphase interdiffusion zone. For the calculation of driving forces, we need to consider the growth of the $MSi_2$ phase between Si and MSi, as shown in Fig. 4.2.5. Free energies calculated from the details available in Refs. [12, 18, 35, 36] are listed in Table 4.2.3. Calculated $D_{Si}^*$ are plotted using Arrhenius equation are shown in Fig. 4.2.6. The calculated activation energies of $HfSi_2$ and $ZrSi_2$ are 430 ± 36 and 348 ± 34 kJ/mol, respectively.

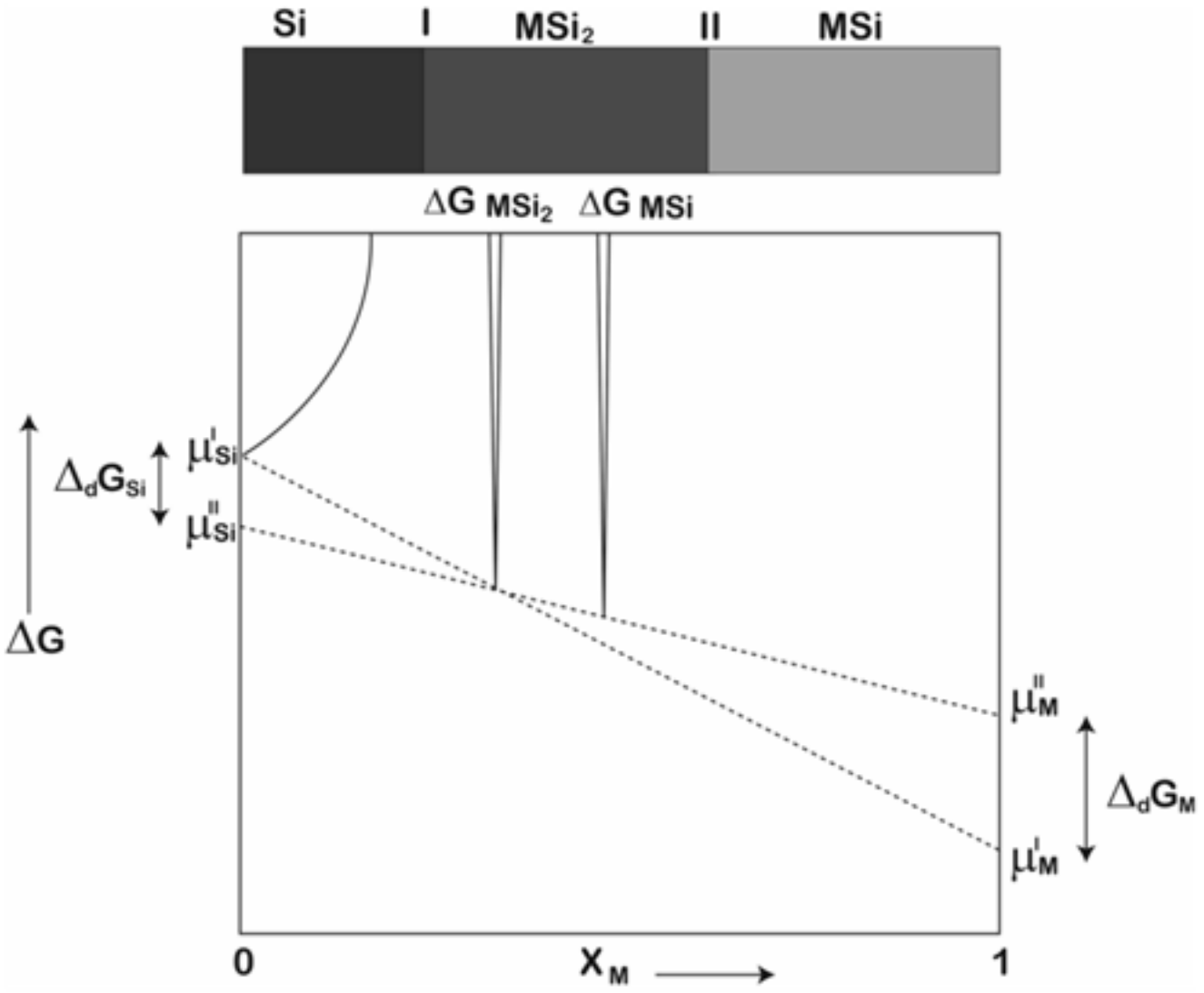

Fig. 4.2.5: Schematic diagram of the driving force calculation for the growth of MSi2 from neighboring phases.

| T ($^oC$) | $\Delta G_{si}$ (diamond) J/mol | $\Delta G$ ($HfSi_2$) J/mol | $\Delta G$ (HfSi) J/mol | $\Delta_d G_{Si}$ J/mol |
|---|---|---|---|---|
| **1275** | -59806.3 | -127919.2 | -151724.1 | -20645.8 |
| **1250** | -58329.1 | -126435.94 | -149893.4 | -21332.5 |
| **1200** | -55410.1 | -123507.6 | -146271.6 | -22675.0 |
| **1175** | -53968.51 | -122062.8 | -144480.7 | -23392.7 |

**(a)**

| T ($^oC$) | $\Delta G_{si}$ (diamond) J/mol | $\Delta G$ ($ZrSi_2$) J/mol | $\Delta G$ (ZrSi) J/mol | $\Delta_d G_{Si}$ J/mol |
|---|---|---|---|---|
| **1250** | -58329.1 | -306867.3 | -308440.38 | -245407 |
| **1225** | -56863.61 | -303516.3 | -305278.8 | -243136 |
| **1200** | -55410.1 | -300189.7 | -302141.5 | -240887 |
| **1150** | -52539.2 | -293610.6 | -295941.1 | -236424 |

**(b)**

**Table 4.2.3 : (a) Free energy $\Delta G_{si}$ (diamond), $\Delta G$ ($HfSi_2$), $\Delta G$ (HfSi) and driving force for Si diffusion, $\Delta_d G_{Si}$ in $ZrSi_2$ (b) $G_{si}$ (diamond), $\Delta G$ ($ZrSi_2$), $\Delta G$ (ZrSi) and driving force for Si diffusion $\Delta_d G_{Si}$ in $HfSi_2$ at the temperatures of our intereset.**

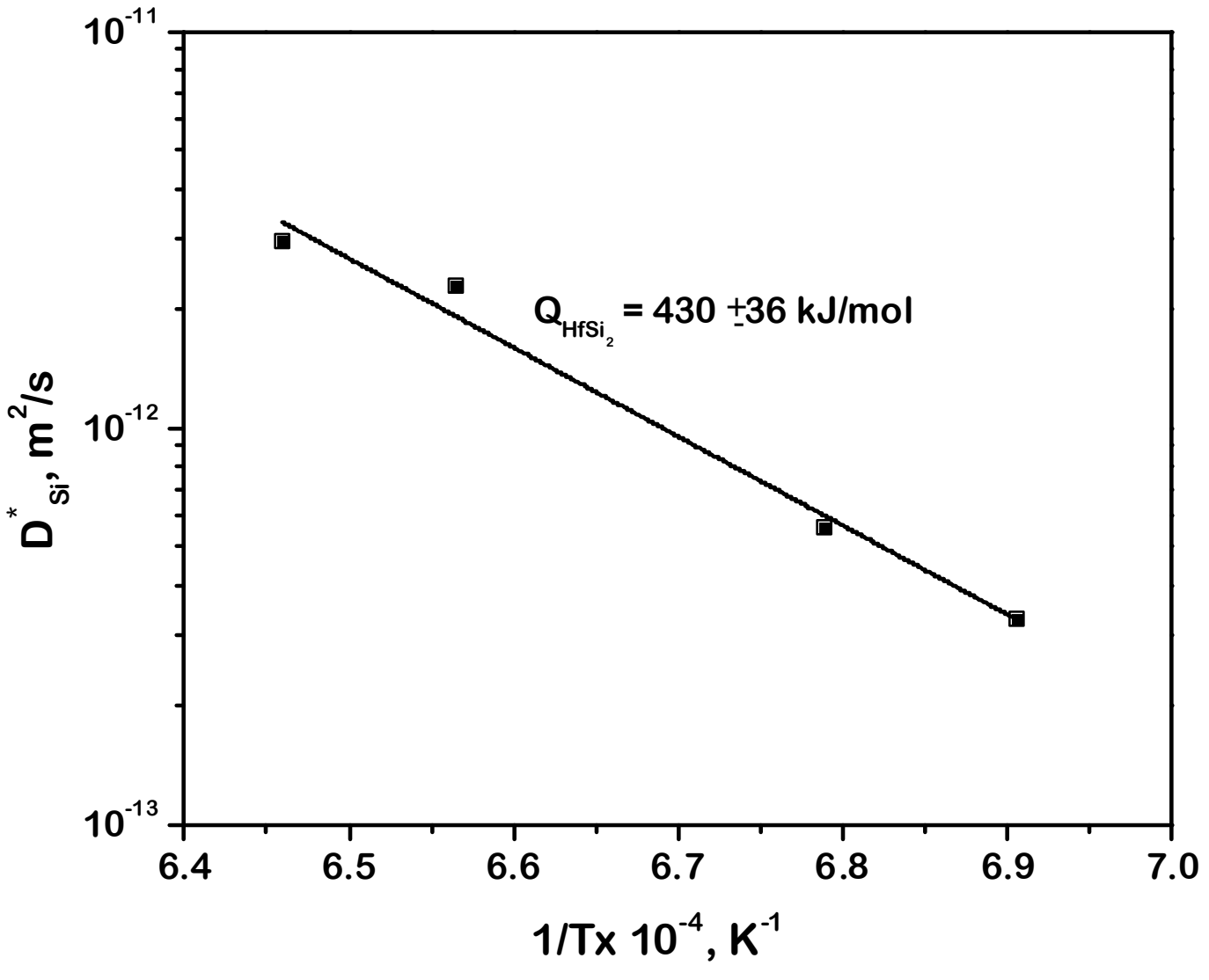

(a)

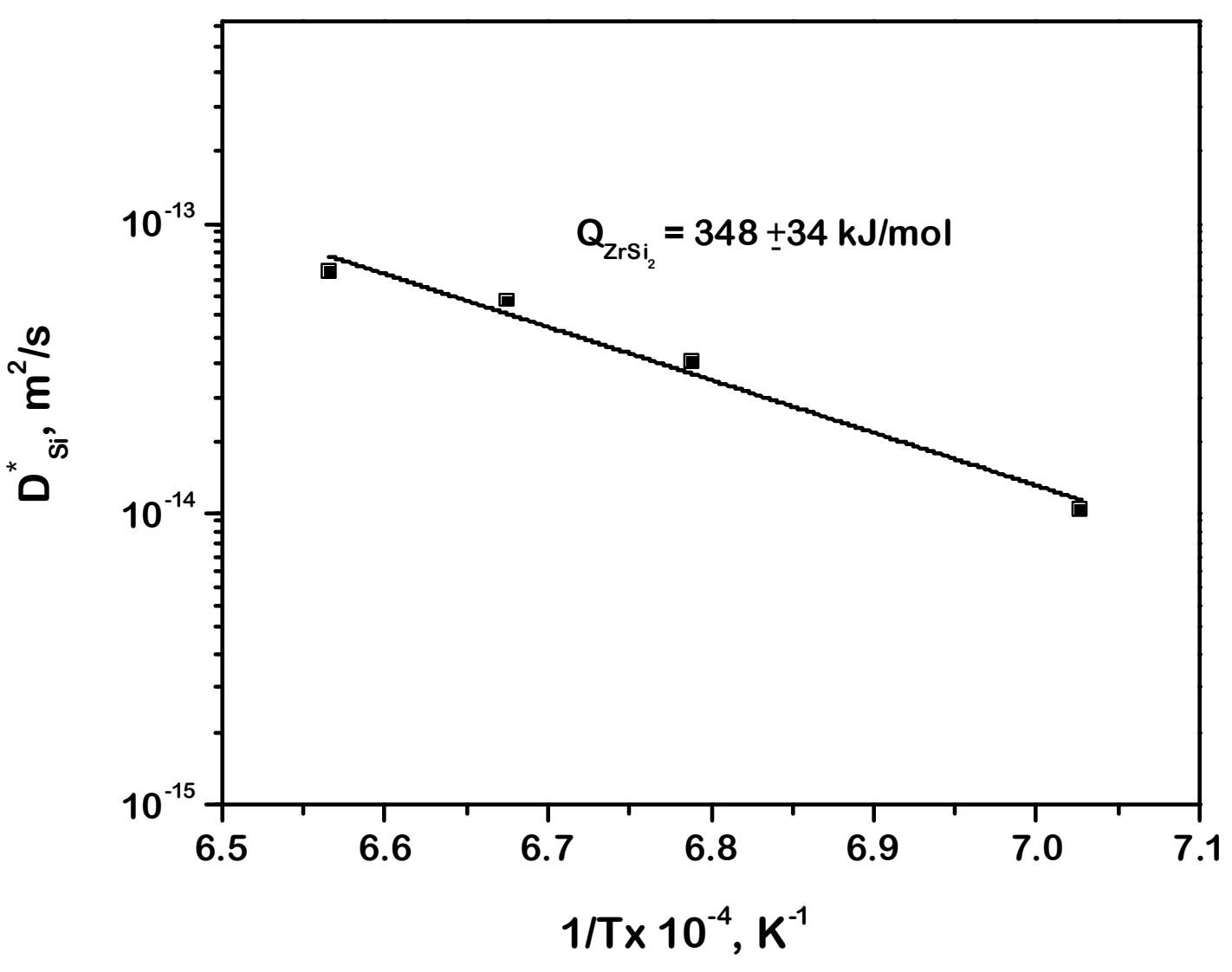

(b)

Fig 4.2.6: Arrhenius plot of the tracer diffusion coefficients of Si in (a) $HfSi_2$ and (b) $ZrSi_2$.

### 4.2.3. Atomic mechanism of diffusion

Diffusion in an ordered phase is a complex process. Different concentration of defects on different sublattices makes it even more difficult to understand. Although diffusion coefficients and activation energies are being calculated in many systems, the lack of data on defects makes it difficult to understand the process in the atomic level. Extensive theoretical and experimental studies are available only in few phases, such as $MoSi_2$ [37,38], β-NiAl [39] and γ'-$Ni_3Al$ [40]. Based on these studies and calculated parameters in $HfSi_2$ and $ZrSi_2$, we can qualitatively discuss the diffusion process and possible defects present in the disilicides. Both the phases have orthorhombic C49 (oC12) crystal structure with 12 atoms per unit cell, as shown in the Fig. 4.2.7.

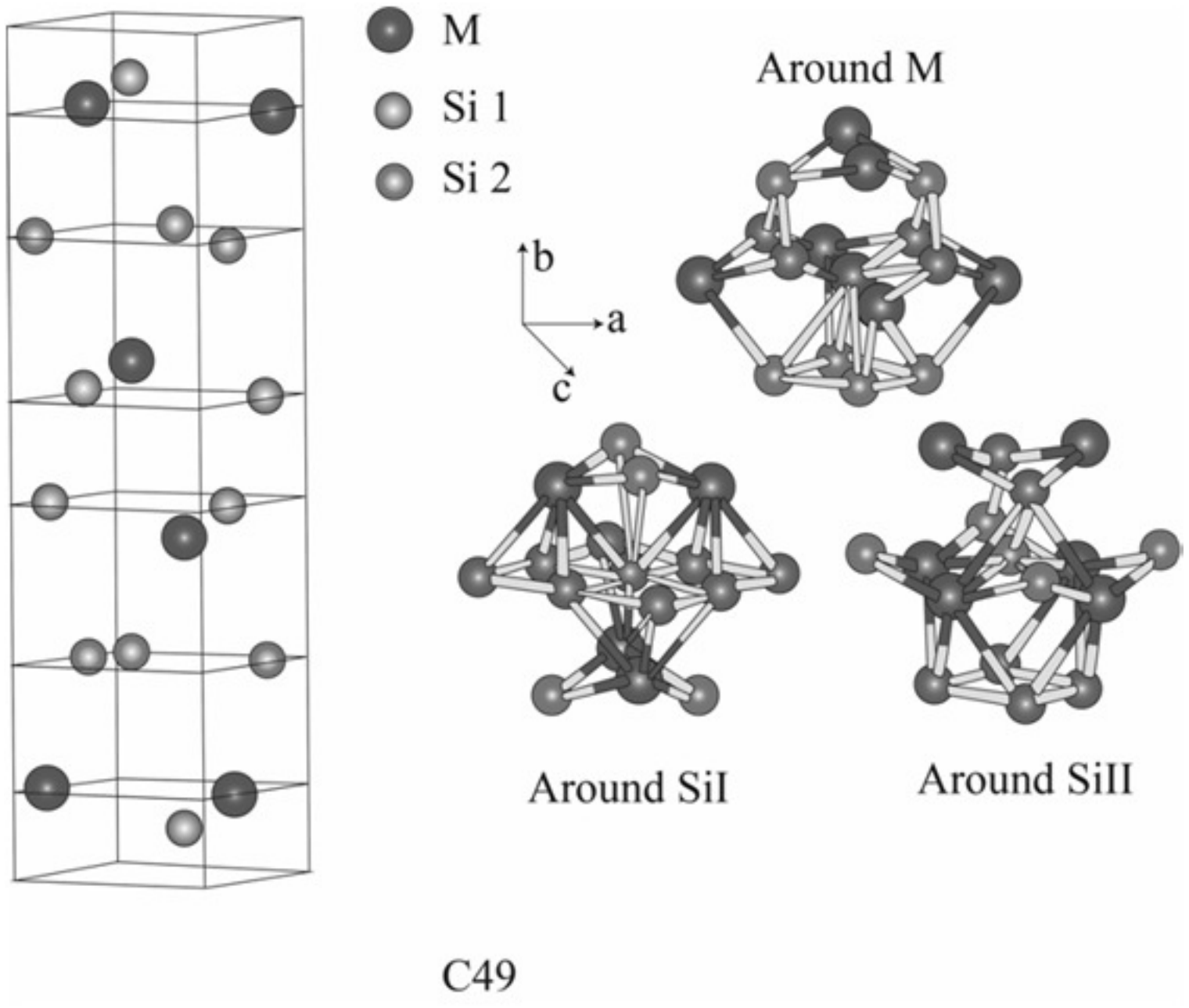

Fig 4.2.7: (a) The crystal structure of C49 (oC12). First and second nearest neighbors of Ti, SiI and SiII are shown.

Metals (M) have only one sublattice whereas Si has two, designated by SiI and SiII. M is surrounded by 10 Si (4 SiI and 6 SiII) and 6 M. SiI is surrounded by 12 Si (8 SiI and 4 SiII). SiII is surrounded by 10 Si (6 SiI and 4 SiII) and 6 M. Hence, both the species can diffuse via its own sublattice if vacancies are present. Negligible diffusion of M compared to Si indicates that vacancies are mainly present on the Si sublattice. Similar defect concentrations were found experimentally in $MoSi_2$ [41].

### 4.2.4. Conclusion

Bulk diffusion couple experiments are conducted between the group IVB elements Hf and Zr with Si. Only two phases, $MSi_2$ and MSi are found to grow in the interdiffusion zone. The thickness of the disilicide is much higher than monosilicide. Absence of other phases indicates much lower growth rate of these phases compared to the phases found in the interdiffusion zones. Time dependent experiments indicate the diffusion controlled growth of the phases. The integrated diffusion coefficient for $HfSi_2$, HfSi and $ZrSi_2$ are calculated at different temperatures. The same is not calculated for the ZrSi phase to avoid error associated in the calculation because of very thin layer. The grain morphologies revealed by acid etching show uniform columnar grains throughout the disilicide phase in both the systems. Based on the understanding in other systems, it indicates the presence of the Kirkendall marker plane at the Si/$MSi_2$ interface. The tracer diffusion coefficients of Si in disilicide are calculated using necessary thermodynamic parameters and activation energies. Si is the only diffusing species since vacancies are present mainly on the Si sublattice.

# Chapter 5

# Growth mechanism of tantalum silicides by interdiffusion

## 5.1. Introduction

Transition metal silicides are widely used in the microelectronics industry as interconnects and contacts [1] because of their low contact resistance and thermal stability, which are produced by the metal–silicon reactive diffusion process in the solid state. The presence of the oxide layer on Si might complicate the growth process. Refractory metals are proposed as the diffusion barrier layer in Cu interconnects between the Cu interconnect lines and the $SiO_2$/Si substrate [2–5]. Immiscibility with Cu even at elevated temperature makes them suitable [6]; however, silicides grow at the barrier metal and the substrate interface, which influences the performance of the product. They are also suitable for high-temperature applications because of their high strength, low density, high melting point and resistance to oxidation and creep [7, 8]. $Ta_5Si_3$ has a superior oxidation resistance compared to other 5:3 silicides.

A few interdiffusion studies on both thin-film and bulk conditions are available [8–10]. The interdiffusion process in the thin-film condition is complicated because of the stress generated during the deposition of films. Nonequilibrium phases might grow and sometimes a few phases might not grow because of nucleation problems. Since we

---

This chapter is written based on
S. Roy and A. Paul, Growth mechanism of tantalum silicides by interdiffusion, Philosophical Magazine, 92 (2012) 4215-4229.

are interested, at this point, in understanding the growth mechanism of the phases and the diffusion mechanism of the species, we shall consider only the interdiffusion process in the bulk condition. Christian et al. [8] studied the growth of the $Ta_5Si_3$ phase in Ta/$TaSi_2$ couples in the temperature range of 1000–1400 °C. $TaSi_2$ was deposited by the chemical vapor deposition process. They calculated the integrated diffusion coefficients of the $Ta_5Si_3$ phase using the relation developed by Wagner [11] and found the activation energy of 271 kJ/mol. Milanese et al. [9] studied the growth of the $TaSi_2$ and $Ta_5Si_3$ phases in Ta/Si couples in the temperature range of 1250–1350 °C. They calculated the parabolic growth constants and the average interdiffusion coefficients. The activation energies of diffusion were estimated as 560 and 450 kJ/mol for the $TaSi_2$ and $Ta_5Si_3$ phases, respectively. The difference between the activation energy values for the $Ta_5Si_3$ phase in these two articles is notable. Moreover, it will be explained later that the calculation methodology followed by Milanese et al. is questionable. Further, there is no knowledge on the relative mobilities of the species without which it is not possible to understand the atomic mechanism of the diffusion of the species. The growth mechanism of the phases has also not been studied.

Therefore the aim of this study is to conduct bulk diffusion couple experiments to calculate first the integrated and the tracer diffusion coefficients of the species. Afterwards, the growth mechanism of the phases will be discussed. In the end, the problem with the calculation procedure followed by Milanese et al. will be explained.

## 5.2. Results: calculation of diffusion parameters

Fig. 5.1 shows the back-scattered electron image (BSE) of the interdiffusion zone developed at 1250 °C in a Ta/Si diffusion couple. It can be seen that the $TaSi_2$ phase grows mainly in the interdiffusion zone; however, a very thin layer of the $Ta_5Si_3$ phase is also visible. The thicknesses of the $TaSi_2$ and $Ta_5Si_3$ phases are found to be of 52–145 μm and 0.9–4 μm, respectively, in the temperature range of 1200–1275 °C. Another two phases $Ta_2Si$ and $Ta_3Si$, which are present in the Ta–Si phase diagram [12], should have also grown in the interdiffusion zone, but these were not detected. It is possible that the growth rates of these two phases are much lower compared to the phases found. After etching, duplex morphology is seen in the $TaSi_2$ phase. Fine grains are found in the $Ta_5Si_3$ phase side whereas; relatively coarse grains can be seen in the Si side. Since the $Ta_5Si_3$ phase has a very fine structure because of a very thin layer, it is possible that the part of the $TaSi_2$ phase, which grows in this phase, also has a fine structure. On the other hand, when the same phase grows from Si single crystal, it has a coarse structure. The presence of duplex morphology indicates the location of the Kirkendall marker plane because the phase grows differently from two different interfaces [13-16]. Even this plane is also identified by the presence of a line of pores.

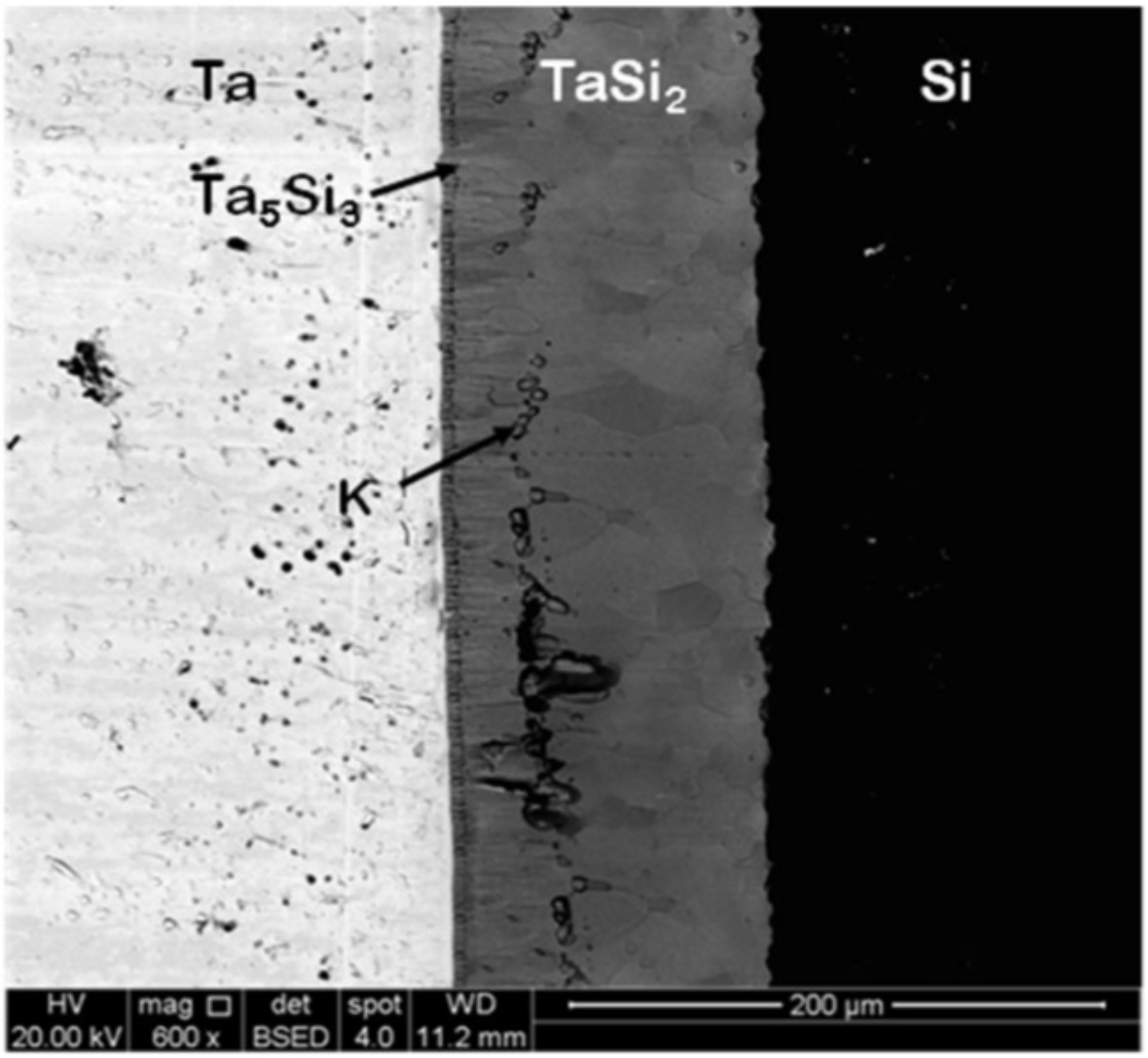

Fig 5.1:BSE image of the etched interdiffusion zone in the Ta/Si diffusion couple annealed at 1250 °C for 9 hours. K indicates the location of the Kirkendall marker plane.

Time-dependent experiments conducted by Milanese et al. [9] show the parabolic growth, that is, the diffusion controlled growth of the phases. However, as will be discussed later, the calculation procedure of the diffusion parameters was not followed correctly. We have used the well-established Wagner's relation to calculate the integrated diffusion coefficients, $\tilde{D}_{int}$ ($m^2/s$) since the phases have a very narrow homogeneity range. $\tilde{D}_{int}$ is the interdiffusion coefficient integrated over the unknown composition range. From the layer thickness data in Ta/Si diffusion couples, we calculated the diffusion parameters only in the $TaSi_2$ phase since the $Ta_5Si_3$ phase is not thick enough to calculate these parameters with sufficient accuracy. The Wagner's relation is expressed as Eqn 2.34. The molar volume of the product phases are calculated as $V_m^{TaSi_2} = 8.71$ and $V_m^{Ta_5Si_3} = 9.48$ cm$^3$/mol from the lattice parameter data available in the literature [10, 17].

When only a single-phase layer grows in the interdiffusion zone, $\tilde{D}_{\text{int}}$ can be expressed as

$$\tilde{D}_{\text{int}}^{\beta} = \frac{\left(N_i^{\beta} - N_i^{-}\right)\left(N_i^{+} - N_i^{\beta}\right)}{N_i^{+} - N_i^{-}} \frac{\Delta x_{\beta}^{2}}{2t} \tag{5.1}$$

In the Ta/Si couples, the layer thickness of the $Ta_5Si_3$ phase is very small compared to the thickness of the $TaSi_2$ phase. So Eq. 5.1 can be used to calculate the integrated diffusion coefficients of the $TaSi_2$ phase using their phase layer thickness and ignoring $Ta_5Si_3$ without introducing any significant error. The annealing temperatures, layer thickness and the integrated diffusion coefficients are listed in Table 5.1. The calculated $\tilde{D}_{\text{int}}$ values are shown in Fig. 5.2 with respect to the Arrhenius equation

$$\tilde{D}_{\text{int}}^{\beta} = \tilde{D}_{\text{int}}^{o} \exp\left(-\frac{Q}{RT}\right) \tag{5.2}$$

where $\tilde{D}_{\text{int}}^{o}$ (m$^2$/s) is the pre-exponential factor, Q (J/mol) is the activation energy and R is the universal gas constant (8.314 J/mol.K). The activation energy is estimated as 550 ± 70 kJ/mol.

In line compounds, we cannot calculate the intrinsic or the ratio of intrinsic diffusion coefficients because we cannot determine the slope of the composition profile and the partial molar volumes of the species, $V_i$. However, we can determine the ratio of the tracer diffusion coefficients using the relation developed by van Loo in Eqn. 4.2.2[18]

| **Temperature ( $^{o}$C)** | **Layer Thickness ($\mu$m)** | $\tilde{D}_{int}^{TaSi_2}$ **x $10^{-14}$** | $\frac{D_{Si}^*}{D_{Ta}^*}$ **in the $TaSi_2$ phase** |
|---|---|---|---|
| 1200 | 52 | 0.9 | 1.3 |
| 1225 | 70 | 1.7 | 1.2 |
| 1250 | 121 | 5.0 | 1.1 |
| 1275 | 145 | 7.2 | 1.1 |

(a)

| **Temperature ( $^{o}$C)** | **Layer Thickness ($\mu$m)** | $\tilde{D}_{int}^{Ta_5Si_3}$ **x $10^{-16}$** | $\frac{D_{Si}^*}{D_{Ta}^*}$ **in the $Ta_5Si_3$ phase** |
|---|---|---|---|
| 1200 | 3.1 | 0.3 | 2.9 |
| 1250 | 7.4 | 1.4 | 5.8 |
| 1300 | 12.2 | 3.5 | 8.9 |
| 1350 | 19.6 | 7.4 | 10.8 |

(b)

Table 5.1: Thickness of the phase layers, the integrated Diffusion coefficients and ratio of the tracer diffusivities $\frac{D_{Si}^*}{D_{Ta}^*}$ in the **(a)** $TaSi_2$ and **(b)** $Ta_5Si_3$ phases.

Note here that we have ignored the vacancy wind effect proposed by Manning [19], since the structure factors required are not known in these complex crystal structures. The ratios of tracer diffusivities, $D_{Si}^* / D_{Ta}^*$, are listed in Table 5.1.

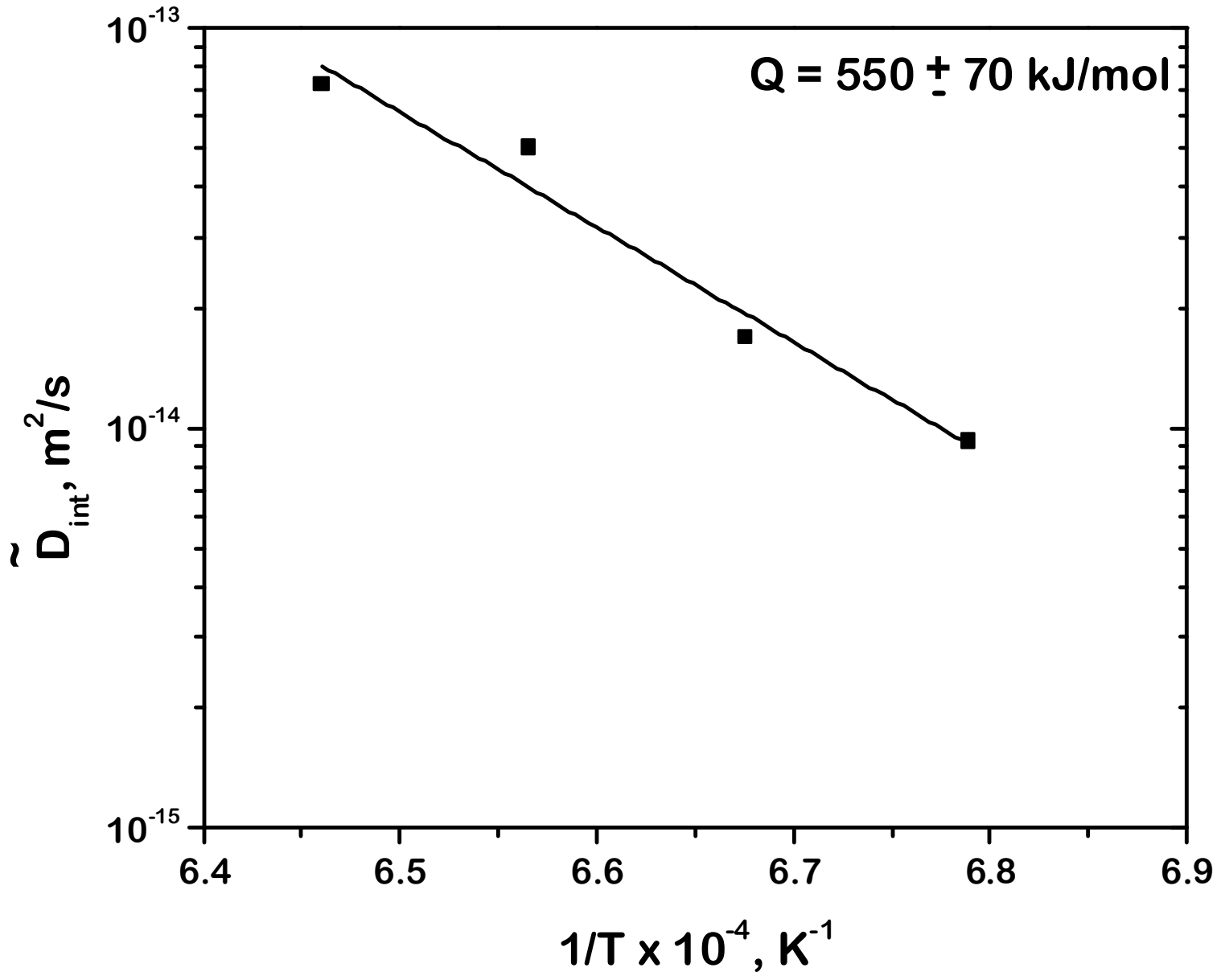

Fig 5.2: Arrhenius plot of the integrated diffusion coefficients calculated for the $TaSi_2$ phase.

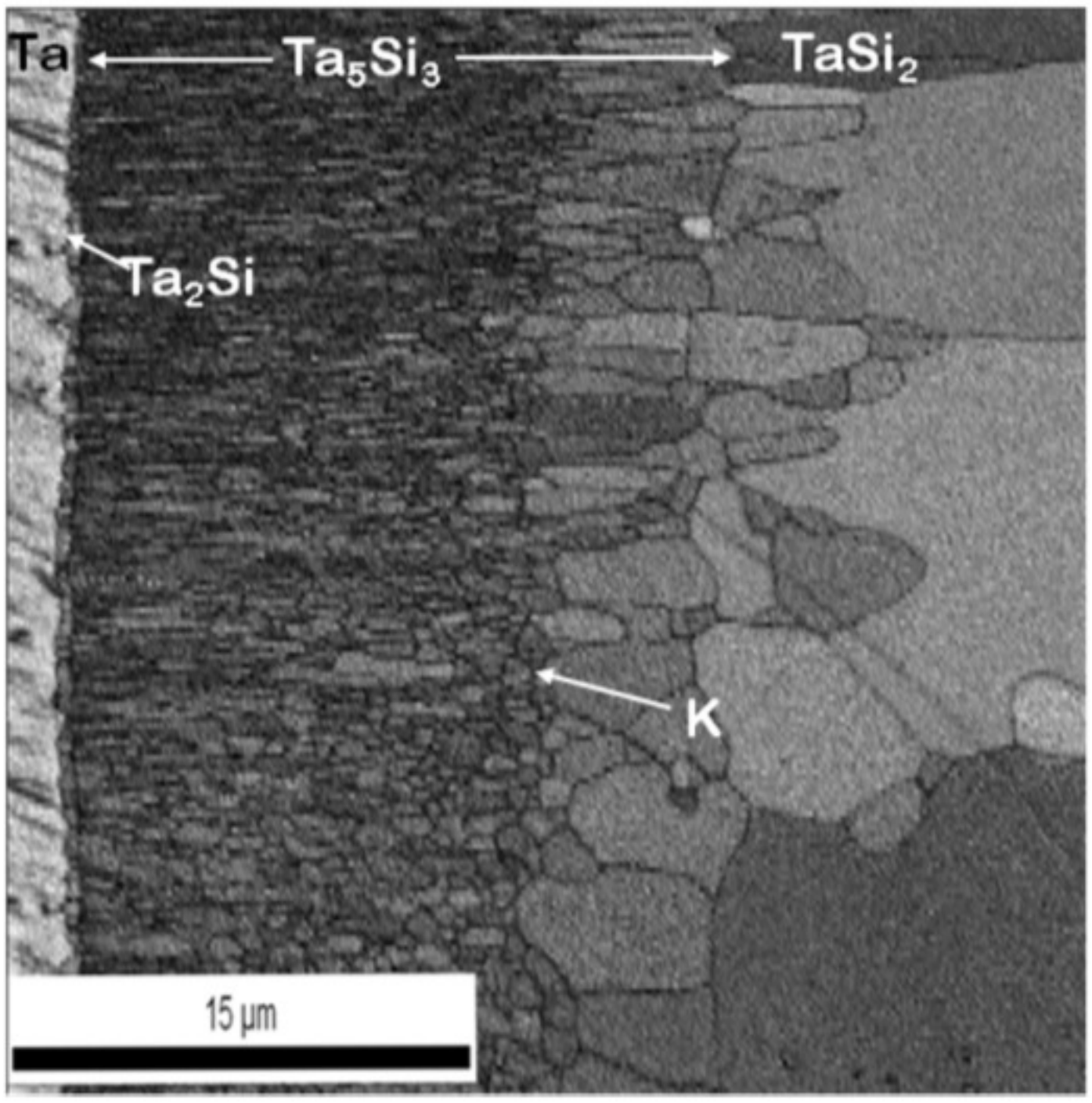

Fig 5.3: EBSD image of the $Ta_5Si_3$ phase in the Ta/$TaSi_2$ diffusion couple annealed at 1350 °C for 16 hours. K indicates the location of the Kirkendall marker plane.

It can be seen that the ratios calculated are more or less the same at different temperatures in the range of 1.1-1.3. Since the ratio does not change much with the temperature, it indicates that the tracer diffusion coefficients of different species change almost at the same rate with the annealing temperature. Since the layer thickness of the $Ta_5Si_3$ phase is very small in the Ta/Si diffusion couple, we conducted incremental diffusion couple experiments between $TaSi_2$ and Ta so that the $Ta_5Si_3$ phase as grows a reasonably thick layer in the interdiffusion zone. It should be noted here that 0.9 μm of the phase layer was already present before the annealing of the incremental couples. The thickness of the phase layer varied in the range of 3.1–19.6 μm (after correcting for the initial layer that was already present before the annealing of the incremental couples). The interdiffusion zone developed at 1350 °C after 16 h of annealing and is shown in Fig. 5.3. A very thin layer of $Ta_2Si$ is present at the $Ta/Ta_5Si_3$ interface and this was not detectable in the Ta/Si diffusion couple. This indicates that the growth rate of this phase layer is much lower compared to other phases found. In this couple, inert particles were not used; however, the location of the Kirkendall plane can be detected from the presence of duplex morphology. Since this phase is very much thicker than the $Ta_2Si$ phase, we can again use Eq. 5.1 for the calculation of the integrated diffusion coefficients. The initial layer thickness that was already present before the start of the annealing of the incremental diffusion couple is considered. This is calculated from $\left[\Delta x_\beta^2 - \left(\Delta x_\beta^o\right)^2\right]/2t$ instead of $\Delta x_\beta^2/2t$, where $\Delta x_\beta^o$ is the initial layer thickness. The integrated diffusion coefficients calculated are listed in Table 5.1 and shown in Fig. 5.4 with respect to the Arrhenius equation. The activation energy is estimated as 410 ± 39 kJ/mol. Later, the ratio of tracer diffusion coefficients are calculated using Eq. 4.2.2 and are listed in Table. 5.1. The initial layer thickness that was already present before the start of the annealing process is deducted from the left hand side sublayer, since the right hand side sublayer from the marker plane grows only after the start of the interdiffusion process in the incremental couple. This argument will be clear once we discuss the growth mechanism of the phases in the next section. It can be seen that Si has a faster diffusion rate; however, the ratio increases with an increase in the annealing temperature. This indicates that the rate of change of the tracer diffusion coefficients of the species with temperature is different for different elements.

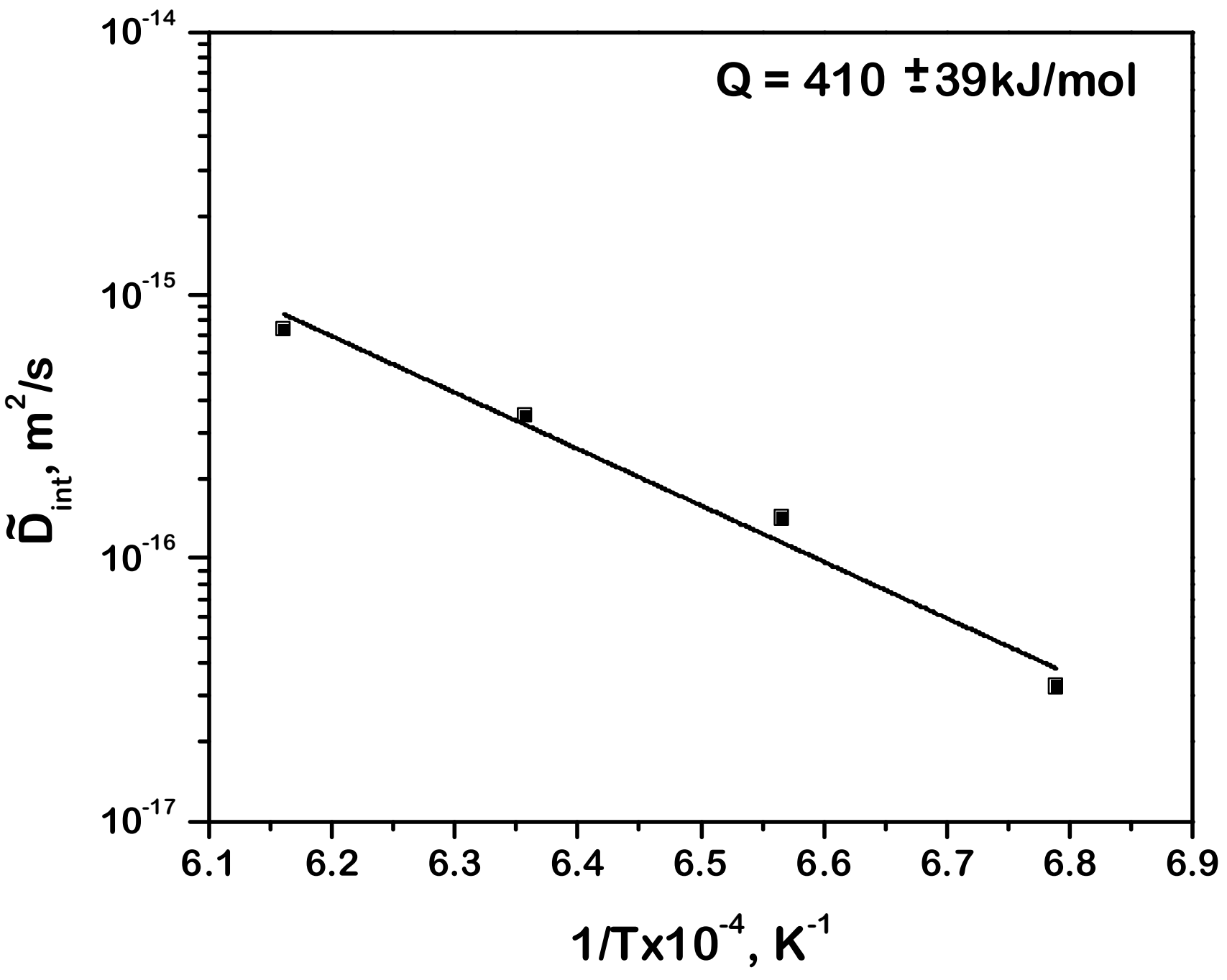

Fig 5.4: Arrhenius plot of the integrated diffusion coefficient calculated in the $Ta_5Si_3$ phase.

The estimated activation energies for integrated diffusion coefficients are in very good agreement with the activation energy for interdiffusion coefficients calculated by Milanese et al. However, they followed a method developed by Bulcagila and Anselmi-Tamburini [20], which is questionable. It is not possible to compare the results since different diffusion parameters are calculated; however, the problem can be understood from the relations used by Milanese et al.

$$\tilde{D}_i = k_i^{II}\left[\frac{v_i^2(v_{i-1}-v_{i+1})^2}{(v_{i-1}-v_i)^2\,(v_i-v_{i+1})^2}\,\frac{1+v_i}{v_i}\,\frac{\left|\Delta G_i^0\right|}{RT}\right]^{-1} \tag{5.3a}$$

$$\tilde{D}_i = \frac{1}{1+v_i}\overline{D}_{Ta,i} + \frac{v_i}{1+v_i}\overline{D}_{Si,i} \tag{5.3b}$$

where, $k_i^{II}$ is the parabolic growth constant of the second kind, $\tilde{D}_i$ is the interdiffusion coefficient of the $i^{th}$ phase, $\overline{D}_{Ta,i}$ and $\overline{D}_{Si,i}$ are the self diffusion coefficients of Ta and Si, respectively in the $i^{\text{th}}$ phase. $\Delta G_i^0$ is the Gibbs free energy change for the formation of 1 mol of phase $A_{v_i}B$, $v_i$ is the mole fraction of A of the of the $i^{\text{th}}$ phase $A_{v_i}B$, $v_{i-1}$ and $v_{i+1}$ are the mole fraction of element A of the adjacent phases in the phase diagram. The problems with the above equations are the following: As discussed before [21], it should be noted that to calculate $\tilde{D}_i$ as can be seen in the Matano-Boltzmann relation or the relations proposed by others [18], one needs the concentration gradient. However, in the relations above (Eq. 5.3a) the concentration gradient term is missing. This is almost impossible to determine it in line compounds and Wagner [11] introduced the concept of the integrated diffusion coefficient, which is used in this study. $\tilde{D}_i$ is directly related to the self diffusion coefficients (Eq. 5.3b), which is not correct. It should be noted that intrinsic or interdiffusion coefficients are related to the tracer diffusion coefficients and not to the self diffusion coefficients. Further, interdiffusion happens under the thermodynamic driving force, which is not considered in Eq. 5.3b.

It is not necessary to consider the thermodynamic parameter to calculate the interdiffusion, integrated or intrinsic diffusion coefficients. These parameters can be directly calculated from the composition profiles. The unnecessary introduction of this parameter (in Eq. 5.3a) is rather a source of error. The different free energy values determined in different articles will give different diffusion parameters. Moreover, Milanese et al. calculated two different interdiffusion coefficients in the $Ta_5Si_3$ phase considering total 2 (what we actually find in the interdiffusion zone) and 4-phase layers (according to the phase diagram) in the interdiffusion zone. However, it should be noted that the diffusion coefficients are actually material constants and should have the same value irrespective of the total number of phase layers grown in the interdiffusion zone.

## 5.3. Discussion

### 5.3.1 Atomic mechanism of diffusion

The atomic mechanism of diffusion in intermetallic compounds is complex since atoms cannot jump randomly like in solid solution. Till date only a very few systems have been studied extensively, such as Ni aluminides [22-24] to understand the diffusion mechanism. The advantage of considering these phases is their relatively simple crystal structure. The knowledge on defect concentrations is also available. In most of the phases, the knowledge on defect concentration is not available. However, diffusion rates determined experimentally indicate the kinds of defects present in the structure. Based on the available studies, it is already proven that different sublattices can have different concentrations of vacancies. Even at stoichiometric compositions, a very high concentration of antisites could be present. In many phases, the nearest neighbor bonds are not present between minor elements; however, significant diffusion is still possible because of presence of antisites.

$TaSi_2$ has a hexagonal C40 structure (hP9), as shown in Fig. 5.5a [25, 26]. As listed in Table 5.2, Ta atoms are surrounded by 6 Si atoms on the same plane and 4 Si atoms on two adjacent planes. On the other hand, Si atoms are surrounded by 3 Ta and 3 Si atoms on the same plane and 2 Ta and 2 Si atoms on two adjacent planes. That means that Ta atoms are surrounded by only Si atoms and Si atoms are surrounded by both Si and Ta atoms in the nearest neighbor positions. This indicates that even if there are no Si antisites present, the diffusion of this element is possible if vacancies are present on the same sublattice. On the other hand, in the absence of Ta antisites, Ta cannot diffuse because it cannot exchange positions with the vacancies on the next neighbor Si sublattice. After exchanging positions, it cannot rest on a wrong sublattice. Ta can diffuse only if Ta antisites are present. So, it is not unusual to find Si as faster diffusing species through this phase. However, note that Ta has a comparable diffusion rate with respect to Si since the ratio of tracer diffusivities, $D^*_{Si} / D^*_{Ta}$ is determined in the range of 1.1−1.3.

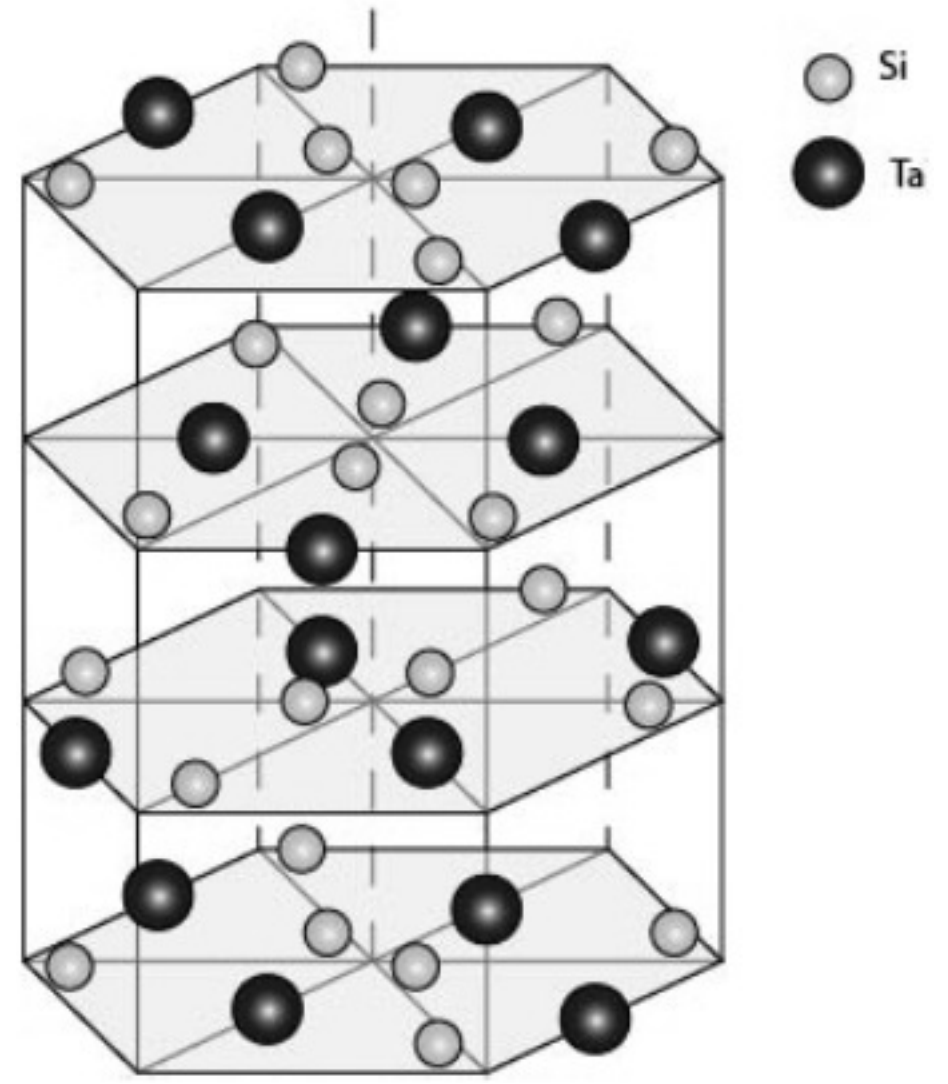

(a)

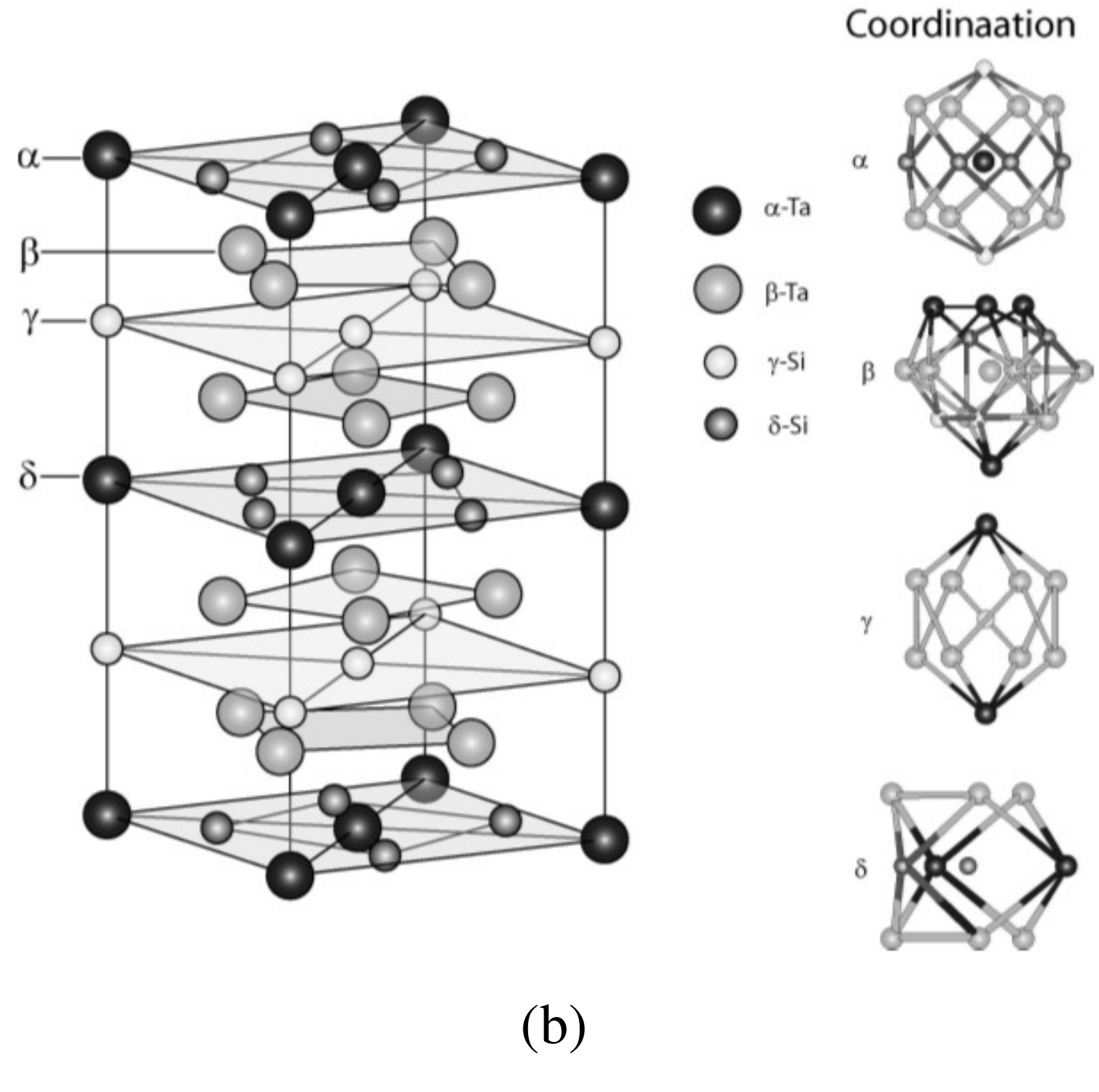

(b)

Fig 5.5 (a): Crystal Structure of the $TaSi_2$ phase, C40 (hp9). **(b)**: Crystal Structure of the α - $Ta_5Si_3$ phase, tI32 (D8l) showing α, β, γ and δ sublattices and their nearest neighbors.

This indicates the presence of a high concentration of Ta antisites. In the $Au_2Bi$ and $AuSb_2$ phases, the minor element had a higher diffusion rate than the major elements [27]. The $Ta_5Si_3$ phase has two different stable crystal structures in different temperature range. Above 2160 °C, the $W_5Si_3$-type $D8_m$ $\beta -$ $Ta_5Si_3$ phase is stable. Below this temperature, the $Cr_5B_3$-type $D8_l$ $\alpha -$ $Ta_5Si_3$ phase is stable. In the temperature range of our study, the $\alpha -$ phase is stable. It has basically a tetragonal structure with 32 atoms (tI32) in the unit cell, as shown in Fig. 5.5b [17]. It has 4 sublattices $\alpha, \beta, \gamma$ and $\delta$ occupied by Ta1, Ta2, Si1 and Si2. As listed in Table 5.2, Ta1 site is surrounded by 8 Ta and 6 Si atoms, Ta2 is surrounded by 11 Ta and 5 Si atoms. Si1 is surrounded by 10 Ta atoms and Si2 is surrounded by 8 Ta and only 1 Si atom. So Ta atoms are surrounded by many Ta and Si atoms. That means these atoms can exchange positions with their own sublattice depending on the availability of vacancies on the $\alpha$ and $\beta$ sublattices. On the other hand, Si has only one Si nearest neighbor. So it is expected that if only thermal vacancies are present, then the diffusion rate of Si should be lower than the diffusion rate of Ta.However, according to the data calculated in this phase, Si has a higher diffusion rate than Ta and the ratio of tracer diffusivities $D^*_{Si} / D^*_{Ta}$ increases with an increase in temperature in the range of 2.9−10.8. This further indicates that the activation energies for diffusion of these two elements are not the same. Further, there must be a high concentration of Si antisites present in the structure.

### 5.3.2 Growth mechanism of the phases

As it is already shown in Fig. 5.1, $TaSi_2$ has much higher thickness as compared to the $Ta_5Si_3$ phase. Moreover, duplex morphology, that is, the marker plane, is present in the $TaSi_2$ phase.

| Atoms in $TaSi_2$ crystal structure | Nearest neighbour | | |
|---|---|---|---|
| | Same plane | Adjacent planes | Total |
| Ta | 6 Si | 4 Si | 10 Si |
| Si | 3 Ta, 3 Si | 2 Ta, 2 Si | 5 Ta, 5 Si |

(a)

| Atoms in $Ta_5Si_3$ crystal structure | Nearest neighbour | | |
|---|---|---|---|
| | Ta | Si | Total |
| α - Ta | 8 β-Ta | 2 γ-Si, 4 δ -Si | 14 |
| β - Ta | 7 β -Ta, 4 α-Ta | 3 γ-Si, 2δ -Si | 16 |
| γ - Si | 2 α-Ta, 8 β-Ta | 0 | 10 |
| δ -Si | 2 α-Ta, 6 β-Ta | 1 δ-Si | 9 |

(b)

Table 5.2: Nearest neighbours in the (a) $TaSi_2$ and (b) $Ta_5Si_3$ phases

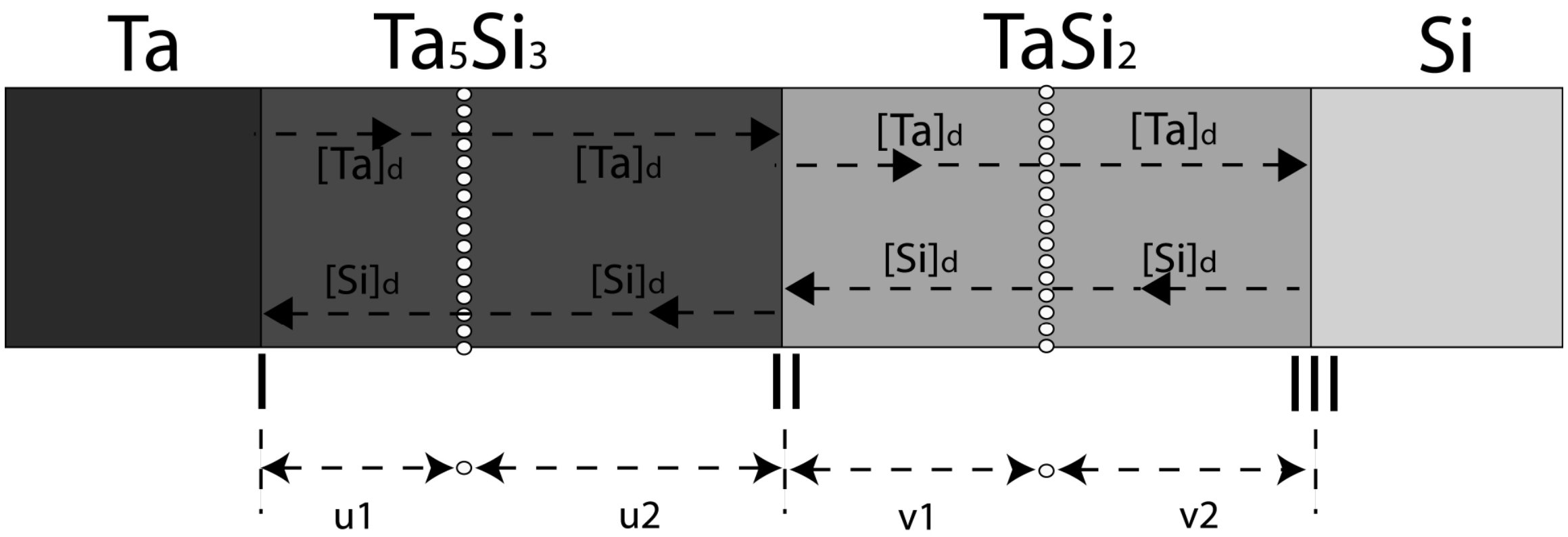

(a)

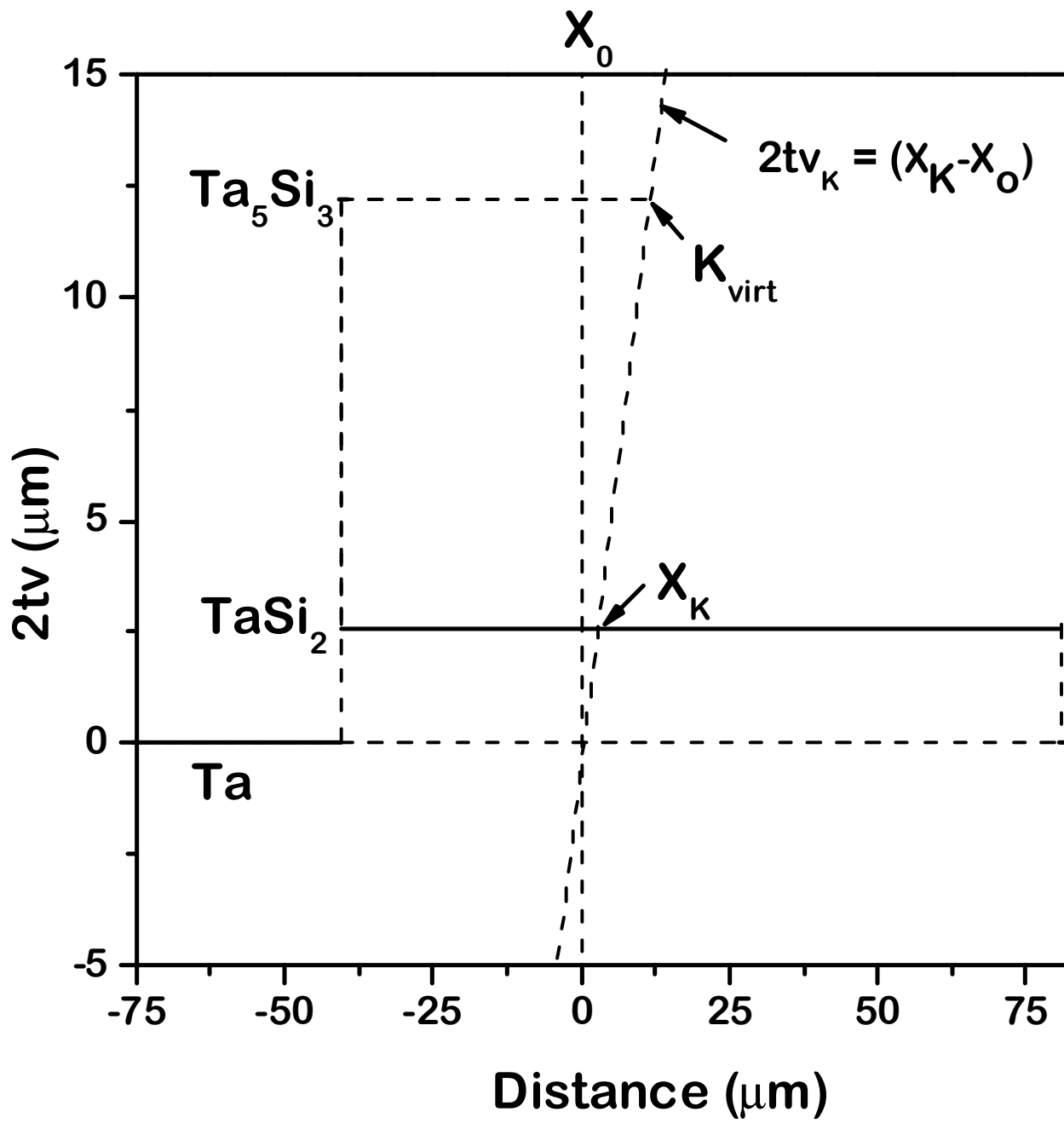

(b)

Fig 5.6: (a): Schematic representation of the dissociation and reaction at different interfaces are shown in the Ta/Si diffusion couple. $[Ta]_d$ and $[Si]_d$ are the diffusing species. (b): Velocity diagram constructed for Ta/Si diffusion couple at 1250 $^oC$. K is the Kirkendal marker plane in the $TaSi_2$ phase and $K_{virt}$ is the location of the virtual Kirkendal marker plane in the $Ta_5Si_3$ phase.

To gain an idea of the growth mechanism of the phases, let us consider the schematic representation, as shown in Fig. 5.6a of the Ta/Si diffusion couple at 1250 °C. From our calculation we have seen that both the elements could diffuse in both the phases. Since the other phases ($Ta_2Si$, $Ta_3Si$) were not detectable, we ignore the presence of these phases. Both the phases grow together and we need to examine the growth with respect to the diffusion of elements and the reaction/dissociation simultaneously. From the schematic representation, it must be clear that Ta diffuses from interface I through the $Ta_5Si_3$ phase and reacts with the $TaSi_2$ phase at interface II to produce the $Ta_5Si_3$ phase. At interface II, Si dissociates from the $TiSi_2$ phase to produce the $Ta_5Si_3$ phase. Dissociated Si diffuses through the $Ta_5Si_3$ phase and reacts with Ta at interface I to produce the $Ta_5Si_3$ phase. At the same time, Ta dissociates from $Ta_5Si_3$ to produce the $TaSi_2$ phase at interface II. Dissociated Ta diffuses through the $TaSi_2$ phase to react with Si and produce the $TaSi_2$ phase. So, it must be clear that very complex reactions and dissociation processes happen at interface II [13-16]. The $Ta_5Si_3$ phase grows by consuming the $TaSi_2$ phase and, at the same time, the $Ta_5Si_3$ phase gets consumed because of the growth of the $TaSi_2$ phase. The growth of the phases is diffusion controlled and the reaction/dissociation process is not the rate-limiting step. However, the reaction/dissociation schemes explain the growth process. Let us first consider that both the sublayers could grow from interface II, so that both the phases have two sublayers. That means we are considering the presence of the Kirkendall marker plane in both the phases. It should be noted here that the diffusion parameters are the material constants. This means that the data calculated from the incremental diffusion couple can be used to understand the growth process in the Ta/Si diffusion couple. We consider the thickness of the sublayers in the $Ta_5Si_3$ phase as $u_1$ and $u_2$ and the thickness of the sublayers in the $TaSi_2$ phase as $v_1$ and $v_2$. The integrated diffusion coefficients of the phases with the thickness of the sublayers at 1250 °C using Eq. 2.34 are

$$\tilde{D}_{int}^{Ta_5Si_3} = 1.41\times10^{-16} m^2/\mathrm{s} = \frac{1}{t}\left[0.117u^2 + 0.068uv\right]\times10^{-12} \qquad (5.4a)$$

$$\tilde{D}_{int}^{TaSi_2} = 5.02\times10^{-14} m^2/s = \frac{1}{t}\left[0.111v^2 + 0.057uv\right]\times10^{-12} \qquad (5.4b)$$

Further, the ratio of the tracer diffusivities at 1250 °C can be related as

$$\left.\frac{D^*_{Si}}{D^*_{Ta}}\right|_{Ta_5Si_3} = 5.8 = \left[\frac{\frac{3}{8}u1}{\frac{1}{3}v\frac{9.48}{8.71} + \frac{5}{8}u2}\right] \qquad (5.5c)$$

$$\left.\frac{D^*_{Si}}{D^*_{Ta}}\right|_{TaSi_2} = 1.1 = \left[\frac{\frac{3}{8}u\frac{8.71}{9.48} + \frac{2}{3}v1}{\frac{1}{3}v2}\right] \qquad (5.5d)$$

where $u_1 + u_2 = u$ and $v_1 + v_2 = v$ (5.5e)

The calculated values of integrated diffusion coefficients and the ratio tracer diffusion coefficients at 1250 °C, as listed in Table 1, are used to derive these relations. Considering annealing time of 9 h, we solve the equations to get the values as $u_1$ = 64.41, $u_2$ = − 63.51, $u_3$ = 42 and $u_4$= 79 μm. Note here the negative value of $u_2$. To explain the growth process further, we need to draw the velocity diagram. The velocity of the phase, $\beta$ can be determined by [28]

$$v_\beta = \frac{V_{Ta}}{V_m} \frac{\frac{D_{Si}}{D_{Ta}} - 1}{\left(\frac{V_{Ta} D_{Si}}{V_{Si} D_{Ta}}\right) N_{Ta} + N_{Si}} \frac{\tilde{D}_{\text{int}}}{\Delta x_\beta} \tag{5.6}$$

In a line compound, the partial molar volumes, $V_i$ are not known and we consider $V_i = V_m$. So we can write that the ratio of intrinsic diffusion coefficients $\frac{D_{Si}}{D_{Ta}}$ is equal to the $\frac{D^*_{Si}}{D^*_{Ta}}$ (see **Eq. 4.2.2**). From Eq. 5.6, we calculate the velocity values as $1.87\text{x}10^{-10}$ *m/s* and as $4.02\text{x}10^{-11}$ *m/s* for the $Ta_5Si_3$ and the $TaSi_2$ phases, respectively. These are drawn with respect to 2tv in Fig. 5.6b. Following, it is possible to draw the straight line $v_K = (x_K - x_o)/2t$ passing through initial contact plane, $x_o$ [13]. We plot the velocity diagram as *2tv* vs. *x* plot. Since the line $2tv_K = (x_K - x_o)$ intersects once with the velocity line of the $TaSi_2$ phase, it indicates the presence of one marker plane in this phase only. When we extend the velocity line of the $Ta_5Si_3$ phase to $2tv_K = (x_K - x_o)$, the intersection point indicates the location of the virtual Kirkendall marker plane of this phase. This is basically the marker plane that would be present in this phase if the $Ta_5Si_3$ phase layer would not get consumed from the interface II. In fact the negative value of the sublayer thickness, $u_2$ indicates that this sublayer of the $Ta_5Si_3$ phase is consumed by the $TaSi_2$ phase. The total layer thickness of the $Ta_5Si_3$ phase is $(u_1+u_2)$ = 0.9 µm. It indicates that not only $u_2$ but part of the sublayer grows from the $Ta/Ta_5Si_3$ interface also gets consumed. This is the reason we do not have any marker plane in this phase and we should not therefore expect duplex morphology in this phase.

## 5.4. Conclusion

In this manuscript, we determined the required diffusion parameters to understand the atomic mechanism of diffusion and the growth mechanism of the phases. The ratios of the diffusivities required for the analysis were not available and we determined these values at the Kirkendall marker planes. The location of the marker planes were detected from the duplex morphology developed during the interdiffusion process. Although Milanese et al. [9] did not determine the ratio of diffusivities, they determined the interdiffusion coefficients. However, they followed a questionable method, as discussed before.

Therefore, following Wagner's approach [11], we first estimated the integrated diffusion coefficients and then calculated the ratio of diffusivities using the relation developed by van Loo [18]. The outcome of this manuscript can be summarized as

- Activation energies estimated for diffusion in this study is very close to the data calculated by Milanese et al. [9]. Christian et al. [8] calculated the activation energy in the $Ta_5Si_3$ phase. However, their estimated data (271 kJ/mol) is very less than the data estimated in our study (410±39 kJ/mol). The reason for this difference is not clear to us. In both the studies, incremental diffusion couple experiments were followed but the techniques were different. Christian et al. deposited the $TaSi_2$ phase by CVD method and then annealed at the desired temperatures. This might be the reason to find the difference; however, it is difficult for us to make any further comment at this point. Further, the activation energies calculated in this system is much higher than the values estimated in other refractory metal-silicon systems [20, 29, 30]. This could be because of high activation energy for the formation of defects in this system compared to other systems.

- The growth rate of the $Ta_5Si_3$ phase is much lower than that of the $TaSi_2$ phase. This is the reason to find almost two orders of magnitude lower integrated diffusion coefficient of the $Ta_5Si_3$ phase compared to the $TaSi_2$ phase.
- The locations of the marker planes were identified from the duplex morphology in the phase layers. The calculated ratios of tracer diffusivities calculated at these planes indicate that Si is the faster diffusing species in both the phases. It is rather surprising to find comparable diffusion rates of the species in the $TaSi_2$ phase, since Ta-Ta bonds are not present. This indicates the presence of a high concentration of Ta antisites. The higher diffusion rate of Si in the $Ta_5Si_3$ phase is also unusual if we consider the nearest neighbors of the atoms as explained in section 4.1. Similar unusual behavior was found in the Nb–Si [29] and Mo–Si [30] systems. These indicate the structural defects present in the crystals. It is also unusual to find that $\frac{D^*_{Si}}{D^*_{Ta}}$ is lower in the Si rich phase, $TaSi_2$ compared to the Si lean phase, $Ta_5Si_3$.
- From our analysis of the growth, it is understood that the $Ta_5Si_3$ phase could not grow with reasonable thickness because of the high growth rate of the $TaSi_2$ phase in the Ta/Si diffusion couple, which consumes most of the $Ta_5Si_3$ phase.

- The velocity diagram and the negative value of the thickness of one of the sublayers calculated in the $Ta_5Si_3$ phase in the Ta/Si diffusion couple indicate that the Kirkendall marker plane should not be present in this phase. This further indicates that duplex morphology should not be present in this phase. On the other hand, the positive values of both the sublayers in the $TaSi_2$ phase indicate the presence of the Kirkendall marker plane and the duplex morphology in this phase.

# Chapter 6

# Diffusion in tungsten silicides

## 6.1. Introduction

Tungsten–silicon is one of the important refractory metal–silicon systems to study for its potential use in integrated circuits [1]. $WSi_2$ layer, which has low electrical resistivity and good thermal stability, is grown by the reactive diffusion process. Alloys in this system could also be useful in high temperature applications because of their excellent strength at high temperatures. $WSi_2$ is also considered as protective coating on W based alloys because of its excellent oxidation resistance [2-4]. This is most probably the only metal-silicon system which is studied extensively, both in thin film [3, 5-12] and bulk conditions [13-15] to understand the diffusion controlled growth of the phases. In all the studies till date, mainly the growth kinetics and the activation energies for the growth are calculated. However, it is necessary to calculate the diffusion parameters to understand the diffusion mechanism. To author`s knowledge, the integrated diffusion coefficients in the $W_5Si_3$ phase were estimated, in three different studies. However, the relative mobilities of the species were not estimated without which it is not possible to understand the atomic mechanism of diffusion. The Kirkendall marker experiments were conducted earlier in the thin film condition, and Si was reported to be the faster diffusing species in

---

This chapter is written based on
S. Roy and A. Paul, Diffusion in tungsten silicides, Intermetallics 37 (2013) 83-87.

the $WSi_2$ phase [12]. However, study in this condition might not be suitable since stress developed during deposition could play an important role on diffusing species and the growth of the phases. Further, the ratio of the diffusivities was not calculated, which indicates the diffusion rates of both the species.

So the bulk diffusion couple experiments are conducted in this study to estimate first the integrated diffusion coefficients at different temperatures. Following this, the activation energies for diffusion are calculated. Ratios of tracer diffusion of elements are estimated at the Kirkendall marker plane. The atomic mechanism of diffusion with the help of crystal structures are then discussed.

## 6.2. Results and discussion

### 6.2.1 Diffusion in the $WSi_2$ phase

Mainly the $WSi_2$ phase was found to grow in the W/Si diffusion couples. The SEM image of the interdiffusion zone annealed at 1225 °C for 9 hours is shown in Fig. 6.1a. The composition was confirmed by spot analysis in an electron probe microanalyzer. The thicknesses of the phase layer at different temperatures are listed in Table 6.1a. $W_5Si_3$ was present at the $W/WSi_2$ interface but with negligible thickness. Since the product phases grow as line compounds, we cannot estimate the interdiffusion coefficients. Therefore, the integrated diffusion coefficients, $\tilde{D}_{int}$ , which is the interdiffusion coefficient integrated over the unknown composition range of the phase is calculated using the relation developed by Wagner [16]. This is expressed as Eqn 5.1 (for the only one product phase β in the interdiffusion zone)

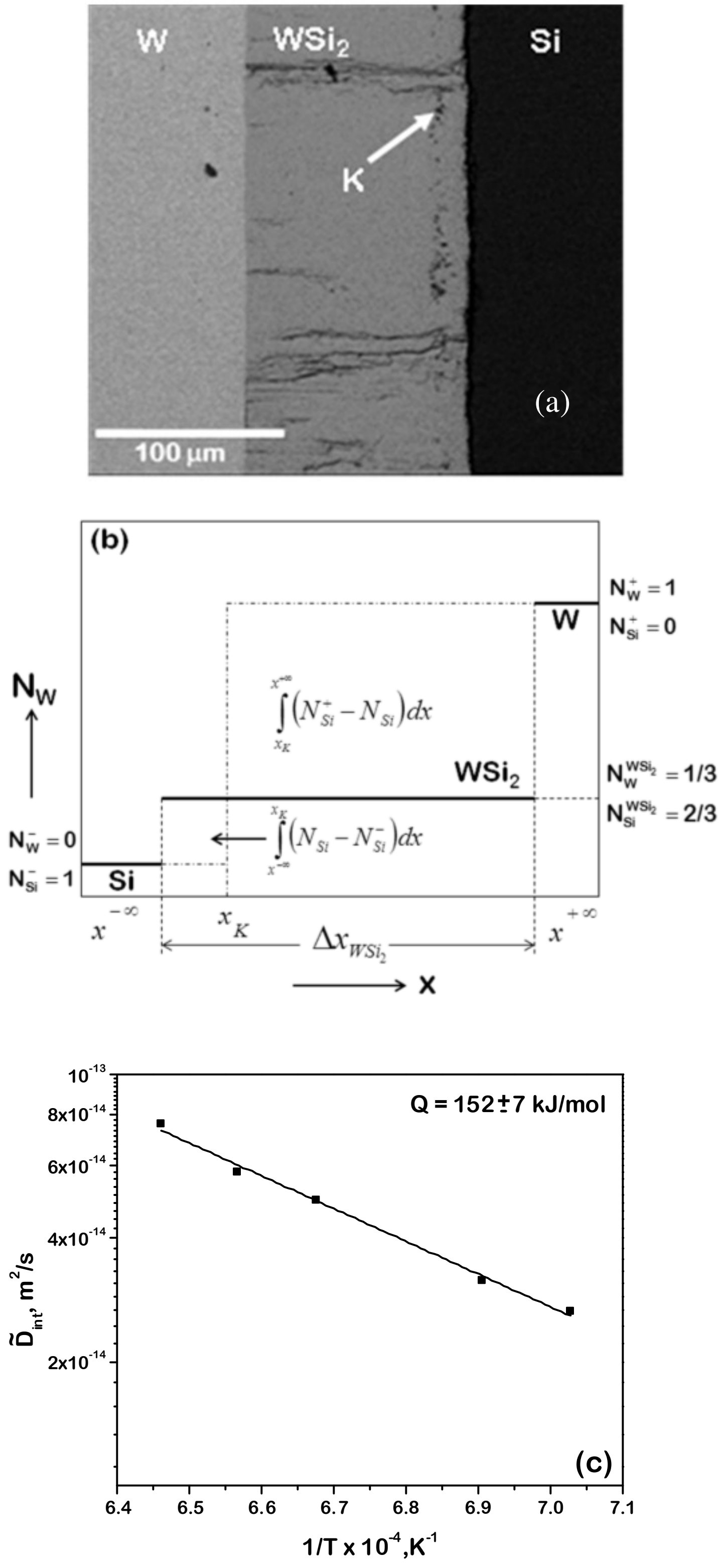

Fig 6.1**:** (a) BSE image of the interdiffusion zone in the W/Si diffusion couple annealed at 1498 (1225 °C) for 9 hours, (b) Schematic profile explain the calculation. (c) Arrhenius plot of the integrated diffusion coefficients calculated for the $WSi_2$ phase.

| Temperature ( $^{o}$C) | Layer Thickness (μm) | $\tilde{D}_{int}^{WSi_2}$ x $10^{-14}$ ($m^2/s$) |
|---|---|---|
| 1150 | 88 | 2.65 |
| 1175 | 96 | 3.16 |
| 1225 | 118 | 4.77 |
| 1250 | 130 | 5.79 |
| 1275 | 149 | 7.61 |

**(a)**

| Temperature ( $^{o}$C) | Layer Thickness (μm) | $\tilde{D}_{int}^{W_5Si_3}$ x $10^{-17}$($m^2/s$) |
|---|---|---|
| 1150 | 2.1 | 0.89 |
| 1250 | 3.7 | 2.78 |
| 1300 | 6.8 | 9.40 |
| 1350 | 10 | 20.3 |

**(b)**

Table 6.1: Thickness of the phase layers and the integrated Diffusion coefficients in the (a) $WSi_2$ and (b) $W_5Si_3$ phases.

Note that this particular relation for one product phase in the interdiffusion zone could be used since $W_5Si_3$ phase has negligible thickness in the W/Si couples. Since the product phase is a line compound, we need only the thickness of the product phase (as listed in Table 6.1a) to calculate the $\tilde{D}_{int}$ at different temperatures, as explained with the help of a schematic profile in Fig. 6.1b. The estimated values are listed in Table 6.1a and plotted in Fig. 6.1c with respect to the Arrhenius equation, expressed as Eqn 5.2. We cannot compare the diffusion coefficients calculated in this study with the results published previously reporting the parabolic growth constant of the phase. However, when a single line compound grows in the interdiffusion zone, it can be understood from the Eqn. 5.1 that the activation energies calculated from the integrated diffusion coefficient and the parabolic growth constant are the same. The activation energy is estimated as 152±7 kJ/mol in our study from the integrated diffusion coefficients. This value is reported in the range of 159-209 kJ/mol [3, 8, 11, 14]. The Kirkendall marker experiments failed since the couples did not join, when inert $TiO_2$ or $Y_2O_3$ particles were used at the interface. However, it is well known fact that the position of the Kirkendall plane still can be recognized based on the line of pores present in the interdiffusion zone [17], as it is found in this couple in Fig. 6.1a. In line compounds, it is not possible to determine the ratio of intrinsic diffusion coefficients because of unknown partial molar volumes. However, the ratio of tracer diffusion coefficients can be determined at the Kirkendall marker plane using the relation developed by van Loo [17], which expressed as Eqn 4.2.2. Note that we have neglected the vacancy wind effect proposed by Manning [18], since the data on structure factor is not available in this complex crystal structure. The contribution of the vacancy wind effects are discussed previously based on the experimental studies [18-21]. The areas expressed by integrals in Eq. 4.2.2 are shown in Fig. 6.1b. The total thickness of the phase layer at 1225 °C is 118 μm. The thickness of the sublayers on the Si side of the Kirkendall marker plane is 15 μm and on the W side is 103 μm. Following we estimated $\frac{D_{Si}^*}{D_W^*} = 13.7$ at the Kirkendall marker plane at this temperature. In the ordered phases, atoms cannot jump randomly and the diffusion process is very complex. Ni-aliminides [22-24] have been studied extensively to understand the atomic mechanism of diffusion based on the information on different kinds of defects. Based on these studies, it is now known that different concentration of vacancies could be present on different sublattices and antisites defects could be present in relatively high concentration even at the stoichiometric composition. Further, in the similar compound (same crystal structure) in different systems concentration of defects could be different, leading to different rates of diffusion [22]. Similar behavior was found in silicides also, where relative mobilities of the species could be very different in different systems [25-29]. Although, details on the defects are not available in the W-Si system, diffusion mechanism still can be discussed based on the mechanism understood in different aluminides. As shown in Fig. 6.2, $WSi_2$ has tetragonal C11b structure (tI6) [30-31].

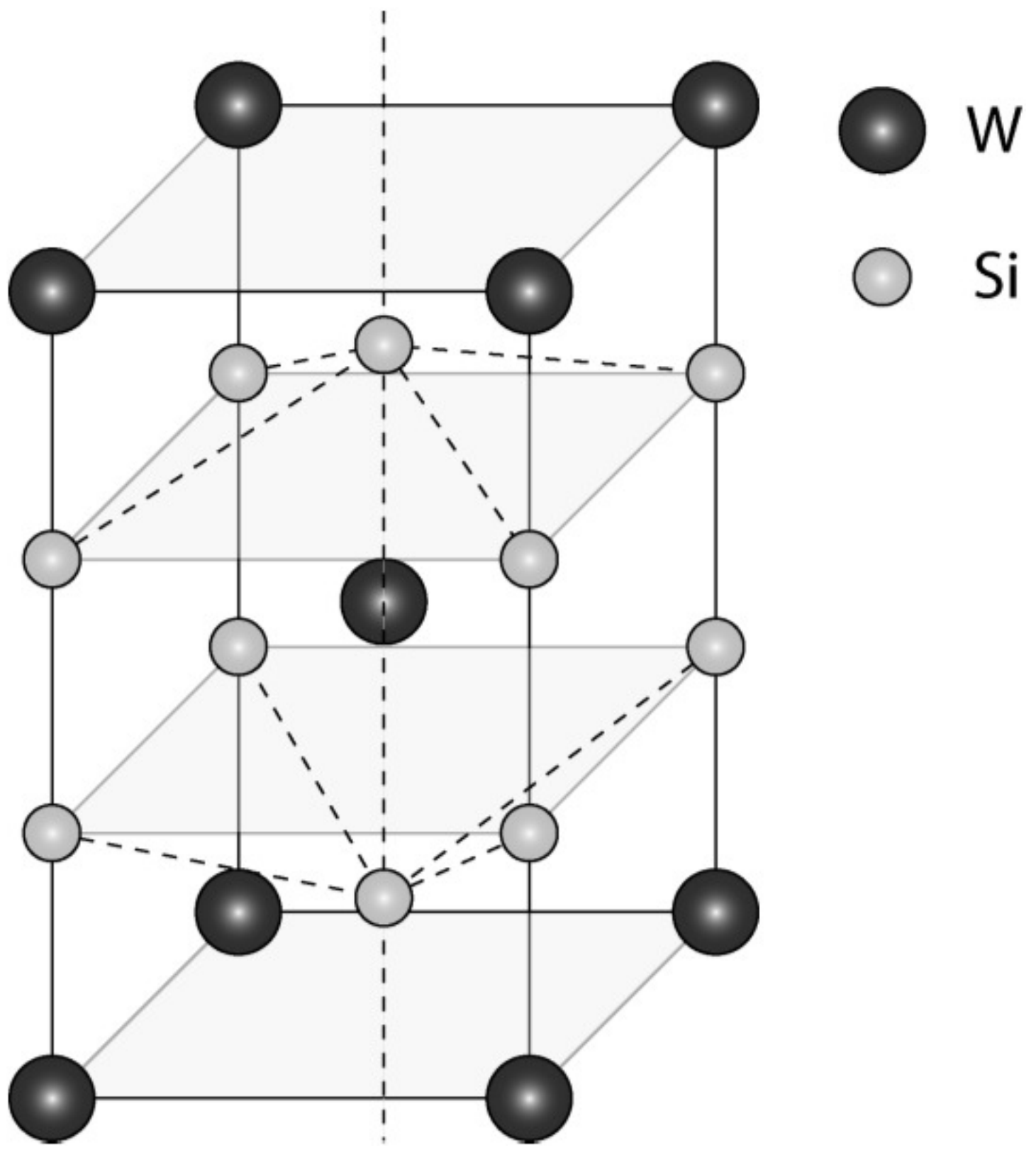

Fig 6.2: Crystal Structure of the $WSi_2$ phase with C11b (tI6) crystal structure.

W atoms are surrounded by 10 Si atoms. Si atoms, on the other hand, are surrounded by 5 W and 5 Si. So W atoms are surrounded by only Si atoms and Si atoms are surrounded by both Si and W atoms. This indicates that Si can diffuse even if no Si antisites are present. Presence of the antisites will increase the diffusion rate. On the other hand, W cannot diffuse easily since it is surrounded by only Si atoms. Exchanging position with the vacancies on the Si sublattice is not possible since then W will be at the wrong sublattice. W can diffuse only if W antisites are present. This could be the reason to find higher diffusion rate of Si compared to W. Note here that we are considering the antisite defect assisted diffusion. Without the presence of antisites also diffusion of elements is possible, as explained in Refs. [23, 32, 33]. The sublayer in the Si side of the Kirkendall plane grows because of diffusion of W. In this phase, Si has around 13.7 times higher diffusion rate compared to W. In $NbSi_2$ [25] and $TaSi_2$ [29] the ratio $\left(D_{Si}^{*} / D_{M}^{*} ; M = \text{metal species}\right)$ was found to be 4.8±1.4 and 1.2±0.1, respectively. On the other hand, in $VSi_2$ [26] and $MoSi_2$ [27], the same were estimated as infinity. This might indicates the presence of Nb antisites in the $NbSi_2$ phase and even higher concentration of Ta antisites in the $TaSi_2$ pahse. On the other hand V and Mo antisites might be absent in the $VSi_2$ and $MoSi_2$ phases. Unusual behavior was noticed in some other $AB_2$ compounds. In the $Au_2Bi$ and $AuSb_2$, unlike the silicides, the minor elements had higher diffusion rates [34].

### 6.2.2 Diffusion in the $W_5Si_3$ phase

Since the thickness of the $W_5Si_3$ phase was negligible in the W/Si couple, experiments in $WSi_2$/W incremental couples were conducted. Interdiffusion zone of the couple annealed at 1623 K (1350 °C) for 16 hours is shown in Fig. 6.3a.

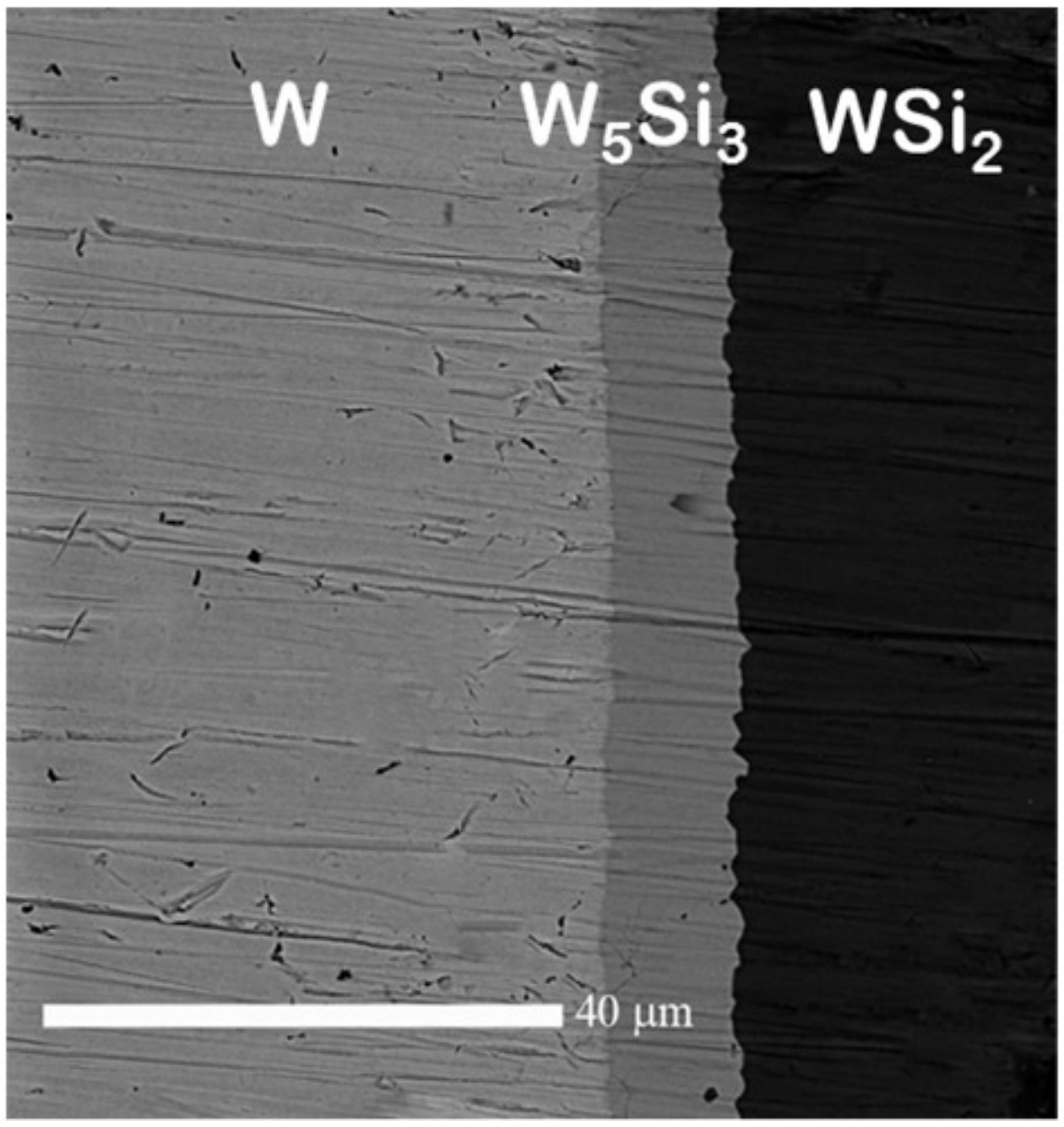

**(a)**

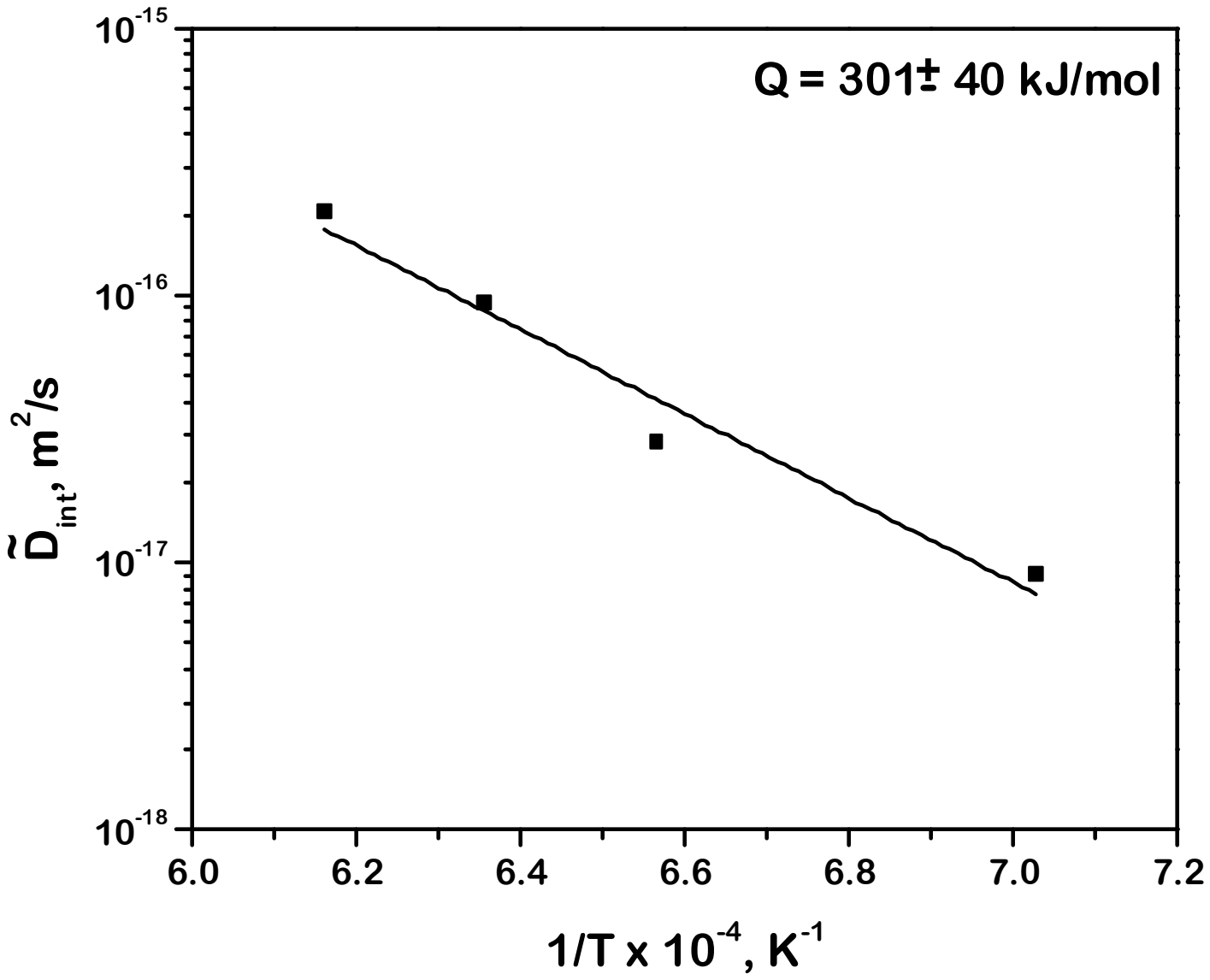

**(b)**

Fig 6.3: BSE image of the interdiffusion zone in the W/$WSi_2$ diffusion couple annealed at 1623 (1350 °C) for 16 hours. (b) Arrhenius plot of the integrated diffusion coefficients calculated for the $W_5Si_3$ phase.

The thicknesses of the phase layer at different temperatures are listed in Table 6.1b. Following similar method, as explained for the $WSi_2$ phase, $\tilde{D}_{int}$ at different temperatures is estimated using Eq. 5.1. These are listed in Table 1b and plotted in Fig. 3b with respect to Arrhenius equation. The activation energy is estimated as 301±40 kJ/mol. The values found in literature are in the range of 289 – 360 kJ/mol [13-15]. We could not locate the marker plane in our experiments. However, Lee et al. [13] have shown the location of this plane in their W/$WiSi_2$ diffusion couple at 1400 °C, although they did not calculate the ratio of tracer diffusion coefficients. Using Eq. 4.2.2, we estimated the value as $\frac{D^*_{Si}}{D^*_W} = 11.9$. To understand the diffusion mechanism based on this calculation, let us consider the crystal structure of this phase. It has tetragonal $D8_m$ structure with 32 atoms (tI32) in the crystal, as shown in Fig. 6.4 [35]. It has 4 sublattices α, β, γ and δ, occupied by W1, W2, Si1 and Si2. W1 is surrounded by 10 W and 4 Si. W2 is surrounded by 9 W and 6 Si. Si1, on the other hand, is surrounded by 8 W and 2 Si. Si2 is surrounded by 10 W and only 2 Si. That means both W and Si have many nearest neighbor bonds of the same elements. However, since Si has higher diffusion rate, it is expected that the concentration of vacancies on the Si sublattice must be higher than the vacancies present on the W sublattice. Moreover, the relatively smaller size of Si might be another reason to find higher diffusion rate of this species. In the $Ta_5Si_3$, this ratio $\left(D^*_{Si}/D^*_M ; M = \text{metal species}\right)$ of diffusivities varies in the range of 2.9–10.8. In the $Nb_5Si_3$ and the $Mo_5Si_3$ phases, this ratio was estimated as 31±15 [25] and 103±40 [27], respectively.

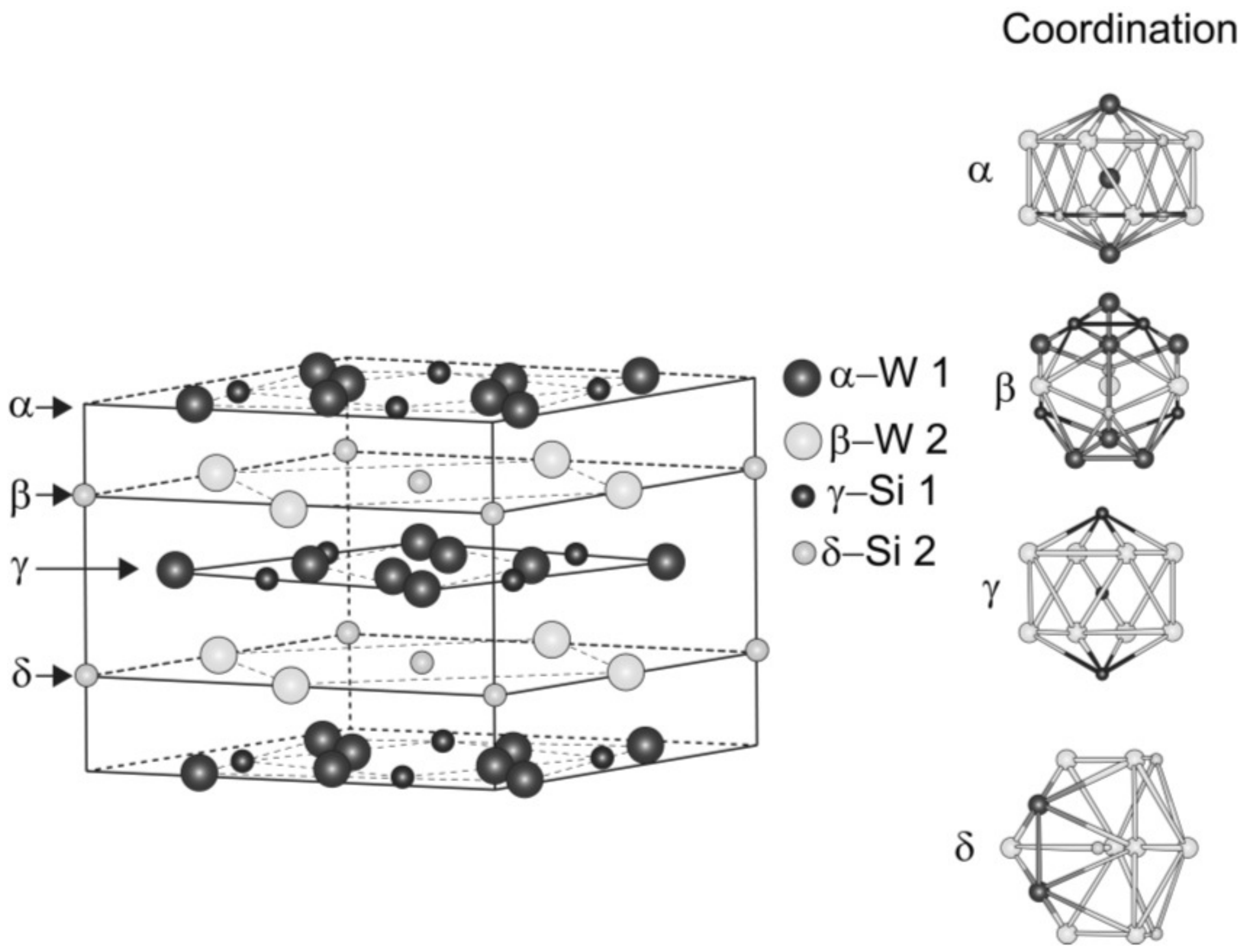

Fig 6.4: Crystal Structure of the $W_5Si_3$ phase, tI32 (D8m), showing α,β,γ and δ sublattices.

## 3. Conclusion

Bulk diffusion couple experiments were conducted to understand the diffusion mechanism in the W-Si system. In the W/Si diffusion couples, only the $WSi_2$ phase grows with measurable thickness in the interdiffusion zone. Si has higher diffusion rate compared to W in this phase, which is rather expected. The presence of W antisites is apparent since it diffuses through the phase layer. W/$WSi_2$ incremental diffusion couple experiments were conducted to study the growth of the $W_5Si_3$ phase. In this phase also Si has higher diffusion rate compared to W. This indicates the presence of high concentration of vacancies on the Si sublattice. Further discussion on the atomic mechanism of diffusion is not possible because of lack of knowledge on bond strengths or the concentration of defects on different sublattices.

# Chapter 7

# Comparison of silicides in different groups in the periodic table

## 7.1. Introduction

As explained in the previous chapter, it is already known from the experimental analysis and theoretical calculations that diffusion on atomic level in ordered phases is mainly assisted with two types of defects, vacancies and antisites. Moreover, different concentrations of the defects are typically present on different sublattices. Often, this is the reason of very different rates of components' diffusion in a phase. The aim of this chapter is to systemize the diffusion behavior of elements with respect to the atomic numbers in different groups. We are considering mainly the diffusion behavior in $MSi_2$ phase in group IVB, and $MSi_2$ and $M_5Si_3$ phases and VB and VIB refractory metal-silicon system. The atomic mechanism of diffusion is then analyzed accounting for the atomic arrangements in the crystal. Following, the trends in diffusion behavior of elements in different systems are examined and they are related to the different concentrations of defects in different systems based on the estimated diffusion parameters as explained in the previous chapter.

---

This chapter is written based on

S. Roy, S. Prasad, S. Divinski and A. Paul, Diffusion pattern in MSi2 and M5Si3 silicides in group VB (M = V, Nb, Ta) and VIB (M = Mo, W) refractory metal-silicon systems Philosophical Magazine 94 (2014) 1508-1528

## 7.2. Results and discussion

In the beginning, we compare the diffusion parameters in the phases of our interest group wise. Following we shall discuss the atomic mechanism of diffusion and possible defects present on different sublattices.

### 7.2.1 Comparison of the growth of the phases in different groups

As already discussed on the growth of the phases in different systems from group IVB refractory metal–silicon systems, we have only one phase that is $MSi_2$, in which the integrated diffusion coefficients and the ratio of diffusivities are estimated.

It can be seen in Fig. 7.1 that in all the systems, the marker plane is located at the $M/MSi_2$ interface. Therefore, Si diffusion rate is much higher compared to the metal species.

In the group VB refractory metal silicon systems, as already discussed, we have studied the Ta/Si system. For the sake of comparison, we consider the results already available in the V-Si [1, 2] and Nb-Si systems [3]. All the phases present in the phase diagram are grown in the interdiffusion zone as shown in Fig. 7.2a [1, 3]. Experiments were conducted in the temperature range of 1150-1300 $^{o}C$ for 16 hrs at the interval of 50 $^{o}C$ in this range. Etched samples reveal continuous morphology of the $VSi_2$ phase. Based on the previous analysis on several systems, we know that the marker plane should be present at any of the interfaces of $VSi_2$ sides. Since, Si is the faster diffusing species in most of the disilicides phase; the position of the Kirkendall plane must be present at the $V/VSi_2$ interface. This indicates that the diffusion rate of V is negligible compared to that of Si. Similar behavior was found in the Mo-Si system [4], which could be validated based on tracer diffusion experiments [5-8].

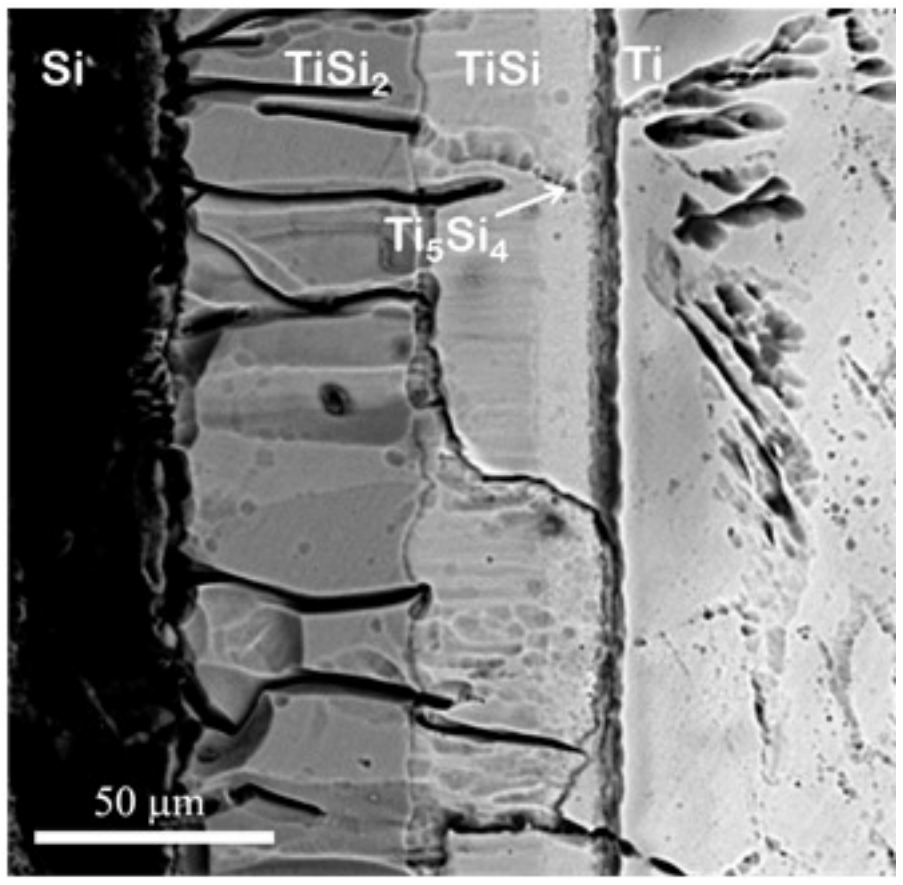

(a)

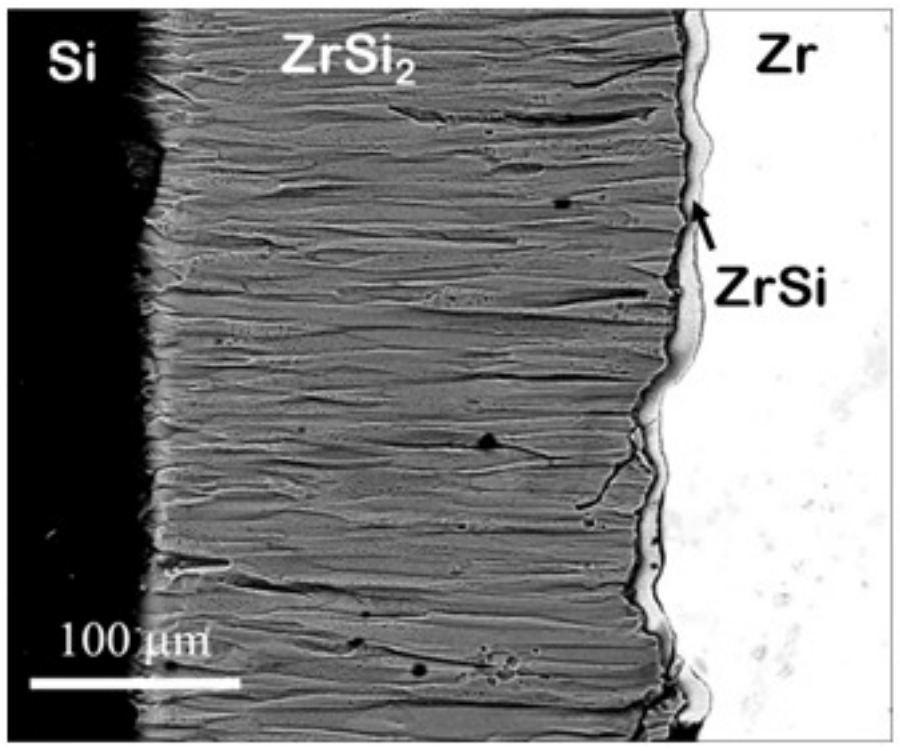

(b)

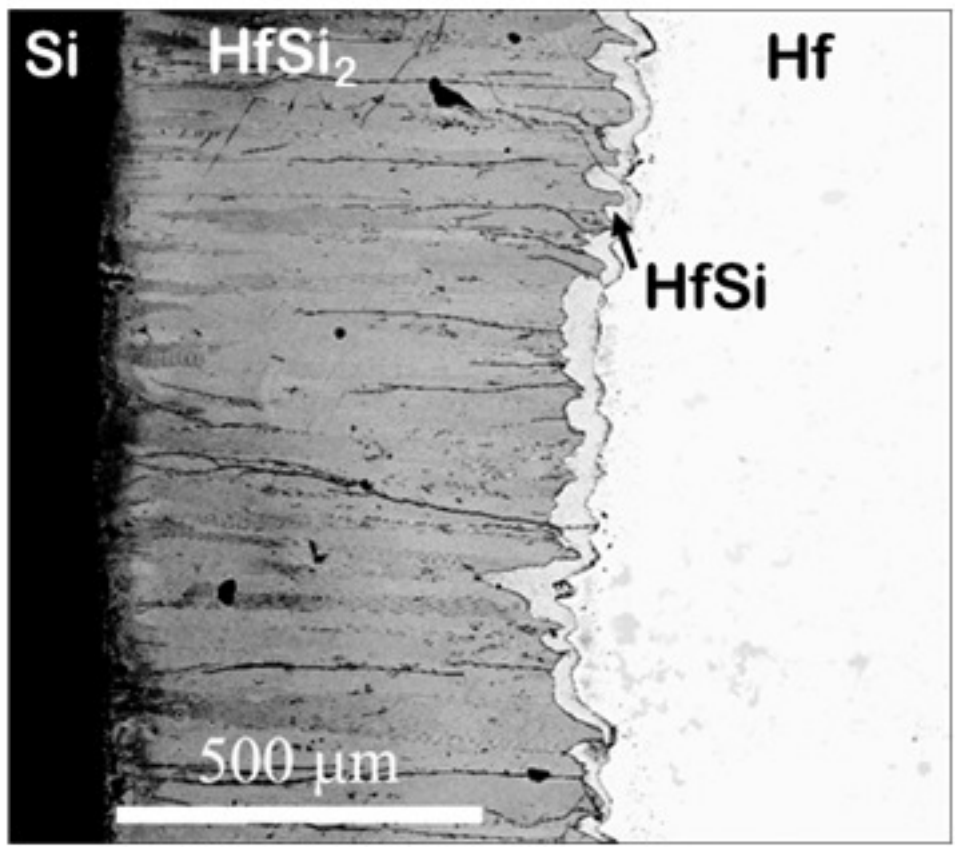

(c)

Fig 7.1: Scanning electron micrograph of etched (a)Ti/Si diffusion couple annealed at 1200 $^oC$ for 16 hrs (b) Zr/Si diffusion couple annealed at 1200 $^oC$ for 16 hrs (c) Hf/Si diffusion couple annealed at 1250 $^oC$ for 16 hrs

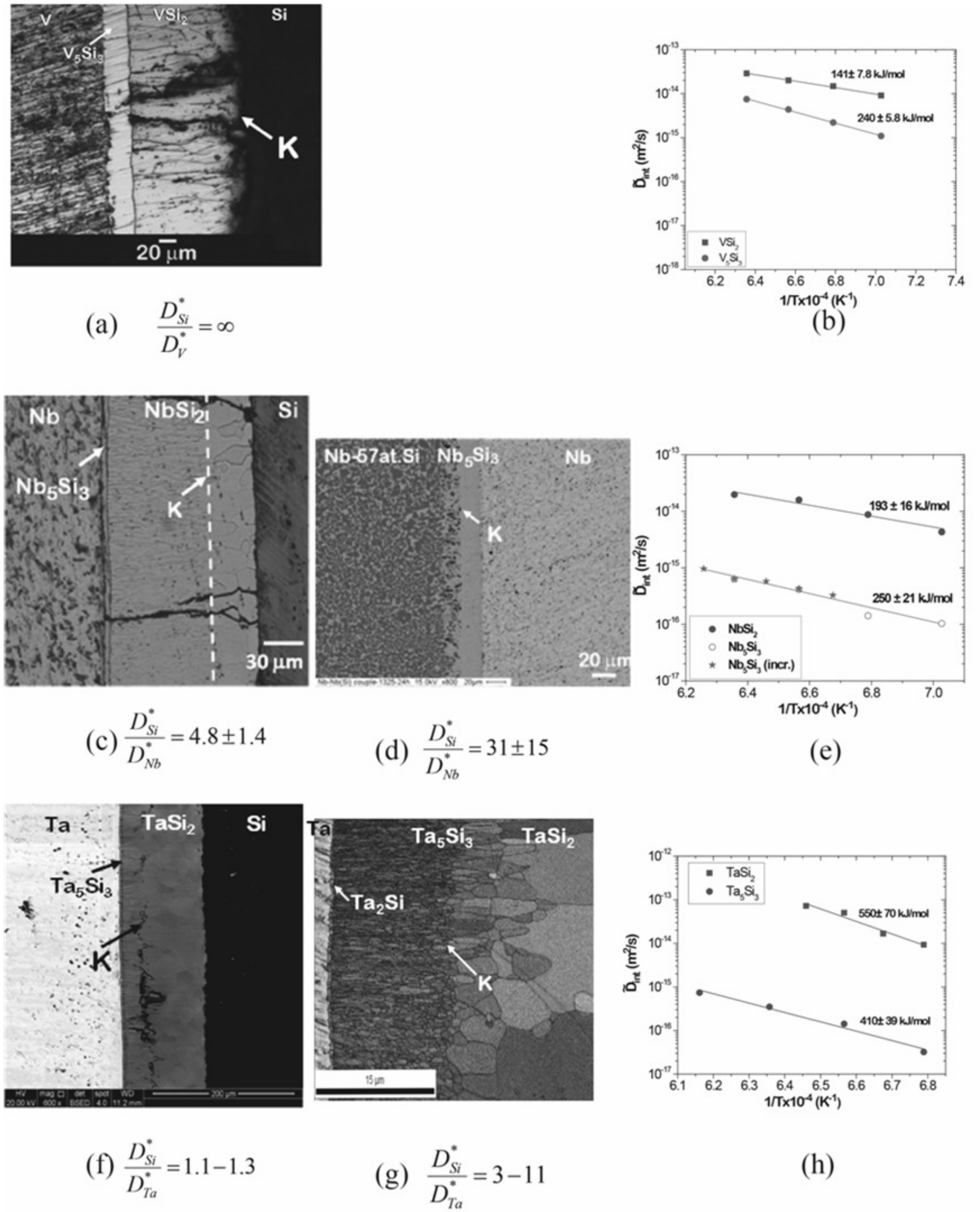

Fig 7.2: (a) Scanning electron micrograph of etched V/Si diffusion couple annealed at 1200°C for 16 hrs (b) Arrhenius plot of the integrated diffusion coefficients calculated for $VSi_2$ and $V_5Si_3$ phases (c) Nb/Si diffusion couple annealed at 1250°C for 24 hrs (d) Incremental diffusion couple of Nb57at%Si/Si annealed at 1325°C for 24 hrs **[3]** (e) Arrhenius plot of the integrated diffusion coefficients calculated for $NbSi_2$ and $Nb_5Si_3$ phases (f) Scanning electron micrograph of etched Ta/Si diffusion couple annealed at 1250°C for 9 hrs (g) Incremental diffusion couple of $TaSi_2$/Si annealed at 1350°C for 16 hrs (h) Arrhenius plot of the integrated diffusion coefficients calculated for $MSi_2$ and $M_5Si_3$ phases.

The integrated diffusion coefficients, $\tilde{D}_{\text{int}}$ for the phases of interest in this manuscript are shown in Fig. 7.2b. As expected $\tilde{D}_{\text{int}}$ values are higher for $VSi_2$ compared to $V_5Si_3$. On the other hand, the activation energies, Q is lower for $VSi_2$ compared to $V_5Si_3$. Since Si is much faster diffusing species through the $VSi_2$ phase, the calculation of $D^*_{Si}/D^*_V$ gives the value of infinity. In the Nb-Si system, only two phases, $NbSi_2$ and $Nb_5Si_3$ are present and both the phases are grown in the interdiffusion zone, as shown in Fig. 7.2c [3]. These were annealed in the temperature range of 1150-1300 ºC at the interval of 50 ºC. $Nb_5Si_3$ grows as thin layer compared to $NbSi_2$. Samples were etched to reveal the grain morphology. Duplex morphology in $NbSi_2$ indicates the location of the Kirkendall marker plane. To grow the $Nb_5Si_3$ phase with reasonable thickness, incremental diffusion couple experiments of (Nb-57 at.% Si)/Nb were conducted in the range of 1150-1300 ºC. $TiO_2$ particles were used to locate the Kirkendall marker plane, as can be seen in Fig. 7.2d. $\tilde{D}_{\text{int}}$ calculated for both the phases are shown in Fig. 7.2e. As expected, in the case of the $Nb_5Si_3$ phase, $\tilde{D}_{\text{int}}$ values are the same when calculated using the thickness from diffusion couple prepared with pure elements or incremental couple. Further, the activation energy for diffusion in $Nb_5Si_3$ is higher than $NbSi_2$. Interestingly, the marker plane moves inside the $NbSi_2$ phase compared to the $VSi_2$ [1, 2] phase, which indicates comparable diffusion rates of Nb. The ratio of diffusivities is calculated as $\frac{D^*_{Si}}{D^*_{Nb}} = 4.8 \pm 1.4$ in $NbSi_2$ and $\frac{D^*_{Si}}{D^*_{Nb}} = 31 \pm 15$ in $Nb_5Si_3$. Very high error in the calculated data in the $Nb_5Si_3$ phase is apparent. In fact, it is very common to find a high error in the calculation from a diffusion couple, when the ratios of diffusivities fall outside the range of 0.1-10. For the sake of comparison, we show the interdiffusion zone in the Ta-Si system again in Fig. 7.2f. Since $Ta_5Si_3$ phase grows with very small thickness, incremental couples of Ta/$TaSi_2$ were prepared after removing Si as explained in the experimental method. In this, another phase $Ta_2Si$ appeared along with a relatively thick layer of $Ta_5Si_3$. Duplex morphologies in the phases indicate the locations of the marker plane. It can be seen that the marker plane move further inside in the $TaSi_2$ phase compared to that in the $NbSi_2$ phase. This indicates that relative mobilities of the metal component increases compared to Si. The ratio of diffusivities is found to be $\frac{D^*_{Si}}{D^*_{Ta}} = 1.1 - 1.3$ in this phase. The ratio of diffusivities, $\frac{D^*_{Si}}{D^*_{Ta}}$ vary in the range of $3 - 11$ in the $Ta_5Si_3$ phase. In this phase also $\frac{D^*_{Si}}{D^*_{Ta}}$ decreases compared to the relative mobilities in $Nb_5Si_3$. This indicates relatively higher diffusion rate of the metal component compared to Si. $\tilde{D}_{int}$ determined for both the phases are presented in Fig. 7.2h. Surprisingly, unlike other systems, $TaSi_2$ has higher activation energy compared to the $Ta_5Si_3$ phase.

In the group VIB refractory metal silicon systems, I have studied the Cr-Si and W-Si systems, whereas, results on Mo-Si systems were taken from the literature for the sake of comparison. We have, at present, conducted only one experiment in the Cr-Si system, as shown in Fig. 7.3a. There are four phases present in this system [9] out of which $CrSi_2$ and $Cr_5Si_3$ grow with higher thickness. Presence of uniform grain morphology in the $CrSi_2$ phase reveals the presence of the Kirkendall plane at the Si/$CrSi_2$ interface. Therefore, it indicates that the $CrSi_2$ phase grows by the diffusion of Si. Cr has negligible diffusion rate through the phase.

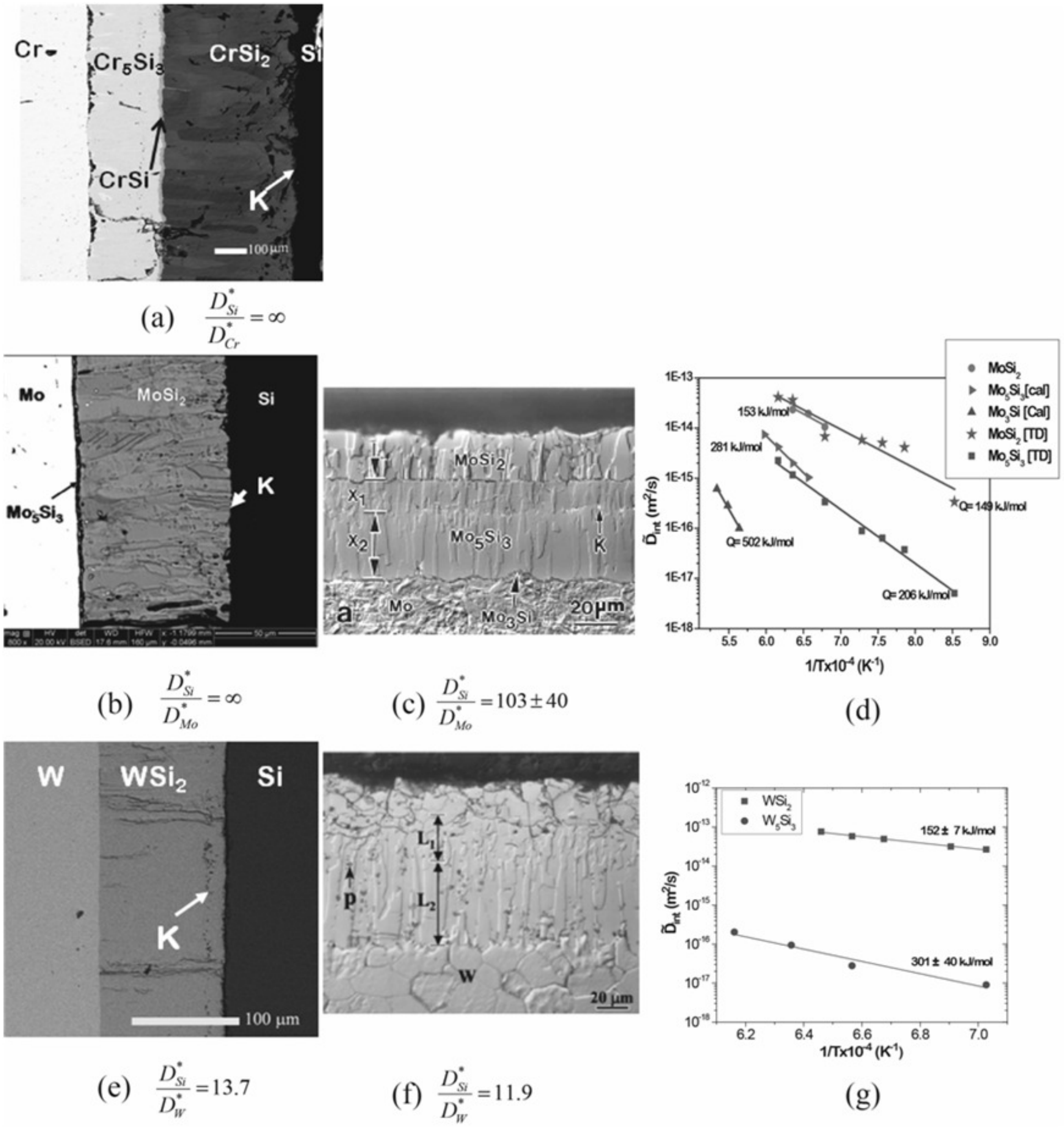

(a) $\frac{D^*_{Si}}{D^*_{Cr}} = \infty$ (b) $\frac{D^*_{Si}}{D^*_{Mo}} = \infty$ (c) $\frac{D^*_{Si}}{D^*_{Mo}} = 103 \pm 40$ (d) (e) $\frac{D^*_{Si}}{D^*_{W}} = 13.7$ (f) $\frac{D^*_{Si}}{D^*_{W}} = 11.9$ (g)

Fig 7.3: (a) Scanning electron micrograph of etched Cr/Si diffusion couple annealed at 1250°C for 16 hrs (b) Mo/Si diffusion couple annealed at 1300°C for 16 hrs [4](c) Optical micrograph of Mo/$MoSi_2$ diffusion couple annealed at 1350°C for 20 hrs [10] (d) Arrhenius plot of the integrated diffusion coefficients calculated for $MoSi_2$ and $Mo_5Si_3$ phases (e) Scanning electron micrograph of W/Si diffusion couple annealed at 1225°C for 9 hrs (f) Optical micrograph of Mo/$MoSi_2$ diffusion couple annealed at 1400°C for 50 hrs [13] (f) Arrhenius plot of the integrated diffusion coefficients calculated for $WSi_2$ and $W_5Si_3$ phases.
*TD stands for the results of Tortorici and dayananda [11]

In the Mo-Si system, three phases are present; however, mainly two phases could be detected in the interdiffusion zone, as shown in Fig. 7.3b [4]. $MoSi_2$ phase grows with much higher thickness compared to $Mo_5Si_3$. As shown in Fig. 7.3c, incremental diffusion couple experiments were conducted by Yoon et al. [10] to study the growth of the $Mo_5Si_3$ phase. However, they calculated only the growth constants of the phase. Prasad and Paul [4] calculated the integrated diffusion coefficients from the growth kinetics data reported by them. These are plotted along with the results reported by Prasad and Paul [4] and Toritorici and Dayananda [11, 12], as shown in Fig. 7.3d. It can be seen that the activation energy for diffusion in $Mo_5Si_3$ phase is higher compared to the $MoSi_2$ phase. Again, like in the case of $VSi_2$ phase, continuous grains were found after etching in the $MoSi_2$ phase. Based on this, Prasad and Paul [4] and Totorici and Dayananda [11, 12] predicted the position of the marker plane at the $MoSi_2$/Si interface. This indicates that Si has much higher diffusion rate compared to Si through in $MSi_2$ phase. The analysis of the diffusion couple gives the value of the ratio of the tracer diffusivity $\left(\frac{D_{Si}^*}{D_{Mo}^*}\right)$ in $MSi_2$ phase as infinite. In fact tracer diffusion studies conducted in this phase also support this prediction [5, 7, 8]. Si or Ge have few orders of magnitude higher diffusion rate compared to that of Mo. Based on the Kirkendall marker experiment conducted by Yoon et al. [10] the ratio of the tracer diffusivity in $Mo_5Si_3$ is calculated as $\left(\frac{D_{Si}^*}{D_{Mo}^*}\right)_{Mo_5Si_3} = 103 \pm 40$. In the W-Si system, two phases are present and both are in the interdiffusion zone, as shown in Fig. 7.3e, mainly the $WSi_2$ phase is visible; however, $W_5Si_3$ was found as very thin layer.

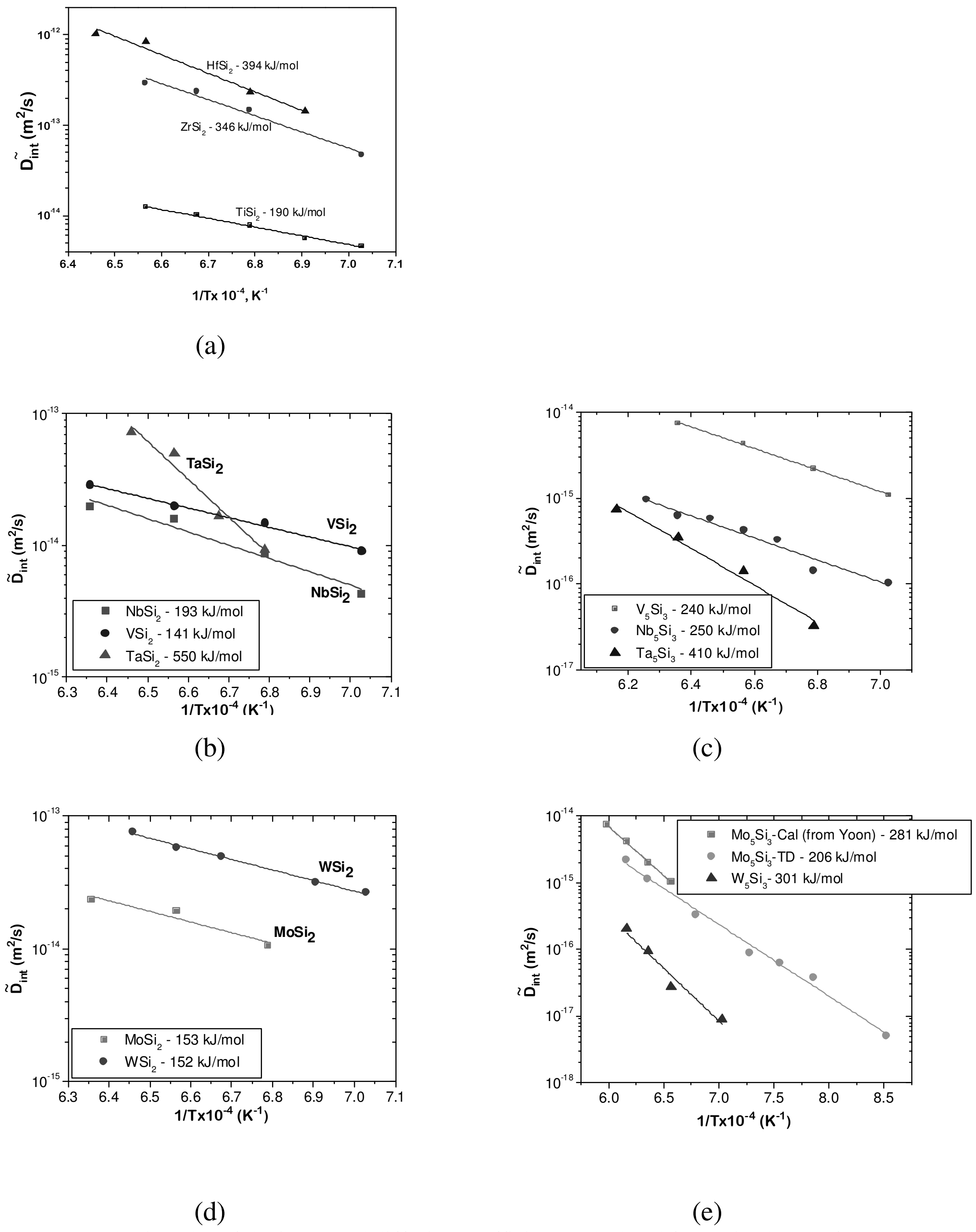

Fig 7.4: (a) Arrhenius plot of the integrated diffusion coefficients calculated for $MSi_2$ phases of group IVB elements (b) Integrated diffusion coefficients calculated for $MSi_2$ phases of group VB elements (c) Integrated diffusion coefficients calculated for $M_5Si_2$ phases of group VB elements (d) Integrated diffusion coefficients calculated for $MSi_2$ phases of group VIB elements (e) Integrated diffusion coefficients calculated for $M_5Si_2$ phases of group VIB elements.

Therefore, the incremental couple experiments were conducted for the calculation of diffusion coefficients in this phase, as shown in Fig. 7.3f. The Kirkendall marker plane in the $WSi_2$ phase was detected from the presence of line of pores, whereas, incremental diffusion couple experiment with inert marker conducted by Lee et al. [13] was considered for the calculation of relative mobilities of the species in the $W_5Si_3$ phase. Although, $WSi_2$ and $MoSi_2$ have the same crystal structure, it can be seen that the marker plane is found inside the $WSi_2$ phase, whereas the Kirkendall plane was found at the $MoSi_2$/Si interface. This indicates the change in diffusion of metal species because of change in the system. Integrated diffusion coefficients estimated are shown in Fig. 7.3g. Like other systems (except Ta-Si), activation energy for disilicide ($WSi_2$) is less than the activation energy for 5:3 silicide ($W_5Si_3$). The ratio of tracer diffusivities, $\frac{D_{Si}^*}{D_W^*}$, is determined to be 13.7 in $WSi_2$ and 11.9 in $W_5Si_3$ from the marker experiment by Lee et al.[13]. In both the phases, the ratio decreases compared to the Mo-Si system, which indicates increase in diffusion of metal species compared to Si. $\tilde{D}_{\text{int}}$ determined for different systems are compared in Fig. 7.4a-e. It can be seen from Fig. 7.4a that in the group IVB refractory metal silicon system that $\tilde{D}_{\text{int}}$ of the disilicides increases with the increase in atomic number. Even the activation energy also increases in similar fashion. Again the trend is the same for group VB refractory metal silicon systems. It can be seen from Fig. 7.4b that with the increasing the atomic number of refractory metals, activation energy increases for the disilicides. For $TaSi_2$, it increases drastically. Even it has higher $\tilde{D}_{\text{int}}$ that is higher growth rate compared to other phases in the higher temperature range. As shown in Fig. 7.4c, in the $M_5Si_3$ phase $\tilde{D}_{\text{int}}$ decreases and activation energy increases with the increase in the atomic number of the refractory element. On the other hand, as shown in Fig. 7.4d, the activation energy for disilicides formed by interaction of Si with group VIB refractory elements are more or less the same with higher $\tilde{D}_{int}$ for $WSi_2$ compared to $MoSi_2$. In the case of 5:3 silicide, as shown in Fig. 7.4e, the activation energy increases and $\tilde{D}_{int}$ decreases with the increase in the atomic number of refractory elements in both the groups.

### 7.3. Atomic mechanisms of diffusion

The previous discussion reveals definite patterns in the ratio of atomic diffusivities of the two components for silicides of the subgroups IVB, VB and VIB components as a function of the atomic number of the refractory metals. These data allows to shed light onto atomistic mechanisms of diffusion of the two components in the silicides under consideration and to reveal some trends in the diffusion behavior as a function of the atomic number in the refractory metal component.

#### 7.3.1 $MSi_2$ disilicides

In the group IV refractory metal silicon systems, all the disilicides have orthorhombic structure. However, there is difference in the atomic arrangements and number of atoms in the unit cells for $TiSi_2$ with other two compounds *i.e.* $ZrSi_2$ and $HfSi_2$. $TiSi_2$ has oF24, C54 structure, as shown in Fig. 7.5a [14]. It can be seen that Ti is surrounded by 4 Ti and 10 Si. On the other hand, Si is surrounded by 5Ti and 9 Si. ZrSi2 and HfSi2 have oC12, C49 structure, as shown in Fig. 7.5b [14]. In this Metal (M) components are surrounded by 6 M and 10 Si.

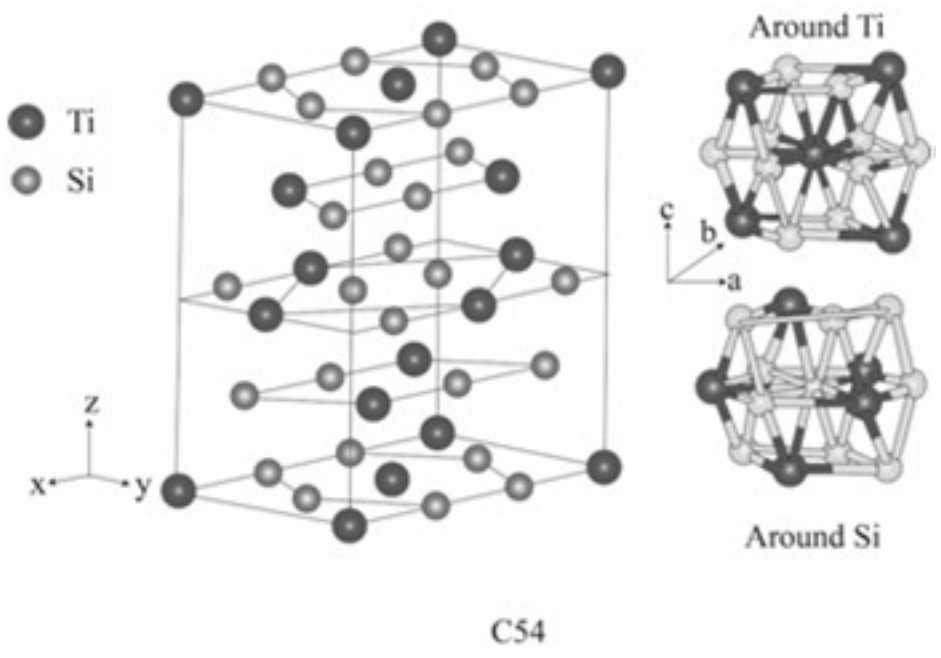

C54, oF24

Ti is surrounded by 4 Ti and 10 Si
Si is surrounded by 5 Ti and 9 Si

(a)

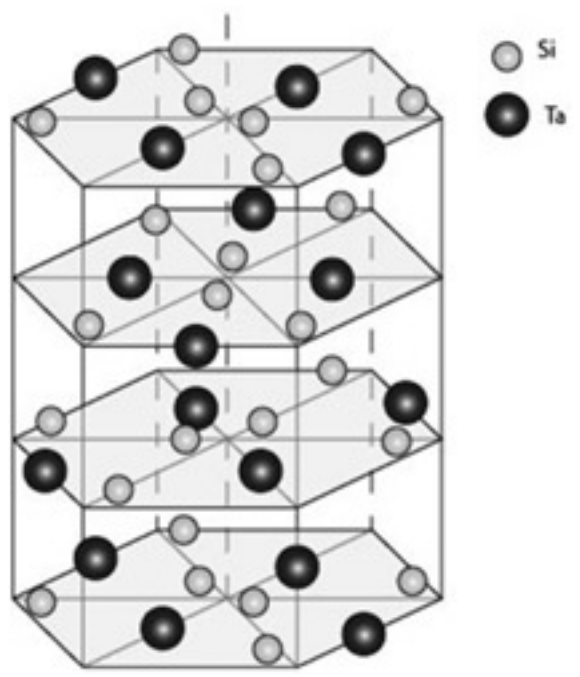

C40, hP9

Ta is surrounded by 5 Si
Si is surrounded by 5 Si and 5 Ta

(c)

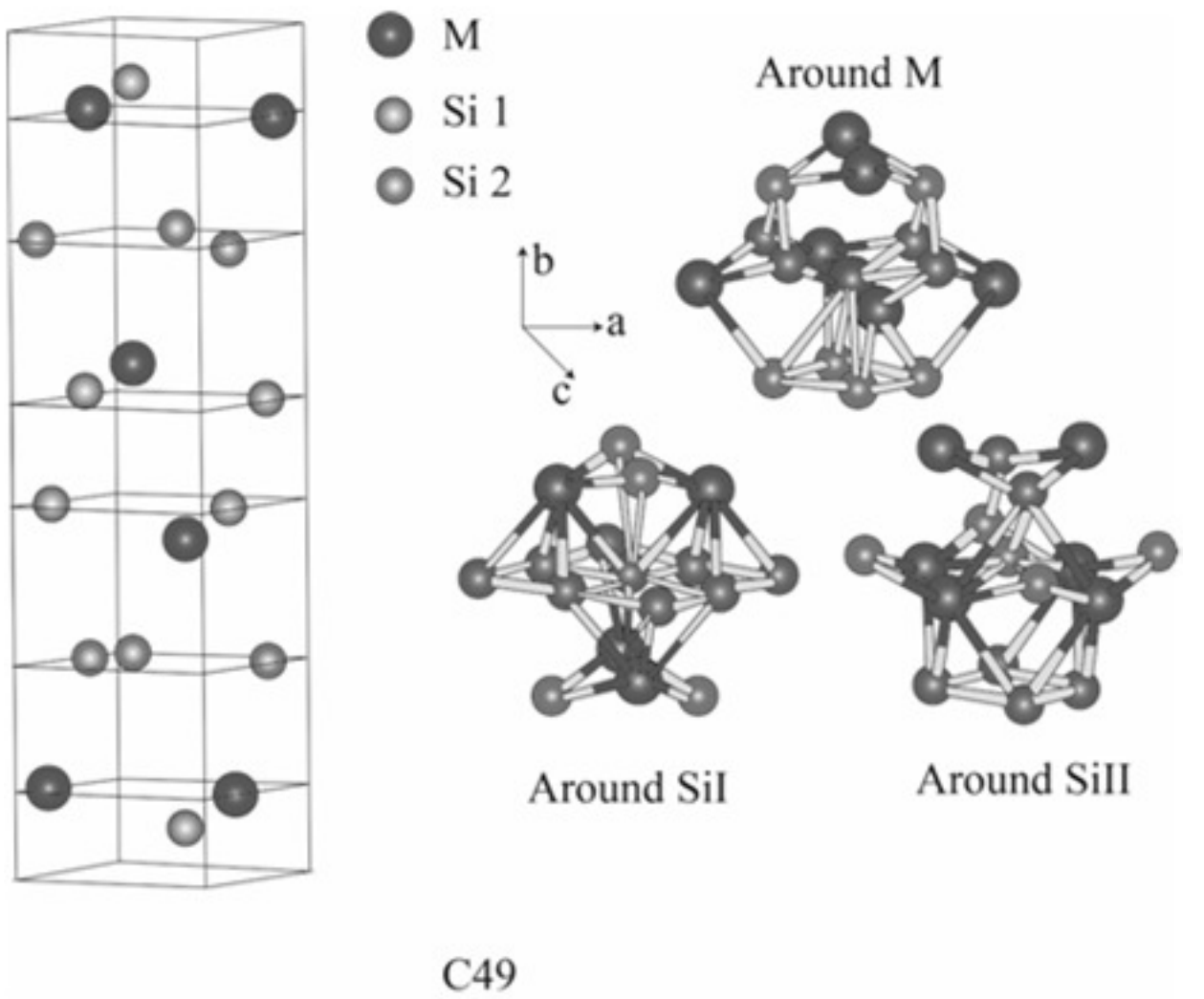

C49 oC12

M is surrounded by 10 Si (4 Si-I and 6 Si-II) and 6 M
Si-I is surrounded by 12 Si (8 Si-I and 4 Si-II)
Si-II is surrounded by 10 Si (6 Si-I and 4 Si-II) and 6 M

(b)

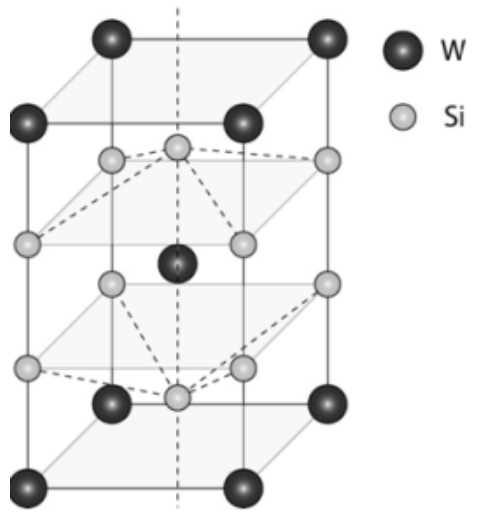

C11b, tI6

W is surrounded by 10 Si
Si is surrounded by 5 Si and 5 W

(d)

Fig 7.5: Crystal structures and the nearest neighbours of the species of (a) $TiSi_2$ phase (b) $ZrSi_2$ and $HfSi_2$ phases (c) $VSi_2$, $NbSi_2$,$TaSi_2$ and $CrSi_2$ phases (d) $MoSi_2$ and $WSi_2$ phases

| Group IVB | | | Group VB | | | Group VIB | | |
|---|---|---|---|---|---|---|---|---|
| Ti | $TiSi_2$<br>1478 °C | oF24 (C54)<br>a = 8.236Å<br>b =4.773 Å<br>c = 8.523Å<br>$V_m$ = 8.41 $cm^3/mol$ | V | $VSi_2$<br>1677 | hP9 (C40)<br>a = 4.58 Å<br>b = -<br>c = 6.37 Å<br>$V_m$ = 7.73 $cm^3/mol$ | Cr | $CrSi_2$<br>1490 °C | hP9 (C40)<br>a = 4.428 Å<br>b = -<br>c = 6.369 Å<br>$V_m$ = 7.23 $cm^3/mol$ |
| | TiSi<br>1570 °C | oP8 (B27)<br>a = 6.540Å<br>b = 3.630Å<br>c = 4.990Å<br>$V_m$ = 8.92 $cm^3/mol$ | | $V_5Si_3$<br>2010 °C | tI32 ($D8_m$)<br>a = 7.135 Å<br>b = -<br>c = 4.842 Å<br>$V_m$ = 7.9 $cm^3/mol$ | | | |
| | $Ti_5Si_4$<br>1920 °C | tP36<br>a = 6.713 Å<br>b = -<br>c = 12.171Å<br>$V_m$ = 9.18 $cm^3/mol$ | Nb | $NbSi_2$<br>1935 °C | hP9 (C40)<br>a = 4.785 Å<br>b = -<br>c = 6.576 Å<br>$V_m$ = 8.7 $cm^3/mol$ | Mo | $MoSi_2$<br>2020 °C | tI6 ($C11_b$)<br>a = 3.206 Å<br>b = -<br>c = 7.874 Å<br>$V_m$ = 8.1 $cm^3/mol$ |
| Zr | $ZrSi_2$<br>1620 °C | oC12 (C49)<br>a = 3.698Å<br>b = 14.761 Å<br>c = 3.664Å<br>$V_m$ = 10.03 $cm^3/mol$ | | $Nb_5Si_3$<br>2515 °C | tI32 ($D8_l$)<br>a = 6.55 Å<br>b = -<br>c = 11.86 Å<br>$V_m$ = 12.06 $cm^3/mol$ | | $Mo_5Si_3$<br>2180 °C | tI32 ($D8_m$)<br>a = 9.642 Å<br>b = -<br>c = 4.909 Å<br>$V_m$ = 8.6 $cm^3/mol$ |
| Hf | $HfSi_2$<br>1541 °C | oC12 (C49)<br>a = 3.667Å<br>b = 14.550 Å<br>c = 3.649Å<br>$V_m$ = 9.77 $cm^3/mol$ | Ta | $TaSi_2$<br>2040 °C | hP9 (C40)<br>a = 4.784 Å<br>b = -<br>c = 6.568 Å<br>$V_m$ = 8.71 $cm^3/mol$ | W | $WSi_2$<br>2433 °C | tI6 ($C11_b$)<br>a = 3.12 Å<br>b = -<br>c = 7.84 Å<br>$V_m$ = 7.66 $cm^3/mol$ |
| | HfSi<br>2145 °C | oP8 (B27)<br>a = 6.855Å<br>b = 3.700 Å<br>c = 5.220Å<br>$V_m$ = 9.97 $cm^3/mol$ | | $Ta_5Si_3$<br>2550 °C | tI32 ($D8_l$)<br>a = 6.516 Å<br>b = -<br>c = 11.873 Å<br>$V_m$ = 9.48 $cm^3/mol$ | | $W_5Si_3$<br>2594 °C | tI32 ($D8_m$)<br>a = 9.60 Å<br>b = -<br>c = 4.972 Å<br>$V_m$ = 7.67 $cm^3/mol$ |

Table 7.1: Details on phases, crystal structures, lattice parameters, molar volumes, $V_m$ and melting points of the phases are listed for the group IVB, V and group VIB refractory components based silicides.

In this Si has two different sublattices and these are designated as SiI and SiII. SiI is surrounded by 12 Si. On the other hand, SiII is surrounded by 6 M and 10 Si. Therefore in both the structures, both the components are surrounded by M and Si. These can diffuse on the condition that vacancies are present on their sublattices. Since only Si diffuse in the disilicides, it indicates that vacancies are mainly present only on the Si sublattices. The negligible diffusion rate of the metal components indicates that vacancies are not present on their sublattice. Further metal antisites must be negligible also. To compare the results estimated in different compounds, we need to plot the data with respect to $\tilde{D}_{int}$ vs. $T_m/T$, where $T_m$ is the melting point and T is annealing temperature at which the diffusion parameters are estimated. Melting points of the phases are listed in Table 7.1. It is shown in Fig. 7.6a. It can be seen that $\tilde{D}_{int}$ increases with the increase in atomic number in the temperature normalized plot. Since Si is the dominant diffusing component, it indicates that concentration of vacancies increases with the increase in the atomic number of the metallic component.

As shown in Fig. 7.5c, $VSi_2$ [15, 16], $NbSi_2$ [17], and $TaSi_2$ [18, 19] have the same hexagonal-based hP9 (C40) atomic structure in which each metallic atom has 5 Si atoms as nearest neighbors, while Si atoms are surrounded by 5 Si and 5 refractory metal atoms. Assuming the vacancy diffusion mechanism and the anti-site type of disorder, the experimentally observed significantly larger diffusion rate of Si atoms predicts straightforwardly the dominant creation of vacancies on the Si sublattice.

This fact was proven for the $MoSi_2$ phase [7] based on experimental data of tracer diffusion coefficients [5, 8] and vacancy concentration measurements on different sublattices [20]. Experiments indicate a nearly constant value of the integrated interdiffusion coefficient at a given absolute temperature, e.g. it amounts to $1.48\times10^{-14}$, $0.87\times10^{-14}$, and $0.90\times10^{-14}$ $m^2/s$ at 1473 K in $VSi_2$, $NbSi_2$, and $TaSi_2$, respectively. However, this temperature corresponds to a sequentially reduced homologous temperature, $T/T_m$, (T is the temperature of interest and $T_m$ is the melting point of the phase of interest), which are e.g. 0.76, 0.67, and 0.64 at 1473 K following the same sequence. Thus, at a constant homologous temperature the atomic mobilities of metal species are significantly increased with an increasing atomic number of the refractory metal, Fig. 7.6b. Simultaneously the ratio of the components' diffusivities decreases from the very large values of the orders of magnitude for $VSi_2$ to a value only slightly larger than unity for $TaSi_2$. We conclude that the diffusion rates of refractory metals are increased with the increasing atomic number and this trend correlates with an increased atomic radius of the atoms. These facts can satisfactory be explained in terms of vacancy-mediated diffusion jumps of metallic atoms over the Si sublattice, as listed in Table 7.2, and sequentially increased attractive interaction of the oversized refractory metal anti-sites with the Si vacancies. Note that whereas vanadium has significantly smaller atomic radius among the other elements of the VB subgroup, about 1.35 Å, Nb and Ta have similar radii, respectively 1.45 and 1.46 Å, that correlates with similar relative diffusivities of Nb and Ta with respect to that of Si atoms and vanishingly small diffusion rate of the V atoms.

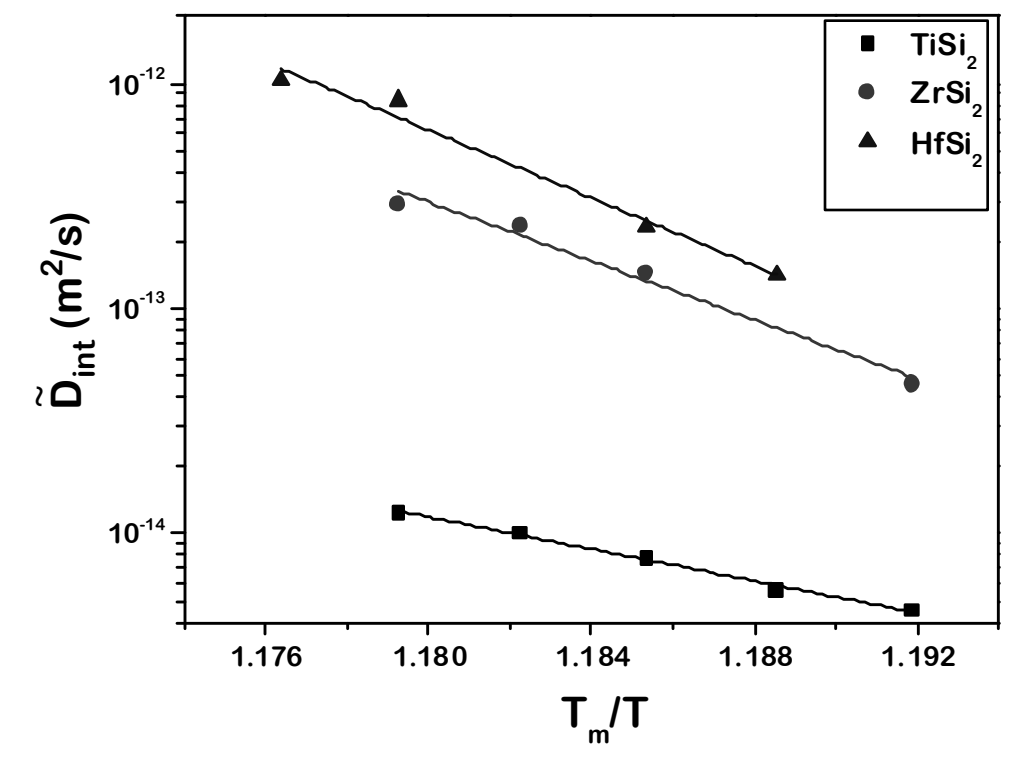

(a)

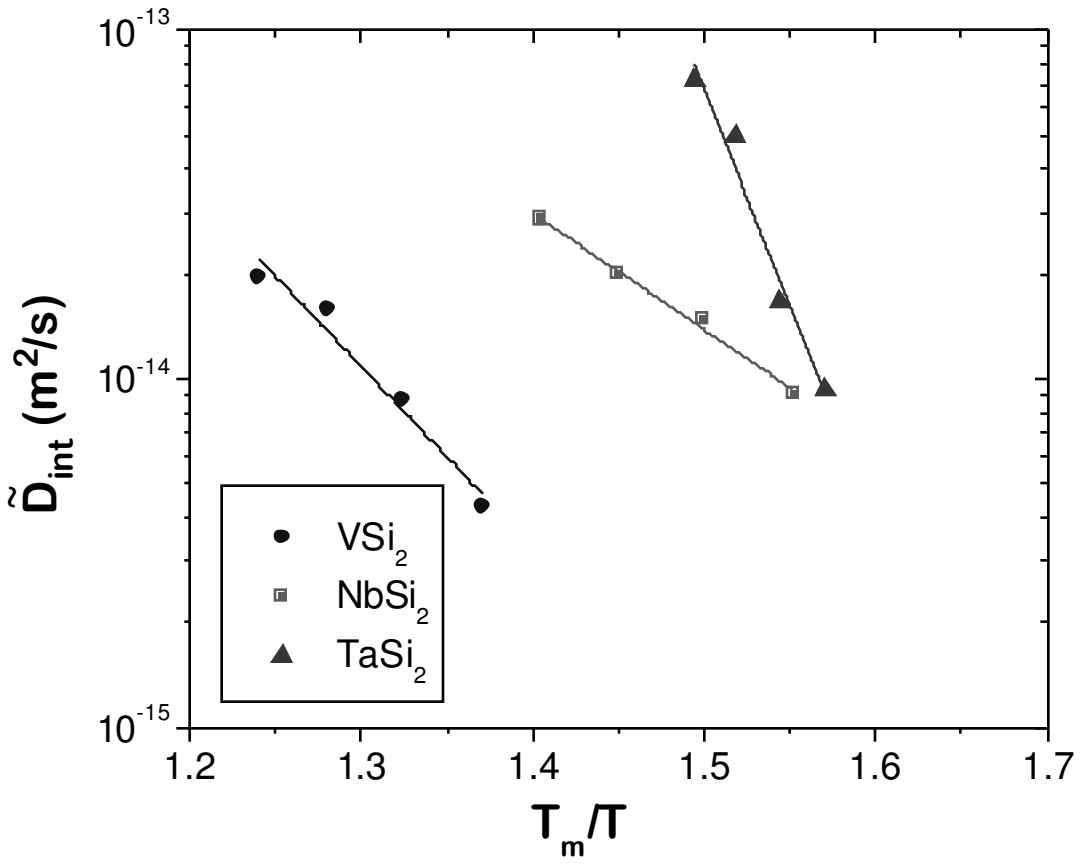

(b)

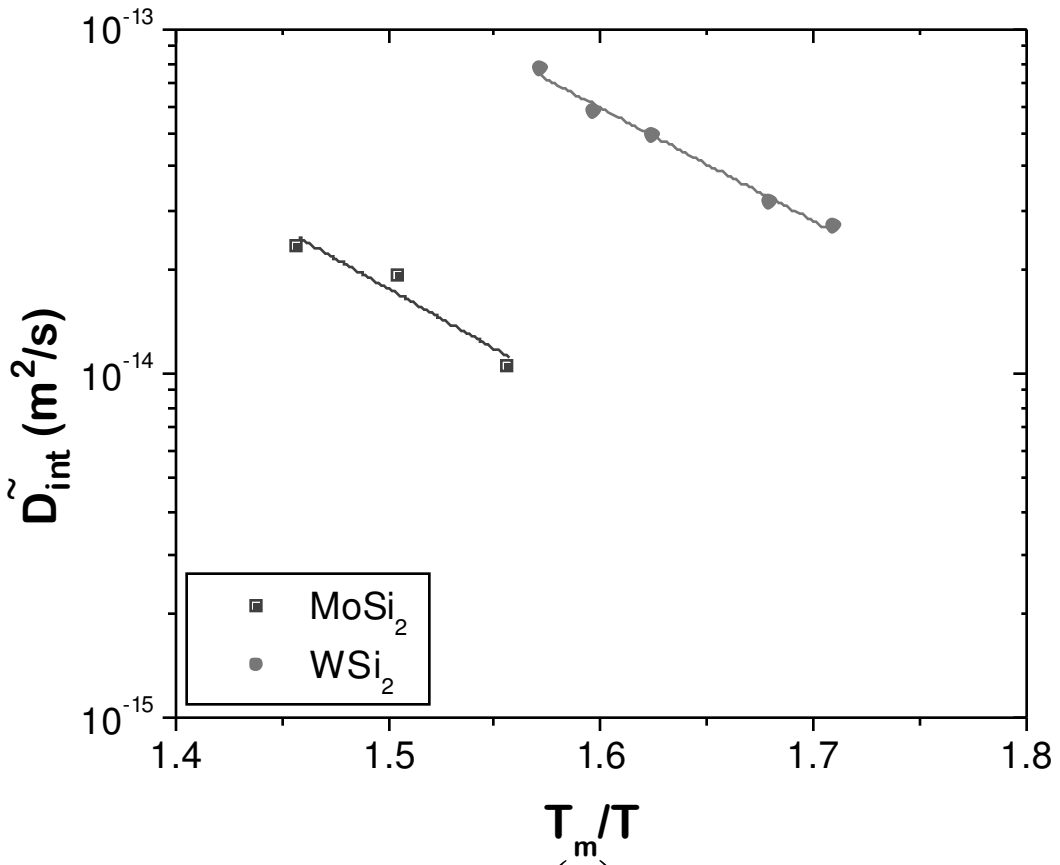

(c)

Fig 7.6 : Integrated diffusion coefficients are plotted with respect to homologous temperature $MSi_2$ phases of (a) group IVB and (b) group VB and (c) VIB refractory metal silicides.

Similar trends are found for the silicides of the subgroup VIB refractory elements which reveal the hP9 structure for CrSi2 and tI6 [21-23] structure for $MoSi_2$ and $WSi_2$, as shown in Fig. 7.5d and 5e. The nearest neighbor in the hP9 is already discussed. In the case of tI6, the metallic atoms are surrounded by 10 Si atoms, whereas the Si atoms have 5 Si and 5 metallic atoms as nearest neighbors. Again, the large W atoms reveal relatively faster diffusivities with respect to the Si atoms in comparison to smaller Mo atoms, which are almost immobile in the $MoSi_2$ phase. Again, the integrated interdiffusion coefficient increases with increasing atomic number if measured at the same homologous temperature, Fig. 7.6c. Thus, the analysis predicts the diffusion mechanism of Cr, Mo and W atoms as anti-sites on the Si sublattice with significant attractive interaction between the W atoms and Si vacancies.

| | **Atomic Number** | **Element Symbol** | **Atomic Radius [Å]** | **Ionic Radius [Å]** | **Covalent Radius [Å]** | **Van-der-Waals Radius [Å]** | **"Crystal" Radius [Å]** |
|---|---|---|---|---|---|---|---|
| | 14 | Si | 1.11 | 1.10 | 1.11 | 2.10 | 0.40 |
| Group VB | 23 | V | 1.71 | 1.35 | 1.25 | | 0.68 |
| | 41 | Nb | 1.98 | 1.45 | 1.37 | | 0.78 |
| | 73 | Ta | 2.00 | 1.45 | 1.38 | | 0.78 |
| Group VIB | 24 | Cr | 1.66 | 1.4 | 1.27 | | |
| | 42 | Mo | 1.90 | 1.45 | 1.45 | | 0.79 |
| | 74 | W | 1.93 | 1.35 | 1.46 | | 0.74 |

Table 7.2: Details of size of the components used in this study are listed.

A hypothetical alternative description of the refractory metals diffusion in disilicides would correspond to next nearest neighbor jumps of metallic atoms over their own sublattice. However, such an explanation would fail to describe the observed significant enhancement of the metal atom diffusion rate with increasing atomic size due to a potentially increasing energetic penalty for the corresponding next nearest neighbor jumps during passing of the saddle-point configuration made of Si atoms.

One may also argue about a potential contribution of cyclic diffusion mechanisms like an updated to the hP9 structure version of the well-known 6-jump-cycle mechanism (originally suggested for B2 compounds [24]). In this case, the refractory atoms would sequentially jumps between their own and the Si sublattices. It may be argued that this mechanism is energetically unfavorable in view of the need to create two metallic anti-sites and one Si anti-site just after half passage of the corresponding atomic jumps. Again, the above reasoning for the next nearest neighbor jumps with respect to the dependence of the relative diffusivity as a function of the atomic number would completely applicable, too, and bring to an explicit contradiction with the experimental data.

A detailed analysis of the correlation effects for Si diffusion via simple vacancy jumps over the Si sublattice provided a perfect agreement with the experiment for the $MoSi_2$ system [7] that additionally indicates a minor (if any) effect of correlative jumps, at least, for Si diffusion. Recently, the correlation effects for both, Mo and Si diffusion in $MoSi_2$ were re-analyzed using the Monte-Carlo simulation approach [8] instead of a more powerful semi-analytical approach in Ref. [7] which combined Monte Carlo calculations with application of the partial correlation factors' concept from LeClair [25]. Again, the vacancy-mediated jumps of Mo anti-sites over Si sublattice agree well with the experimental facts and provide little support for cyclic diffusion mechanisms in $MoSi_2$.

### 7.4.2 $M_5Si_3$ silicides.

These silicides offer a surplus of the fraction of refractory metal atoms over Si atoms in the atomic cell that has a strong effect on the connectivity of the corresponding sublattices, shown in Fig. 7.7.

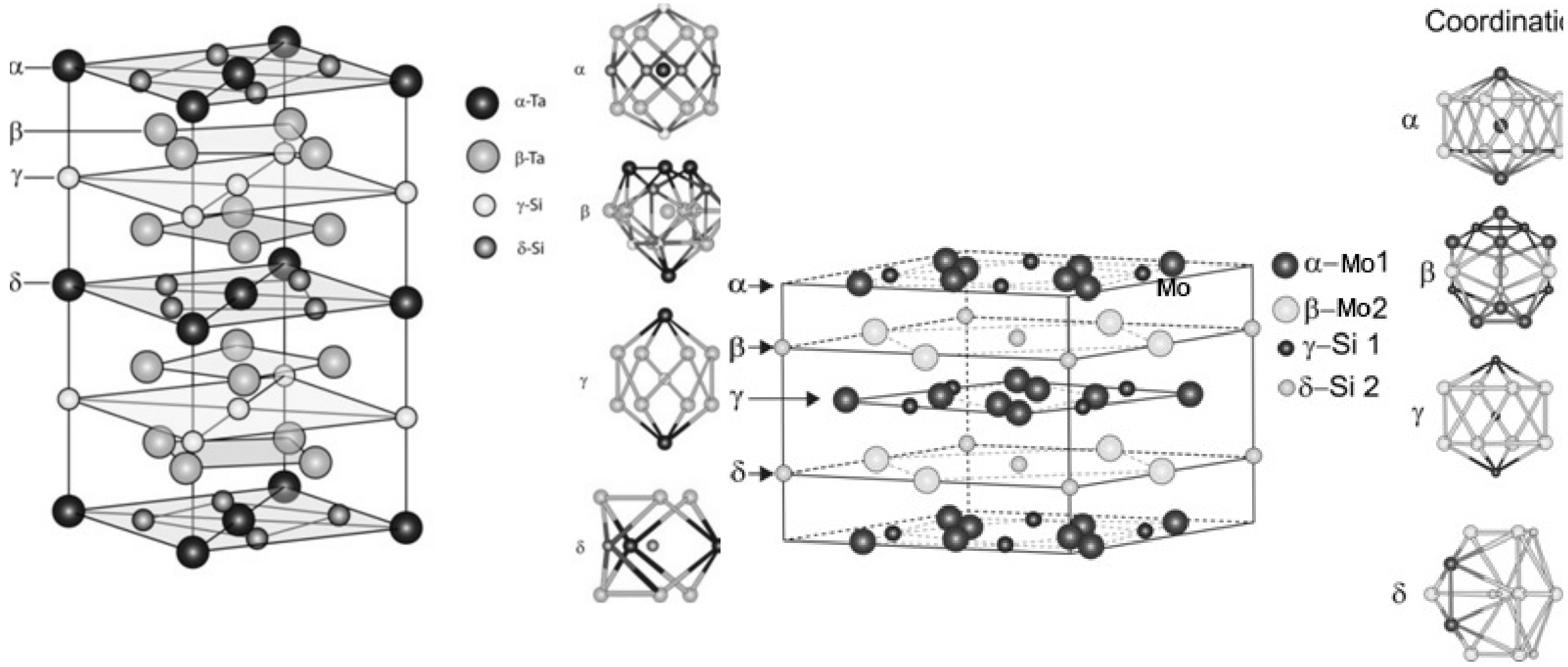

D8l, tI32

αTa is surrounded by 8 Ta and 6 Si
βTa is surrounded by 11 Ta and 5 Si
γ Si is surrounded by 10 Ta
δ Si is surrounded by 8 Ta and 1 Si

(a)

D8m, tI32

α W is surrounded by 10 W and 4 Si
β W is surrounded by 9 W and 6 Si
γ Si is surrounded by 8 W and 2Si
δ Si is surrounded by 10 W and 2Si

(b)

Fig 7.7: Crystal structures and the nearest neighbours of the species of $M_5Si_2$ phase (a) group VB elements (Nb, Ta) (b) group VIB elements (Mo and W).

Alternatively to the case of $MSi_2$ disilicides, the metallic atom sublattice is connected with respect to the next neighbor jumps of a metallic vacancy, whereas the Si atoms have mainly M atoms as nearest neighbors and only one to two neighboring Si atoms [16, 17], in dependence on the given position, are available.

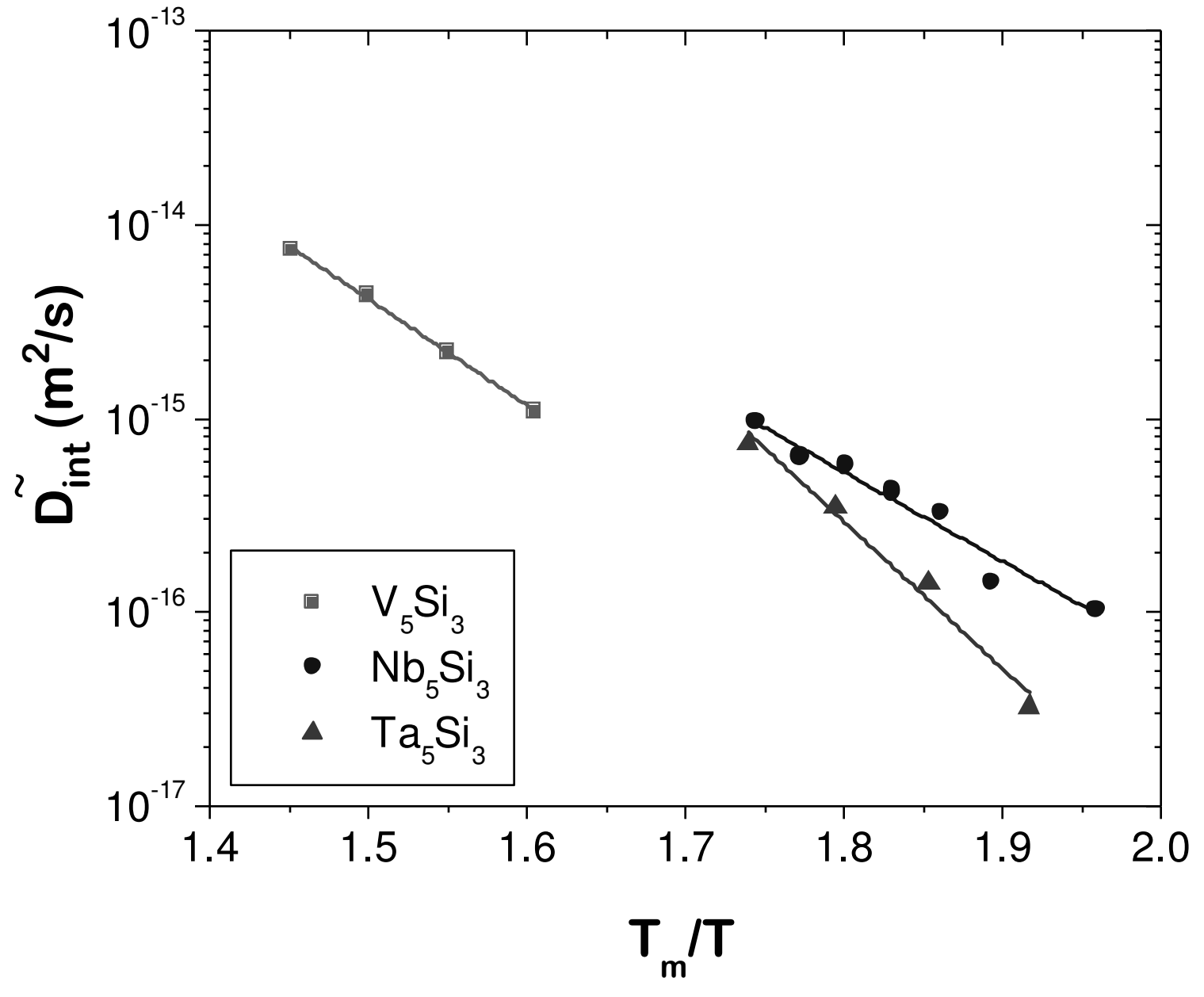

(a)

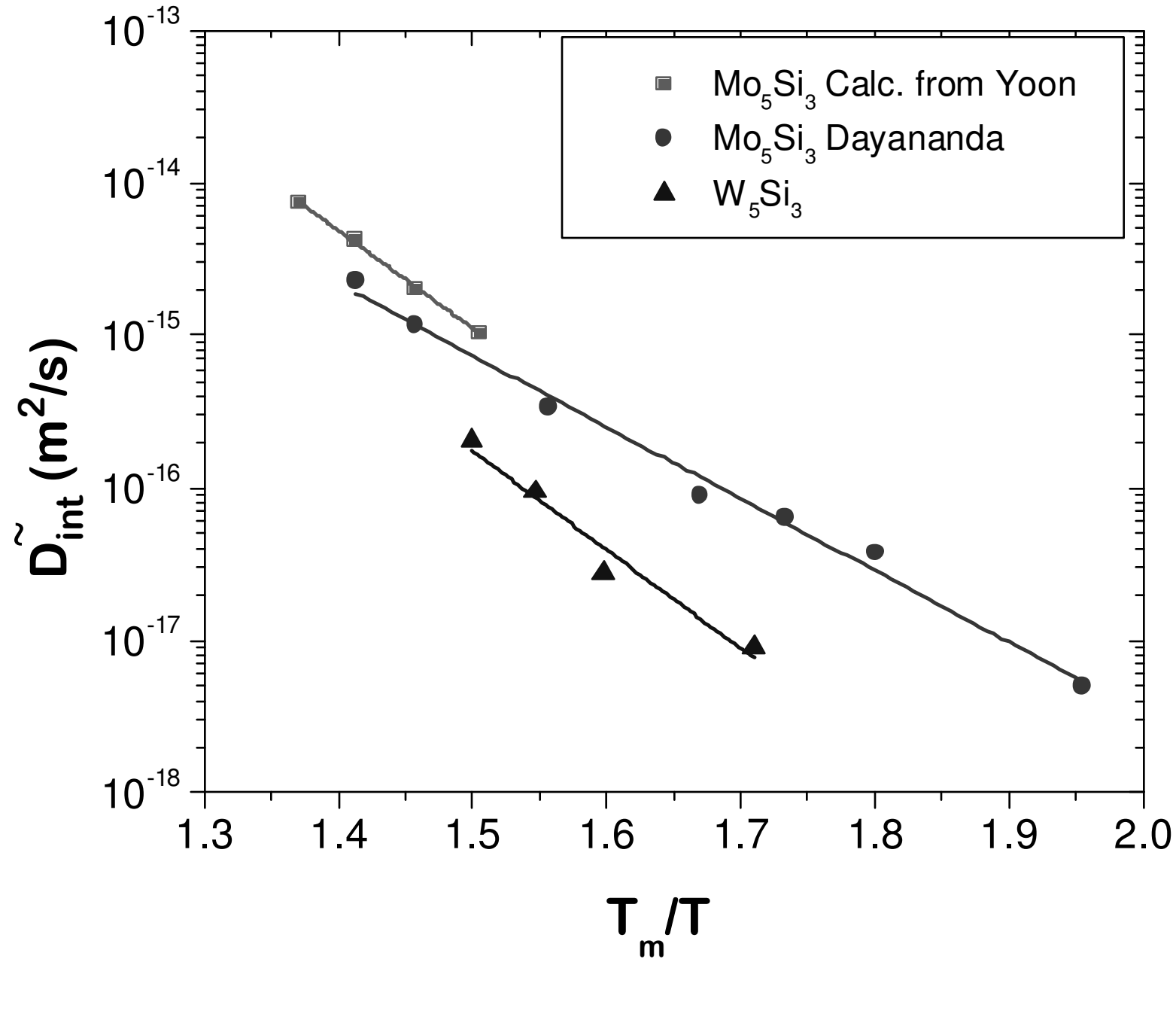

(b)

Fig 7.8 : Integrated diffusion coefficients are plotted with respect to homologous temperature $M_5Si_2$ phases of group VB and (b) group VIB refractory metal silicides.

The integrated interdiffusion coefficients are not much different in $V_5Si_3$, $Nb_5Si_3$ and $Ta_5Si_3$, if they are determined at the same homologous temperature, e.g. they are equal to $1.1\times10^{-16}$, $3\times10^{-16}$ and $5\times10^{-16}$ $m^2/s$ at $T_m/T = 1.8$, respectively, Fig. 7.8a. The ratio of components' diffusivities is somewhat decreased with increasing atomic number in this row. These facts can be understood in terms of nearest neighbor jumps of metallic atoms over own sublattice using corresponding vacancies. Si atoms will also diffuse over refractory metal sublattice as anti-sites. The increase of atomic size of the refractory metal with atomic number for the VB silicides induces a slight decrease of the absolute interdiffusion coefficient which is still determined by mobility of Si atoms. Smaller Si atoms have presumably lower barriers for jumps into the M vacancies providing fast diffusion rates. Additionally the so-called anti-structure bridge mechanism [18] could provide significant contribution for Si atoms, similarly to Al diffusion in $Ni_3Al$ as it was discussed in [26]. An increase of the atomic size of the metallic component in the $M_5Si_3$ silicides hinders effectively these jumps increasing corresponding barriers. As a result, the mobilities of the two components approach one another in $Ta_5Si_3$ with respect to that in $Nb_5Si_3$. Similar trends are observed for the $M_5Si_3$ silicides of the subgroup VIB, Fig. 7.8b.

**7.4. Conclusion**

We have considered group IVB, VB and VIB refractory metal silicides to show an interesting pattern in diffusion behavior with the change in atomic number in their respective groups. Mainly $MSi_2$ and $M_5Si_3$ (M:Si) silicides are considered for discussion, which have similar crystal structures in a group. Following important conclusions are drawn for this study:

- Except Ta-silicides, $MSi_2$ have lower activation energy for integrated diffusion compared to $M_5Si_3$. In the Ta-Si system, $TaSi_2$ has higher activation energy compared to $Ta_5Si_3$.
- Location of the Kirkendall marker plane is detected by the presence of uniform or duplex grain morphology and line of pores in the phases. In the $Nb_5Si_3$, both inert particles ($TiO_2$) and morphology indicates the location.
- In the IVB M-Si systems, marker planes are present at the Si/$MSi_2$ interface. That means Si has much higher diffusion rate compared to metal component. The increase in diffusion rate with the increase in atomic number of the metal species indicate that vacancies increase on the Si sublattice.
- In the group VB M-Si systems, the Kirkendall marker plane moves systematically from the $MSi_2$/Si interface towards inside the $MSi_2$ phase, with the increase in atomic number of the metal species i.e. in the sequence of $VSi_2$, $NbSi_2$ and $TaSi_2$. Since the marker plane is located at the $MSi_2$/Si interface in the V-Si system, it indicates few orders of magnitude higher diffusion rate of Si compared to V. On the other hand, the marker plane in the $TaSi_2$ is close to the initial contact plane (Matano plane), which indicates comparable diffusion rates of both the species.
- Similar behavior is found in the group VIB M-Si systems, where marker plane is located at the $MoSi_2$/Si interface and with the increase in atomic number of the metal species that is in W-Si system, the marker plane moves inside $WSi_2$.

- Therefore, with the increase in atomic number of the refractory metal, the relative mobilities measured by the ratio of the tracer diffusion coefficients, $\frac{D^*_{Si}}{D^*_M}$, decreases in both the phases. This indicates relative increase in rate of diffusion of the metal components compared to Si. Simultaneously the integrated interdiffusion coefficients increase with increasing atomic number of the refractory metal if determined at the same homologous temperature.
- Different behavior is found in the $M_5Si_3$ phase, where $\frac{D^*_{Si}}{D^*_M}$ decreases with the increasing atomic number of the metal species in both the groups, too, although the interdiffusion coefficients decrease with the increasing atomic number of the refractory metal at a given homologous temperature.
- Since no M-M nearest neighbors are present in $MSi_2$, diffusion of M must be possible because of presence of M anti-sites. High diffusion rate of Si is expected because of presence of vacancies mainly on the Si sublattice, which is determined experimentally in the $MoSi_2$ phase [3]. Further, the relative increase in diffusion rate of M compared to Si with the increase in atomic number of M in both the groups indicates higher attraction of M and vacancies because of size effect.
- In $M_5Si_3$, only one or two Si-Si bonds are present compared to many M-M bonds; however, Si has much higher diffusion rate. This fact may be interpreted in terms of Si diffusion over M sublattice via M vacancies. A strong contribution of the anti-structure bridge mechanism for diffusion of Si atoms is predicted.

# Summary

The solid state interdiffusion in refractory metals and single crystal Si are studied in details by diffusion couple technique. The wide range of application and importance of silicides in various devices are the motivation for these studies. The refractory metals which are considered and compared in these M/Si systems (M = metal) are group IVB elements Ti, Zr and Hf, group VB elements V, Nb and Ta, group VIB elements Cr, Mo and W. Out of these systems, the Ti, Zr, Hf, Ta and W M-Si systems have been studied in this thesis. The results of V, Nb and Mo systems, which were studied before in our group, are compared with the result of present study and a systematic pattern in diffusion behavior is found with increasing atomic number of the elements, group-wise in the periodic table. These studies help to understand the growth kinetics of the phases formed, diffusion parameters, relative mobilities of the species and presence and behavior of the different defects in the crystal structure. The different diffusion parameters calculated in present study are, parabolic growth constants, integrated diffusion coefficients, activation energy for diffusion, ratio of tracer diffusivities and tracer diffusivity of the species.

In group IVB, Ti/Si system all the equilibrium phases are found to grow in the interdiffusion zone whereas in Zr/Si and Hf/Si systems only $MSi_2$ and MSi are found, although there are many phases present in the both systems. The thickness of the $MSi_2$ phase is much higher than MSi phase in both the system. Therefore the growth rate of $MSi_2$ is found to be much faster than other phases. In group VB and VIB, in all M/Si systems, mainly $MSi_2$ and $M_5Si_3$ phases are found to grow. The growth rate of $M_5Si_3$ phases is found to be negligible compared to $MSi_2$. Because of this reason the $M_5Si_3$ phases are studied by incremental diffusion couple. Analysis shows that Si is the faster diffusing element in all the silicides in the present study.

The Ti-Si system was studied to understand the growth mechanism of the phases and the diffusion mechanism of the species. All the five equilibrium phases are found to grow in Ti/Si diffusion couple up to 1225 $^{o}$C. The maximum stability temperature for the $Ti_3Si$ phase is found to be higher than the temperature reported in the Ti-Si phase diagram. The layer thickness of TiSi phase does not change significantly with increasing temperature. On the other hand, time dependent experiments show all the phases grow by diffusion controlled process. Therefore in the case of multiphase growth, the study just on the parabolic growth constant can draw a wrong conclusion on the growth mechanism of the phases. The calculation of the diffusion parameters is always reliable in discussing the diffusion mechanism. Uniform and continuous grain morphology in the $TiSi_2$ phase indicates the location of the Kirkendall marker plane at the Si/$TiSi_2$ interface. Therefore, $TiSi_2$ grows mainly because of the diffusion of Si. The Si tracer diffusivity in $TiSi_2$ phase is calculated by the relation developed by van Loo with help of the thermodynamic data from literature. The grain morphology developed in the interdiffusion zone is explained with imaginary diffusion couples. In the $TiSi_2$ phase, Ti atoms are surrounded by 4 Ti and 10 Si. Similarly, Si atoms are surrounded by 5 Ti and 9 Si. Therefore both the elements can diffuse through their own sublattice if the vacancies are present. Since the phase grows because of Si diffusion, vacancies must be present mainly on the Si sublattice. Negligible diffusion rate of Ti indicates the low concentration of Ti antistes and vacancies on the Ti sublattice.

Similar bulk diffusion couple experiments are conducted between other two group IVB elements Hf and Zr with Si. Unlike Ti/Si system, in these two systems only two phases, $MSi_2$ and MSi are found to grow in the interdiffusion zone. The thickness of the disilicide is much higher than monosilicide. Absence of other phases indicates much lower growth rate of these phases compared to the phases found in the interdiffusion zones. Time dependent experiments indicate that the growth of the phases is diffusion controlled. The integrated diffusion coefficient for $HfSi_2$, HfSi and $ZrSi_2$ are calculated at different temperatures. The same is not calculated for the ZrSi phase to avoid error associated in the calculation because of very thin layer. The grain morphologies revealed by acid etching show uniform columnar grains throughout the disilicide phase in both the systems. That is because Si is the dominant diffusing species in the disilicide phase and the disilicide phase grow on the single crystal Si. Based on the understanding in other systems; it affirms the presence of the Kirkendall marker plane at the Si/$MSi_2$ interface. The tracer diffusion coefficients of Si in disilicide are calculated using available thermodynamic parameters and activation energies. As Si is the only diffusing species, we can conclude that vacancies are present mainly in the Si sublattice.

In Ta/Si system, after etching, the duplex grain morphology is found in the interdiffusion zone. This duplex grain morphology clearly shows the location of the Kirkendall marker plane inside the $TaSi_2$ phase. The $Ta_5Si_3$ phase was grown by incremental diffusion couple technique. The grain morphology in $Ta_5Si_3$ phase is revealed by EBSD to locate the Kirkendall marker plane. We have estimated the integrated diffusion coefficients and then calculated the ratio of tracer diffusivities. The activation energy for diffusion in $TaSi_2$ phase is found to be much higher that of compared to other refractory metal disilicides. The growth rate of the $Ta_5Si_3$ phase is found much lower than that of the $TaSi_2$ phase. This is the reason to find almost two orders of lower magnitude of integrated diffusion coefficient of the $Ta_5Si_3$ phase compared to the $TaSi_2$ phase. The calculated ratios of tracer diffusivities indicate that Si is the faster diffusing species in both the phases. In $TaSi_2$ phase the Kirkendall marker plane is found very near to the Matano plane but toward the Si side. This clearly shows the higher diffusivity of Si in this phase with comparable diffusivity of Ta. Although Si is the higher diffusing species, Ta-Ta bonds are not present in the crystal structure. This indicates the presence of a high concentration of Ta antisites. The higher diffusion rate of Si in the $Ta_5Si_3$ phase is also unusual. This observation indicates that the structural defects present in the crystals have a strong influence on their diffusion behavior. From our analysis of the growth, it is understood that the $Ta_5Si_3$ phase could not grow with reasonable thickness because of the high growth rate of the $TaSi_2$ phase in the Ta/Si diffusion couple, because of the consumption of the $Ta_5Si_3$ phase. The velocity diagram and the negative value of the thickness of one of the sublayers calculated in the $Ta_5Si_3$ phase in the Ta/Si diffusion couple indicate that the Kirkendall marker plane should not be present in this phase. This further indicates that duplex morphology should not be present in this phase. On the other hand, the positive values of both the sublayers in the $TaSi_2$ phase indicate the presence of the Kirkendall marker plane and the duplex morphology in this phase.

In the W/Si diffusion couples, only the $WSi_2$ phase is found to grow with measurable thickness in the interdiffusion zone. Si has higher diffusion rate compared to W in this phase, which is rather expected from the crystal structure. The presence of W antisites is apparent since it diffuses through the phase layer. $W/WSi_2$ incremental diffusion couple experiments were conducted to study the growth of the $W_5Si_3$ phase. In this phase also Si has higher diffusion rate compared to W. This indicates the presence of high concentration of vacancies on the Si sublattice.

Group IVB, VB and VIB refractory metal silicides show an interesting pattern in diffusion behavior with increase in atomic number in their respective groups. After studying individually different systems, we compared the results of the similar phases in these groups. In this comparison we also include the results of V/Si, Nb/Si and Mo/Si systems, which had been studied in our group before. Mainly $MSi_2$ in group IVB and $MSi_2$ and $M_5Si_3$ (M:Si) silicides in group VB and VIB are considered for discussion, which have similar crystal structures in a group.

In group IVB all the disilicides show continuous grains through the phase. From the previous studies it is obvious that the Kirkendall marker plane is present in the $Si/MSi_2$ interface. That also indicates the diffusivity of metal component is negligible and the disilicide phase grow mainly by Si diffusion. The $TiSi_2$ have C54 crystal structure and $ZrSi_2$ and $HfSi_2$ both have C49 crystal structure. Both components (M and Si) have reasonable numbers of nearest neighbor of the same component. Therefore, both the components can easily diffuse on the condition that vacancies are present in both the sublattices. However these group IVB disilicides grow by Si diffusivity only. Therefore high concentration of Si vacancies must be present in the crystal structure.

In the Group VB M-Si systems, The Kirkendall marker plane moves systematically from the $MSi_2/Si$ interface to inside the $MSi_2$ phase, with the increase in atomic number of the metal species i.e. in the sequence of $VSi_2$, $NbSi_2$ and $TaSi_2$. Since the marker plane is located at the $MSi_2$/Si interface in the V-Si system, it indicates few orders of higher diffusion rate of Si compared to V. On the other hand, the marker plane in the TaSi2 is close to the initial contact plane (Matano plane), which indicates comparable diffusion rates of both the species. Similar behavior is found in the group VIB M-Si systems, where marker plane is located at the $MoSi_2$/Si interface and with the increase in atomic number of the metal species, the marker plane moves inside $WSi_2$. Therefore, with the increase in atomic number of the refractory metal, the relative mobilities measured by the ratio of the tracer diffusion coefficients, $\frac{D_{Si}^*}{D_M^*}$ decreases in both the phases. This indicates relative increasing rate of diffusion of the metal species compared to Si. Similar behavior is found in the $M_5Si_3$ phase, where $\frac{D_{Si}^*}{D_M^*}$ decreases with the increasing atomic number of the metal species in both the groups. Since no M-M bonds are present in $MSi_2$, diffusion of M should be possible because of presence of M antisites. High diffusion rate of Si is expected because of presence of vacancies mainly on the Si sublattice, which is determined experimentally in the $MoSi_2$ phase. Further, the relative increase in diffusion rate of M compared to Si with the increase in atomic number of M in both the groups indicates higher attraction of M and vacancies because of increasing atomic size. In $M_5Si_3$, only one or two Si-Si bonds are present compared to many M-M bonds; however, Si has much higher diffusion rate. This indicates that vacancies must be present mainly on the Si sublattice. In this phase also the diffusion rate of M component increases relatively compared to Si with the increase in atomic number in both the groups indicating higher attraction of M and vacancies because of increase in atomic size.

## Publications:

1. S. Roy and A. Paul, *Growth mechanism of tantalum silicides by interdiffusion, Philosophical Magazine* 92 (2012) 4215-4229.

2. S. Roy and A. Paul, *Diffusion in tungsten silicides*, Intermetallics, 37 (2013) 83-87.

3. S. Roy, S. Divinski and A. Paul, *Reactive diffusion in the Ti-Si system and the significance of the parabolic growth constant*, Philosophical Magazine, 94 (2014) 683-699.

4. V. Baheti, S. Roy, R. Ravi and A. Paul, *Study on diffusion mechanism and the phase boundary compositions in the Co-Ta system*, Intermetallics, 33 (2013) 87-91.

5. S. Roy, S. Prasad, S. Divinski and A. Paul, *Diffusion pattern in* $MSi_2$ *and* $M_5Si_3$ *silicides in group VB (M = V, Nb, Ta) and VIB (M = Mo, W) refractory metal-silicon systems*, Philosophical Magazine, Philosophical Magazine 94 (2014) 1508-1528.

6. S. Roy and A. Paul, *Growth of hafnium and zirconium silicides by reactive diffusion, Materials Chemistry and Physics* 143 (2014) 1309-1314.

   S. Roy, S. Prasad and A. Paul, *An overview of the interdiffusion studies in Mo-Si and W-Si systems*, Defects and Diffusion Forum: Advanced Diffusion Processes and Phenomena 354 (2014) 79-84.

7. S. Roy and A. Paul, *Reactive diffusion in the Re-Si system*, Journal of Phase Equilibria and Diffusion 35 (2014) 631-635.